\documentclass[iop,twocolappendix,numberedappendix]{emulateapj}
\usepackage{amsmath,natbib,graphicx}
\usepackage{epsf}
\usepackage{epstopdf}
\usepackage{epsfig}
\usepackage{color}
\DeclareGraphicsExtensions{.jpg,.pdf,.png,.eps,.ps}
\usepackage{scrextend}
\usepackage{hyperref}
\usepackage{multirow}

\begin{document}

\title{Spitzer catalog of {\it Herschel}-selected ultrared dusty, star-forming galaxies}




\author{Jingzhe Ma$^{1}$, Asantha Cooray$^{1}$, Hooshang Nayyeri$^{1}$, Arianna Brown$^{1}$, Noah Ghotbi$^{1}$, Rob Ivison$^{2,3}$, Ivan Oteo$^{3}$, Steven Duivenvoorden$^{4}$, Joshua Greenslade$^{5}$, David Clements$^{5}$, Julie Wardlow$^{6}$, Andrew Battisti$^{7}$, Elisabete da Cunha$^{7}$, Matthew L. N. Ashby$^{8}$, Ismael Perez-Fournon$^{9}$, Dominik Riechers$^{10}$, Seb Oliver$^{4}$, Stephen Eales$^{11}$, Mattia Negrello$^{11}$, Simon Dye$^{12}$, Loretta Dunne$^{11}$, Alain Omont$^{13}$, Douglas Scott$^{14}$, Pierre Cox$^{15}$, Stephen Serjeant$^{16}$, Steve Maddox$^{11}$, Elisabetta Valiante$^{11}$}


\altaffiltext{1}{Department of Physics \& Astronomy, University of California, Irvine, CA 92617, USA}
\altaffiltext{2}{European Southern Observatory, Karl-Schwarzschild-Strasse 2, 85748, Garching, Germany}
\altaffiltext{3}{Institute for Astronomy, University of Edinburgh, Royal Observatory, Blackford Hill, Edinburgh EH9 3HJ, UK}
\altaffiltext{4}{Astronomy Centre, Department of Physics and Astronomy, University of Sussex, Brighton BN1 9QH, UK}
\altaffiltext{5}{Astrophysics Group, Imperial College, Blackett Laboratory, Prince Consort Road, London SW7 2AZ, UK}
\altaffiltext{6}{Department of Physics, Lancaster University, Lancaster, LA1 4YB, UK}
\altaffiltext{7} {Research School of Astronomy and Astrophysics, Australian National University, Canberra, ACT 2611, Australia}
\altaffiltext{8}{Center for Astrophysics $|$ Harvard \& Smithsonian, 60 Garden Street, Cambridge, MA 02138, USA}
\altaffiltext{9}{Instituto de Astrofisica de Canarias, E-38200 La Laguna, Tenerife, Spain}
\altaffiltext{10}{Cornell University, Space Sciences Building, Ithaca, NY 14853, USA}
\altaffiltext{11}{School of Physics \& Astronomy, Cardiff University, Queen's Buildings, The Parade, Cardiff CF24 3AA, UK}
\altaffiltext{12}{School of Physics and Astronomy, University of Nottingham, University Park, Nottingham, NG7 2RD, UK}
\altaffiltext{13}{Institut d'Astrophysique de Paris, Centre national de la recherche scientifique (CNRS) \& Universit\'e Pierre et Marie Curie (UPMC), 98 bis boulevard Arago, F-75014 Paris, France}
\altaffiltext{14}{Department of Physics and Astronomy, University of British Columbia, 6224 Agricultural Road, Vancouver, BC V6T 1Z1, Canada}
\altaffiltext{15}{Institut d'astrophysique de Paris, Sorbonne Universit, CNRS, UMR 7095, 98 bis bd Arago, 7014 Paris, France}
\altaffiltext{16}{School of Physical Sciences, The Open University, Milton Keynes, MK7 6AA, UK}


\begin{abstract}
The largest {\it Herschel} extragalactic surveys, H-ATLAS and HerMES, have selected a sample of ``ultrared'' dusty, star-forming galaxies (DSFGs) with rising SPIRE flux densities ($S_{500} > S_{350} > S_{250}$; so-called ``500 $\mu$m-risers'') as an efficient way for identifying DSFGs at higher redshift ($z > 4$). In this paper, we present a large {\it Spitzer} follow-up program of 300 {\it Herschel} ultrared DSFGs. We have obtained high-resolution ALMA, NOEMA, and SMA data for 63 of them, which allow us to securely identify the {\it Spitzer}/IRAC counterparts and classify them as gravitationally lensed or unlensed. Within the 63 ultrared sources with high-resolution data, $\sim$65\% appear to be unlensed, and $\sim$27\% are resolved into multiple components. We focus on analyzing the unlensed sample by directly performing multi-wavelength spectral energy distribution (SED) modeling to derive their physical properties and compare with the more numerous $z \sim 2$ DSFG population. The ultrared sample has a median redshift of 3.3, stellar mass of 3.7 $\times$ 10$^{11}$ $M_{\odot}$, star formation rate (SFR) of 730 $M_{\odot}$yr$^{-1}$, total dust luminosity of 9.0 $\times$ 10$^{12}$ $L_{\odot}$, dust mass of 2.8 $\times$ 10$^9$ $M_{\odot}$, and V-band extinction of 4.0, which are all higher than those of the ALESS DSFGs. Based on the space density, SFR density, and stellar mass density estimates, we conclude that our ultrared sample cannot account for the majority of the star-forming progenitors of the massive, quiescent galaxies found in infrared surveys. Our sample contains the rarer, intrinsically most dusty, luminous and massive galaxies in the early universe that will help us understand the physical drivers of extreme star formation. 
\end{abstract}


\section{Introduction}\label{sec:intro}

It has become clear that observing at UV/optical wavelengths is insufficient to probe the total star formation history of the Universe as a large fraction of star formation is obscured by dust (e.g., \citealt{Madau2014,Gruppioni2017}). Wide-area infrared (IR) surveys have revolutionized our understanding of obscured star formation by discovering a large number of dusty, star-forming galaxies (DSFGs; also known as ``submillimeter galaxies'' or SMGs; see \citealt{Casey2014} for a review), which make up the bulk of the cosmic infrared background (e.g., \citealt{Dole2006}). Some of the DSFGs represent the rarest and most extreme starbursts at high redshift (with star formation rates, SFRs $>$ 10$^{3}$ $M_{\odot}$yr$^{-1}$, and number densities $<$ 10$^{-4}$ Mpc$^{-3}$; \citealt{Gruppioni2013}), which still pose challenges to galaxy formation and evolution models (e.g., \citealt{Baugh2005,Narayanan2010,Hayward2013,Bethermin2017}). The discovery and detailed characterization of this population is required to understand the most extreme obscured star formation, which is only made possible now by deep and large-area surveys at far-infrared (FIR) and sub-mm/mm wavelengths with e.g., the {\it Herschel Space Observatory} \citep{Pilbratt2010,Eales2010,Oliver2012}, the South Pole Telescope (SPT; \citealt{Carlstrom2011,Vieira2010,Mocanu2013}), the Atacama Cosmology Telescope (ACT; \citealt{Marriage2011,Marsden2014,Gralla2019}), and the {\it Planck} Satellite \citep{Planck2011,Planck2014,Canameras2015,Planck2016}.

The largest {\it Herschel} extragalactic surveys, {\it Herschel} Astrophysical Terahertz Large Area Survey (H-ATLAS; \citealt{Eales2010}) and {\it Herschel} Multitiered Extragalactic Survey (HerMES; \citealt{Oliver2012}) covering a total area of $\sim$1300 deg$^2$, have revealed a large number of DSFGs. While most of them are $z$ $\sim$ 1-2 starburst galaxies (e.g., Casey et al. 2012a,b), selecting those with ultrared colors is extremely efficient for identifying a tail extending towards higher redshift ($z$ $>$ 4). A well-defined population of ``ultrared''  DSFGs using the {\it Herschel} SPIRE bands $S_{500} > S_{350} > S_{250}$ (``500 $\mu$m-risers'') has been established (e.g., \citealt{Cox2011,Dowell2014,Ivison2016,Asboth2016}). Based on their colors, these are likely to be $z\gtrsim4$, dusty and rapidly star-forming ($>500$\,M$_{\odot}$yr$^{-1}$) galaxies. These systems are believed to be the progenitors of massive elliptical (red and dead) galaxies identified at $z \sim 3$ (e.g., Oteo et al. 2016). Spectroscopic confirmation of a sub-sample of 26 sources based on CO rotational lines, an indicator of the molecular gas that fuels the prodigious star formation in these galaxies, has verified the higher redshifts compared to general DSFG samples (e.g., \citealt{Cox2011,Combes2012,Riechers2013,Riechers2017,Fudamoto2017,Donevski2018,Zavala2018,Pavesi2018}). Meanwhile, relatively wide and shallow surveys with SPT have discovered a large number of gravitationally lensed DSFGs at $z$ $>$ 4 (Vieira et al. 2013; Spilker et al. 2016; Strandet et al. 2016), including the highest-redshift DSFG discovered so far at $z$ = 6.9 \citep{Strandet2017,Marrone2018}.

To fully characterize the {\it Herschel}-selected ultrared DSFGs, we have been conducting a multi-wavelength observational campaign to probe as many regimes of their spectral energy distributions (SEDs) as possible. \cite{Ivison2016} searched the 600 deg$^2$ H-ATLAS survey and initially selected 7961 high-redshift DSFG candidates. A subset of 109 DSFGs were further selected for follow-up observations with SCUBA-2 \citep{Holland2013} or LABOCA \citep{Siringo2009} at longer wavelengths (850/870 $\micron$). \cite{Asboth2016} identified 477 ``500 $\micron$-risers'' from the 300 deg$^2$ HerMES Large Mode Survey (HeLMS), and 188 of the brightest 200 ($S_{500}$ $>$ 63 mJy) sources were followed up with SCUBA-2 \citep{Duivenvoorden2018}. The addition of these longer wavelength data to the three {\it Herschel}/SPIRE bands better constrains the photometric redshifts and FIR luminosities. 

The nature of the {\it Herschel}-selected ultrared sources, which can be gravitationally lensed sources, blends (multiple components including mergers at the same redshift or just projection effect from sources at different redshifts), or unlensed intrinsically bright DSFGs, requires high-resolution data to confirm. The ultrared sample provides a good opportunity for studying the gas, dust, and stellar properties in detail of starburst galaxies at $z\gtrsim4$ out to the epoch of reionization at multiple wavelengths. In particular, the ultrared sources that are not lensed, blended, or otherwise boosted are of great interest, because they may represent the most extreme galaxies in the early universe. 

Gas and dust properties of DSFGs have been routinely studied with high-resolution interferometers such as the Atacama Large Millimeter/submillimeter Array (ALMA) and the Northern Extended Millimeter Array (NOEMA). An observational campaign is being conducted with ALMA and NOEMA on a sub-sample (63 so far) of the {\it Herschel} ultrared sources that have SCUBA-2/LABOCA data to further pinpoint their locations, reveal their morphologies, and confirm their redshifts (\citealt{Fudamoto2017, Oteo2017}). 

Observations of the stellar populations at optical or near-IR (NIR) are needed in order to place this population in the context of galaxy formation and evolution and provide a complete picture of their physical properties. {\it Spitzer}/IRAC is the only currently available facility that probes the rest-frame optical stellar emission of these sources and the only way to constrain their stellar masses and thus their specific SFRs (sSFRs), which are a critical diagnostic for the star formation mode of these galaxies. In this work, we present a follow-up study of 300 {\it Herschel} ultrared sources from H-ATLAS and HeLMS with {\it Spitzer}/IRAC in combination with multi-wavelength ancillary data. The {\it Spitzer} data allow us to constrain the stellar masses of a statistical sample of DSFGs at $z > 4$ in a consistent manner for the first time and provide the first constraint on the stellar mass density and evolution of this population. 

The paper is organized as follows. We describe the {\it Spitzer}/IRAC observations, source detection and photometry in Section \ref{sec:observations} along with the multi-wavelength ancillary data. In Section \ref{sec:catalog}, we introduce the cross-identification methods and describe the photometric catalog of the cross-matched {\it Spitzer} counterparts. We perform panchromatic SED modeling to derive their physical properties. The results of the SED fitting are presented in Section \ref{sec:sed}. We discuss the redshift distribution, multiplicity, unlensed fraction, SFR surface density, space density, SFR density, and stellar mass density in Section \ref{sec:discussion}. Section \ref{sec:summary} summarizes our conclusions and future plans. 

Throughout this paper, we adopt a concordance $\Lambda$CDM cosmology with $H_0$ = 70 kms$^{-1}$Mpc$^{-1}$, $\Omega$$_{\Lambda}$ = 0.7, and $\Omega$$_{\rm m}$ = 0.3. We use a \cite{Chabrier2003} initial mass function (IMF) and AB system magnitudes.

\begin{figure}
\includegraphics[width=8.7cm]{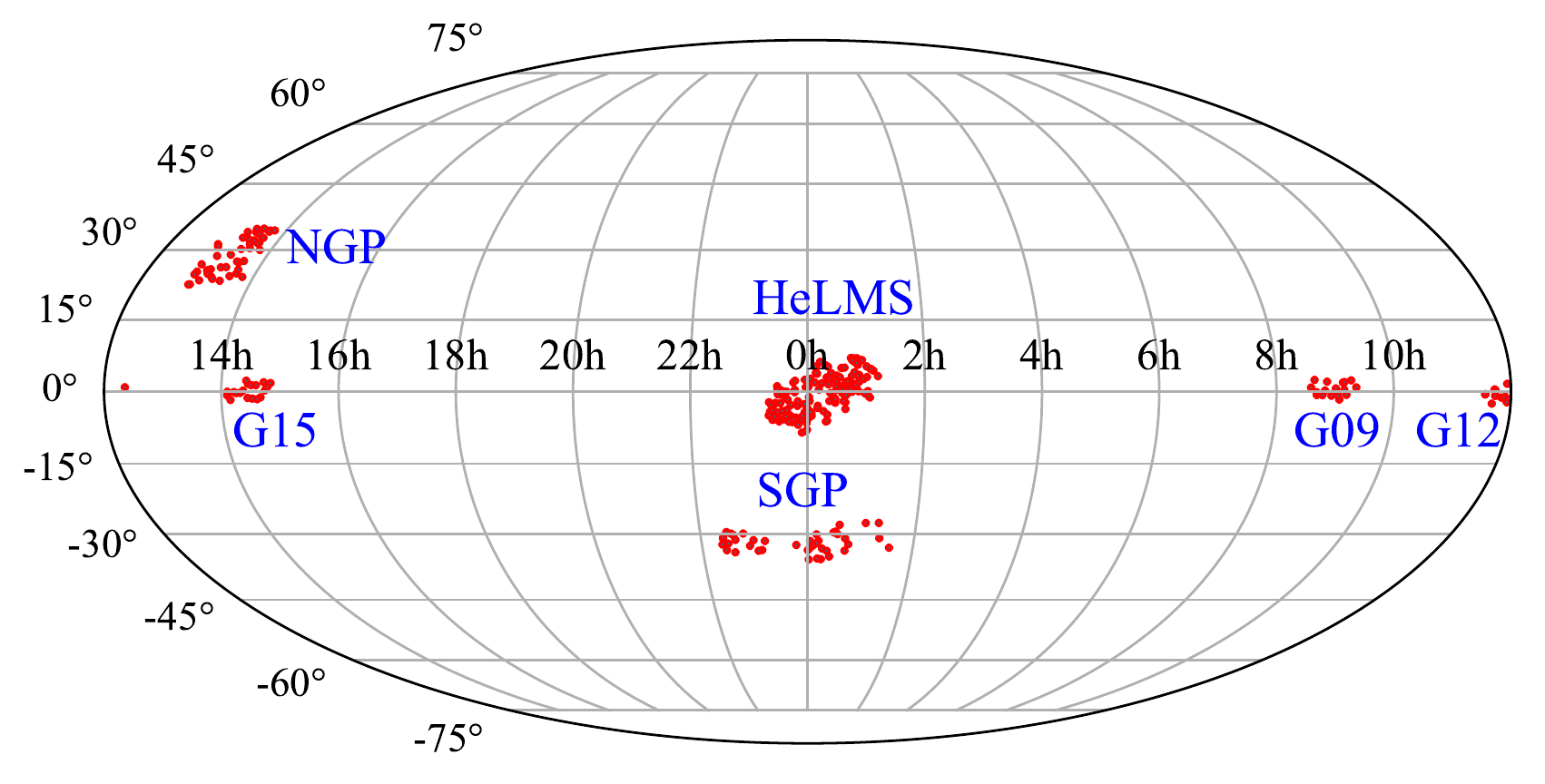} 
\caption{The locations of the 300 {\it Herschel}-selected ultrared sources on the all-sky map, including the GAMA09, GAMA12, GAMA15, NGP, SGP, and HeLMS fields. }
\label{allskymap}
\end{figure}

\section{Sample selection, {\it Spitzer} observations, and ancillary data}
\label{sec:observations}

\subsection{{\it Herschel}/SPIRE at 250, 350, and 500 $\micron$}
\label{sec:herschel}

The H-ATLAS observations were performed in the parallel-mode with both SPIRE \citep{Griffin2010} and PACS \citep{Poglitsch2010}. The survey covers three equatorial fields at right ascensions (R.A.) of 9, 12, and 15 hr (namely the GAMA09, GAMA12, and GAMA15 fields), the North Galactic Pole (NGP) field in the north, and the South Galactic Pole (SGP) field in the south. The total survey area is about 600 deg$^2$. For SPIRE maps, sources were extracted and flux densities were measured based on a matched-filter approach which mitigates the effects of confusion (e.g., \citealt{Chapin2011}). The depth of the PACS data ($1\sigma$ $\sim$ 45 mJy) is insufficient to detect our FIR-rising sources. The detailed descriptions of the H-ATLAS observations and source extraction can be found in \cite{Valiante2016}, \cite{Ivison2016}, and \cite{Maddox2018}. \cite{Ivison2016} selected the ultrared sample based on the following criteria: $3.5 \sigma$ detection threshold at $S_{500} > 30$ mJy and the color selection with $S_{500}/S_{250}$ $\geq$ 1.5 and  $S_{500}/S_{350}$ $\geq$ 0.85. 

The HeLMS field covers an effective area of  $\sim$274 deg$^2$ of the equatorial region spanning 23$^h$14$^m$ $<$ R.A. $<$ 1$^h$16$^m$ and -9$^{\circ}$ $<$ Dec. $<$ +9$^{\circ}$. The observations were conducted with the SPIRE instrument \citep{Griffin2010}. Sources were extracted using a map-based search method \citep{Dowell2014,Asboth2016} that combines the information in the 250, 350, and 500 $\micron$ maps simultaneously. {\it Herschel}/SPIRE maps have pixel scales of 6$\arcsec$, 8$\arcsec$ (8.333$\arcsec$ for HeLMS maps), and 12$\arcsec$ at 250, 350, and 500 $\micron$, which are one-third of the FWHM beam sizes of 18$\arcsec$, 25$\arcsec$, 36$\arcsec$, respectively. We refer the reader to \cite{Asboth2016} and \cite{Duivenvoorden2018} for more information on the HeLMS observations and detailed description of the source extraction and photometry. The HeLMS ultrared sample was selected with flux densities $S_{500} > S_{350} > S_{250}$ and a 5$\sigma$ cut-off $S_{500}$ $>$ 52 mJy \citep{Asboth2016}. 

Figure \ref{allskymap} demonstrates the distributions of the 300 {\it Herschel} ultrared sources on the all-sky map for which we have obtained {\it Spitzer} data. These include all the H-ATLAS sources in \cite{Ivison2016} and additional 31 sources from Ivison et al. in prep., and 161 HeLMS sources (with a flux density cut $S_{500}$ $>$ 64 mJy) from \cite{Asboth2016}. The {\it Herschel}/SPIRE photometry of these sources at 250, 350, and 500 $\micron$ are shown in Table \ref{tab:catalog}.

\begin{table*}
\centering
\caption{Catalog of {\it Herschel}-selected ultrared sources}  
\begin{tabular}{lllc}
\hline
\hline
Column   & Parameter    & Description &    Units  \\
\hline
1          & IAU\_name    & Survey name +  {\it Herschel} sourcename   &                \\
2          & ID    &  Nickname or short name in \citep{Ivison2016} and \citep{Duivenvoorden2018}  & \\
3           & $z$  & Photometric redshift from FIR SED fitting (2 decimal points; \citealt{Ivison2016} and &\\
             &         &  \citealt{Duivenvoorden2018}) or spectroscopic redshift (3 decimal points; see Table \ref{tab:specz}) &\\
4          & Lensed?  &  Lensed (y), unlensed (n), yn (both lensed and unlensed components), unknown (x) & \\
5            & RA\_IRAC  & {\it Spitzer}/IRAC counterpart position: right ascension (J2000) & degree \\
6           &  Dec\_IRAC & {\it Spitzer}/IRAC counterpart position: declination (J2000)& degree \\

7            &  MAG\_AUTO$_{3.6\micron}$  & Kron-like elliptical aperture magnitude at 3.6 $\micron$ & AB mag\\
8           &  MAGERR\_AUTO$_{3.6\micron}$  &  AUTO magnitude uncertainty at 3.6 $\micron$ & AB mag\\
9          &  MAG\_AUTO$_{4.5\micron}$  & Kron-like elliptical aperture magnitude at 4.5 $\micron$ & AB mag\\
10            &  MAGERR\_AUTO$_{4.5\micron}$  &  AUTO magnitude uncertainty at 4.5 $\micron$ & AB mag\\
 
11            &  $S_{3.6\micron}$  & Flux density at 3.6 $\micron$ &$\mu$Jy\\
12            &  $S_{3.6\micron}$\_err  & Flux density uncertainty at 3.6 $\micron$ & $\mu$Jy\\
13            &  $S_{4.5\micron}$  & Flux density at 4.5 $\micron$ & $\mu$Jy\\
14            &  $S_{4.5\micron}$\_err  & Flux density uncertainty at 4.5 $\micron$ & $\mu$Jy\\
           
15            & RA\_H  & {\it Herschel}/SPIRE position: right ascension (J2000)& degree \\
16            & Dec\_H & {\it Herschel}/SPIRE position: declination (J2000)& degree \\
17            &  $S_{250\micron}$  & Flux density at 250 $\micron$ &mJy\\
18            &  $S_{250\micron}$\_err  & Flux density uncertainty at 250 $\micron$ & mJy\\
19            &  $S_{350\micron}$  & Flux density at 350 $\micron$ & mJy\\
20            &  $S_{350\micron}$\_err  & Flux density uncertainty at 350 $\micron$ & mJy\\
21            &  $S_{500\micron}$  & Flux density at 500 $\micron$ & mJy\\
22            &  $S_{500\micron}$\_err  & Flux density uncertainty at 500 $\micron$ & mJy\\
           
23            & RA\_S  & SCUBA2 position: right ascension (J2000)& degree \\
24            & Dec\_S &SCUBA2 position: declination (J2000)& degree \\
25            &  $S_{850\micron}$  & Flux density at 850 $\micron$ (SCUBA2) or at 870 $\micron$ (LABOCA)  &mJy\\
26            &  $S_{850\micron}$\_err  & Flux density uncertainty at 850 $\micron$ (SCUBA2) or at 870 $\micron$ (LABOCA) & mJy\\
\hline
\end{tabular}
\tablecomments{A machine-readable table is available online.}
\label{tab:catalog}
\end{table*}

\subsection{{\it Spitzer} observations and source photometry}

A total of 300 {\it Herschel} ultrared sources were followed up by the {\it Spitzer} snapshot imaging program PID 13042 (PI: A. Cooray) using IRAC \citep{Fazio2004}. Images were taken at 3.6 and 4.5 $\micron$ with 30 second exposure per frame and a 36-position dither pattern in each band for each source (one source per Astronomical Observation Request (AOR)), totaling 1080 s integration per band. We obtained the post- basic calibrated data (pBCDs) from the {\it Spitzer} Science Center (SSC; pipeline version S19.2), which have been reduced with the Mosaicing and Point-source Extraction ({\sc mopex}\footnote{https://irsa.ipac.caltech.edu/data/SPITZER/docs/dataanaly-sistools/tools/mopex/}) package, including background matching, overlap correction and mosaicking. The pBCDs, in most cases, are of good quality for our purpose of source detection and photometry. When necessary, we also downloaded and re-processed the artifact-corrected BCDs with MOPEX (version 18.5.0; \citealt{Makovoz2006}) to generate improved final mosaics. The IRAC mosaics (one mosaic per source/AOR) have a resampled pixel scale of 0.6$\arcsec$ pixel$^{-1}$ and an angular resolution of $\sim$ 1.9$\arcsec$.

We use SExtractor (version 2.8.6; \citealt{Bertin1996}) in dual-image mode to perform source detection and photometry. SExtractor is well-suited for this task, because the relatively sparse mosaics have sufficient source-free pixels available for robust sky background estimation. We use the co-added 3.6 and 4.5 $\mu$m image as the detection image. The dual-image approach ensures that the photometry is measured in identical areas in both bands, yielding accurate source colors. We obtained the Kron-like elliptical aperture magnitudes, MAG\_AUTO, and corresponding flux densities for the following analyses. To estimate the accuracy of the astrometry, we compared the positions of bright IRAC sources with their 2MASS counterparts \citep{Skrutskie2006}. The astrometric discrepancy is small, about one-fourth (0.15$\arcsec$) of the mosaic pixel.

\subsection{SCUBA-2/LABOCA at 850/870 $\micron$}

For the H-ATLAS ultrared DSFGs, \cite{Ivison2016} obtained the 850 $\micron$ continuum imaging with SCUBA-2 and/or 870 $\micron$ continuum imaging with LABOCA. The FWHM of the main beam of SCUBA-2 is 13.0$\arcsec$ at 850 $\micron$, and the astrometry accuracy $\sigma$ is 2-3$\arcsec$. LABOCA has a FWHM resolution of 19.2$\arcsec$ and the positional uncertainty is estimated to be about 1-2$\arcsec$. They measured the 850 or 870 $\micron$ flux densities via several methods (brightest pixel values and aperture photometry) and performed template fitting together with the three {\it Herschel}/SPIRE bands to derive photometric redshifts and FIR luminosities. \cite{Duivenvoorden2018} presented the SCUBA-2 observations of the HeLMS sources and extracted flux densities by taking the brightest pixel values within the 3$\sigma$ of positional uncertainty in the SCUBA-2 map. Photometric redshifts and integrated properties (e.g., FIR luminosities) are derived from the {\sc eazy} code \citep{Brammer2008} using representative FIR/sub-mm templates. We refer the reader to \cite{Ivison2016} and \cite{Duivenvoorden2018} for detailed descriptions of the SCUBA-2/LABOCA observations, flux density measurements, and FIR SED fitting. We adopt the peak-value photometry and photometric redshifts from both works (listed in Table \ref{tab:catalog}). A total of 261 out of the 300 {\it Spitzer} follow-up sources have SCUBA-2/LABOCA data, which are used in the following SED modeling.

\subsection{High-resolution submm/mm data}

\subsubsection{Continuum data at 870 $\micron$, 1.1mm, 1.3 mm, and 3 mm}

A subset (21) of the H-ATLAS ultrared sources in \cite{Ivison2016} were selected for continuum imaging with NOEMA (PIs: R.J. Ivison, M. Krips; PIDs: W05A, X0C6, W15ET) at 1.3 and/or 3 mm and ALMA (PI: A. Conley; PID: 2013.1.00499.S) at 3 mm as described in detail in \cite{Fudamoto2017}. They observed 17 with NOEMA and 4 with ALMA, based on their accessibility and high photometric redshifts. The synthesized FWHM beam sizes are 1-1.5$\arcsec$ for NOEMA and 0.6-1.2$\arcsec$ for ALMA. Precise positions were determined from the continuum images for 18 ultrared sources, and the remaining 3 sources lack secure detection. 

\cite{Oteo2017} presented the high-resolution ($\sim$ 0.12$\arcsec$) ALMA continuum imaging at 870 $\micron$ (PI: R.J. Ivison; PIDs: 2013.1.00001.S, 2016.1.00139.S) for a sample of 44  equatorial and southern ultrared sources from both the H-ATLAS and HerMES surveys. Thirty-one of them are in our {\it Spitzer} sample and thus included in our following analysis. Additional 18 H-ATLAS ultrared sources whose ALMA data were released after \cite{Oteo2017} are also included in this work. 

Five HeLMS ultrared sources (HELMS\_RED\_1, 2, 4, 10, 13) were observed with SMA at 1.1 mm in the compact array configuration (PI: D. Clements; PID: 2013A-S005). The reduced maps have an average rms noise level of 2.2 mJy/beam and the beam FWHM sizes are typically 2.5$\arcsec$. The flux densities at 1.1 mm are given in \cite{Duivenvoorden2018} and included in the SED fitting below. The details of the observations and data reduction will be presented in Greenslade et al. in prep. Additional data with MUSIC/CSO \citep{Sayers2014} and ACT \citep{Su2017} were obtained for 5 HeLMS sources (HELMS\_RED\_1, 3, 4, 6, 7). We list the sources with submm/mm flux density measurements from SMA, MUSIC, and ACT in the Appendix (Table \ref{tab:moredata}). \cite{Duivenvoorden2018} summarized the observations obtained so far as part of the still on-going observational campaign. 

We analyze in this paper a total of 63 {\it Herschel}-selected ultrared sources from H-ATLAS and HeLMS that have high-resolution positions from various observations as described above. The positions and flux densities of the high-resolution sub-sample are listed in Table \ref{tab:highres}. The original 63 ultrared sources are resolved into 86 individual submm/mm sources as seen at the high resolution (see further discussion in Section \ref{sec:catalog} and Section \ref{sec:multiplicity}). These sources are classified as lensed or unlensed based on the submm/mm morphology and the presence or absence of low-redshift foreground galaxies (further discussion in Section \ref{sec:catalog} and Section \ref{sec:unlensed}).

\begin{table*}
\centering
\caption{High-resolution sample followed up with ALMA, NOEMA, and SMA}
\begin{tabular}{lllc}
\hline
\hline
Column   & Parameter    & Description &    Units  \\
\hline
1              &  ID               & Sourcename &   \\
2              &  Highres\_data         & High-resolution data from: ALMA (1), NOEMA (2), SMA (3) etc. & \\
3              & Lensed?     &  Lensed (y), unlensed (n), or unknown (x) according to the high-resolution interferometry data & \\
4             &  RA\_IRAC  & {\it Spitzer}/IRAC counterpart position: right ascension (J2000) & degree \\
5           &  Dec\_IRAC & {\it Spitzer}/IRAC counterpart position: declination (J2000)& degree \\
6            &  MAG\_AUTO$_{3.6\micron}$  & Kron-like elliptical aperture magnitude at 3.6 $\micron$ & AB mag\\
7           &  MAGERR\_AUTO$_{3.6\micron}$  &  AUTO magnitude uncertainty at 3.6 $\micron$ & AB mag\\
8          &  MAG\_AUTO$_{4.5\micron}$  & Kron-like elliptical aperture magnitude at 4.5 $\micron$ & AB mag\\
9            &  MAGERR\_AUTO$_{4.5\micron}$  &  AUTO magnitude uncertainty at 4.5 $\micron$ & AB mag\\
10            &  $S_{3.6\micron}$  & Flux density at 3.6 $\micron$&$\mu$Jy\\
11            &  $S_{3.6\micron}$\_err  & Flux density uncertainty at 3.6 $\micron$ & $\mu$Jy\\
12            &  $S_{4.5\micron}$  & Flux density at 4.5 $\micron$ & $\mu$Jy\\
13            &  $S_{4.5\micron}$\_err  & Flux density uncertainty at 4.5 $\micron$ & $\mu$Jy\\
14            &  $S_{250\micron}$  & Flux density at 250 $\micron$; de-blended if multiple components &mJy\\
15            &  $S_{250\micron}$\_err  & Flux density uncertainty at 250 $\micron$ & mJy\\
16            &  $S_{350\micron}$  & Flux density at 350 $\micron$; de-blended if multiple components & mJy\\
17            &  $S_{350\micron}$\_err  & Flux density uncertainty at 350 $\micron$ & mJy\\
18            &  $S_{500\micron}$  & Flux density at 500 $\micron$; de-blended if multiple components & mJy\\
19            &  $S_{500\micron}$\_err  & Flux density uncertainty at 500 $\micron$ & mJy\\
20            &  $S_{850\micron}$  & Flux density at 850 $\micron$ (SCUBA2) or at 870 $\micron$ (LABOCA); de-blended  &mJy\\
              &              &          if multiple components  &\\
21            &  $S_{850\micron}$\_err  & Flux density uncertainty at 850 $\micron$ (SCUBA2) or at 870 $\micron$ (LABOCA) & mJy\\ 
\hline
\end{tabular}
\tablecomments{A machine-readable table is available online.}
\label{tab:highres}
\end{table*}

\subsubsection{Spectroscopic redshifts from spectral scans}

\cite{Fudamoto2017} conducted spectral scans at the 3-mm atmospheric window for 21 H-ATLAS ultrared sources with NOEMA (Co-PIs: R.J. Ivison, M. Krips; PIDs: W05A, X0C6) and ALMA (PI: A. Conley; PID: 2013.1.00499.S). They obtained 8 secure redshifts via detections of multiple CO lines and 3 redshifts via a single CO line detection. One of the SGP sources, SGP-354388, was confirmed to be the core of a protocluster that lies at $z$ = 4.002 via detections of CO, [CI], and H$_2$O lines with ALMA and ATCA \citep{Oteo2018}. Four HeLMS ultrared sources have spectroscopic redshifts confirmed with ALMA and CARMA (\citealt{Asboth2016,Duivenvoorden2018}). Table \ref{tab:specz} lists all the spectroscopically confirmed {\it Herschel}-selected ultared DSFGs at $z \gtrsim 4$ in this work or in the literature.

\begin{table*}
\centering
\caption{Spectroscopically confirmed {\it Herschel}-selected ultrared DSFGs at $z \gtrsim 4$. } 
\begin{tabular}{lllccc}
\hline
\hline
IAU name                               & Nickname         & $z_{\rm spec}$             &  $L_{\rm IR}$                        & Lensed? & Reference\\
      					   &				&			& (10$^{13}$ $L_{\odot}$)	&   			&  \\
\hline
HATLAS J084937.0+001455 &G09-81106        &4.531           &        2.7                                       &   No        & This work; a, g  \\
HATLAS J090045.4+004125 &G09-83808        &6.027           &       3.2                                        & Yes                  &This work; a\\
HATLAS J133337.6+241541 &NGP-190387     &4.420	     &		3.1		  		   &Yes	     &This work; a \\
HATLAS J134114.2+335934 &NGP-246114     &3.847            &        2.0                                      &No                     & This work; a \\
HATLAS J133251.5+332339 &NGP-284357     &4.894         & 		2.5				   & No		     &This work; a\\
HATLAS J000306.9-330248 &SGP-196076    &   4.425	&	2.5				   & No		     & This work; a, b \\ 
HATLAS J000607.6-322639 &SGP-261206     &4.242	&	4.4					   & Yes		     & This work; a\\
HATLAS J004223.5-334340 &SGP-354388     & 4.002	&	4.8					  & No		     &This work; c\\
HerMES J004409.9+011823 &HELMS\_RED\_1   & 4.163	&	8.9					  & Yes		     &This work; d\\
HerMES J005258.9+061319 &HELMS\_RED\_2   & 4.373	&	8.3					  & Yes		      &This work; d\\
HerMES J002220.8-015521 &HELMS\_RED\_4 & 5.161	&	5.8					  & Yes		      & This work; e \\
HerMES J002737.4-020801 &HELMS\_RED\_31 &  3.798	&	3.4					  & Yes		      &This work; e \\
\hline
HATLAS J142413.9+022304 & ID141            &4.243          &     8.5                                            & Yes                 & h\\
HerMES J043657.5-543809 &ADFS-27     & 5.655		&	2.4					  &No			&f\\
HLS        J091828.6+514223 &   ---                & 5.243        & 11                                              & Yes                  &i\\
HerMES J104050.6+560654 & LSW102     & 5.29 		 & ---    					  & Yes               &j, k, m\\
HerMES J170647.7+584623 & HFLS3       &  6.337           &  2.9					  & Yes          & k, l \\
HerMES J170817.3+582844 & HFLS1      & 4.286              & 5.6                                            & No                  &j, k\\
HerMES J172049.0+594623  & HFLS5     & 4.44               & 2.8 				           & No		& j, k\\
\hline
\end{tabular}
\tablecomments{a.\cite{Fudamoto2017}, b. \cite{Oteo2016}, c. \cite{Oteo2018}, d. \cite{Duivenvoorden2018}, e. \cite{Asboth2016}, f. \cite{Riechers2017}, g. \cite{Zavala2018}, h. \cite{Cox2011}, i. \cite{Combes2012}, j. \cite{Dowell2014}, k. \cite{Riechers2013}, l. \cite{Cooray2014}, m. Wardlow et al. in prep.}
\label{tab:specz}
\end{table*}

\section{Cross-identification and {\it Spitzer} catalogs}
\label{sec:catalog}

Before we cross-identify the {\it Spitzer} counterparts to the {\it Herschel} ultrared sources, we crossmatch the IRAC data with shallow and medium-depth large-area surveys at optical/NIR wavelengths, e.g., the Sloan Digital Sky Survey (SDSS; \citealt{York2000}) Data Release 12 (DR12; \citealt{Alam2015}) and the VISTA Kilo-degree Infrared Galaxy (VIKING) Survey \citep{Edge2013}, to identify low-redshift galaxies that can potentially magnify a higher redshift ultrared source via gravitational lensing. Figure \ref{cutouts} in Appendix shows the 60$\arcsec$ $\times$ 60 $\arcsec$ {\it Spitzer}/IRAC cutouts centered on the {\it Herschel} positions for all the ultrared sources in our {\it Spitzer} program. The 3$\arcsec$-radius cyan circles denote the positions of the SDSS sources in the field, and the yellow circles show the positions of the VIKING sources. 

For the ultrared sources with high-resolution submm/mm detections in Table \ref{tab:highres}, we can use these data to pinpoint the locations and securely identify the {\it Spitzer}/IRAC counterparts. Within this high-resolution sub-sample of 63 original ultrared sources, 23 of them are gravitationally lensed sources with clear lensing signatures like rings or arcs as shown in \cite{Oteo2017}, and rest of them that do not show any lensing features are likely unlensed, intrinsically bright ultrared DSFGs (more discussion on lensed/unlensed fraction in Section \ref{sec:unlensed}). For sources that are classified as unlensed at sub-mm and successfully cross-identified in {\it Spitzer}/IRAC, we can directly extract the {\it Spitzer}/IRAC source flux densities. We list the flux densities at 3.6 and 4.5 $\mu$m as well as de-blended SPIRE, SCUBA-2/LABOCA, and ALMA flux densities of the unlensed sample in Table \ref{tab:unlensed}, which will be used in multi-wavelength SED fitting to derive their physical properties. The unlensed sample is the focus of our SED analysis in Section \ref{sec:sed}. 

For lensed sources, however, a significant fraction of the emission seen at {\it Spitzer}/IRAC is likely due to the foreground lensing galaxies, and higher-resolution optical/NIR imaging is required to perform source/lens de-blending. Therefore we can only place an upper limit on the IRAC photometry for the background lensed ultrared sources until we are able to de-blend the source/lens photometry. For a few lensed ultrared sources, high-resolution {\it HST} imaging is being acquired and the de-blending results will be presented in Brown et al. in prep. 

Within the high-resolution sample of 63 original {\it Herschel}-selected ultrared sources, 17 ultrared sources are resolved into multiple components. A total of 86 individual submm sources were identified from the 63 ultrared sources as a result of the multiple components (more discussion on multiplicity in Section \ref{sec:multiplicity}). Corresponding {\it Spitzer}/IRAC counterparts are identified based on the high-resolution positions from ALMA, NOEMA, and SMA. We use the probabilistic de-blender XID+ \citep{Hurley2017}, which is a new prior-based source extraction tool in confusion-dominated maps, with the positional priors from the high-resolution interferometry data to disentangle the SPIRE flux densities over the sub-components. XID+ is developed and has been tested on SPIRE maps using a probabilistic Bayesian framework which includes prior information, and uses the Bayesian inference to obtain the full posterior probability distribution function (PDF) on flux estimates. The de-blended flux densities are shown in Table \ref{tab:highres}. The SCUBA-2 or LABOCA flux densities are split among sub-components assuming the same relative ratios between the individual components derived from the ALMA 870 $\micron$ flux densities \citep{Oteo2017}. We caution though that there is inconsistency between the ALMA and SCUBA-2 photometry that could be partly due to the small field of view of ALMA Band 7 \citep{Oteo2017}. We use the SCUBA-2/LABOCA flux densities in the following SED fitting in Section \ref{sec:sed}.

For the ultrared sources that currently lack high-resolution interferometry data, we use the SCUBA-2 or {\it Herschel} positions and cross-match with any IRAC sources that are within 2$\sigma$ positional uncertainties of SCUBA-2 or {\it Herschel}. For sources with robust SCUBA-2 detections (i.e., S/N $>$ 3), we combine a statistical positional accuracy of $\sigma_{\rm pos}$ = 0.6 $\times$ FWHM/(S/N) \citep{Ivison2007} and the JCMT pointing accuracy of 2$\arcsec$-3$\arcsec$. For sources with low S/N or no SCUBA-2 data, we use the {\it Herschel} positions for cross-matching (positional uncertainty $\sigma_{\rm H}$ $\sim$ 6$\arcsec$; \citealt{Asboth2016}). If the SCUBA-2/{\it Herschel} positions are on top of or in close proximity to an IRAC source that is identified as a SDSS or VIKING low redshift object, the ultrared source is likely lensed by the foreground galaxy and most often the background emission in IRAC is blended with that from the foreground lens. We provide a catalog of the IRAC counterpart candidates identified as the closest IRAC source in search radius in Table \ref{tab:catalog} but defer counterpart confirmation and robust assessment of lensing till we obtain more high-resolution interferometry data.

\begin{table*}
\scriptsize
\centering
\caption{Unlensed sample}
\begin{tabular}{lccccccccc}
\hline
\hline
Sourcename                &     $z_{\rm FIR}$  & $z_{\rm \footnotesize{MAGPHYS}}$  & $S_{\rm 3.6\mu m}$ & $S_{\rm 4.5\mu m}$& $S_{\rm 250\mu m}$& $S_{\rm 350\mu m}$ &$S_{\rm 500\mu m}$& $S_{\rm 850\mu m}^{\rm SCUBA-2}$&   $S_{\rm 870\mu m}^{\rm ALMA}$    \\
                                    & 	     &     & ($\mu$Jy)                 &		 ($\mu$Jy)      &   (mJy)                     & (mJy)                          & (mJy)                       &   (mJy)                                                & (mJy)                         \\
\hline
G09-47693                  & 3.12$^{+0.39}_{-0.33}$   &2.73$^{+1.01}_{-0.71}$ & 35.21 $\pm$ 9.06    & 45.30 $\pm$ 10.24   &  27.4 $\pm$ 7.3       &  34.4 $\pm$ 8.1         &  45.4 $\pm$ 8.6        &  12.5 $\pm$ 4.0                                 &  5.5 $\pm$ 0.4  \\
G09-47693a               &                   ---                   &2.51$^{+0.91}_{-0.69}$   & 13.66 $\pm$ 5.64      & 17.05 $\pm$ 6.28     &  11.9 $\pm$ 8.1       &  20.8 $\pm$ 9.0         &  38.0 $\pm$ 10.1        &   5.1 $\pm$ 1.6                               &   2.24 $\pm$ 0.21        \\
G09-47693b               &                    ---                      &2.79$^{+1.33}_{-0.96}$   & 21.55 $\pm$ 7.09      & 28.25 $\pm$ 8.09     &  11.3 $\pm$ 7.9       &  8.1 $\pm$ 6.0          &  7.9 $\pm$ 6.0          & 7.4 $\pm$ 2.4                                       &  3.25 $\pm$ 0.29      \\
G09-51190                 &  3.83$^{+0.58}_{-0.48}$    & 3.31$^{+1.32}_{-0.81}$ &  28.96 $\pm$ 8.25       &  33.18 $\pm$ 8.83    &  28.5 $\pm$ 7.6       &  39.5 $\pm$ 8.1         &  46.6 $\pm$ 8.6        & 28.3 $\pm$ 7.3                                   & 9.7 $\pm$ 0.5 \\ 
G09-51190a               &                    ---                    & 2.73$^{+1.02}_{-0.69}$ & 28.96 $\pm$ 8.25      &  33.18 $\pm$ 8.83    &  21.9 $\pm$ 6.4       &  27.6 $\pm$ 7.8          &  41.1 $\pm$ 7.3        & 10.4 $\pm$ 2.9                                   & 3.55 $\pm$ 0.21     \\
G09-51190b               &         ---                		     & 4.69$^{+1.58}_{-1.54}$  &  $<$ 7.84                   &   $<$ 6.50                &  11.4 $\pm$ 6.2     &     14.4 $\pm$ 7.7          &  6.2 $\pm$ 5.9       & 17.9 $\pm$ 4.6                                    & 6.15 $\pm$ 0.48 \\ 
G09-59393                & 3.70$^{+0.35}_{-0.26}$     & 3.42$^{+1.20}_{-0.86}$ & $<$ 15.07         &10.19 $\pm$ 5.08      &  24.1 $\pm$ 7.0      & 43.8 $\pm$ 8.3           & 46.8 $\pm$ 8.6         & 23.7 $\pm$ 3.5                                     & 12.4 $\pm$ 0.4 \\
G09-59393a              &          	---      		    & 3.15$^{+1.01}_{-0.87}$ & $<$ 7.30       & 10.19 $\pm$ 5.08      &  23.7 $\pm$ 4.8      & 41.7 $\pm$ 5.2           & 36.3 $\pm$ 6.8         & 16.4 $\pm$ 2.4                                     & 7.33 $\pm$ 0.48 \\
G09-59393b               &          	---			    & 5.07$^{+1.43}_{-1.28}$ &   $<$ 7.77               &   $<$ 8.55                &   3.6 $\pm$ 3.3       &   2.7 $\pm$ 2.9            &  6.1 $\pm$ 5.4          & 7.3 $\pm$ 1.1                                    & 3.25 $\pm$ 0.64  \\ 
G09-62610                &  3.70$^{+0.44}_{-0.26}$    & 3.30$^{+1.23}_{-0.85}$ & 12.98 $\pm$ 5.55       & 12.51 $\pm$ 5.51         & 18.6 $\pm$ 5.4      & 37.3 $\pm$ 7.4           & 44.3 $\pm$ 7.8         & 19.5 $\pm$ 4.9                                   &  13.6 $\pm$ 0.7 \\
G09-62610a               &        	---			  & 3.11$^{+1.08}_{-0.84}$  & 2.51 $\pm$ 2.46    &  2.47 $\pm$ 2.44     &    9.1 $\pm$ 4.4       &   18.0 $\pm$ 6.1          &  23.8 $\pm$ 8.5       &    6.2 $\pm$ 1.6                                   &      4.31 $\pm$ 0.35      \\ 
G09-62610b               &        	---			  & 4.30$^{+1.47}_{-1.21}$   & $<$ 7.02   &   3.75 $\pm$ 3.05      &   3.6 $\pm$ 3.4        &   7.8 $\pm$ 6.9           & 9.5 $\pm$ 8.5           &   8.0 $\pm$ 2.0                                   &   5.57 $\pm$ 0.28         \\ 
G09-62610c              &              ---                          &3.55$^{+1.45}_{-1.01}$   &  10.47 $\pm$ 4.97   &   6.29 $\pm$ 3.89     &  3.8 $\pm$ 3.4         &  8.6 $\pm$ 6.4           &  8.0 $\pm$ 7.5            &   5.3 $\pm$ 1.3                                  & 3.72 $\pm$ 0.50   \\ 
G09-64889               &   3.48$^{+0.48}_{-0.40}$   & 3.12$^{+1.18}_{-0.81}$ & 11.24 $\pm$ 5.21     & 10.57 $\pm$ 5.09       & 20.2 $\pm$ 5.9      & 30.4 $\pm$ 7.7           & 34.7 $\pm$ 8.1          & 15.1 $\pm$ 4.3                                  & 7.91 $\pm$ 0.35 \\
G09-79552               &  3.59$^{+0.34}_{-0.26}$    & 3.24$^{+1.14}_{-0.85}$  &7.37 $\pm$ 4.16       & 10.47 $\pm$ 4.92       & 16.6 $\pm$ 6.2      & 38.1 $\pm$ 8.1           & 42.8 $\pm$ 8.5          & 17.0 $\pm$ 3.6                                  & 12.7 $\pm$ 0.6\\
G09-80620               & 4.01$^{+0.22}_{-0.78}$     & 3.17$^{+1.16}_{-0.83}$ & 11.97 $\pm$ 5.50        & 19.46 $\pm$ 6.92        & 13.5 $\pm$ 5.0      & 25.3 $\pm$ 7.4           & 28.4 $\pm$ 7.7         &  13.2 $\pm$ 4.3                                & 8.4 $\pm$ 0.7 \\ 
G09-80620a	        &            ---		          & 1.99$^{+1.10}_{-0.67}$  &7.85 $\pm$ 4.43       & 14.00 $\pm$ 5.83        & 4.5 $\pm$ 3.9      & 7.3 $\pm$ 5.8              & 7.5 $\pm$ 6.8         &  10.3 $\pm$ 3.4                                 & 5.08 $\pm$ 0.97\\ 
G09-80620b        	 &         	---	                    & 2.88$^{+1.19}_{-0.85}$& 4.12 $\pm$ 3.26       &  5.46 $\pm$ 3.73         &  5.5 $\pm$ 4.5       & 6.7 $\pm$ 5.7           &  12.8 $\pm$ 7.7               & 2.9 $\pm$ 0.9                                  & 1.45 $\pm$ 0.29\\ 
G09-80658               &  4.07$^{+0.09}_{-0.72}$    & 3.19$^{+1.14}_{-0.79}$ & 20.00 $\pm$ 6.85      & 31.62 $\pm$ 8.57        & 17.8 $\pm$ 6.4      & 31.6 $\pm$ 8.3          & 39.5 $\pm$ 8.8          & 17.6 $\pm$ 4.1                                 & 10.7 $\pm$ 0.7\\
G09-80658a           &           ---				  & 3.75$^{+1.36}_{-0.92}$ & 20.00 $\pm$ 6.85      & 31.62 $\pm$ 8.57        & 5.9 $\pm$ 4.7      & 12.0 $\pm$ 7.2          & 20.3 $\pm$ 10.2          & 13.2 $\pm$ 3.1                                  & 4.08 $\pm$ 0.40 \\
G09-80658b           &          ---				  & 4.26$^{+1.45}_{-1.17}$ &  $<$ 5.94       & $<$ 10.02          &  3.0 $\pm$ 2.3     &  3.4 $\pm$ 2.5           &  8.0 $\pm$ 5.5                &  4.4 $\pm$  1.0                                    & 1.35 $\pm$ 0.13 \\ 
G09-81106                & 4.531 				 & 4.23$^{+1.32}_{-1.06}$  &3.22 $\pm$ 2.81       &  4.68 $\pm$ 3.35       &  14.0 $\pm$ 6.0       & 30.9 $\pm$ 8.2          & 47.5 $\pm$ 8.8           & 30.2 $\pm$ 5.2                                    & 28.4 $\pm$ 0.8 \\
G09-81271               &  4.62$^{+0.46}_{-0.38}$   & 4.32$^{+1.39}_{-1.02}$  &4.52 $\pm$ 3.36        & 3.54 $\pm$ 3.00        &  15.0 $\pm$ 6.1       & 30.5 $\pm$ 8.2          & 42.3 $\pm$ 8.6           & 29.7 $\pm$ 3.7                                   & 20.5 $\pm$ 0.7\\
G09-87123               &  4.28$^{+0.52}_{-0.34}$    & 3.85$^{+1.25}_{-0.97}$  &7.78 $\pm$ 4.28       & 13.36 $\pm$ 5.57      &  10.4 $\pm$ 5.8       & 25.3 $\pm$ 8.2          & 39.2 $\pm$ 8.7           & 20.7 $\pm$ 4.6                                   & 6.63 $\pm$ 0.47 \\
G09-100369            &  3.79$^{+0.61}_{-0.46}$     & 3.42$^{+1.23}_{-0.89}$ & 3.75 $\pm$ 3.10      & 6.48 $\pm$ 3.95        &  15.4 $\pm$ 5.5       & 17.3 $\pm$ 7.6          & 32.3 $\pm$ 8.0           & 13.2 $\pm$ 3.6                                    & 3.74 $\pm$ 0.25 \\
G09-101355             & 4.20$^{+0.70}_{-0.39}$     & 3.74$^{+1.32}_{-1.02}$ & $<$ 10.86      & 10.22 $\pm$ 4.90        &  9.5 $\pm$ 5.5         & 14.6 $\pm$ 7.9           & 33.4 $\pm$ 8.3           & 13.5 $\pm$ 4.9                                 & 7.6 $\pm$ 0.7 \\
G09-101355a         &           ---				 & 3.99$^{+1.50}_{-1.21}$ &   $<$ 4.68     & 4.24 $\pm$ 3.16        &  3.6 $\pm$ 3.1         & 7.2 $\pm$ 4.4           & 15.6 $\pm$ 7.9            & 8.5 $\pm$ 3.1                                  & 4.78 $\pm$ 0.60 \\
G09-101355b         &          ---				  & 2.92$^{+1.13}_{-0.86}$ & $<$ 6.18       & 5.99 $\pm$ 3.75       &  9.0 $\pm$ 4.2        & 11.1 $\pm$ 4.7         & 12.7 $\pm$ 8.0            & 5.0 $\pm$ 1.8                                   & 2.80 $\pm$ 0.42\\
NGP-101333          &  3.53$^{+0.34}_{-0.27}$     & 3.23$^{+1.24}_{-0.80}$ & 12.75 $\pm$ 5.26     & 13.11 $\pm$ 5.66     &  32.4 $\pm$ 7.5         & 46.5 $\pm$ 8.2          & 52.8 $\pm$ 9.0           & 24.6 $\pm$ 3.8  &---\\
NGP-111912          & 3.27$^{+0.36}_{-0.26}$       & 2.81$^{+1.03}_{-0.72}$ & 36.92 $\pm$ 9.27      & 47.05 $\pm$ 10.42  &  25.2 $\pm$ 6.5         & 41.5 $\pm$ 7.6           & 50.2 $\pm$ 8.0          & 14.9 $\pm$ 3.9  &---\\  
NGP-136156          & 3.95$^{+0.06}_{-0.57}$      & 3.25$^{+1.21}_{-0.79}$ & 7.92 $\pm$ 4.15       & 10.23 $\pm$ 4.95     &  29.3 $\pm$ 7.4         & 41.9 $\pm$ 8.3          & 57.5 $\pm$ 9.2           & 23.4 $\pm$ 3.4  &---\\
NGP-246114          &3.847     				& 3.91$^{+1.43}_{-0.94}$ & 17.12 $\pm$ 5.80      & 17.88 $\pm$ 5.86      & 17.3 $\pm$ 6.5         & 30.4 $\pm$ 8.1          &  33.9 $\pm$ 8.5          & 25.9 $\pm$ 4.6  &---\\
NGP-252305         &4.34$^{+0.43}_{-0.38}$       & 3.95$^{+1.35}_{-0.93}$ & 6.32 $\pm$ 3.89         & 7.90 $\pm$ 4.33       &  15.3 $\pm$ 6.1         & 27.7 $\pm$ 8.1          & 40.0 $\pm$ 9.4          & 24.0 $\pm$ 3.5  &---\\
NGP-284357         & 4.894  				 & 4.51$^{+1.33}_{-1.07}$ & 4.62 $\pm$ 3.32        & 3.43 $\pm$ 2.87       &  12.6 $\pm$ 5.3         & 20.4 $\pm$ 7.8          & 42.4 $\pm$ 8.3          & 28.9 $\pm$ 4.3 &---\\
SGP-72464           & 3.06$^{+0.21}_{-0.19}$     & 2.74$^{+1.00}_{-0.73}$ & 26.76 $\pm$ 8.31      & 28.51 $\pm$ 8.52     &  43.4 $\pm$ 7.6         & 67.0 $\pm$ 8.0          & 72.6 $\pm$ 8.9          & 20.0 $\pm$ 4.2                                   & 16.9 $\pm$ 0.4\\
SGP-93302b          & 3.91$^{+0.27}_{-0.22}$     & 2.84$^{+0.90}_{-0.79}$  & $<$ 8.13        &4.52 $\pm$ 3.66        &  18.3 $\pm$ 8.1         &   33.4 $\pm$ 14.9      &   19.5 $\pm$ 16.3      &9.4 $\pm$ 0.9                                      & 9.95 $\pm$ 0.90\\
SGP-135338         & 3.06$^{+0.33}_{-0.26}$     & 3.14$^{+0.94}_{-0.84}$  & $<$ 3.30                   &  $<$ 6.94                  &  32.9 $\pm$ 7.3        & 43.6 $\pm$ 8.1          & 53.3 $\pm$ 8.8          & 14.7 $\pm$ 3.8                                     & 6.1 $\pm$ 0.4\\
SGP-196076        &  4.425  				 & 4.17$^{+1.36}_{-0.96}$  &12.66 $\pm$ 5.19      & 10.17 $\pm$ 4.91     &  28.6 $\pm$ 7.3         & 28.6 $\pm$ 8.2          & 46.2 $\pm$ 8.6          & 32.5 $\pm$ 4.1                                   & 34.6 $\pm$ 2.3 \\
SGP-196076a     & 4.425    				  & 4.92$^{+1.40}_{-1.19}$ &  5.94 $\pm$ 3.11       &  5.93 $\pm$ 1.81      &   8.6 $\pm$ 6.0       &   11.2 $\pm$ 9.4            & 16.4 $\pm$ 12.2       & 21.3 $\pm$ 2.7                                    & 17.58 $\pm$ 1.03 \\ 
SGP-196076b    & 4.425      				 & 4.35$^{+1.44}_{-1.24}$ &   6.72 $\pm$ 2.08      &   4.24 $\pm$ 1.57      &   9.6 $\pm$ 6.5        &   8.4 $\pm$ 6.1             & 12.0 $\pm$ 8.6         & 9.6 $\pm$ 1.2                                      & 7.90 $\pm$ 0.60 \\ 
SGP-196076c    &   4.425             		 & 2.78$^{+1.17}_{-0.80}$ &     $<$ 6.24                &    $<$ 6.34                 &   5.1 $\pm$ 4.0         &   8.2 $\pm$ 5.9            &  11.5 $\pm$ 8.3      &  1.6 $\pm$ 0.2                                       & 1.33 $\pm$ 0.17  \\ 
SGP-208073        & 3.48$^{+0.40}_{-0.28}$      &3.10$^{+1.19}_{-0.75}$ & 42.70 $\pm$ 9.59         & 50.51 $\pm$ 10.77     &  28.0 $\pm$ 7.4        &  33.2 $\pm$ 8.1         & 44.3 $\pm$ 8.5         & 19.4 $\pm$ 2.9                                     & 13.9 $\pm$ 0.9\\
SGP-208073a     &           	---			        & 4.37$^{+1.38}_{-1.13}$ &  2.84 $\pm$ 2.10      &  3.19 $\pm$ 2.18        &   3.6 $\pm$ 2.7         & 9.8 $\pm$ 8.3            &  12.0 $\pm$ 8.6          & 8.7 $\pm$ 1.3                                      & 6.24 $\pm$ 0.56 \\
SGP-208073b     &            ---				 &3.61$^{+1.42}_{-0.94}$&  28.13 $\pm$ 7.76   & 31.03 $\pm$ 8.58        &  15.9 $\pm$ 9.9        & 9.4 $\pm$ 6.7            &  10.9 $\pm$ 8.1           &9.6 $\pm$ 1.4                                      & 6.91 $\pm$ 0.63 \\
SGP-208073c     &          	---			        &1.87$^{+0.71}_{-0.52}$ & 11.73 $\pm$ 5.23    & 16.29 $\pm$ 6.14      &   9.1 $\pm$ 8.0            & 12.2 $\pm$ 8.0          &  11.8 $\pm$8.2           & 1.2 $\pm$ 0.2                                      &0.83 $\pm$ 0.21 \\
SGP-213813        & 3.49$^{+0.40}_{-0.32}$      &2.98$^{+1.39}_{-0.71}$ & 75.90 $\pm$ 13.25    & 69.53 $\pm$ 12.67   &  23.9 $\pm$ 6.3         & 35.1 $\pm$ 7.6          & 35.9 $\pm$ 8.2          & 18.1 $\pm$ 3.6                                   & 13.9 $\pm$ 0.7\\
SGP-219197        & 2.94$^{+0.25}_{-0.24}$      & 2.61$^{+0.92}_{-0.70}$& 29.18 $\pm$ 8.31      & 36.62 $\pm$ 9.25     &  27.6 $\pm$ 7.4         & 51.3 $\pm$ 8.1         & 43.6 $\pm$ 8.4           & 12.2 $\pm$ 3.7                                  & 9.47 $\pm$ 0.40\\
SGP-317726        & 3.69$^{+0.39}_{-0.30}$      & 3.56$^{+1.20}_{-0.92}$& $<$ 10.37                 & $<$ 8.99                   &  20.4 $\pm$ 6.0        & 35.1 $\pm$ 7.7          & 39.5 $\pm$ 8.0          & 19.4 $\pm$ 3.2                                     & 26.9 $\pm$ 2.9 \\
SGP-354388        &  4.002   				& 4.89$^{+1.45}_{-1.16}$  &5.35 $\pm$ 3.43        & 8.76 $\pm$ 4.58       &  26.6 $\pm$ 8.0         & 39.8 $\pm$ 8.9          & 53.5 $\pm$ 9.8          & 64.1 $\pm$ 10.9                                 & 24.1 $\pm$ 1.7 \\
SGP-354388a      &   4.002       				  & 5.60$^{+1.32}_{-1.22}$&  $<$ 5.35                   &   $<$ 8.76              & 6.2 $\pm$ 4.6        &  16.0 $\pm$ 12.0          &  18.8 $\pm$ 13.1       &  36.7 $\pm$ 6.2                                 & 9.64 $\pm$ 0.33\\
SGP-354388b     & 4.002           			& 4.46$^{+1.40}_{-1.16}$ &  $<$ 5.35                    &   $<$ 8.76               &   6.3 $\pm$ 4.5         &  16.8 $\pm$ 11.7           & 19.3 $\pm$ 16.1        & 13.7 $\pm$ 2.3                                  & 3.61 $\pm$ 0.25 \\
SGP-354388c     &  4.002         				& 4.67$^{+1.40}_{-1.18}$ & $<$ 5.35                  &      $<$ 8.76             &  6.2 $\pm$ 4.6          &  11.7 $\pm$ 8.4           &  13.6 $\pm$ 10.1        & 13.6 $\pm$ 2.3                                  & 3.58 $\pm$ 0.16 \\
SGP-380990        & 2.84$^{+0.22}_{-0.21}$     & 2.68$^{+0.92}_{-0.76}$ &  4.99 $\pm$ 3.45       &  8.59 $\pm$ 4.48       &  14.4 $\pm$ 5.9         & 45.6 $\pm$ 8.2          & 40.6 $\pm$ 8.5          & 7.7 $\pm$1.8 					& 8.93 $\pm$ 0.36\\
SGP-381615        & 2.98$^{+0.29}_{-0.29}$    & 2.97$^{+0.98}_{-0.81}$  & $<$ 4.09                    & $<$ 5.41                   &19.40 $\pm$ 6.6        & 39.1 $\pm$ 8.1           & 34.7 $\pm$ 8.5          & 8.5 $\pm$ 3.6                                      & 6.20 $\pm$ 0.29 \\
SGP-381637       & 3.30$^{+0.28}_{-0.25}$      & 3.04$^{+1.02}_{-0.83}$ &   3.53 $\pm$ 3.01      & 7.49 $\pm$ 4.27       &  18.7 $\pm$ 6.8         & 41.5 $\pm$ 8.4           & 49.3 $\pm$ 8.6          & 12.6 $\pm$ 3.7                                    & 4.45 $\pm$ 0.31 \\
SGP-382394       & 2.96$^{+0.29}_{-0.26}$     &2.73$^{+0.98}_{-0.76}$  &  9.58 $\pm$ 5.00       & 11.94 $\pm$ 5.46      &  15.7 $\pm$ 5.9        & 35.6 $\pm$ 8.1           & 35.9 $\pm$ 8.6          & 8.0 $\pm$ 2.4                                      & 2.11 $\pm$ 0.29\\
SGP-385891       & 3.70$^{+0.29}_{-0.24}$     & 3.32$^{+1.13}_{-0.89}$&     8.44 $\pm$ 4.26     &   13.29 $\pm$ 5.59    &  13.0 $\pm$ 8.2         & 45.6 $\pm$ 9.8          & 59.6 $\pm$ 1.2           &20.5 $\pm$ 3.6                                      & 11.1 $\pm$ 0.7 \\
SGP-385891a   &            	---				& 5.04$^{+1.41}_{-1.25}$ &    $<$ 7.31               &  $<$ 7.81                   &  5.3 $\pm$ 4.7             &  7.6 $\pm$ 5.2         &  20.2 $\pm$ 13.9        &  16.3 $\pm$ 2.9                                        & 5.9 $\pm$ 0.3 \\
SGP-385891b   &           	---				 & 2.43$^{+0.83}_{-0.68}$  &  8.44 $\pm$ 4.26     &  13.29 $\pm$ 5.59      &  12.1 $\pm$ 6.9         &  27.6 $\pm$ 9.0         &  24.7 $\pm$ 14.8        &  4.2 $\pm$ 0.7                                       & 1.5 $\pm$ 0.2 \\
SGP-386447       & 4.89$^{+0.78}_{-0.73}$    & 4.18$^{+1.35}_{-1.10}$ &14.81 $\pm$ 5.75     & 21.52 $\pm$ 7.24      &  10.5 $\pm$ 6.0         & 33.6 $\pm$ 8.4          & 34.5 $\pm$ 8.6          & 34.3 $\pm$ 8.4                                     & 6.5 $\pm$ 0.6 \\
SGP-392029       & 3.42$^{+0.47}_{-0.32}$   & 3.70$^{+1.67}_{-1.27}$  & $<$ 8.01         & 3.29 $\pm$ 2.91        &  18.3 $\pm$ 6.5         & 30.5 $\pm$ 8.3          & 35.3 $\pm$ 8.4          & 13.8 $\pm$ 3.5                                     & 10.8 $\pm$ 0.8 \\
SGP-392029a   &          	---			  & 3.81$^{+1.68}_{-1.27}$  & $<$ 4.74        &$<$ 7.44         &  10.5 $\pm$ 5.8         & 7.7 $\pm$ 5.8            & 5.2 $\pm$ 3.8          & 5.2 $\pm$ 1.3                                        & 3.11 $\pm$ 0.24 \\
SGP-392029b    &          	---				 & 3.28$^{+1.03}_{-0.92}$ & $<$ 3.27       & $<$ 4.56        &  9.8 $\pm$ 5.4        & 18.7 $\pm$ 6.7          &  28.9 $\pm$ 7.0       & 8.6 $\pm$ 2.2                                         & 5.15 $\pm$ 0.28\\ 
SGP-499646       & 4.68$^{+0.49}_{-0.34}$    & 4.37$^{+1.32}_{-1.10}$&   $<$ 4.55       & 4.76 $\pm$ 3.37        &  5.8 $\pm$ 5.9            & 10.8 $\pm$ 8.1         & 41.4 $\pm$ 8.6        & 18.7 $\pm$ 3.0                                        & 14.1 $\pm$ 1.6\\
\hline
\end{tabular}
\label{tab:unlensed}
\end{table*}

\addtocounter{table}{-1}

\begin{table*}
\scriptsize
\centering
\caption{Unlensed sample continued}
\begin{tabular}{lccccccccc}
\hline
\hline
Sourcename                &     $z_{\rm FIR}$  & $z_{\rm \footnotesize{MAGPHYS}}$  & $S_{\rm 3.6\mu m}$ & $S_{\rm 4.5\mu m}$& $S_{\rm 250\mu m}$& $S_{\rm 350\mu m}$ &$S_{\rm 500\mu m}$& $S_{\rm 850\mu m}^{\rm SCUBA-2}$&   $S_{\rm 870\mu m}^{\rm ALMA}$    \\
                                    & 	        &  & ($\mu$Jy)                 &		 ($\mu$Jy)      &   (mJy)                     & (mJy)                          & (mJy)                       &   (mJy)                                                & (mJy)                         \\
\hline
UR72S                 & 3.35                                 &2.84$^{+1.33}_{-0.88}$   &  10.78 $\pm$ 4.84    & 12.20 $\pm$ 5.46     &  35.4 $\pm$ 7.3        & 37.0 $\pm$ 8.2       & 56.0 $\pm$ 8.6       &  	---							&  8.3 $\pm$ 0.9\\
UR72Sa               &           	---	                      & 2.42$^{+1.69}_{-1.00}$ &   10.78 $\pm$ 4.84    &  12.20 $\pm$ 5.46    & 5.9 $\pm$ 4.3          &  7.8 $\pm$ 5.6            &  31.1 $\pm$ 14.4   &      ---                                                       &  3.57 $\pm$ 0.23\\ 
UR72Sb               &            ---		              & 2.39$^{+1.06}_{-0.77}$  &  $<$ 14.53               &   $<$ 16.37                 &  24.1 $\pm$ 7.8        & 28.5 $\pm$ 8.9           &  17.5 $\pm$ 12.0    &     ---                                                        &  0.96 $\pm$ 0.21 \\
UR72Sc               &          	---		             & 2.22$^{+1.42}_{-0.85}$ &  $<$ 14.53                &    $<$ 16.37               &  11.0 $\pm$ 6.3        & 5.1 $\pm$ 3.4             &  7.3 $\pm$ 5.3        &    ---                                                         &  3.72 $\pm$ 0.85 \\
UR73S                    & 3.45    		             & 2.53$^{+1.25}_{-0.82}$   & 48.96 $\pm$ 10.13  & 56.22 $\pm$ 11.44   &  38.6 $\pm$ 7.7      & 39.6 $\pm$ 8.7     &63.7 $\pm$ 9.1         &    ---							 & 7.6 $\pm$0.5 \\ 
UR73Sa                 &         ---  			     & 3.60$^{+1.35}_{-0.98}$  &  48.96 $\pm$ 10.13  &  56.22 $\pm$ 11.44  &  42.3 $\pm$ 5.2       &   37.8 $\pm$ 6.8        &  62.4 $\pm$ 7.5         &   ---                                                          &  2.95 $\pm$ 0.29 \\
UR73Sb                 &         ---  			        & 3.12$^{+1.54}_{-1.11}$ &   $<$ 9.88                 & $<$ 10.02                & 4.7 $\pm$ 3.1           & 19.3 $\pm$ 6.7         & 11.7 $\pm$ 8.2         &    ---                                                         &  4.65 $\pm$ 0.45\\
HELMS\_RED\_10   & 4.62$^{+0.75}_{-0.63}$ & 3.17$^{+1.36}_{-0.63}$ &  43.91 $\pm$ 9.55   & 48.19 $\pm$ 10.54    & 33.6 $\pm$ 5.7        & 53.9 $\pm$ 6.5         & 86.5 $\pm$ 6.9          & 37.9 $\pm$ 4.4                                      & 42.7 $\pm$ 0.9 \\
HELMS\_RED\_10a &           ---                           & 2.37$^{+0.83}_{-0.69}$ &     $<$ 18.64              &   $<$ 21.62               &  16.8 $\pm$ 9.0        &  17.3 $\pm$ 12.4     &   58.0 $\pm$ 17.7   &  3.1 $\pm$ 0.2                                     & 2.00 $\pm$ 0.35 \\
HELMS\_RED\_10b &       	---	              & 4.62$^{+1.31}_{-1.14}$ &  43.91 $\pm$ 9.55      &  48.19 $\pm$ 10.54   &  14.5 $\pm$ 8.7      &  36.1 $\pm$ 13.3       & 22.5 $\pm$ 16.2     &  34.8 $\pm$ 2.6                                   & 22.5 $\pm$ 1.1 \\
HELMS\_RED\_23   &4.20$^{+0.63}_{-0.61}$  & 3.15$^{+1.19}_{-0.81}$  & 30.65 $\pm$ 8.94      &   35.90 $\pm$ 9.18    &  48.2 $\pm$ 6.7          & 87.6 $\pm$ 6.3         & 97.2 $\pm$ 7.4        & 42.1 $\pm$ 4.9                                   & 47.7 $\pm$ 0.9 \\
HELMS\_RED\_23a&            ---                          & 4.67$^{+1.39}_{-1.16}$   &   $<$ 9.32                 &  $<$ 15.90               & 9.3 $\pm$6.6             & 6.8 $\pm$ 5.1           &  20.5 $\pm$ 15.0      & 11.5 $\pm$ 1.3                                   &  9.9 $\pm$ 0.2\\
HELMS\_RED\_23b&              ---                        & 4.92$^{+1.40}_{-1.13}$    & 9.13 $\pm$ 5.35      &17.54 $\pm$ 6.99       &  12.8 $\pm$ 7.7         & 5.8 $\pm$ 4.1            &  13.2 $\pm$ 8.9        &  17.1 $\pm$ 2.0                                  &14.7 $\pm$ 0.8 \\
HELMS\_RED\_23c&               ---                       & 2.59$^{+0.88}_{-0.73}$   &  21.52 $\pm$ 7.16     & 35.9 $\pm$ 9.18      &  25.9 $\pm$ 11.3         & 61.6 $\pm$ 11.0       &  47.4 $\pm$ 22.3      &  13.4 $\pm$ 1.6                                    & 11.5 $\pm$ 0.7\\
HELMS\_RED\_68 & 3.60$^{+0.63}_{-0.64}$  & 3.11$^{+1.08}_{-0.79}$& 10.33 $\pm$ 5.02    & 15.13 $\pm$ 6.02    &  55.4 $\pm$ 5.6           & 73.9 $\pm$ 6.1        & 76.1 $\pm$ 6.5         & 32.7 $\pm$ 3.8                                         & 24.1 $\pm$ 3.1 \\
\hline
\end{tabular}
\tablecomments{$z_{\rm FIR}$ is the photometric redshift derived from FIR SED fitting by \cite{Ivison2016} and \cite{Duivenvoorden2018}. We list the spectroscopic redshift (3 decimals) instead of $z_{\rm FIR}$ whenever available. $z_{\rm MAGPHYS}$ is the photometric redshift derived from {\sc magphys}+photo-$z$ SED fitting in Section \ref{sec:sed}. For non-detections in IRAC, we quote the 3$\sigma$ upper limits. The ALMA flux densities are from \cite{Oteo2017} and this work. The de-blended SPIRE and SCUBA-2/LABOCA flux densities for the individual components are described in Section \ref{sec:catalog}.}  
\label{tab:unlensed}
\end{table*}

\section{SED fitting}
\label{sec:sed}

\begin{figure*}
\centering
\includegraphics[width=17cm]{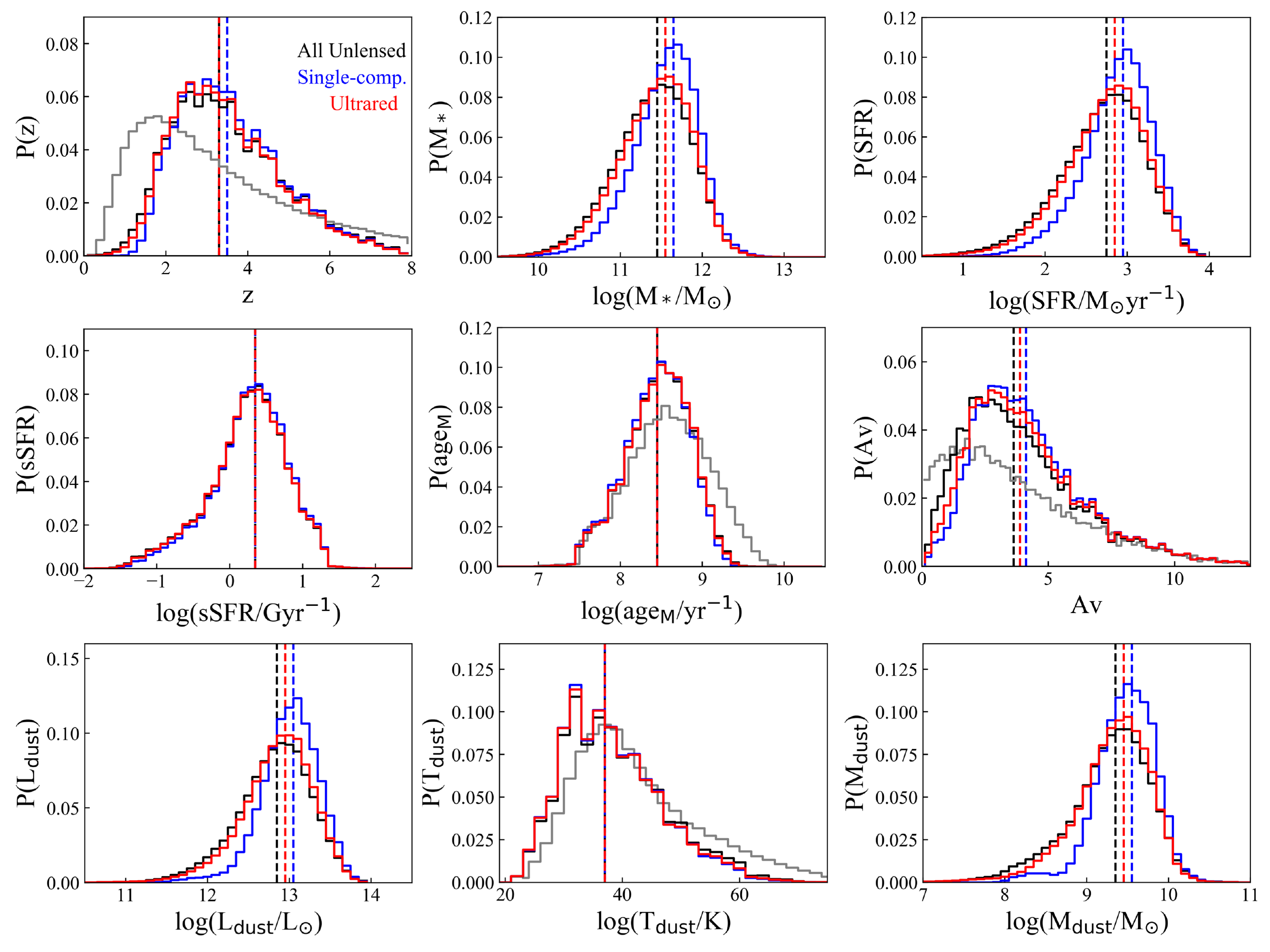}  
\caption{Normalized stacked posterior probability distributions of key physical parameters of the whole unlensed sample (63 DSFGs; black), the unlensed ultrared sub-sample (48 DSFGs; red), and the single-component ultrared sub-sample (31 DSFGs; blue). The median values of the distributions are indicated by the vertical lines with corresponding colors (3 lines are overlapping if only one line is seen). We also plot the prior distributions of $z$, age$_{\rm M}$, $A_{\rm V}$, and $T_{\rm dust}$ in gray for comparison.}
\label{pdf}
\end{figure*}

In this work, we aim to model the multi-wavelength observed SEDs of the cross-identified ultrared sources and derive their physical properties in order to place this population in the context of galaxy formation and evolution. Here we only show the results of the 41 unlensed {\it Herschel} ultrared sources (i.e., 63 individual DSFGs at high resolution), which have been cross-identified at multi-wavelengths. To facilitate comparison (i.e., avoid systematic uncertainties in SED fitting due to different choices of SED codes, SED models and assumptions) with the more abundant DSFGs at $z \sim$ 2, particularly the well-studied ALMA-LESS (ALESS; ALMA follow-up of the LABOCA submillimeter survey in the Extended {\it Chandra} Deep Field South) SMGs at $z$(median) $\sim$ 2.5 \citep{Hodge2013}, we use the same SED code, i.e., the updated version of {\sc magphys} as in \cite{daCunha2015} and \cite{Danielson2017}. {\sc magphys} relies on a self-consistent energy balance argument to combine stellar emission with dust attenuation and dust emission in galaxies \citep{daCunha2008}. The updated version extends the model parameter space to account for properties that are more likely applicable to heavily obscured galaxies at high redshifts. The major SED model components include the stellar population synthesis models of \cite{Bruzual2003}, the delayed-$\tau$ star formation histories, the two-component dust attenuation model \citep{Charlot2000}, dust emission, and radio emission based on the radio-to-FIR correlation. 

We use the spectroscopic redshift as the input redshift whenever available. {\sc magphys} has also been tested and used as a photometric redshift code by leaving the redshift as a free parameter and being simultaneously constrained with all the other physical parameters \citep{daCunha2015,Battisti2019}. The key feature of the photo-$z$ version (referred to as ``{\sc magphys+}photo-$z$'' hereafter; \citealt{Battisti2019}) is the self-consistent incorporation of photometric uncertainty into the uncertainty of all the derived properties. By accounting for this effect, {\sc magphys+}photo-$z$ therefore provides more realistic uncertainties for physical properties. 

We run {\sc magphys+}photo-$z$ with all the multi-wavelength data and compare with the photo-$z$'s derived from the FIR SEDs \citep{Ivison2016,Duivenvoorden2018}. The resultant best-fit parameters and associated uncertainties are determined from the posterior probability distributions. We take the median value as the best-fit parameter and the 16th or 84th percentile range as the $1\sigma$ uncertainty. In order to analyze the overall properties of the unlensed sample and take into account the associated uncertainties, we stack individual probability distributions of the key physical parameters in Figure \ref{pdf}, including the photometric redshift, stellar mass, SFR, specific SFR, mass-weighted stellar population age, V-band dust extinction, total dust luminosity, luminosity-weighted dust temperature, and dust mass. The average (median) properties and associated 16th-84th percentile ranges are listed in Table \ref{tab:sedresults} for each parameter. We display and compare the stacked probability distributions for the following three sub-samples:

$\bullet$ {\it All Unlensed} DSFGs: this is the whole unlensed sample containing 63 individual DSFGs (Table \ref{tab:unlensed}). 

$\bullet$ {\it Unlensed Ultrared} DSFGs: this sub-sample includes all the unlensed DSFGs that satisfy the ultrared selection criteria\footnote{Here we use the selection criteria in \cite{Ivison2016} as mentioned in Section \ref{sec:herschel}.} (48 DSFGs) including multi-component sources. At high resolution, some sub-components no longer meet the ultrared selection criteria so we remove those from this sub-sample. 

$\bullet$ {\it Unlensed Single-component Ultrared} DSFGs: this sub-sample (31 DSFGs) contains the single-component unlensed sources that have FIR photo-$z$'s or the components with spectroscopic redshifts (Table \ref{tab:unlensed}). This is the subset that we can run {\sc magphys} using the fixed-$z$ version (i.e., FIR photo-$z$ or spectroscopic redshift as input) and compare with the results from {\sc magphys}+photo-$z$. As shown later, this sub-sample contains the intrinsically most FIR-luminous and massive DSFGs in our sample, which are likely the best candidates for real $z > 4$ DSFGs. 

We also compare the physical properties of the unlensed ultrared sample with those of the ALESS DSFGs below.

\begin{table*}
\centering
\caption{Average properties of the {\it Herschel} unlensed sample derived from {\sc magphys}+photo-$z$ SED fitting}
\begin{tabular}{lcccc}
\hline
\hline
 Parameter                                    & All Unlensed &     Unlensed Ultrared                 &  Single-comp. Ultrared &ALESS Sample          \\
                                                       &  (63 DSFGs)    &   (48 DSFGs)                                &(31 DSFGs)                          &  (99 DSFGs) \\[0.05in]
\hline
 $z_{\rm phot}$                              & 3.3$^{+1.8}_{-1.0}$  &  3.3$^{+1.6}_{-1.0}$         & 3.5$^{+1.6}_{-1.0}$          & 2.7$^{+1.4}_{-1.1}$ \\[0.04in]
 log($M_{*}$/$M_{\odot}$)             &11.45$^{+0.4}_{-0.5}$ & 11.55$^{+0.4}_{-0.5}$     &11.65$^{+0.3}_{-0.4}$& 10.95$^{+0.6}_{-0.8}$\\[0.04in]
 log(SFR/$M_{\odot}$yr$^{-1}$)    & 2.75$^{+0.5}_{-0.5}$  & 2.85$^{+0.4}_{-0.6}$      &2.95$^{+0.4}_{-0.4}$  & 2.45$^{+0.4}_{-0.5}$ \\[0.04in]
 log(sSFR/Gyr$^{-1}$)                   & 0.35$^{+0.5}_{-0.5}$  & 0.35$^{+0.5}_{-0.5}$      &0.35$^{+0.5}_{-0.4}$  &0.45$^{+0.6}_{-0.6}$\\[0.04in]
 log(age$_M$/yr)  			   & 8.45$^{+0.5}_{-0.3}$   &8.45$^{+0.5}_{-0.3}$       & 8.45$^{+0.5}_{-0.4}$  &8.35$^{+0.5}_{-0.6}$\\[0.04in]
 $A_{V}$           				   &3.6$^{+2.8}_{-1.8}$    & 3.9$^{+2.8}_{-1.8}$            & 4.1$^{+2.5}_{-1.8}$  &1.9$^{+1.2}_{-1.0}$\\[0.04in]
 log($M_{*}/L_{H}$/$M_{\odot}/L_{\odot}$)   &0.28$^{+0.6}_{-0.5}$ & 0.33$^{+0.6}_{-0.5}$ &0.33$^{+0.6}_{-0.4}$ & -0.13$^{+0.4}_{-0.4}$ \\[0.04in]
 log($L_{\rm dust}$/$L_{\odot}$)  & 12.85$^{+0.4}_{-0.4}$  &12.95$^{+0.4}_{-0.5}$   &13.05$^{+0.3}_{-0.3}$  &  12.55$^{+0.3}_{-0.5}$\\[0.04in]
 $T_{\rm dust}$/K                          &37$^{+12}_{-6}$   & 37$^{+10}_{-6}$                         & 37$^{+10}_{-6}$   &43$^{+10}_{-10}$\\[0.04in]
 log($M_{\rm dust}$/$M_{\odot}$)  &9.35$^{+0.4}_{-0.5}$  & 9.45$^{+0.3}_{-0.5}$       &9.55$^{+0.3}_{-0.4}$  & 8.75$^{+0.3}_{-0.4}$  \\[0.04in]
\hline
\end{tabular}
\tablecomments{A machine-readable table of the physical parameters of individual sources is available online. $z_{\rm phot}$: photometric redshift; $M_{*}$: stellar mass; SFR: current SFR defined as the average of the star formation history over the last 10 Myr; sSFR: specific SFR; age$_M$: mass-weighted stellar population age; $A_{V}$: V-band extinction; $M_{*}/L_{H}$: stellar mass to H-band luminosity ratio; $L_{\rm dust}$: total dust luminosity = IR luminosity; $T_{\rm dust}$: luminosity-weighted dust temperature; $M_{\rm dust}$: dust mass. The median values and 16th-84th percentile ranges are determined from the stacked posterior probability distributions shown in Figure \ref{pdf}.}
\label{tab:sedresults}
\end{table*}

\subsection{Photometric redshifts}

{\sc magphys} has been tested as a photometric redshift code on ALESS SMGs (as a preliminary version of {\sc magphys}+photo-$z$ in \citealt{daCunha2015}) by comparing to the spectroscopic redshifts from \cite{Danielson2017}. \cite{daCunha2015} find a good agreement between the {\sc magphys}-based photometric redshifts and the ALESS spectroscopic redshifts, with a small median relative difference of $\Delta$$z$/(1+$z_{\rm spec}$) = -0.005. \cite{Battisti2019} further demonstrated the success of {\sc magphys}+photo-$z$ in estimating redshifts and physical properties based on over 4000 IR-selected galaxies at $0.4 < z < 6.0$ in the COSMOS field with robust spectroscopic redshifts. They achieved high photo-$z$ accuracy (median offset $\Delta z$/(1+$z_{\rm spec}$) $\lesssim$ 0.02), and low catastrophic failure rates ($\eta$ $\lesssim$ 4\%; a catastrophic failure is defined as a source with $\Delta z$/(1+$z_{\rm spec}$) $>$ 0.15) over all redshifts. The median value and uncertainties on the photometric redshift are determined in a self-consistent manner as all the other physical parameters, since the likelihood distributions of the redshift and physical parameters are computed simultaneously. \cite{Battisti2019} also demonstrated that the choice of priors, especially the non-uniform prior for the model redshift distribution (Figure \ref{pdf}) does not introduce significant redshift bias in the results. 

\begin{figure}
\centering
\includegraphics[width=7.5cm]{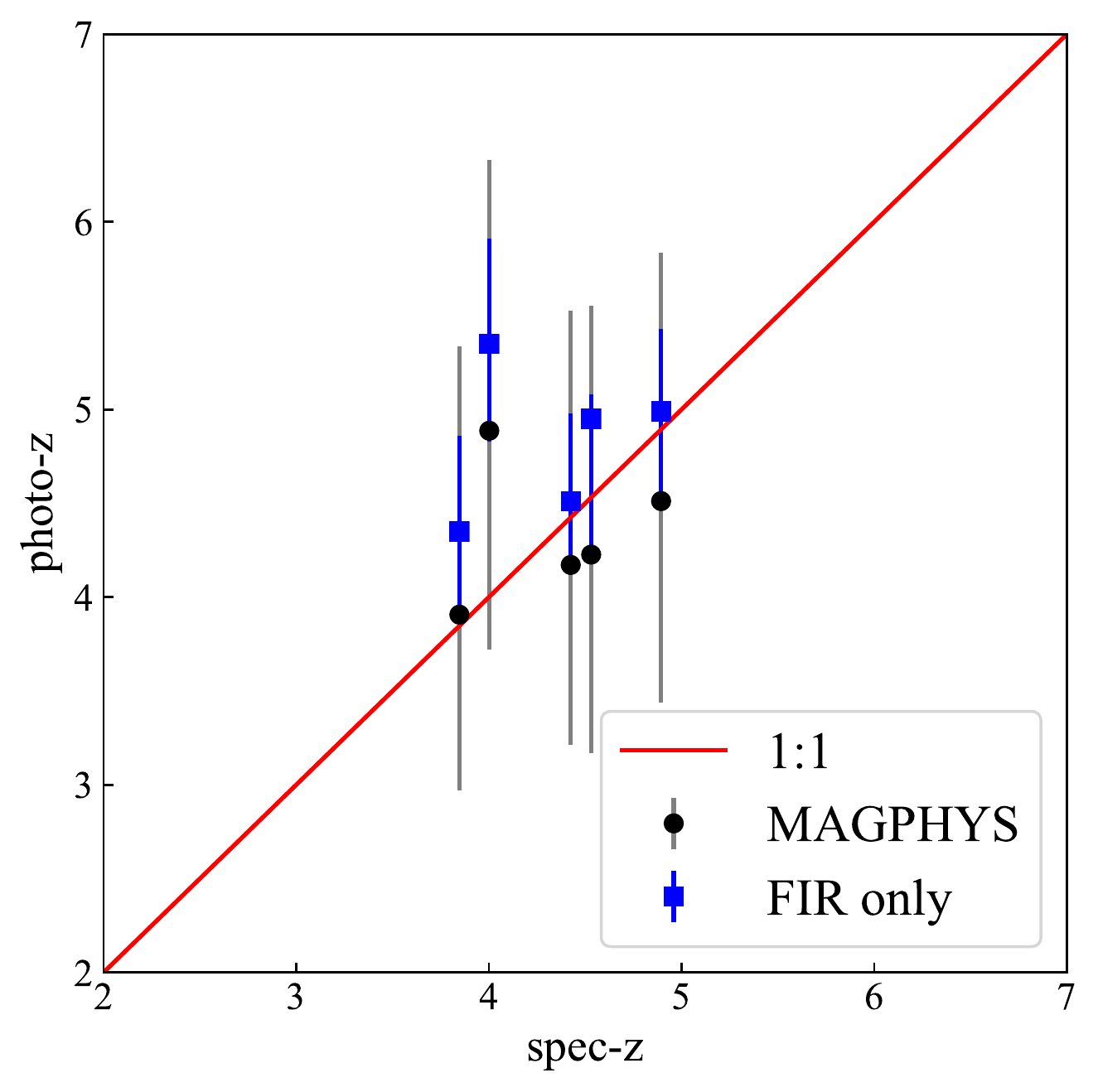} 
\includegraphics[width=7.5cm]{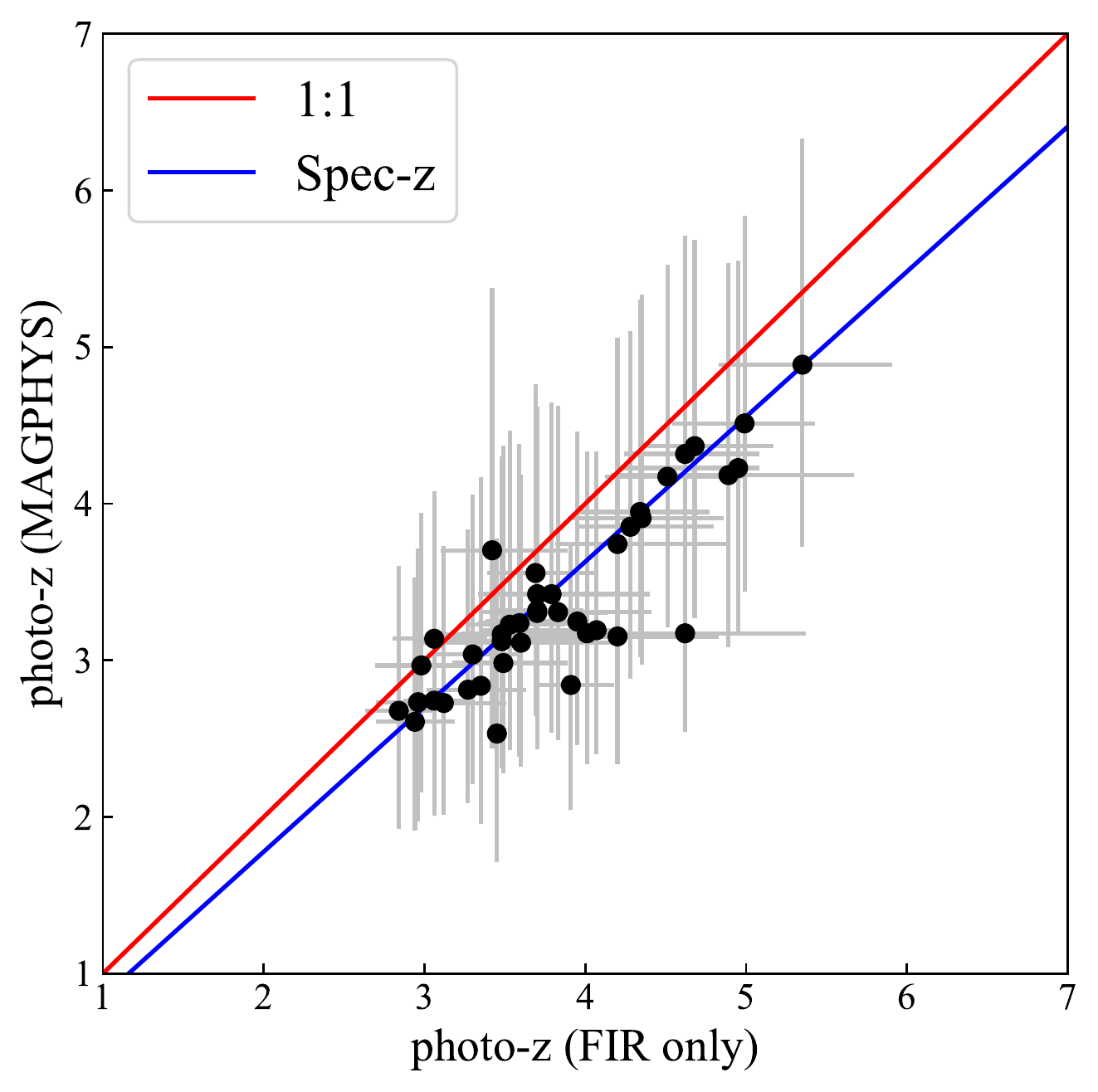} 
\caption{{\it Top}: We have 5 sources for directly comparing photometric redshifts ({\sc magphys} and FIR) with available spectroscopic redshifts. The red solid line represents the 1:1 ratio. {\it Bottom}: Comparison of the photometric redshifts derived from the FIR photometry only and from {\sc magphys}+photo-$z$ multi-wavelength SED fitting for the unlensed sample. The blue line shows the median relative offset ($\Delta$$z$/(1+$z$) = 0.08) of the FIR-derived photometric redshifts from the available spectroscopic redshifts based on reliability tests in \cite{Ivison2016} and \cite{Duivenvoorden2018}. The {\sc magphys}-derived redshifts are systematically lower than the FIR-derived redshifts but are on average more consistent with the expected spectroscopic redshifts.  }
\label{z_magphys}
\end{figure}

We compare the photometric redshifts from {\sc magphys}+photo-$z$ and FIR SED fitting with spectroscopic redshifts based on the 5 sources whose spectroscopic redshifts are available (Figure \ref{z_magphys} top panel). The mean relative offset $\Delta z$/(1+$z_{\rm spec}$) is 0.096 for the FIR method and -0.005 for {\sc magphys}+photo-$z$, while the median relative offset is 0.076 for the FIR method and -0.047 for {\sc magphys}+photo-$z$. The {\sc magphys}-derived redshifts on average (both mean and median) are more accurate than the FIR method. We further make the comparison between the {\sc magphys}-derived redshifts and the FIR-derived redshifts for the whole unlensed sample (Figure \ref{z_magphys} bottom panel). The {\sc magphys}-derived redshifts, with a median value of 3.3, are systematically lower than the FIR-derived redshifts with a median value of 3.7. \cite{Ivison2016} and \cite{Duivenvoorden2018} compared the FIR-derived redshifts with available spectroscopic redshifts of a larger sample and found a median relative offset of $\Delta$$z$/(1+$z_{\rm spec}$) = 0.08. We compare the {\sc magphys} redshifts with the expected true redshifts based on their tests \footnote{The median relative offset between $z_{\rm FIR}$ and $z_{\rm spec}$ is ($z_{\rm FIR}$-$z_{\rm spec}$)/(1+$z_{\rm spec}$) = 0.08. Therefore, $z_{\rm spec}$ = ($z_{\rm FIR}$-0.08)/1.08, which is plotted as the blue line in Figure \ref{z_magphys}.}. Given the additional data and constraining power in the NIR along with the FIR photometry, the {\sc magphys}-derived redshifts are on average more consistent with the expected true redshifts based on this comparison (Figure \ref{z_magphys} bottom panel), although the error bars are larger. The large error bars from {\sc magphys}+photo-$z$ are partly due to the fact that we do not have enough data to constrain the full SED but reflect more realistic uncertainties than the errors for the FIR-only photo-$z$'s, which are based on the templates used. Obtaining more rest-frame UV/optical data would further improve the accuracy and precision \citep{Battisti2019}. 

\subsection{Stellar mass versus SFR}
\label{sec:mstar_sfr}

\begin{figure}
\centering
\includegraphics[width=8.8cm]{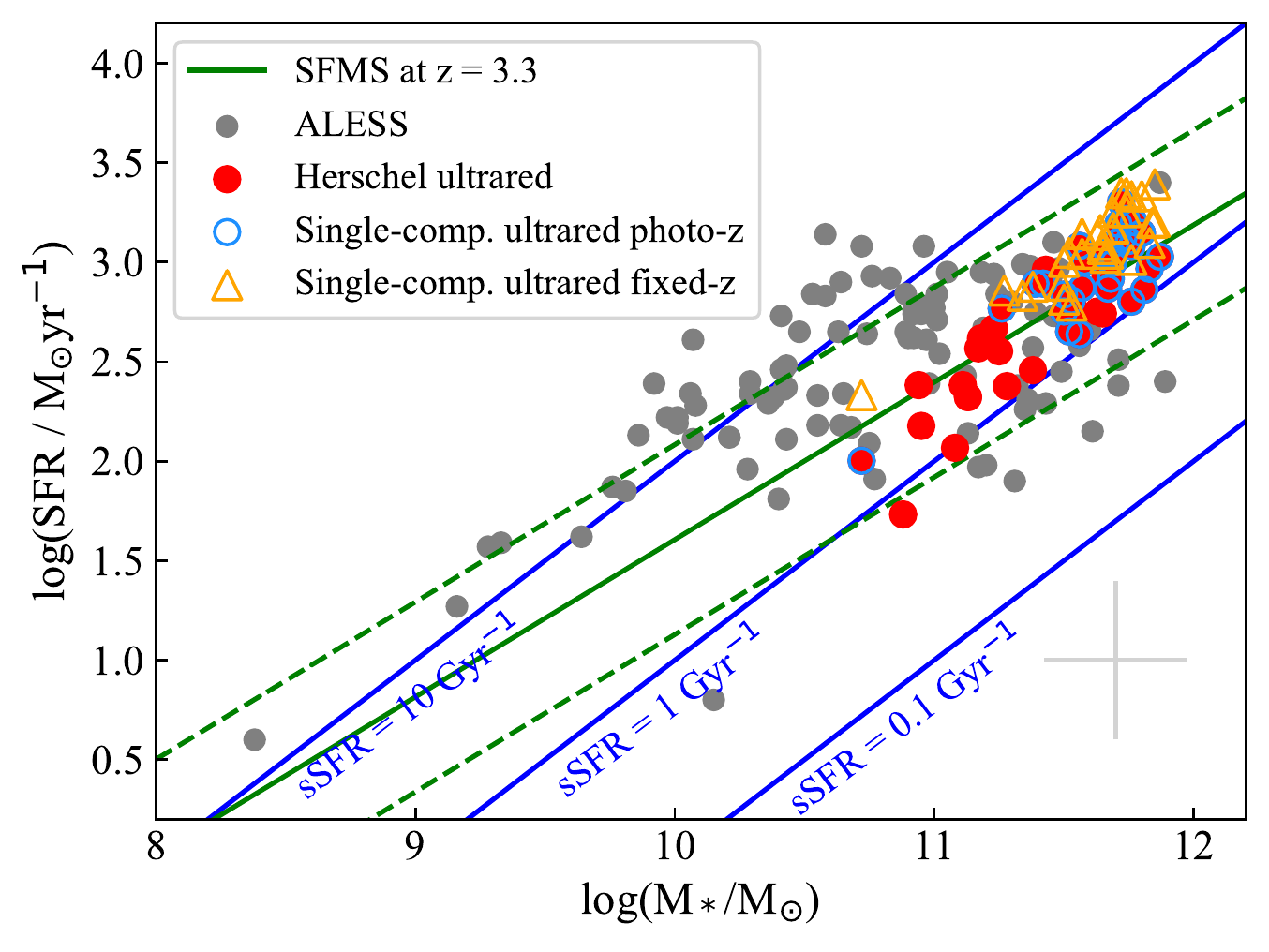}  
\caption{SFR versus stellar mass. The red circles are our unlensed {\it Herschel} ultrared DSFGs while the gray circles denote the ALESS DSFGs at $z \sim2.5$ from \cite{daCunha2015}. The typical error bar derived from {\sc magphys} SED fitting is shown at the lower right corner. The green solid line represents the star-forming main sequence from \cite{Speagle2014} at the median redshift $z$ = 3.3 of our sample, and the dashed lines are three times above or below this relation. The blue open circles mark the single-component ultrared DSFGs with parameters derived from {\sc magphys}+photo-$z$, while the orange triangles are derived from fixing the redshift to the FIR photo-$z$ or known spectroscopic redshift as input. }
\label{sfrmstar}
\end{figure}

Galaxy surveys at low and high redshifts have shown that star-forming galaxies form a power-law relation between their SFR and stellar mass, known as the Òmain-sequence (MS)Ó of star-forming galaxies (e.g., \citealt{Daddi2007,Magdis2010,Speagle2014}). Figure \ref{sfrmstar} shows the stellar masses versus SFRs of the unlensed ultrared DSFGs compared to those of ALESS DSFGs. The green solid line shows the star-forming MS at the median redshift of our sample, $z =3.3$, from \cite{Speagle2014}, and the dashed lines show three times above or below the MS. The wide spread of stellar masses and SFRs of the ALESS DSFGs compared to the star-forming MS, suggests that these DSFGs are not a homogeneous population with some lying significantly ($>$ 3 times) above the MS thus defined as starbursts and some being consistent with the MS ($<$ 3 times) but just at the high-mass end of this relation \citep{daCunha2015}. This bimodal distribution is also in line with theoretical predictions, e.g., simulations by \cite{Hayward2012}. However, the fraction of starbursts at high redshift depends on how we define the normal star-forming MS at these redshifts. Since the sSFR of the MS predicted by \cite{Speagle2014} continues increasing with redshift, this fraction therefore would be lower than if the MS flattens at $z \sim2$ as suggested by some other studies (e.g., \citealt{Weinmann2011,Gonzalez2014}). The star-forming MS is also dependent on the stellar mass, and the slope of the MS is found to be steeper at low masses (log($M_*/M_{\odot}$) $<$ 10.5) and flattens at the high-mass end out to $z$ of 2.5 (e.g. \citealt{Whitaker2014,Leja2015}). A shallower slope at high masses than the one in \cite{Speagle2014} could mean a higher starburst fraction.

The {\it Herschel} ultrared DSFGs on average have higher stellar masses and SFRs than the ALESS DSFGs, with a median stellar mass of (3.7 $\pm$ 0.2) $\times$ 10$^{11}$ $M_{\odot}$ and a median SFR of 730 $\pm$ 30 $M_{\odot}$yr$^{-1}$ (modulo assumptions about the IMF, e.g., \citealt{Romano2017,Zhang2018}). The 16th-84th percentile ranges of the stacked stellar mass and SFR probability distributions are (1.1-8.9) $\times$ 10$^{11}$ $M_{\odot}$ and 180-1800 $M_{\odot}$yr$^{-1}$. Almost all of the ultrared DSFGs have specific SFRs higher than 1 Gyr$^{-1}$. The open blue circles mark the single-component ultrared sources, which are mostly hyper-luminous IR galaxies (HyLIRGs; $L_{\rm IR}$ $\geq$ 10$^{13}$ $L_{\odot}$) and are at the high-mass end of our sample. Since leaving redshift as a free parameter introduces extra degree of freedom and degeneracy in SED fitting than fixing the redshift, here we also compare with the {\sc magphys} results by fixing the input redshift to the FIR-derived photo-$z$ or spectroscopic redshift for the single-component ultrared sample (open triangles). The stellar masses and SFRs from the photo-$z$ and fixed-$z$ versions are generally consistent with each other although the SFRs from the fixed-$z$ version are higher, which is mostly driven by the higher FIR-derived photo-$z$ than the {\sc magphys}-derived photo-$z$ as shown in Figure \ref{z_magphys}. Nevertheless, they are all consistent with and at the high-mass end of the star-forming MS by \cite{Speagle2014}. If the MS slope flattening continues at $z >3$ at the most massive end where our ultrared DSFGs reside, a shallower MS slope and lower sSFR than that in \cite{Speagle2014} would infer a higher starburst fraction, but major mergers, which trigger short-phased enhanced SFRs thus lie significantly above the MS, are not a dominant driver of our ultrared DSFG population.

\subsection{Dust properties}
\begin{figure*}
\centering
\includegraphics[width=8.8cm]{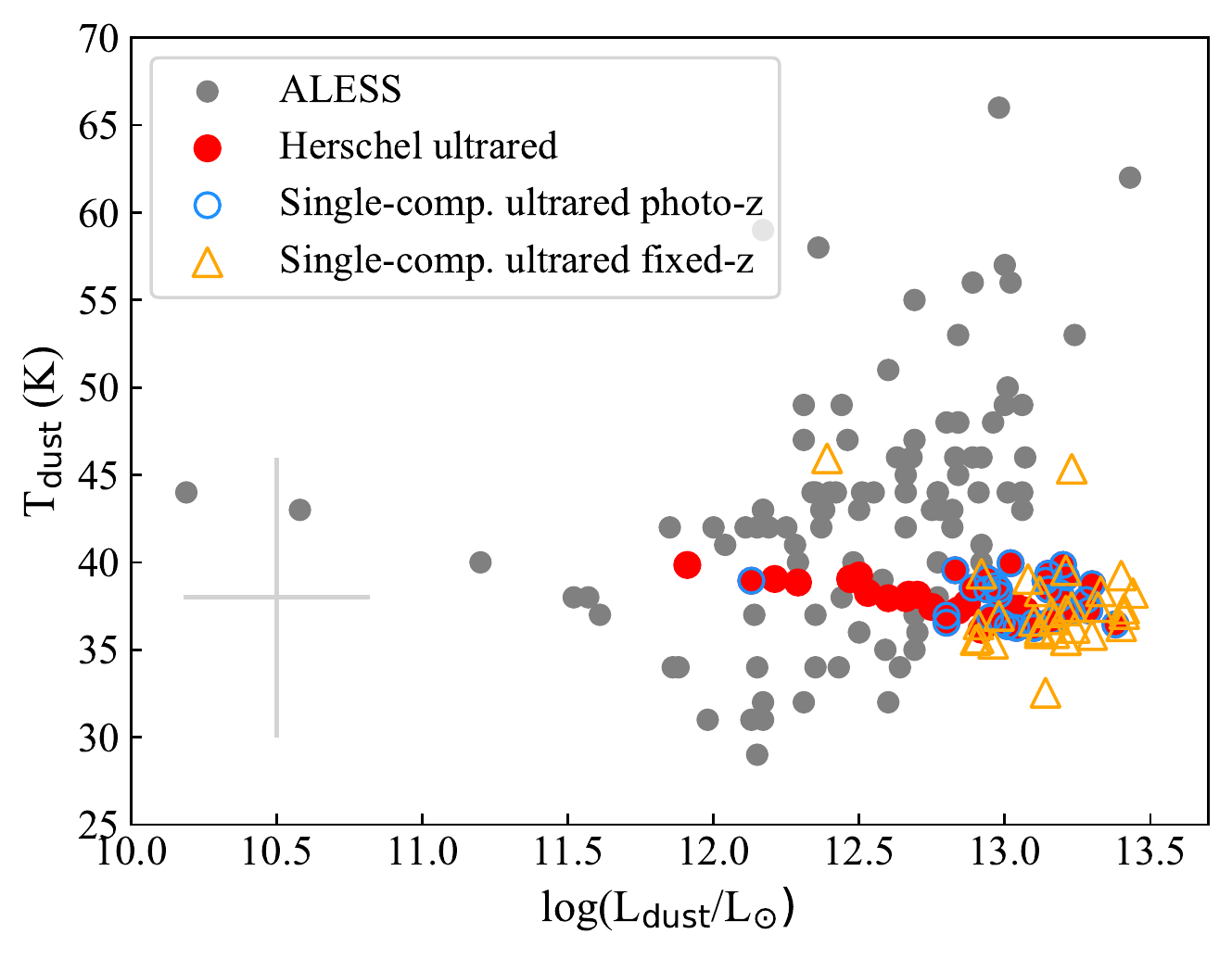} 
\includegraphics[width=8.7cm]{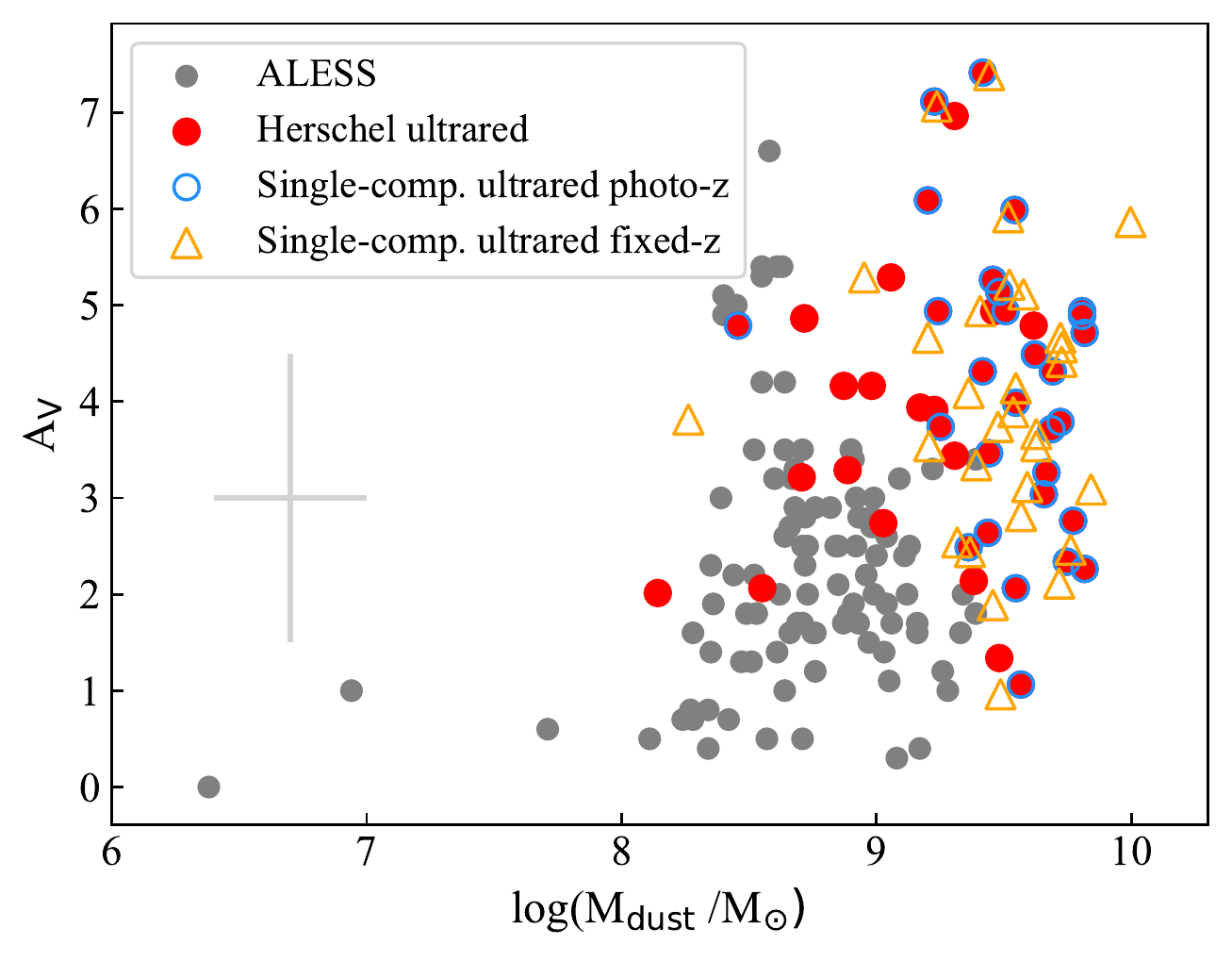}
\caption{{\it Left:} $L_{\rm dust}$ versus $T_{\rm dust}$. $T_{\rm dust}$ is the luminosity-weighted dust temperature from {\sc magphys}. The red circles are the unlensed {\it Herschel} ultrared DSFGs while the gray circles denote the ALESS DSFGs at $z \sim2.5$ from \cite{daCunha2015}. The typical error bar derived from {\sc magphys} SED fitting is shown at the lower left corner. The blue open circles mark the single-component ultrared DSFGs with parameters derived from {\sc magphys}+photo-$z$, while the orange triangles are derived from fixing the redshift to the FIR photo-$z$ or known spectroscopic redshift as input. {\it Right:} Av versus dust mass.}
\label{LdustTd}
\end{figure*}

Figure \ref{LdustTd} shows the comparisons of the total dust luminosity, dust temperature, V-band extinction, and dust mass between the unlensed {\it Herschel} ultrared DSFGs and ALESS DSFGs. The {\it Herschel} ultrared DSFGs have a median total dust luminosity of (9.0 $\pm$ 2.0) $\times$ 10$^{12}$ $L_{\odot}$, a dust mass of (2.8 $\pm$ 0.6) $\times$ 10$^9$ $M_{\odot}$, a luminosity-averaged dust temperature of 38 $\pm$ 2 K, and a V-band extinction of 4.0 $\pm$ 0.3. Again, we show the properties of the single-component ultrared sub-sample derived from both the photo-$z$ version and the fixed-$z$ version. As noted in \cite{daCunha2015} and other SMG studies (e.g., \citealt{Chapman2005,Wardlow2010,Magnelli2012,Swinbank2014}), there exists a correlation between the total dust luminosity and the average temperature of DSFGs. The dust temperatures of the ultrared sample distribute in a narrower range across the dust luminosity although the error bars are large. This narrow range is almost entirely driven by the prior distribution as shown in Figure \ref{pdf}, i.e., $T_{\rm dust}$ is basically not constrained by the data. Although we have FIR-submm/mm data to locate the dust emission peak, there still exists the degeneracy between $T_{\rm dust}$ and redshift. More scatter in $T_{\rm dust}$ can be seen for the single-component sub-sample if we fix the input redshift to the FIR photo-$z$ or spectroscopic redshift, i.e., break the $T_{\rm dust}$-redshift degeneracy. The single-component sub-sample (fixed-$z$ version) shifts to the higher-luminosity end with most of them being HyLIRGs, and the luminosity-temperature correlation is very weak. The V-band extinction values widely spread over the range from $\sim$ 1-7.5 for the whole ultrared sample and the single-component sub-sample and is on average higher than that of ALESS DSFGs by $\sim$2 magnitudes. The high $A_{\rm V}$ values ($A_{\rm V} > 4$) have also been found in other SMGs (e.g., \citealt{Hopwood2011,Ma2015}). The {\it Herschel} ultrared DSFGs, which have significantly higher IR luminosities, shift to the high dust-mass end. The sample selection methods factor into the differences we observe here for the two DSFG samples. The ALESS SMGs are selected based on a single flux density limit ($S_{\rm 870\mu m}$ $>$ 4.2 mJy; \citealt{Swinbank2014}), while the {\it Herschel} ultrared DSFGs are selected by their rising SPIRE flux densities and limited by the sensitivity that {\it Herschel} probes, i.e., the {\it Herschel} ultrared selection naturally chooses higher 870 $\mu$m-flux density sources than the ALESS sample. We also notice that the {\it Herschel} ultrared sample distributes similarly on the $A_{\rm v}$ versus log$L_{\rm dust}$ plane as the $A_{\rm v}$-log$M_{\rm dust}$ plot without a clear correlation. Any correlation may be diluted by the statistical errors on these parameters due to limited sampling of the SEDs. Only by obtaining spectroscopic redshifts and a better sampling of the SEDs can we ultimately break the degeneracy between these parameters (e.g., dust temperature-redshift degeneracy) and investigate potential intrinsic correlations between the physical properties.

\subsection{The average SED of the unlensed ultrared sample}

\begin{figure*}
\centering
\includegraphics[width=11cm]{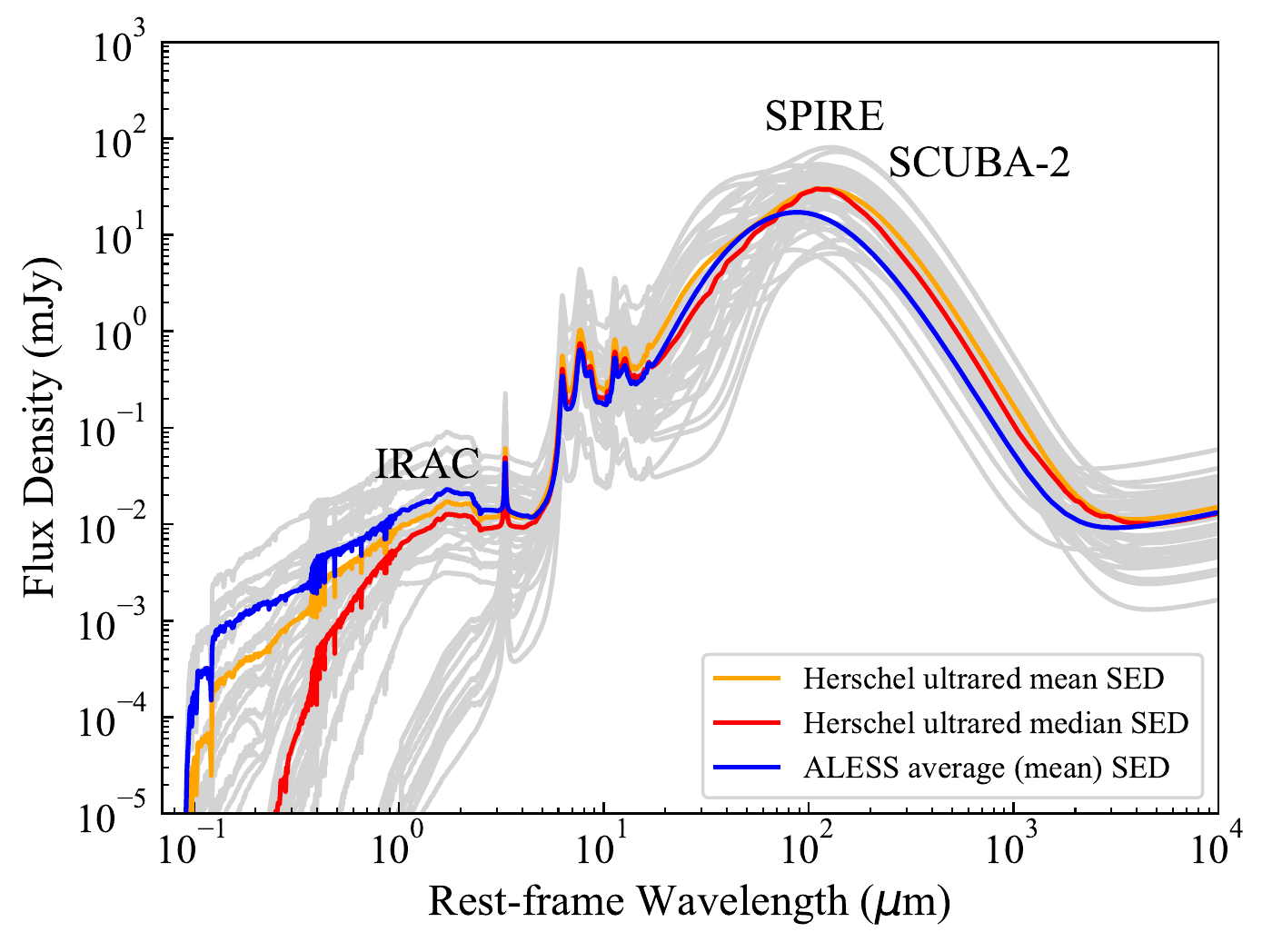} 
\caption{Best-fit SEDs in the rest-frame. The gray curves show the individual best-fit SEDs from {\sc magphys}+photo-$z$ of the 48 {\it Herschel} unlensed ultrared DSFGs and the red curve and orange curve are the median and mean SED of this sample. The average (mean) SED of the ALESS DSFGs at $z \sim2.5$ is overlaid in blue. Our {\it Herschel} ultrared DSFGs on average are more dust obscured and more luminous in the FIR than the $z \sim2.5$ ALESS DSFGs. We do not attempt to further compare the detailed shape of the SED due to the limited filter bands hence uncertainties in our best-fit SEDs, e.g., the rest-frame UV region is not constrained for our ultrared DSFGs.  }
\label{sed}
\end{figure*}

Figure \ref{sed} shows the best-fit SED shifted to the rest-frame wavelength for each unlensed ultrared DSFG (gray) to highlight the intrinsic SED variations of these sources. We generate the median SED (red) and mean SED (orange) of this sample by averaging flux densities of all the best-fit SEDs at each rest-frame wavelength in the same manner as for ALESS DSFGs. The average (mean) SED of the ALESS DSFGs from \cite{daCunha2015} is also overlaid for comparison. The comparison of the average SEDs shows that the {\it Herschel} ultrared DSFGs on average are more luminous at FIR-submm, more dust obscured, and peak at longer wavelengths than ALESS DSFGs at $z \sim 2.5$. We caution though that our current sample of the SED fitting analysis is smaller and our SEDs are not as well-sampled as ALESS DSFGs due to limited photometry (e.g., the rest-frame UV region is not constrained), therefore, we do not attempt to further compare the detailed shape of the SED.

\section{Discussion}
\label{sec:discussion}

\subsection{Redshift distribution}

\begin{figure*}
\centering
\includegraphics[width=8.1cm]{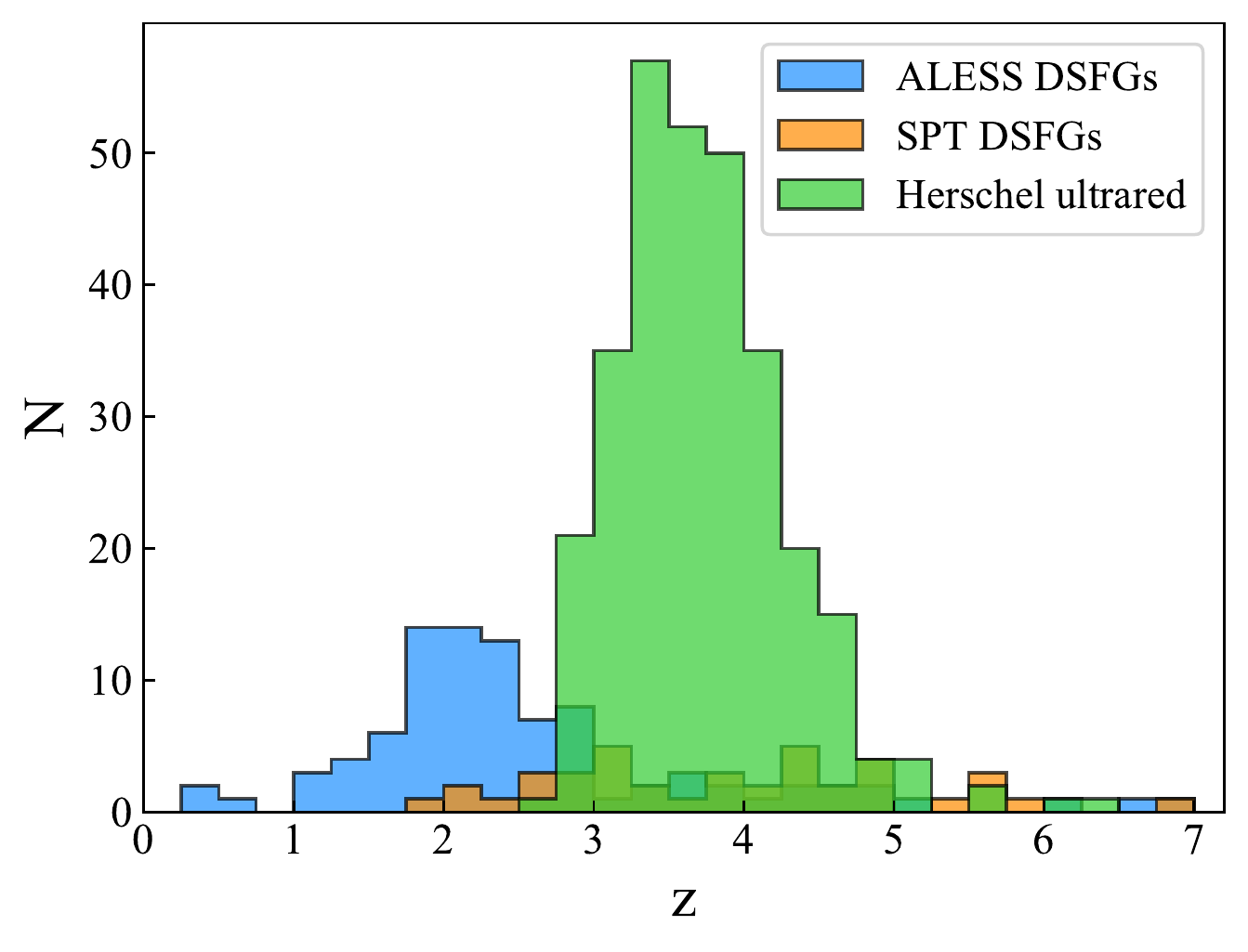} 
\includegraphics[width=8.3cm]{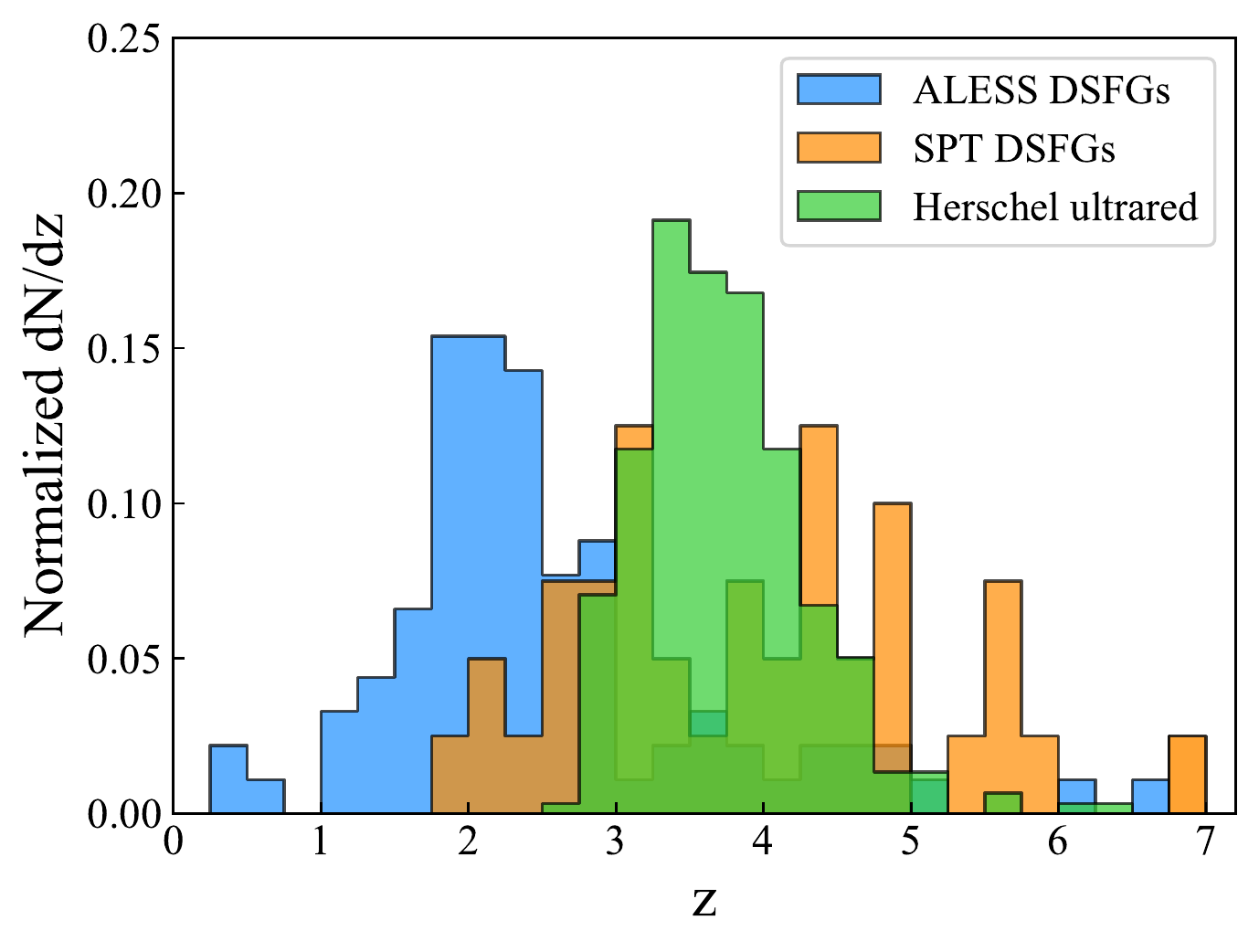}
\caption{{\it Left}: Redshift distributions of the {\it Herschel} ultrared DSFGs (green), ALESS DSFGs (blue; \citealt{Simpson2014}), and SPT DSFGs (orange; \citealt{Strandet2016, Strandet2017}). {\it Right}: Normalized dN/d$z$ (by sample size) for the three samples. }
\label{fig:dndz}
\end{figure*}

The raw photometric redshift distribution of the 300 {\it Herschel} ultrared DSFGs are shown in Figure \ref{fig:dndz} (Left), which has a median redshift of 3.7 $\pm$ 0.7 based on the FIR method as we do not have {\sc magphys}-derived redshifts for all of them. We compare it with the photometric redshift distribution of ALESS DSFGs from \cite{Simpson2014}. \cite{Danielson2017} present spectroscopic redshifts for 52 ALESS DSFGs and the distribution is consistent with the photometric redshift distribution for these sources. Here we use the photometric redshift distribution for comparison because it is more complete although less precise. The median redshift of ALESS DSFGs, 2.5 $\pm$ 0.2, is consistent with that of \cite{Chapman2005}, which is relied on radio-wavelength counterpart identification, although \cite{Simpson2014} shows a higher fraction of high-redshift sources than the earlier work.  We also overlay the spectroscopic redshift distribution of SPT DSFGs from \cite{Strandet2016,Strandet2017} (also \citealt{Weiss2013}), which are almost purely gravitationally lensed sources. The observed median redshift of the SPT sample, 3.9 $\pm$ 0.4, is higher than other samples due to two major selection effects: longer selection wavelengths and gravitational lensing \citep{Bethermin2015}. After correcting for the lensing effect, the median redshift of SPT DSFGs decreases to 3.1 $\pm$ 0.3. Figure \ref{fig:dndz} (Right) demonstrates the normalized dN/d$z$ that scales to the same total number of source in each sample. We do not attempt to correct the raw distribution for any selection effects due to the complicated selection functions for our sample. Spectroscopic observations are required to ascertain the redshift distribution of our ultrared sample especially the ones at $z > 4$.

\subsection{Multiplicity}
\label{sec:multiplicity}

Source blending or multiplicity has always been a concern for 500 $\mu$m-risers due to the relatively large SPIRE beam at 500 $\mu$m. Different multiplicity rates have been reported in the literature, depending upon various selection criteria and instruments used (e.g., \citealt{Cowley2015,Simpson2015,Scudder2016}). Now that we have obtained high-resolution data for a significant subset of ultrared sources, we are able to investigate multiplicity rates and fractional contribution from brightest components. Within our high-resolution sample, $\sim$ 27\% (17 out of 63) {\it Herschel} ultrared sources are resolved into multiple components of two or more. The multiplicity rate is about 39\% (16 out of 41) for the unlensed sample. The brightest components seen in ALMA contribute 41\%-80\% of the total ALMA flux. This fraction is lower, in the range of 15\%-59\%, if we use the SCUBA2/LABOCA flux at similar wavelengths as the total flux as the ALMA observations may not recover the total flux. Based on SMA follow-up observations at 1.1 mm and/or 870 $\mu$m of 36 500 $\mu$m-risers from various {\it Herschel} fields, Greenslade et al. in prep found a multiplicity rate of 33\% with a fractional contribution of 50\%-75\% due to the brightest components. The de-blending results from XID+ suggest that the brightest components contribute to 35\%-87\% of the total 500 $\mu$m flux. \cite{Donevski2018} investigated multiplicity using simulated SPIRE maps and compared the extracted flux of the brightest galaxy to the total flux, resulting in an average brightest-galaxy fraction of 64\%. The average observed brightest-galaxy fraction of our sample is consistent with this prediction.

Multiple components at the same redshift will not have a serious impact on the determination of redshift, and physical properties such as IR luminosity and SFR are determined for the combined system. Several {\it Herschel} ultrared sources have been spectroscopically confirmed as major mergers (e.g., SGP-196076 at $z$ = 4.425 \citep{Oteo2016}, ADFS-27 at $z$ = 5.655 \citep{Riechers2017}) or galaxies in the core of protoclusters (e.g., SGP-354388 at $z$ = 4.002 \citep{Oteo2018}). Detailed SED analysis especially their stellar properties, with additional high-resolution optical/NIR data, for the individual merging systems or protoclusters will be published in forthcoming papers. Multiple components at different redshifts, e.g., blending with foreground objects, will produce composite SEDs that may lead to an intermediate redshift estimate. Once decomposed, the individual components may not satisfy the 500 $\mu$m-riser selection criteria. \cite{Duivenvoorden2018} derived from mock observations that 60\% of the detected HeLMS galaxies pass the selection threshold due to flux boosting partially caused by blending with foreground objects. Our SPIRE de-blending results with XID+, which are based on high-resolution positional priors of mostly H-ATLAS galaxies, suggest that $\sim$20\% of our sources would not pass the selection criteria of 500 $\mu$m-risers if without blending.  

\subsection{Unlensed fraction}
\label{sec:unlensed}

Based on the high-resolution sub-sample, 65\% of the ultrared sources (41 out of 63) do not have clear signatures of lensing thus are classified as unlensed. This fraction is higher, 73\% (63 out of 86), if we count the individual components. We do not exclude the possibility that a small fraction of them might be moderately lensed (a lensing magnification factor of 1 $<$ $\mu$ $<$ 2). This would not significantly affect our derived properties. 

\cite{Donevski2018} compared the observed lensed fractions of 500 $\mu$m-risers in different fields, H-ATLAS \citep{Ivison2016}, HeLMS \citep{Asboth2016}, and HerMES \citep{Dowell2014}, with predicted fractions using the \cite{Bethermin2017} model and applying the same selection criteria in each study. The observed lensed fractions are consistent with the corresponding predicted lensed fractions. Since the HeLMS ultrared sources are selected with a higher flux cut at 500 $\mu$m, the lensed fraction is the highest (75\%), compared to the lowest lensed fraction (28\%) in H-ATLAS. Our sample contains sources from both H-ATLAS and HeLMS therefore the observed lensed fraction (35\%) is in between as expected. This number is more towards the lower end because most of the sources in our high-resolution sample are from H-ATLAS. 

We acknowledge that there could be lenses that we are missing with the current ancillary data, for example, those with small Einstein radii and faint, high-redshift lensing galaxies. High-resolution imaging (e.g., with {\it HST}/{\it JWST}/ALMA) and spectroscopic confirmation are needed to truly address the lensing fraction.  

\subsection{SFR surface density}

\begin{figure}
\centering
\includegraphics[width=8.8cm]{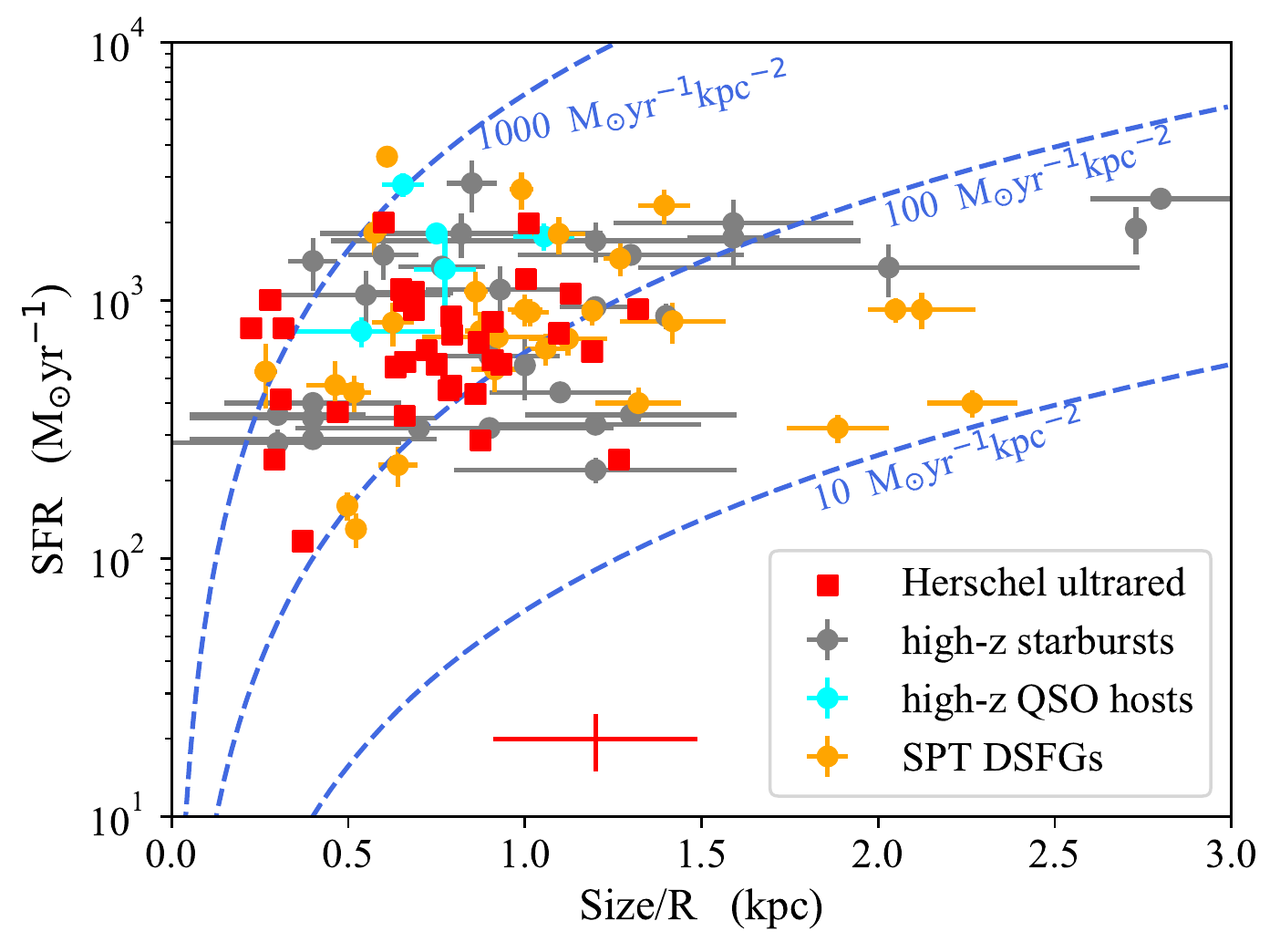} 
\caption{SFR versus dust continnuum size. The dashed lines show constant SFR surface density values. The red squares are our {\it Herschel} ultrared DSFGs with size measurements from \cite{Oteo2017} and this work. The typical error bar is shown as the red cross on the bottom of the plot. The orange circles are SPT DSFGs from \cite{Spilker2016} and \cite{Ma2016}. The gray circles are individual starburst galaxies at $z > 3$ in the literature with dust continuum size measurements. The magenta circles are quasar host galaxies at $z > 3$ in the literature. The literature galaxies are drawn from \cite{Younger2008}, \cite{Walter2009}, \cite{Magdis2011}, \cite{Walter2012}, \cite{Fu2012}, \cite{Bussmann2013}, \cite{Carniani2013}, \cite{Wang2013}, \cite{Cooray2014}, \cite{DeBreuck2014}, \cite{Riechers2014}, \cite{Yun2015}, \cite{Simpson2015}, \cite{Ikarashi2015}, and \cite{Riechers2017}.}
\label{sfrsize}
\end{figure}

Figure \ref{sfrsize} shows the SFR as a function of dust continuum size for our {\it Herschel} ultrared DSFGs as well as other high-redshift ($z > 3$) galaxies and quasar hosts in the literature. All the sources in the literature comparison sample have dust continuum size measurements. The SFR surface density is defined as $\Sigma$$_{\rm SFR}$ = SFR/2$R_{\rm a}R_{\rm b}$, where $R_{\rm a}$ = FWHM$_{\rm major}$/2 and $R_{\rm b}$ = FWHM$_{\rm minor}$/2 are the measured semi-major and semi-minor axes. The sizes and areas are measured by carrying out 2D elliptical Gaussian fitting on the high-resolution ALMA dust continuum images at 870 $\mu$m \citep{Oteo2017}. The dashed lines denote the constant SFR surface densities at 10, 100, and 1000 $M_{\odot}$ yr$^{-1}$ kpc$^{-2}$. Most of our sources are above the $\Sigma$$_{\rm SFR}$ = 100 $M_{\odot}$ yr$^{-1}$ kpc$^{-2}$ curve, and some are at or close to the Eddington limit for radiation-pressure supported starbursts \citep{Thompson2005}. The radiation pressure would drive dusty gaseous outflows, which have been observed in starbursting galaxies (e.g., \citealt{Martin2005,Diamond-Stanic2012,Spilker2018}). Future observations are required to confirm this. The very high $\Sigma$$_{\rm SFR}$ could be explained by either high star-formation efficiency, high gas fraction, or both. Gas-rich mergers are a viable mechanism for triggering the compact, enhanced star formation \citep{Mihos1996}. For example, SGP-196076, one of our unlensed ultrared sources, has been spectroscopically confirmed to be at least two interacting galaxies at $z = 4.425$ that will eventually merge \citep{Oteo2016}. However, major mergers do not seem to be the dominant driver of our ultrared DSFGs as discussed in Section \ref{sec:mstar_sfr}.

\subsection{Space density, SFR density, and stellar mass density}

The space density of the H-ATLAS ultrared DSFGs in the redshift range $4 < z < 6$ is estimated to be $\sim$ 6 $\times$ 10$^{-7}$ Mpc$^{-3}$ after completeness and duty-cycle corrections \citep{Ivison2016}, while this number is about an order of magnitude smaller (7 $\times$ 10$^{-8}$ Mpc$^{-3}$) for the HeLMS sample \citep{Duivenvoorden2018} due to the higher flux cut (i.e., HeLMS $S_{\rm 500}$ $>$ 63 mJy and H-ATLAS $S_{\rm 500}$ $>$ 30 mJy). 

DSFGs at $z > 4$ have been proposed to be the star-forming progenitors of the population of massive, quiescent galaxies at $z \sim 3$ uncovered from NIR surveys (e.g., \citealt{Simpson2014,Toft2014, Straatman2014, Nayyeri2014}). However, the quiescent galaxies represented by the mass-limited (log($M_{*}$/$M_{\odot}$) $>$ 10.6) sample at $3.4 < z < 4.2$ in the ZFOURGE survey, whose star formation is predicted to occur at $z \sim 5$, have a comoving space density of $\sim$ 2 $\times$ 10$^{-5}$ Mpc$^{-3}$ \citep{Straatman2014}. This is more than a factor of 30 higher than the H-ATLAS ultrared sample \citep{Ivison2016} and more than 2 orders of magnitude higher than the HeLMS sample \citep{Duivenvoorden2018}. Based on the space density comparison, \cite{Ivison2016} and \cite{Duivenvoorden2018} conclude that the {\it Herschel} ultrared sample, which is limited by the flux density levels probed by {\it Herschel}, cannot account for the formation of massive, quiescent galaxies at $z \sim 3$. Our {\it Herschel}-selected ultrared sample contains more FIR-luminous and thus rarer DSFGs than the star-forming progenitors of the massive, quiescent galaxies. 

The SFR density (SFRD) of DSFGs can be estimated by summing SFRs of all the sources divided by the comoving volume contained in a redshift range. Although it has become clear that UV/optical is insufficient to probe the total cosmic star formation at $z < 3$, there is no consensus on the significance of the contribution of dust-obscured star formation in DSFGs at $z > 3$ due to lack of complete surveys. The SFRD estimates at $4 < z < 5$ based on different samples vary by 3 orders of magnitude from $\sim$10$^{-4}$ to 10$^{-1}$ $M_{\odot}$yr$^{-1}$Mpc$^{-3}$ (e.g., \citealt{Dowell2014,Rowan-Robinson2016,Bourne2017,Donevski2018,Duivenvoorden2018}), with our HeLMS ultrared sample representing the lower limit of $\sim$10$^{-4}$ $M_{\odot}$yr$^{-1}$Mpc$^{-3}$.       

\begin{figure}
\centering
\includegraphics[width=8.7cm]{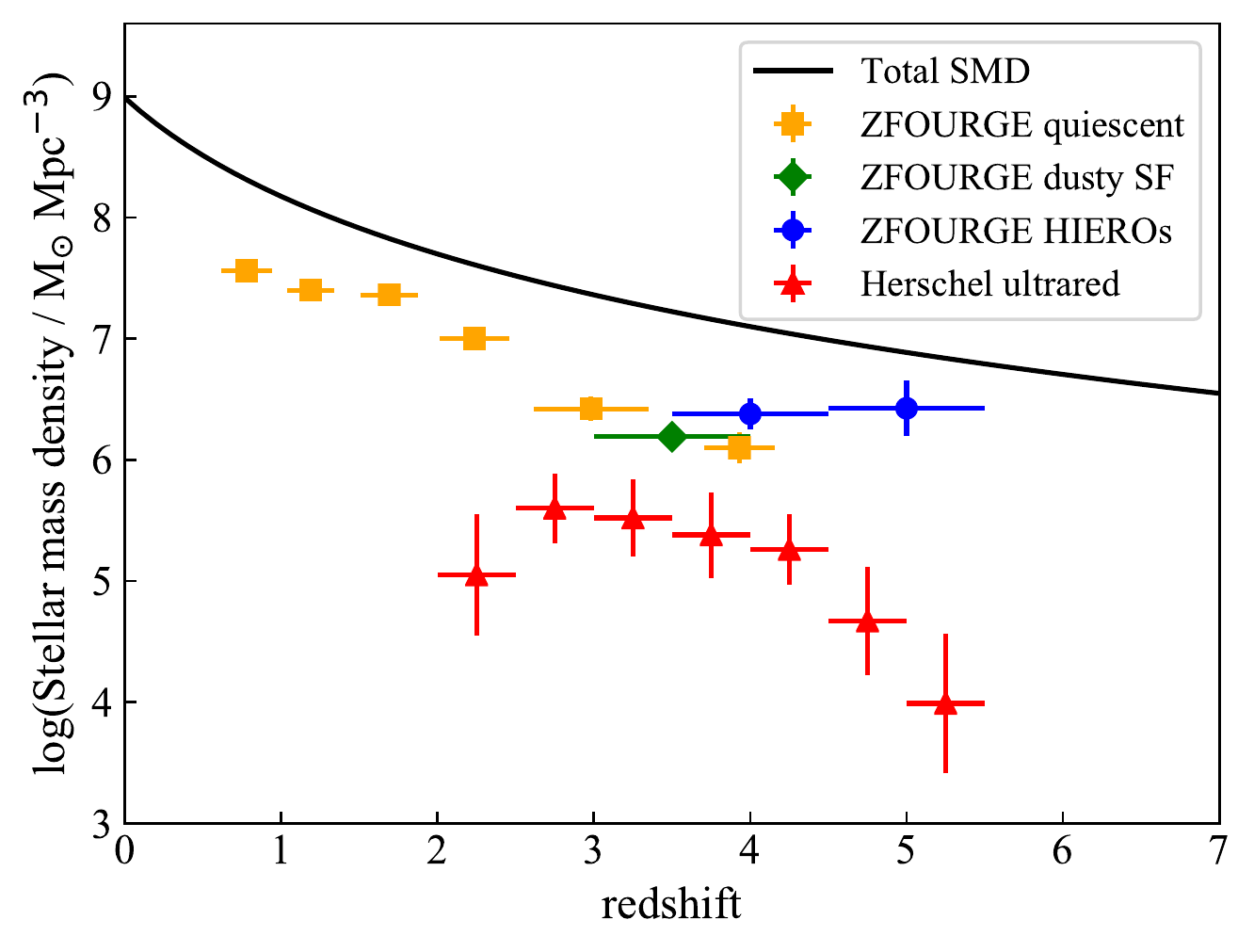}  
\caption{Stellar mass density (SMD) as a function of redshift. The black curve shows a simultaneous fit to the total SMD of the $K_{\rm s}$-selected galaxies at $z < 3.5$ from the UltraVISTA survey and the SMD of the UV-selected samples at $z > 3.5$ \citep{Stark2009,Labbe2010,Gonzalez2011,Lee2012,Muzzin2013}. The orange squares are the $K_{\rm s}$-selected, mass-limited (log($M_{*}$/$M_{\odot}$) $>$ 10.6) quiescent galaxies in the ZFOURGE survey \citep{Straatman2014}. The green diamond shows the $K_{\rm s}$-selected DSFGs at $3 < z < 4$ from the ZFOURGE sample \citep{Spitler2014}. The blue circles are $H$-[4.5] color selected massive DSFGs (so-called `HIEROs' as defined in \citealt{Wang2016}) using the ZFOURGE catalog and applying the same stellar mass cut, i.e., log($M_{*}$/$M_{\odot}$) $>$ 10.6.  }
\label{smd}
\end{figure}

The space density and SFRD of DSFGs at $z > 4$ have been widely explored as summarized above. However, hardly any estimates have been made yet on the stellar mass density contribution of DSFGs at $z > 4$ mainly due to the highly dust-obscured nature of these objects, and therefore the difficulty in detecting the rest-frame optical stellar emission to constrain their stellar masses. Our {\it Spitzer} follow-up sample provides the first attempt to constrain the stellar mass density of ultrared DSFGs at $z > 4$. Figure \ref{smd} shows the comparison of the stellar mass density (SMD) as a function of redshift for different populations. We construct the SMD of our H-ATLAS\footnote{We do not have enough HeLMS ultrared DSFGs to construct SMD yet.} unlensed ultrared DSFGs in 7 redshift bins and have corrected for the H-ATLAS ultrared sample completeness \citep{Ivison2016} and duty-cycle (assuming a starburst duty cycle of $\sim$ 100 Myr). The total SMD curve is a simultaneous fit to the total SMD of the $K_{\rm s}$-selected galaxies at $z < 3.5$ from the COSMOS/UltraVISTA survey and the SMD of the UV-selected samples at $z > 3.5$ (\citealt{Muzzin2013}; see also \citealt{Davidzon2017} and references therein). The UV samples at $z > 3.5$ contain galaxies that are based on drop-out selection methods, e.g., Lyman break galaxies (LBGs), B-dropout etc. \citep{Stark2009,Labbe2010,Gonzalez2011,Lee2012}. The total SMD increases with cosmic time as galaxies build up their stellar masses through star formation. The maximum contribution of our {\it Herschel} ultrared sources to the total SMD is 1.7\% at the $z = 3.25$ bin. The SMD of massive, quiescent galaxies also evolves with redshift, as represented by the $K_{\rm s}$-selected, mass-limited (log($M_{*}$/$M_{\odot}$) $>$ 10.6) ZFOURGE quiescent galaxies \citep{Straatman2014}. The DSFGs at $3 < z < 4$ in the ZFOURGE sample and the $H$-[4.5] color selected `HIEROs' at $z > 3.5$ as defined in \cite{Wang2016}, which are also massive DSFGs, have comparable SMDs as the quiescent galaxies at $3 < z < 4$. However, our {\it Herschel} ultrared DSFGs at $z \sim 5$ have $\sim$2 orders of magnitude lower SMDs than the quiescent sample at $z \sim 3$ and the dusty star-forming HIEROs at similar redshifts, which cannot be reconciled by even increasing the stellar mass limit to log($M_{*}$/$M_{\odot}$) $>$ 11.0. Again, this suggests that our ultrared sample cannot account for the star-forming progenitors of the massive, quiescent galaxies at $z \sim 3$ found in NIR surveys (although likely include the progenitors of the most extremely massive, quenched systems), while other selections, such as HIEROs \citep{Wang2016} and $S_{850\mu m}$- or $S_{870\mu m}$-selected DSFGs at $z > 4$ (e.g., \citealt{Oteo2016b,Michalowski2017}), likely include the majority of the progenitors of the massive, quiescent galaxies at $z \sim 3$.

Nevertheless, our {\it Herschel} ultrared sample contains the intrinsically most FIR-luminous (i.e., HyLIRGs) and massive galaxies in the early universe that are extremely interesting by themselves. These ultrared DSFGs are crucial to understanding the drivers of extreme star formation and assembly and evolution of massive galaxies with cosmic time. 

\section{Summary and Conclusions}
\label{sec:summary}

We have presented a large {\it Spitzer} follow-up program of 300 {\it Herschel}-selected ultrared DSFGs. For a significant subset, we have obtained high-resolution interferometry data such that we can pinpoint the locations and securely cross-identify {\it Spitzer} counterparts, and classify them as lensed or unlensed based on the morphology and the presence or absence of low-redshift foreground galaxies. For the rest of the sample, we have selected {\it Spitzer} counterpart candidates based on the SCUBA-2 positions. We have provided a catalog of all the cross-matched sources, including their positions, {\it Spitzer}/IRAC magnitudes and flux densities, as well as multi-wavelength photometry. In this paper, we have focused on analyzing the unlensed sample by performing {\sc magphys} SED modeling with the multi-wavelength photometry to derive their physical properties and compare with the more abundant $z \sim2$ DSFG population. We have also estimated the stellar mass density as a function of redshift and compared with massive, quiescent galaxies at lower redshifts. Our main results are summarized as follows. 

\begin{enumerate}
\item Within the 63 {\it Herschel} ultrared sources that have high-resolution data, $\sim$65\% (41 out of 63) appear to be unlensed, and about 27\% (17 out of 63) are resolved into multiple components. Some of the de-blended components are no longer 500 $\mu$m-risers. About 20\% of the original ultrared sources would not pass the selection criteria without blending with other sources at lower redshifts. 

\item We run {\sc magphys}+photo-$z$ on the unlensed sample to simultaneously constrain their redshifts and physical properties. The ultrared sample has a median redshift of 3.3, which is lower than the median value of the FIR-derived redshifts, and the 16th-84th percentile range is from 2.3 to 4.9. The {\sc magphys}-based photometric redshifts are more in line with the expected true redshifts based on the test by comparing with spectroscopic redshifts. 

\item We derive the median properties of the whole unlensed sample, the unlensed ultrared sub-sample, and the single-component ultrared sub-sample from stacked probability distributions. The unlensed ultrared sample has a median stellar mass of (3.7 $\pm$ 0.2) $\times$ 10$^{11}$ $M_{\odot}$, a SFR of 730 $\pm$ 30 $M_{\odot}$yr$^{-1}$, a total dust luminosity of (9.0 $\pm$ 2.0) $\times$ 10$^{12}$ $L_{\odot}$, a dust mass of (2.8 $\pm$ 0.6) $\times$ 10$^9$ $M_{\odot}$, and a V-band extinction of 4.0 $\pm$ 0.3. These properties are all higher than those of the ALESS DSFGs. 

\item We estimate the stellar mass densities of our ultrared DSFGs as a function of redshift. The stellar mass density at $z \sim 5$ is significantly lower than that of the massive, quiescent galaxies at lower redshifts from the ZFOURGE survey and HIREOs at similar redshifts. Combined with the comparison of space density and SFR density, we conclude that our ultrared sample cannot account for the majority of the star-forming progenitors of the massive, quiescent galaxies. Our sample is limited by the flux density levels probed by {\it Herschel} thus contains more FIR-luminous and rarer DSFGs than the progenitors of the massive, quiescent galaxies found in NIR surveys. 

\item We have identified a sample of unlensed, intrinsic HyLIRGs. These HyLIRGs are potentially extremely valuable for our understanding of galaxy evolution, because they present the most luminous, massive, and active galaxies in the early universe. Future investigations of their detailed kinematics are needed to understand the physical drivers of such extreme star formation. 

\end{enumerate}

This paper provides a catalog of high-redshift DSFGs for spectroscopic follow-up observations and future JWST observations to probe the mid-IR and rest-frame UV continuum. Our sample contains largely unlensed DSFGs that are especially advantageous because it avoids uncertainties in lens modeling and differential lensing, which will enable us to draw definite conclusions on the connections between stellar, gas and dust emission components.

\section*{acknowledgments}

We thank the anonymous referee for the constructive comments that have greatly improved the manuscript. We thank Maria Strandet and Tao Wang for sharing data with us.
This work is based on observations made with the Spitzer Space Telescope, which is operated by the Jet Propulsion Laboratory, California Institute of Technology under a contract with NASA. This research was supported by NASA grants NNX15AQ06A, NNX16AF38G, and NSF grant AST-131331. JLW acknowledges support from an STFC Ernest Rutherford Fellowship (ST/P004784/2). LD and SJM acknowledge funding from the European Research Council Consolidator Grant: CosmicDust. D.R. acknowledges support from the National Science Foundation under grant number AST-1614213. 

This paper makes use of the following ALMA data: ADS/JAO.ALMA \#2013.1.00001.S and \#2016.1.00139.S. ALMA is a partnership of ESO (representing its member states), NSF (USA) and NINS (Japan), together with NRC (Canada) and NSC and ASIAA (Taiwan) and KASI (Republic of Korea), in cooperation with the Republic of Chile. The Joint ALMA Observatory is operated by ESO, AUI/NRAO and NAOJ. The National Radio Astronomy Observatory is a facility of the National Science Foundation operated under cooperative agreement by Associated Universities, Inc.

The James Clerk Maxwell Telescope is operated by the East Asian Observatory on behalf of The National Astronomical Observatory of Japan, Academia Sinica Institute of Astronomy and Astrophysics, the Korea Astronomy and Space Science Institute, the National Astronomical Observatories of China, and the Chinese Academy of Sciences (Grant No. XDB09000000), with additional funding support from the Science and Technology Facilities Council of the United Kingdom and participating universities in the United Kingdom and Canada.

The Herschel spacecraft was designed, built, tested, and launched under a contract to ESA managed by the Herschel/Planck Project team by an industrial consortium under the overall responsibility of the prime contractor Thales Alenia Space (Cannes), and including Astrium (Friedrichshafen) responsible for the payload module and for system testing at spacecraft level, Thales Alenia Space (Turin) responsible for the service module, and Astrium (Toulouse) responsible for the telescope, with in excess of a hundred subcontractors.

SPIRE has been developed by a consortium of institutes led by Cardiff University (UK) and including University of Lethbridge (Canada); NAOC (China); CEA, LAM (France); IFSI, University of Padua (Italy); IAC (Spain); Stockholm Observatory (Sweden); Imperial College London, RAL, UCL-MSSL, UKATC, University of Sussex (UK); and Caltech, JPL, NHSC, University of Colorado (USA). This development has been supported by national funding agencies CSA (Canada); NAOC (China); CEA, CNES, CNRS (France); ASI (Italy); MCINN (Spain); SNSB (Sweden); STFC, UKSA (UK); and NASA (USA).


\begin{appendix}
\section{{\it Spitzer}/IRAC cutout images of {\it Herschel}-selected ultrared sources }

Figure 10 shows the 60$\arcsec$ $\times$ 60 $\arcsec$ {\it Spitzer}/IRAC cutouts centered on the {\it Herschel} positions (green cross) for the 300 ultrared sources in {\it Spitzer} program PID13042. The white circle shows the 36$\arcsec$ FWHM beam size of SPIRE at 500 $\micron$. The 3$\arcsec$-radius cyan circle denotes the positions of the SDSS sources in the field, and the yellow circle shows the positions of the VIKING sources. The red circles indicate the SCUBA-2 positions and corresponding closest IRAC counterparts (red plus). The magenta circles show the high-resolution positions from ALMA, NOEMA, and/or SMA. 

\begin{table*}
\centering
\caption{Sources with submm/mm photometry from SMA, MUSIC, and ACT}
\begin{tabular}{lcccccccc}
\hline
\hline
Sourcename   & MUSIC                 & SMA                      &  ACT                & MUSIC                   &  ACT                  & MUSIC                & ACT                     & MUSIC  \\
                       &   0.92mm              &  1.1mm                & 1.1mm               &  1.1mm                   & 1.4mm              & 1.4mm                 & 2.0mm                  &   2.0mm \\
                       & 	 (mJy)            &		 (mJy)         &   (mJy)                 & (mJy)                     & (mJy)                &   (mJy)                 & (mJy)                     & (mJy)        \\
\hline
HELMS\_RED\_1 &                          &28.6 $\pm$ 2.3   & 72.32 $\pm$ 6.26 & 35.11 $\pm$ 2.62 & 12.49 $\pm$ 1.74   &                          &                             &                   \\
HELMS\_RED\_2 &                          & 33.9 $\pm$ 2.3   &                              &                              &                                &                          &                             &                   \\
HELMS\_RED\_3 &100.3 $\pm$ 52.8 &                         & 35.32 $\pm$ 6.24 & 24.9 $\pm$ 12.3 & 19.50 $\pm$ 2.56 & 16.0 $\pm$ 6.9 &  6.14 $\pm$ 1.76 &                       \\
HELMS\_RED\_4 & 65.2 $\pm$ 57.3  & 21.3 $\pm$ 1.9 &                             & 32.8 $\pm$ 12.4  &                        		& 19.4 $\pm$ 6.7  &                       & 8.4 $\pm$ 7.4         \\
HELMS\_RED\_6 &                             &                            &                           & 32.3 $\pm$ 9.8   &                                & 8.8 $\pm$ 4.3  &                            & 13.1 $\pm$ 5.7 \\                                                 
HELMS\_RED\_7 &108.7 $\pm$ 35.1 &                           &                           & 62.4 $\pm$ 13.5  &                                & 18.1 $\pm$ 5.1   &                          &                        \\
HELMS\_RED\_10 & 			      &13.3 $\pm$ 2.8 &                            &                             &                                   &                          &                             &                   \\
HELMS\_RED\_13 & 			     &11.5 $\pm$ 1.8 &                            &                             &                                    &                          &                             &                   \\
\hline
Sourcename &  NOEMA                &    NOEMA/ALMA           &                           &                             &                                    &                          &                             &                   \\
                     &  1.3mm                   & 3mm                             &                              &                             &                                    &                          &                             &                   \\
                     &  (mJy)                     & (mJy)                            &                            &                             &                                    &                          &                             &                   \\
\hline
G09-59393          & 4.0 $\pm$ 0.6   &                               &                          &                             &                                 &                             &                           &                            \\
G09-62610          & 5.2 $\pm$ 0.8    &   $<$ 0.18           &                          &                             &                                 &                             &                           &                            \\
G09-81106           &9.7 $\pm$ 1.3    &  0.24 $\pm$ 0.04 &                          &                             &                                 &                             &                           &                            \\
G09-83808           & 19.4 $\pm$ 2.0&   0.66 $\pm$ 0.12 &                          &                             &                                 &                             &                           &                            \\      
NGP-101333        & 10.8 $\pm$ 1.3  &$<$ 0.25              &                          &                             &                                 &                             &                           &                           \\
NGP-111912         & 4.7 $\pm$ 0.9  & $<$ 0.26             &                          &                             &                                 &                             &                           &                            \\
NGP-113609        & 13.0 $\pm$ 2.3  & $<$ 0.26            &                          &                             &                                 &                             &                           &                           \\
NGP-126191        & 12.3 $\pm$ 1.7  & 0.30 $\pm$ 0.11&                          &                             &                                 &                             &                           &                            \\
NGP-136156        &  3.1 $\pm$ 0.8   & $<$ 0.25           &                          &                             &                                 &                             &                           &                           \\
NGP-190387        & 12.2 $\pm$ 1.2  & 0.84 $\pm$ 0.14&                          &                             &                                 &                             &                           &                            \\
NGP-206987        & 9.2 $\pm$ 1.8    & $<$ 0.32            &                          &                             &                                 &                             &                           &                           \\
NGP-246114        & 8.0 $\pm$ 1.5    & 0.42 $\pm$ 0.06&                          &                             &                                 &                             &                           &                            \\
NGP-252305        & 6.5 $\pm$ 0.7   &  $<$ 0.29            &                          &                             &                                 &                             &                           &                           \\
\hline
\end{tabular}
\label{tab:moredata}
\end{table*}

\newpage

\begin{figure*}
\centering
{\includegraphics[width=4.4cm, height=4.4cm]{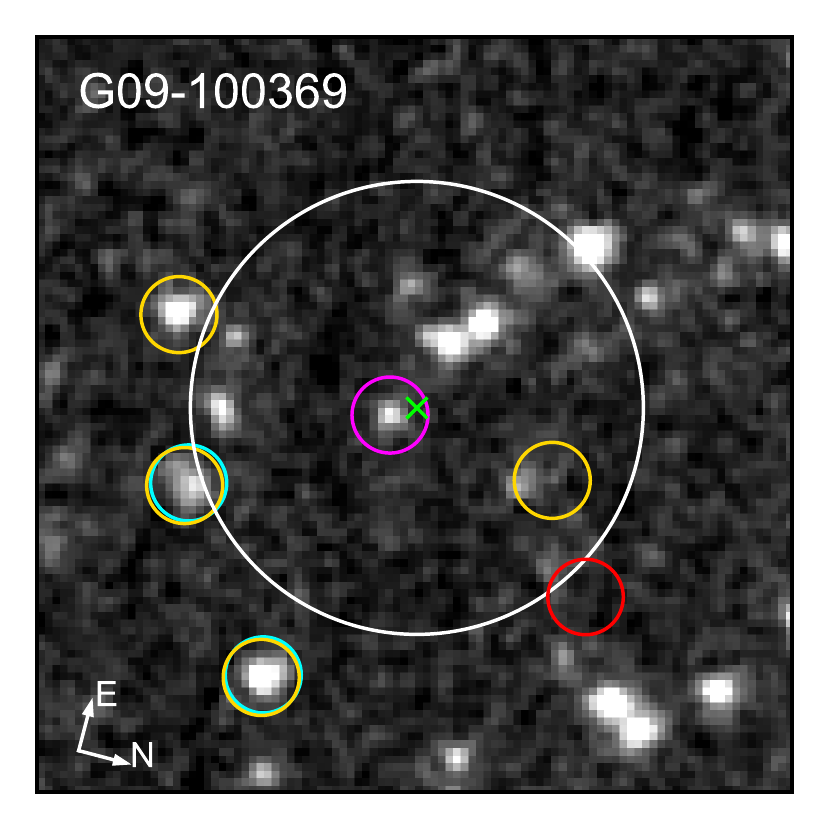}}
{\includegraphics[width=4.4cm, height=4.4cm]{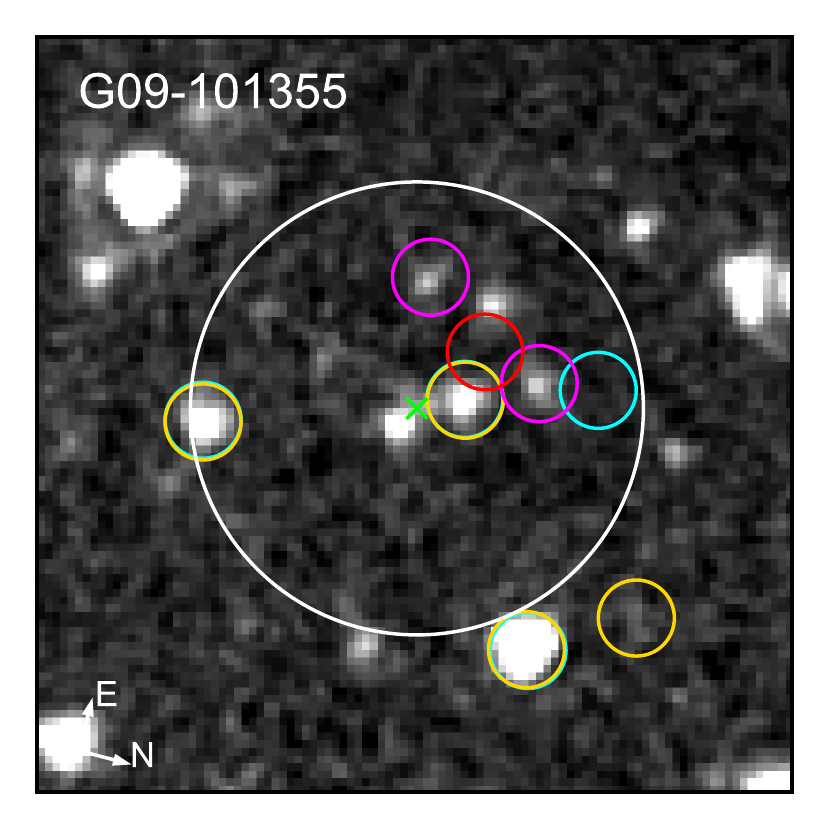}}
{\includegraphics[width=4.4cm, height=4.4cm]{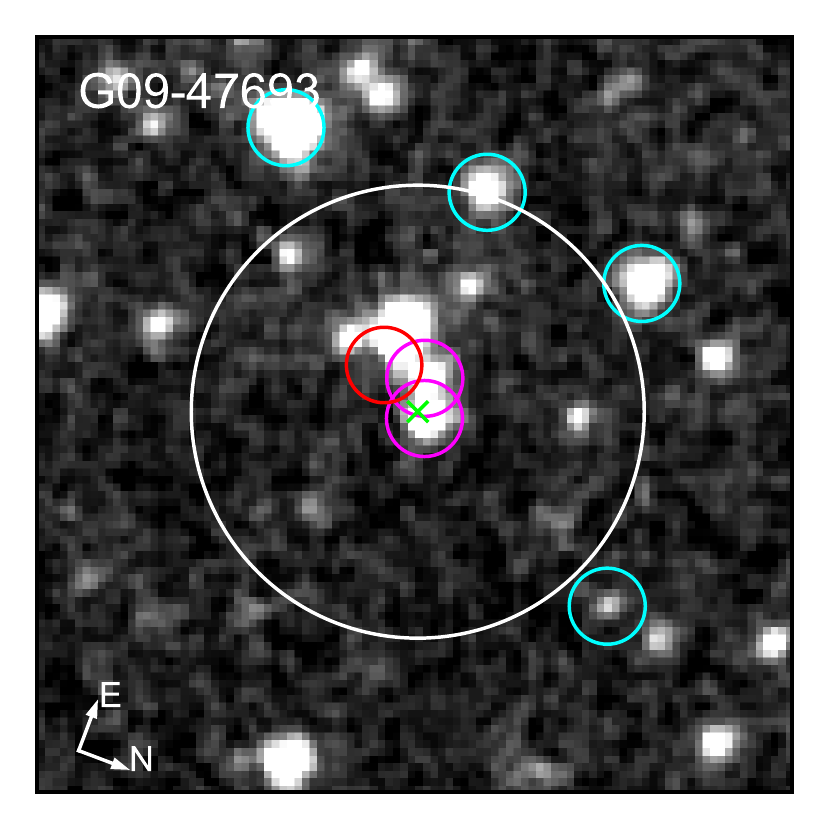}}
{\includegraphics[width=4.4cm, height=4.4cm]{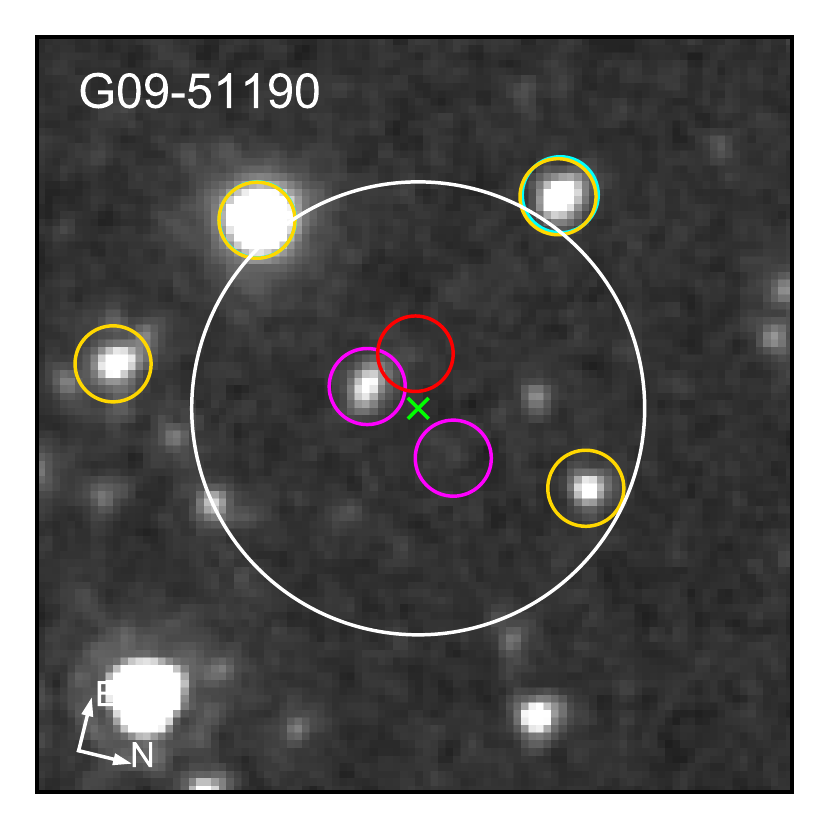}}
{\includegraphics[width=4.4cm, height=4.4cm]{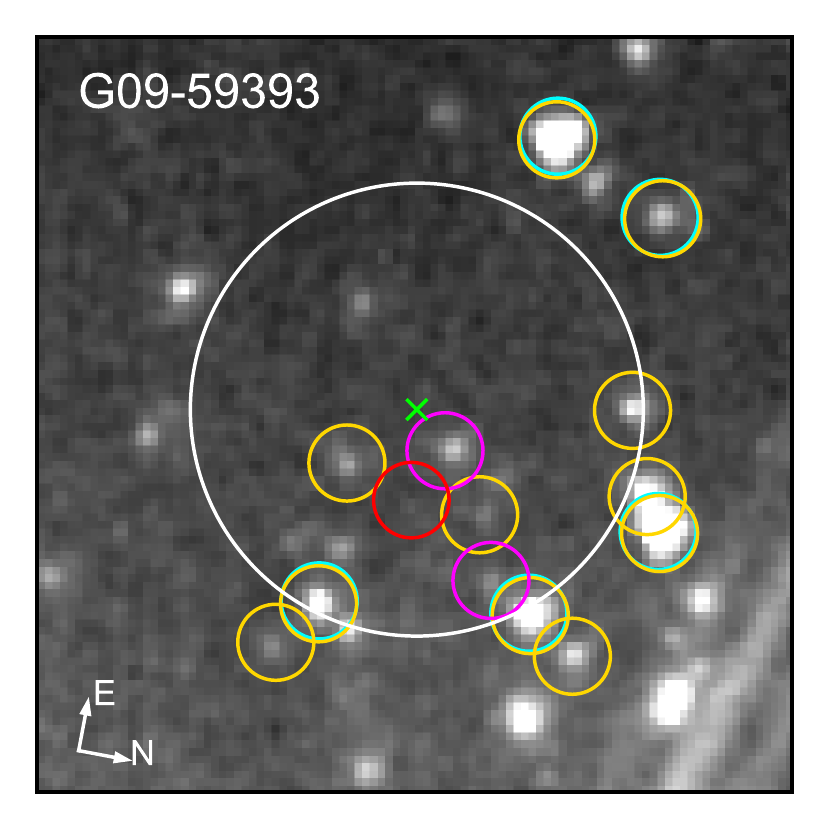}}
{\includegraphics[width=4.4cm, height=4.4cm]{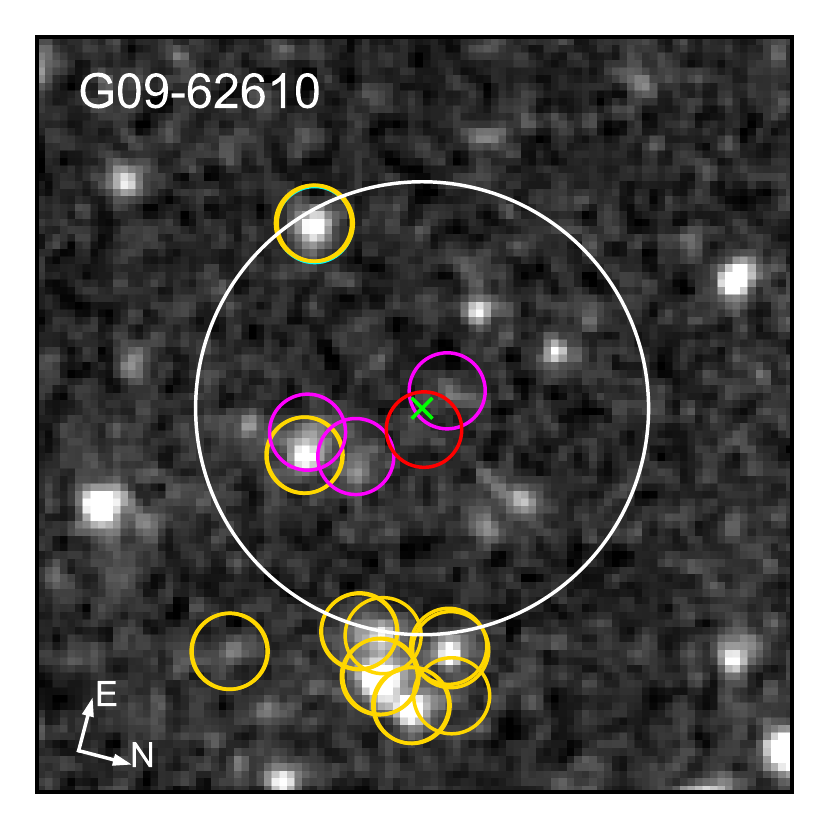}}
{\includegraphics[width=4.4cm, height=4.4cm]{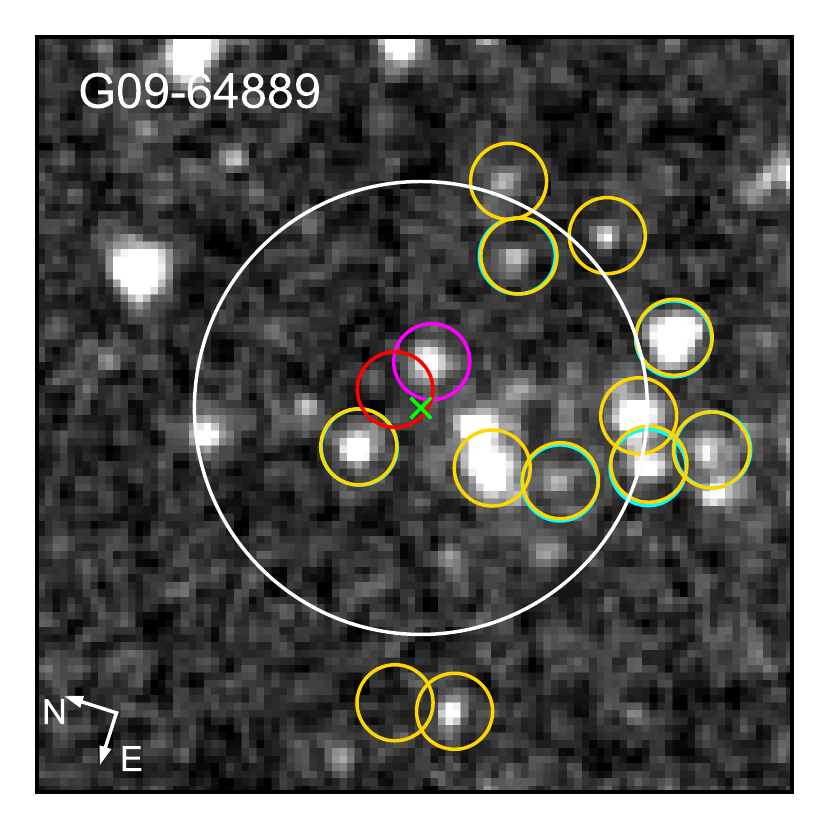}}
{\includegraphics[width=4.4cm, height=4.4cm]{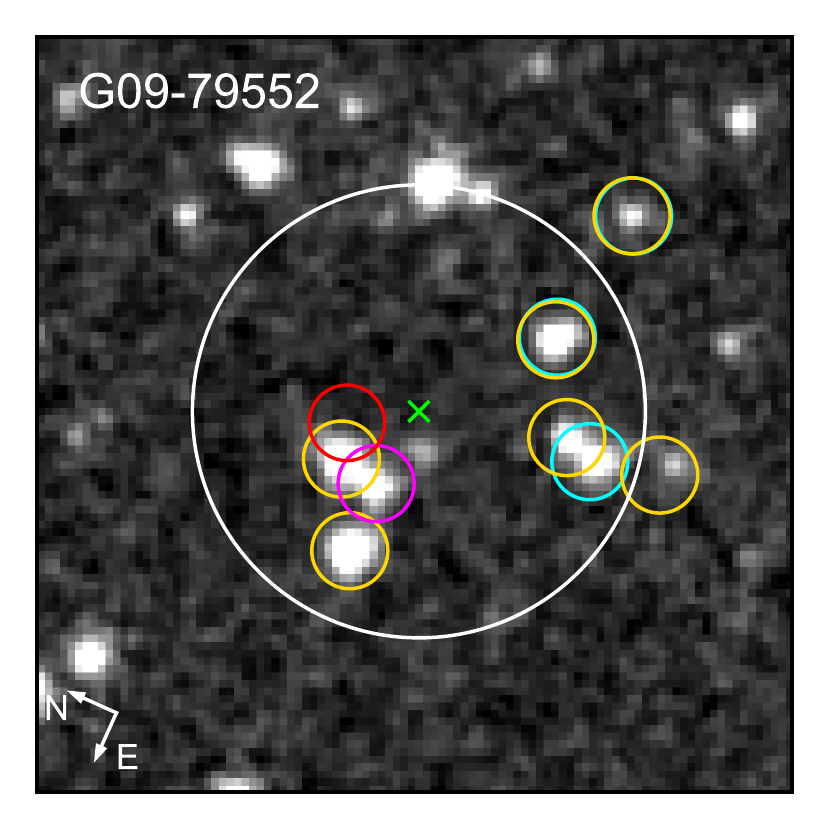}}
{\includegraphics[width=4.4cm, height=4.4cm]{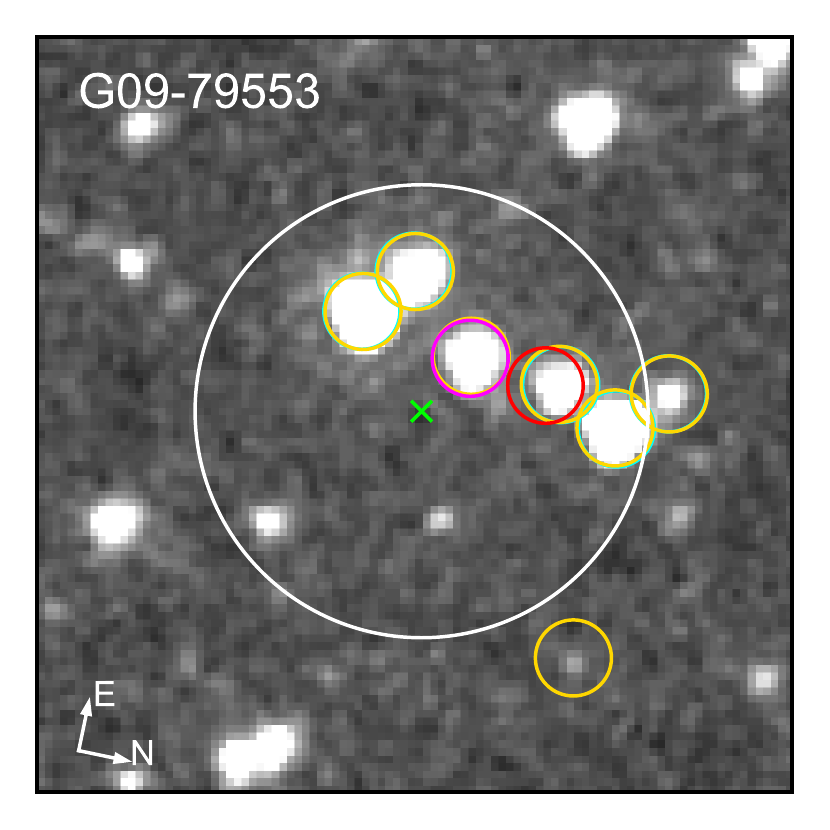}}
{\includegraphics[width=4.4cm, height=4.4cm]{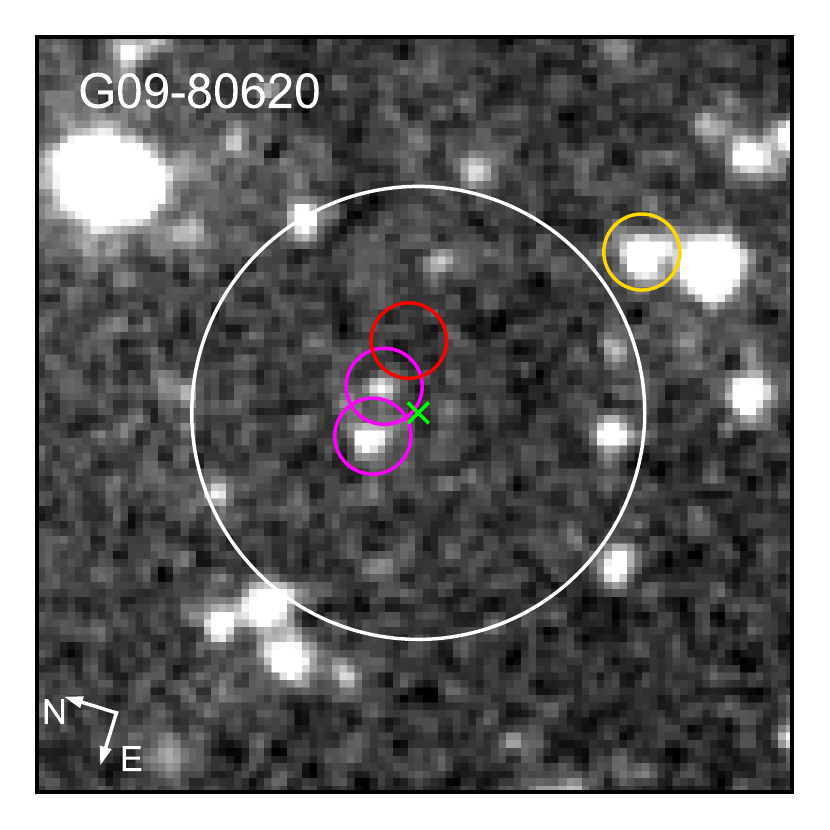}}
{\includegraphics[width=4.4cm, height=4.4cm]{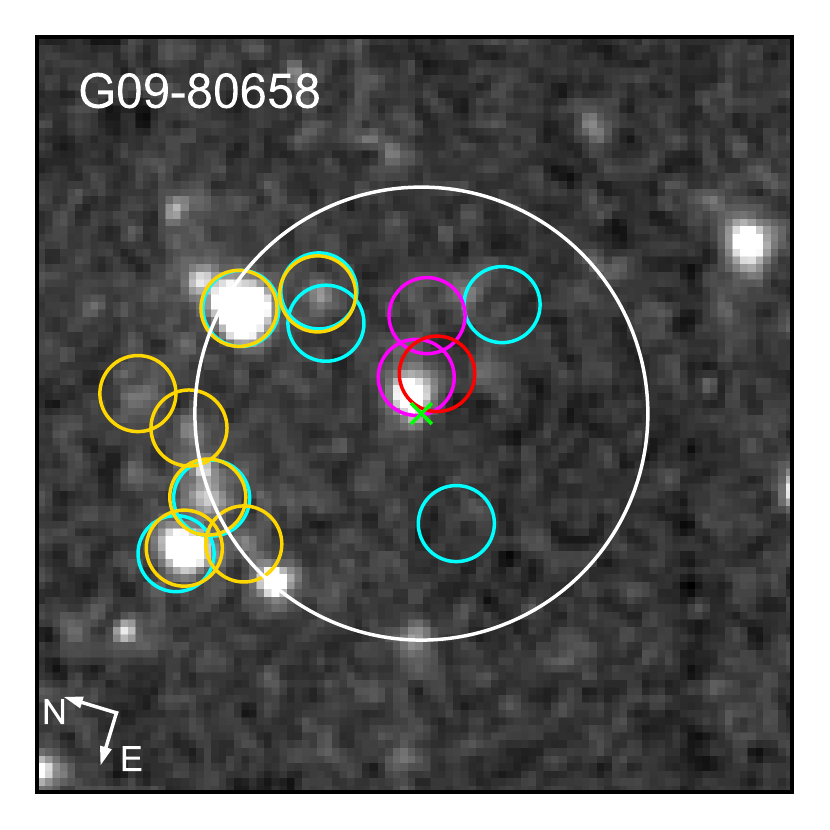}}
{\includegraphics[width=4.4cm, height=4.4cm]{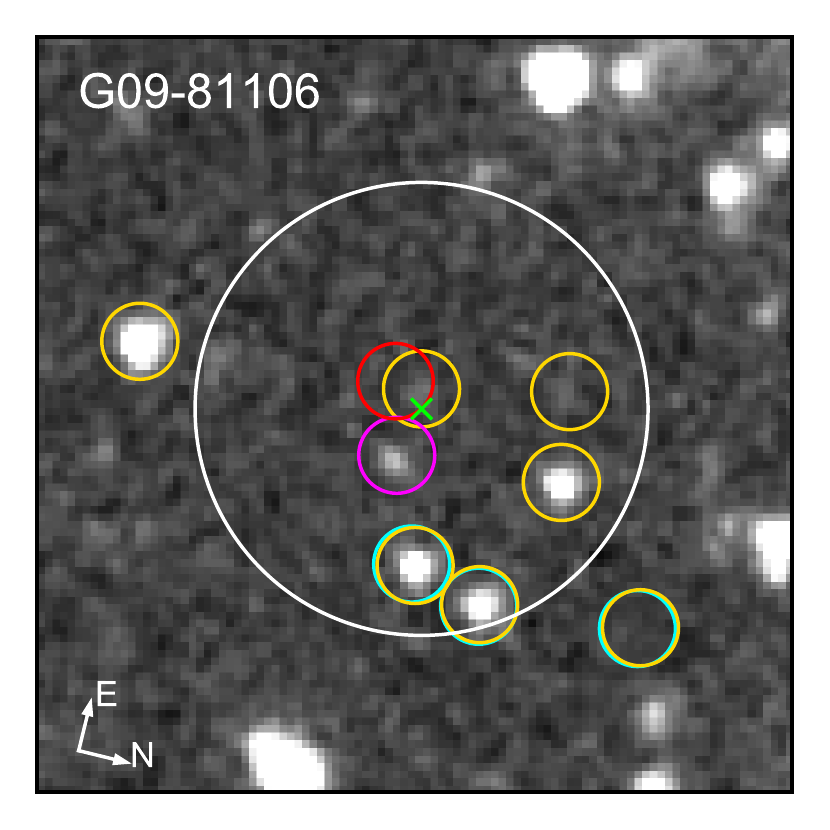}}
{\includegraphics[width=4.4cm, height=4.4cm]{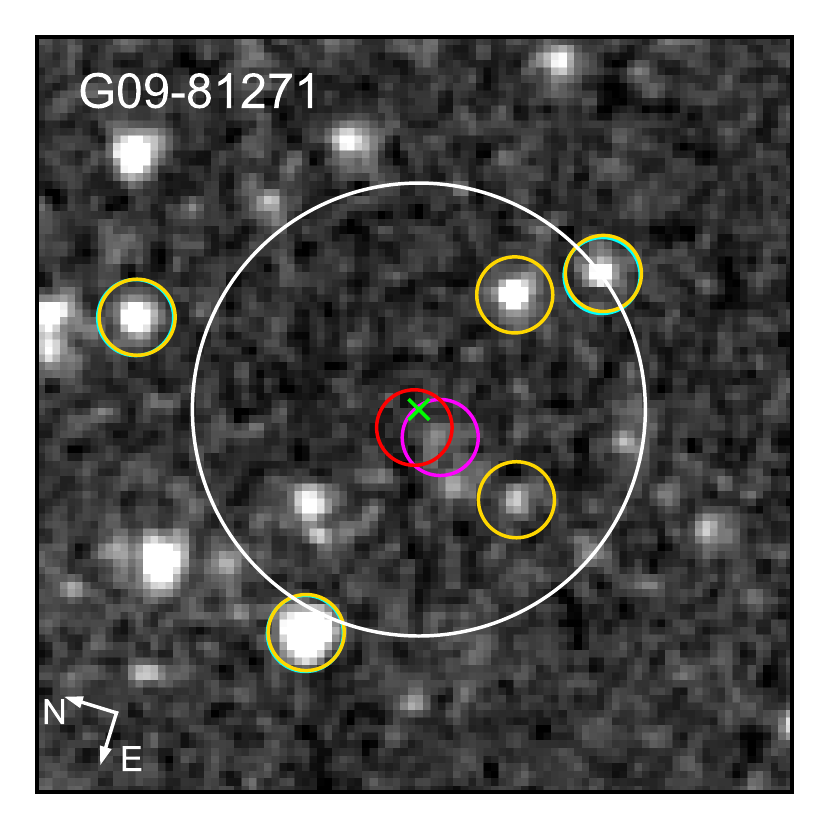}}
{\includegraphics[width=4.4cm, height=4.4cm]{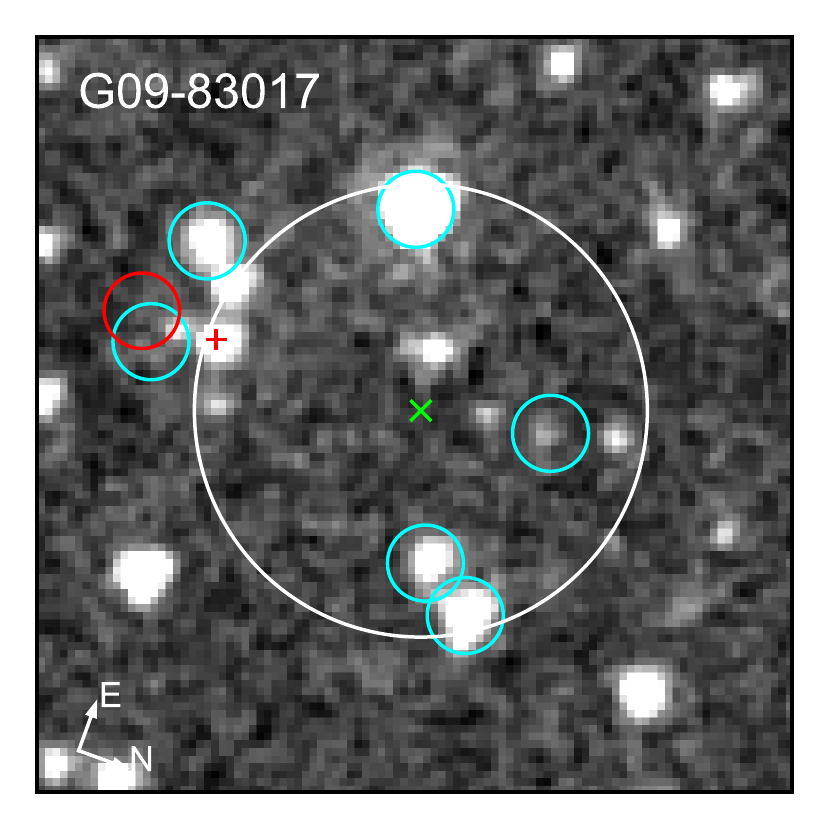}}
{\includegraphics[width=4.4cm, height=4.4cm]{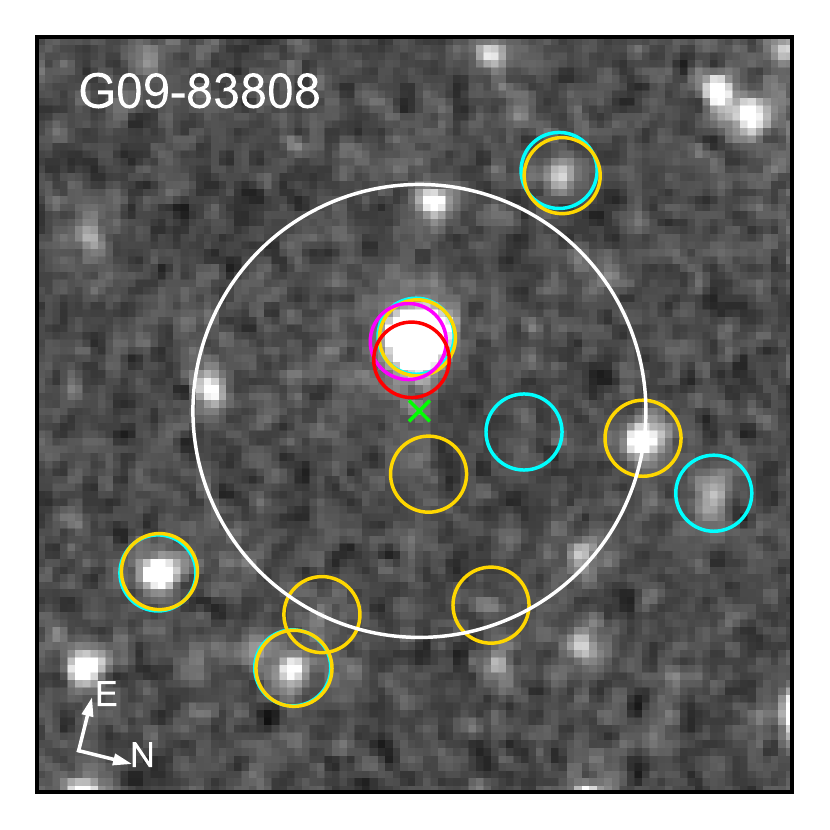}}
{\includegraphics[width=4.4cm, height=4.4cm]{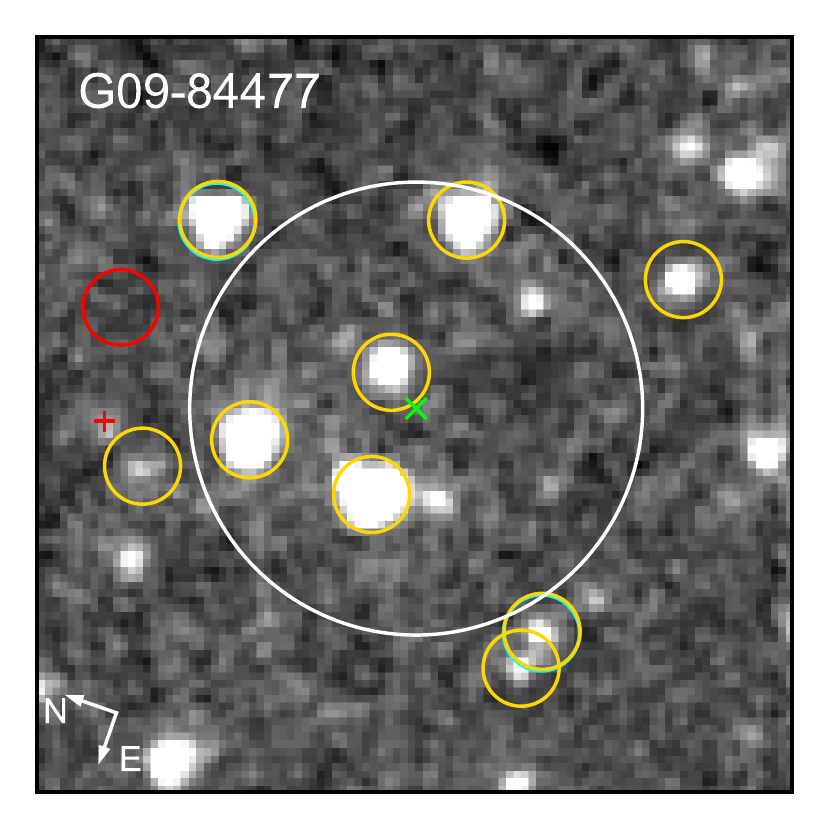}}
{\includegraphics[width=4.4cm, height=4.4cm]{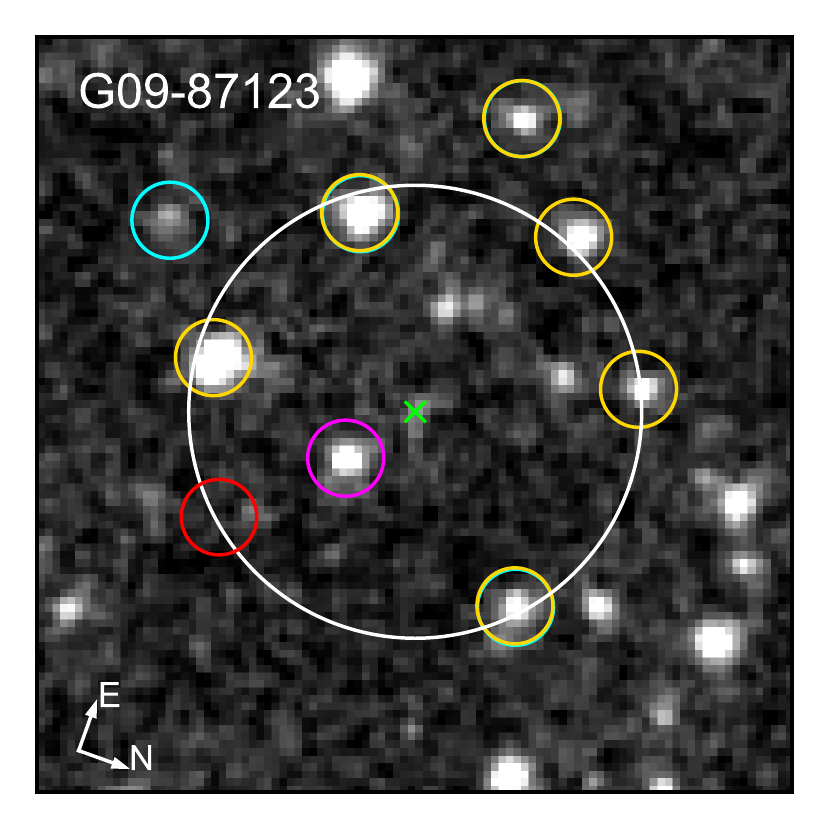}}
{\includegraphics[width=4.4cm, height=4.4cm]{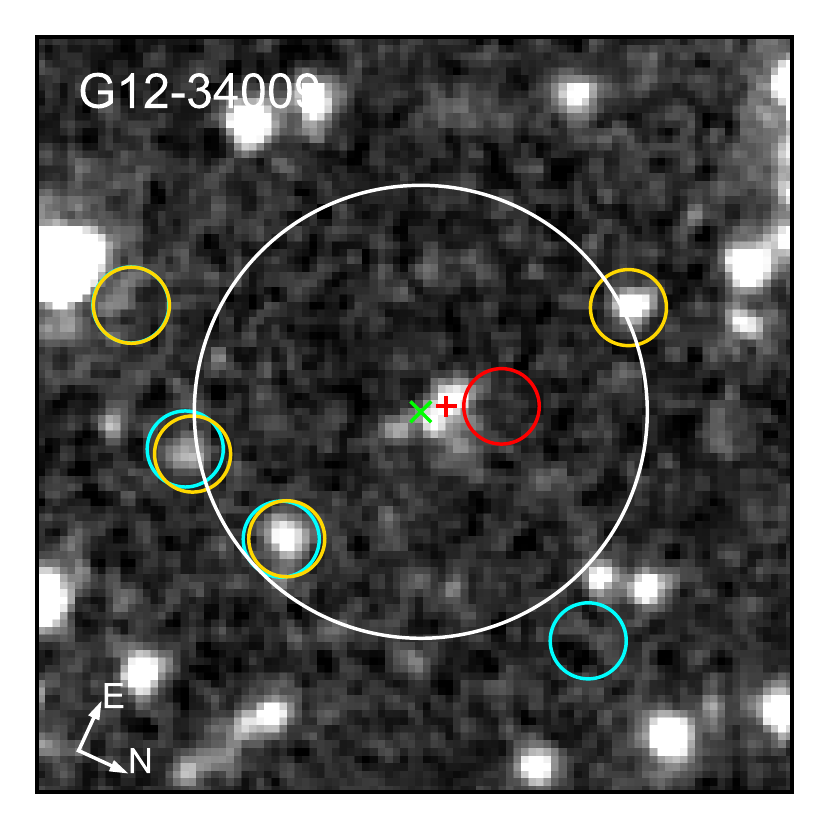}}
{\includegraphics[width=4.4cm, height=4.4cm]{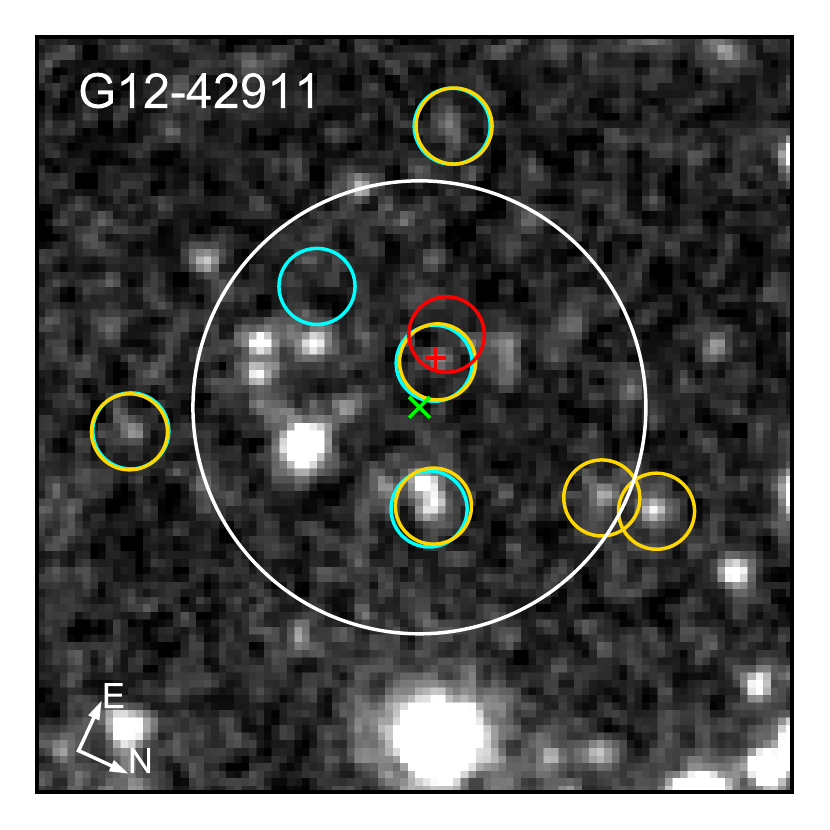}}
{\includegraphics[width=4.4cm, height=4.4cm]{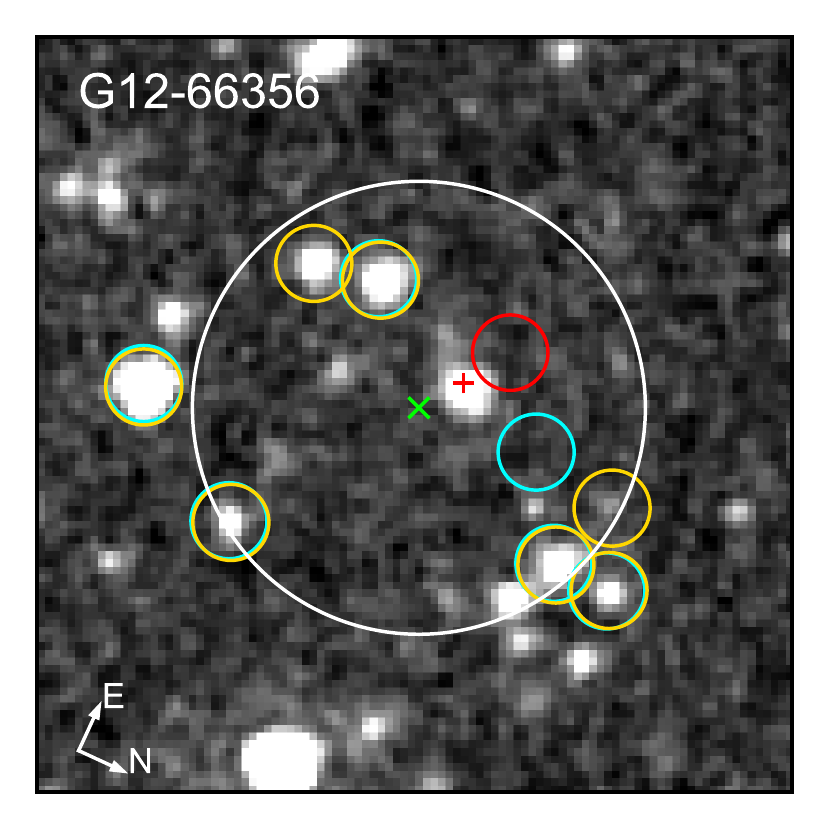}}
\caption{60$\arcsec$ $\times$ 60$\arcsec$ cutouts }
\label{cutouts}
\end{figure*}

\addtocounter{figure}{-1}

\begin{figure*}
\centering
{\includegraphics[width=4.4cm, height=4.4cm]{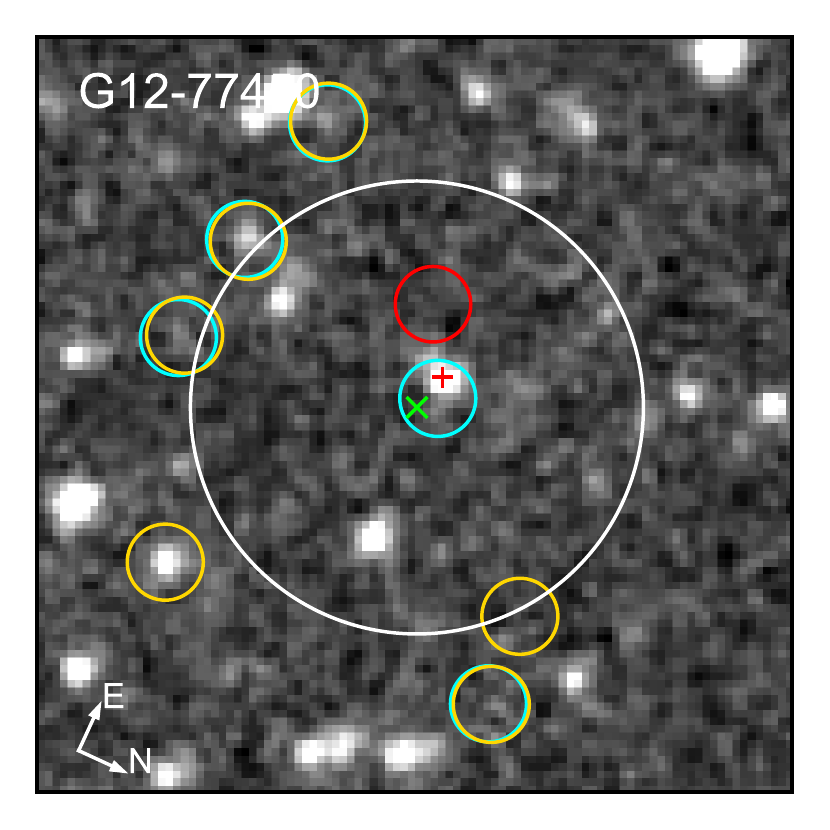}}
{\includegraphics[width=4.4cm, height=4.4cm]{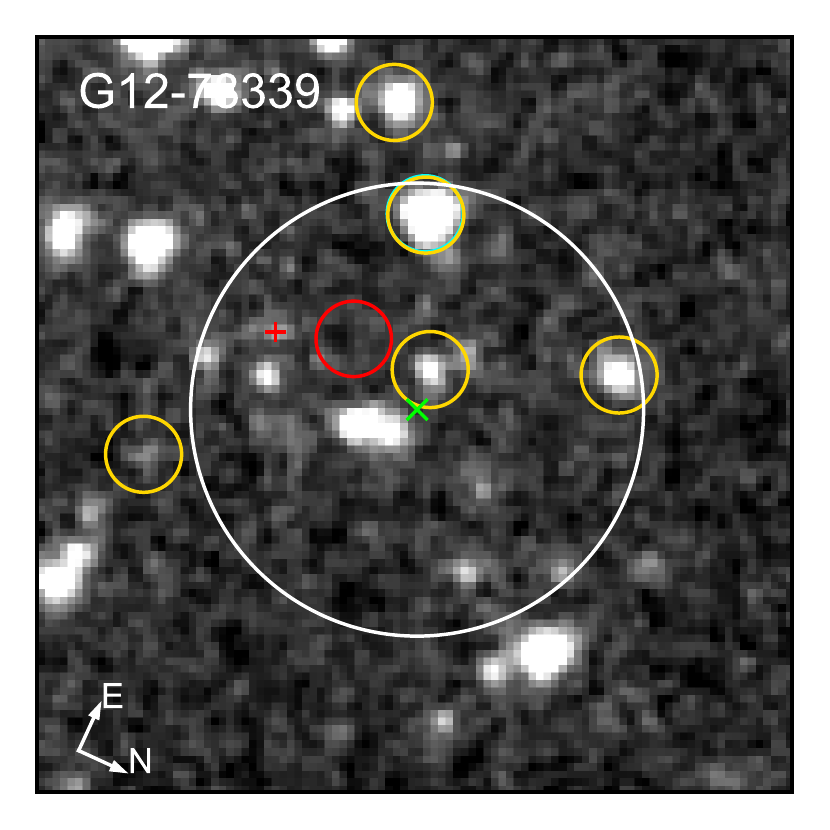}}
{\includegraphics[width=4.4cm, height=4.4cm]{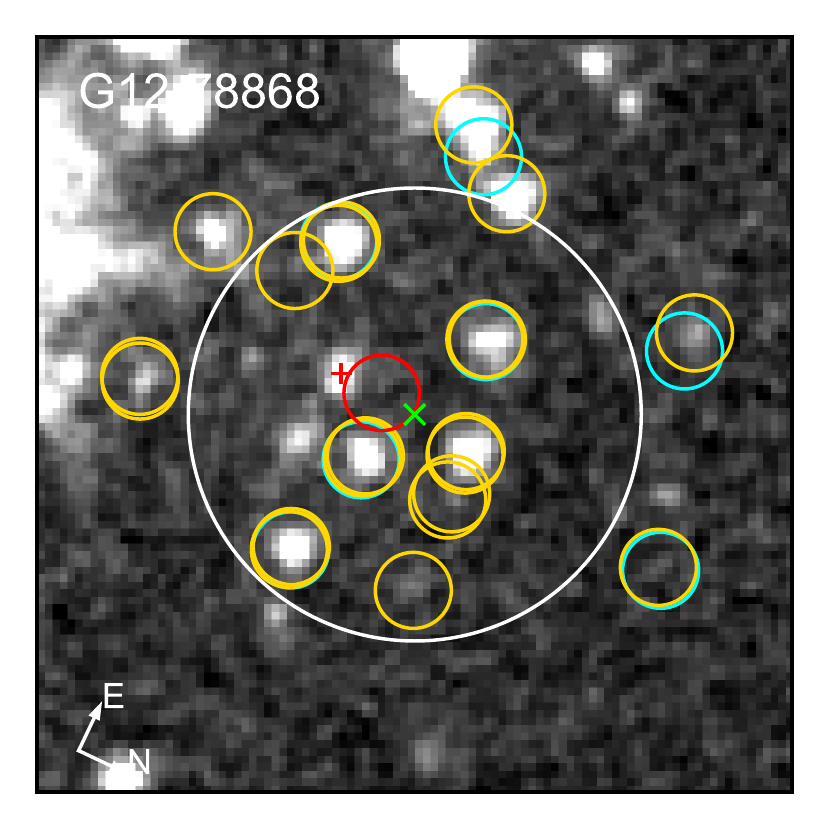}}
{\includegraphics[width=4.4cm, height=4.4cm]{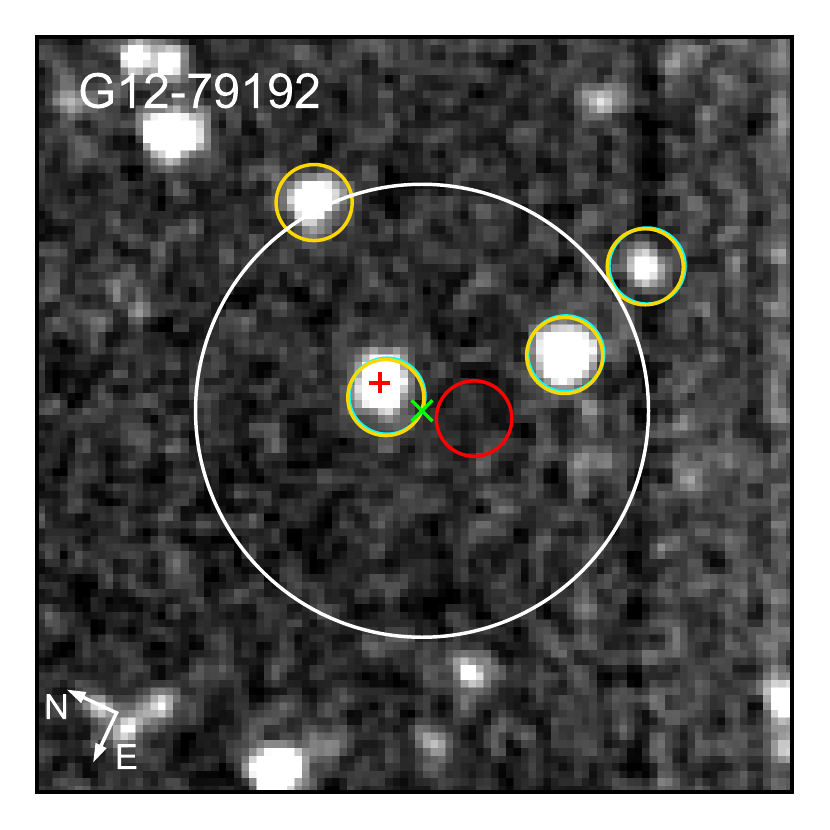}}
{\includegraphics[width=4.4cm, height=4.4cm]{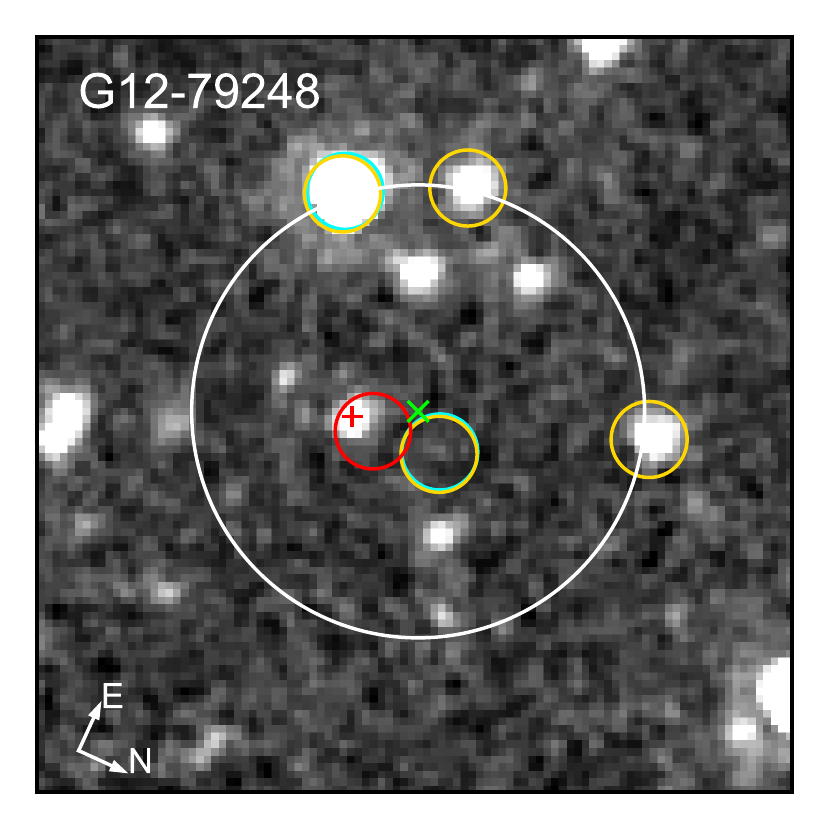}}
{\includegraphics[width=4.4cm, height=4.4cm]{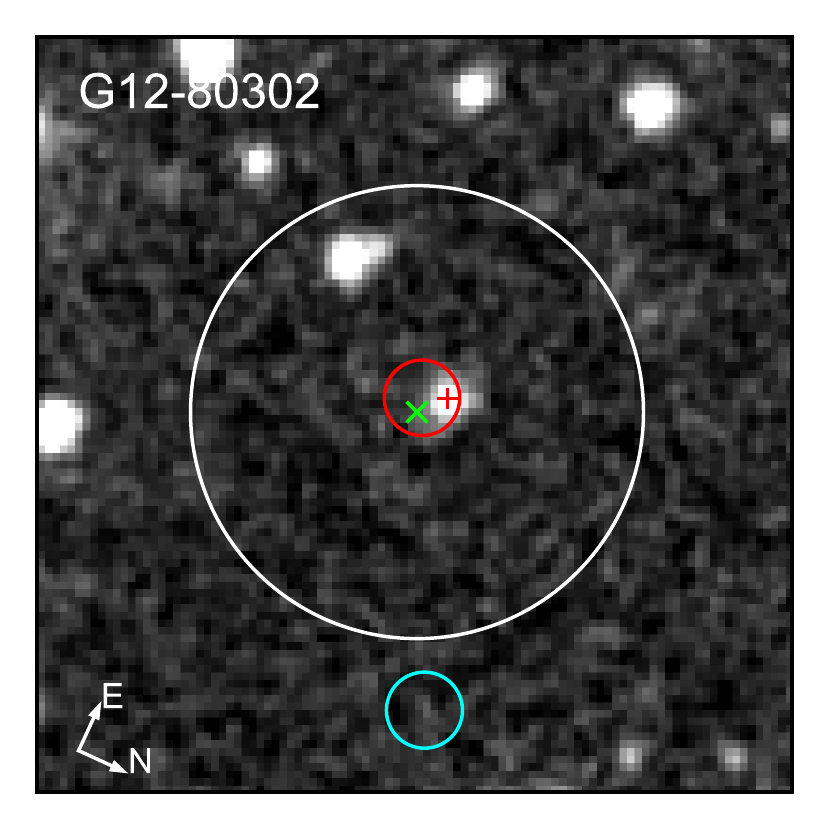}}
{\includegraphics[width=4.4cm, height=4.4cm]{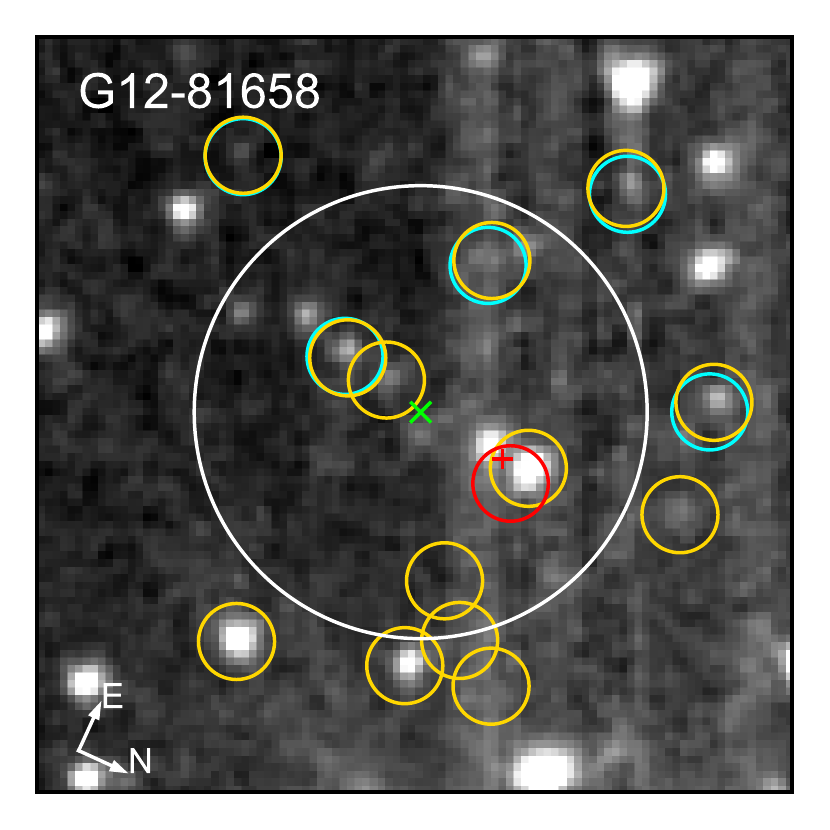}}
{\includegraphics[width=4.4cm, height=4.4cm]{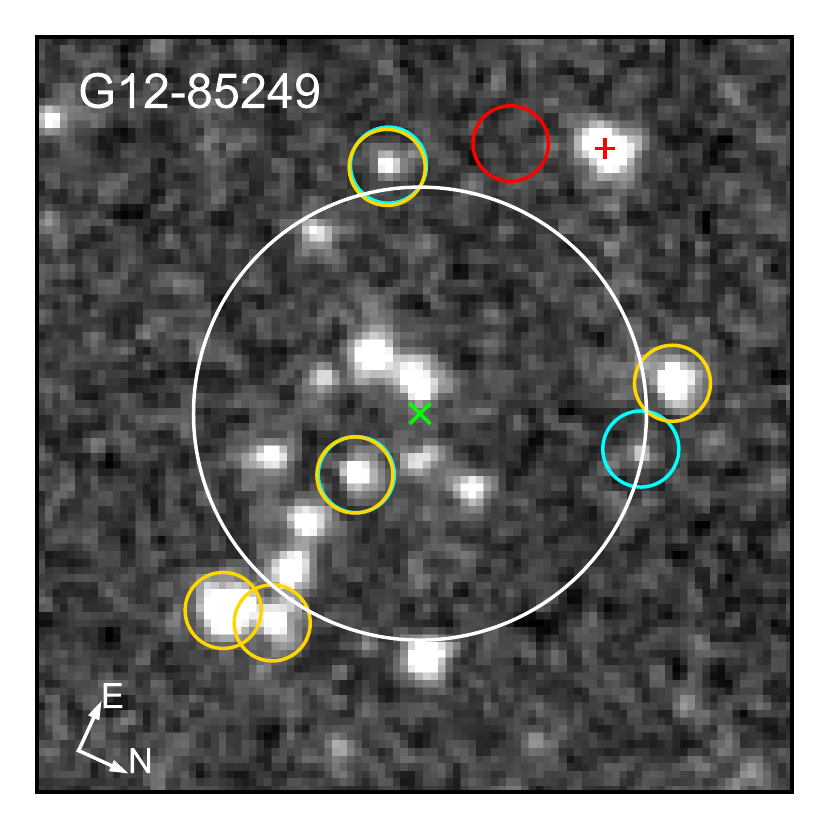}}
{\includegraphics[width=4.4cm, height=4.4cm]{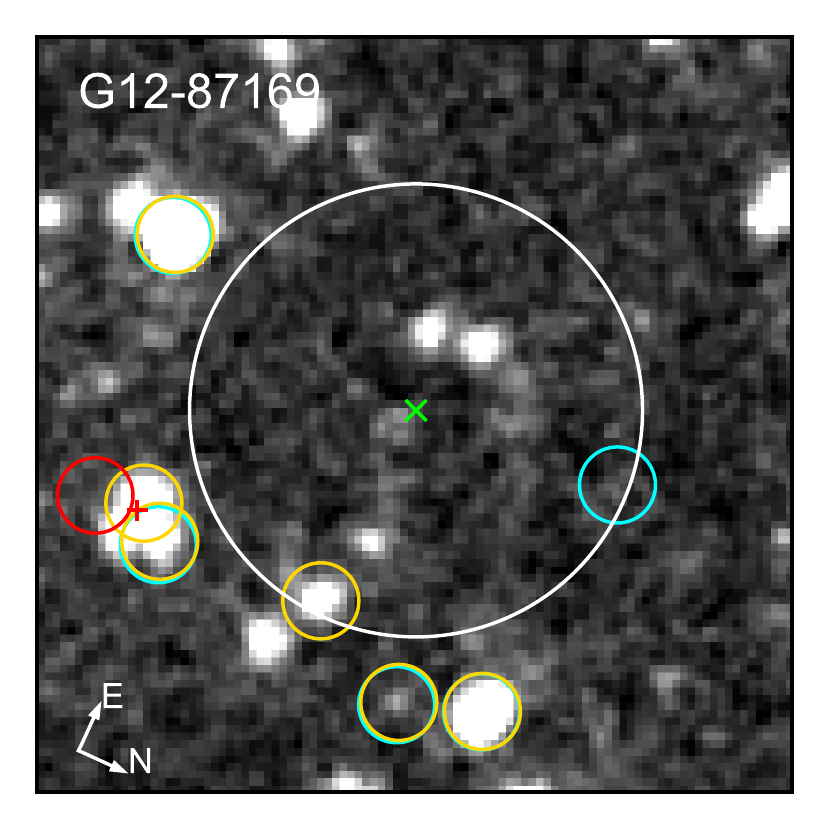}}
{\includegraphics[width=4.4cm, height=4.4cm]{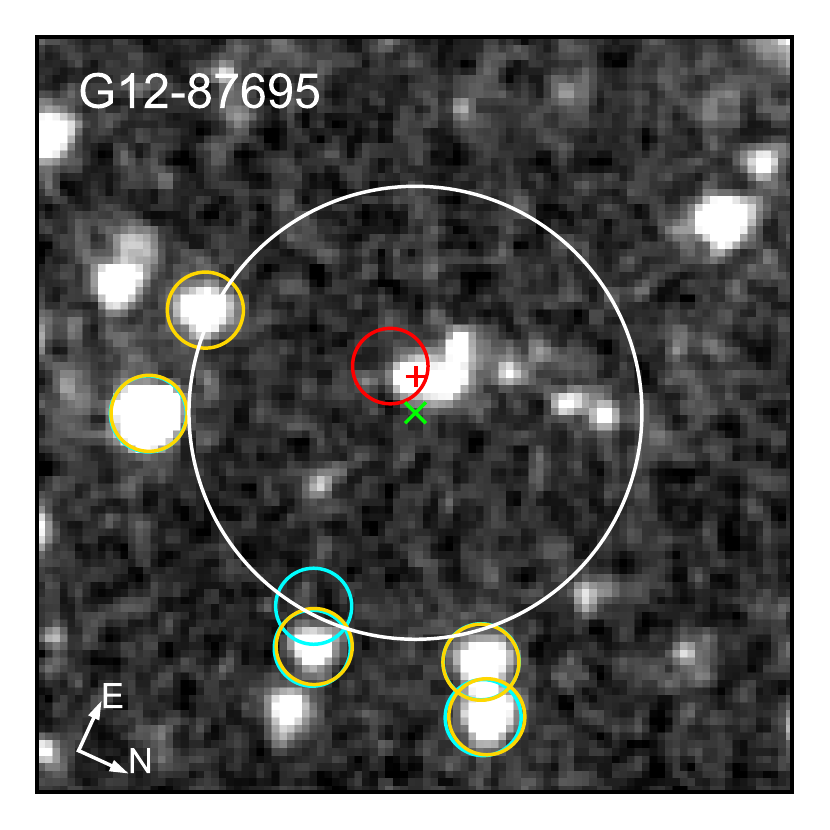}}
{\includegraphics[width=4.4cm, height=4.4cm]{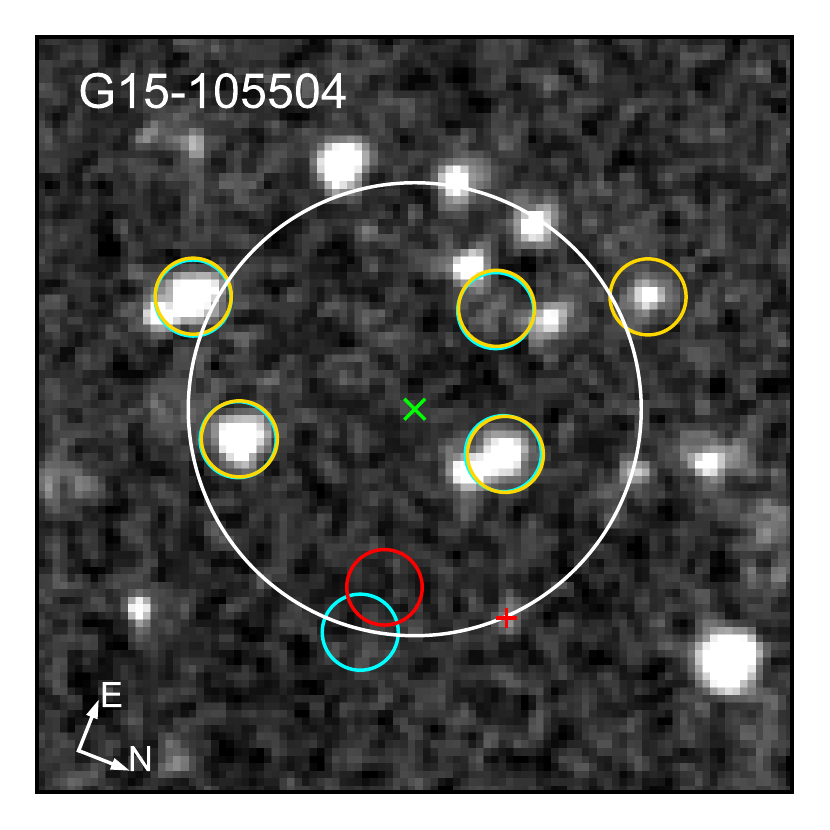}}
{\includegraphics[width=4.4cm, height=4.4cm]{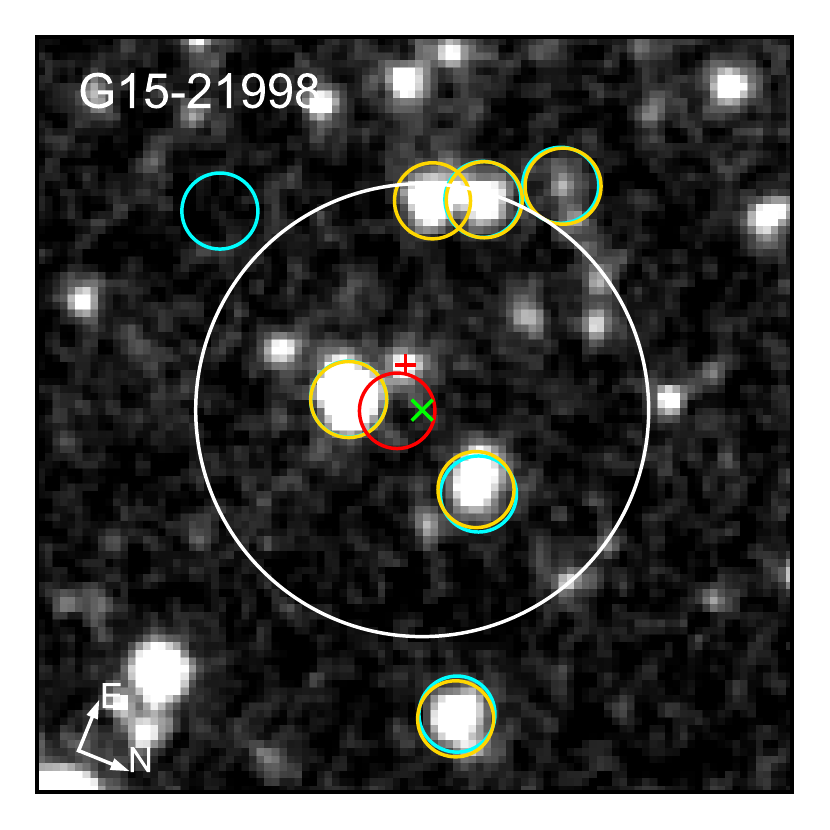}}
{\includegraphics[width=4.4cm, height=4.4cm]{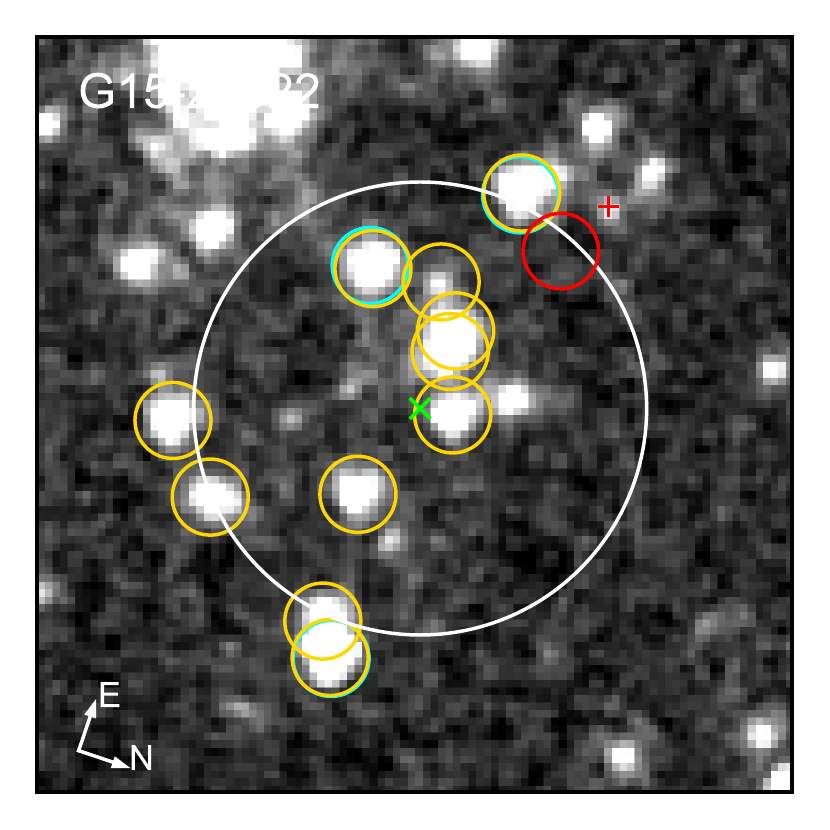}}
{\includegraphics[width=4.4cm, height=4.4cm]{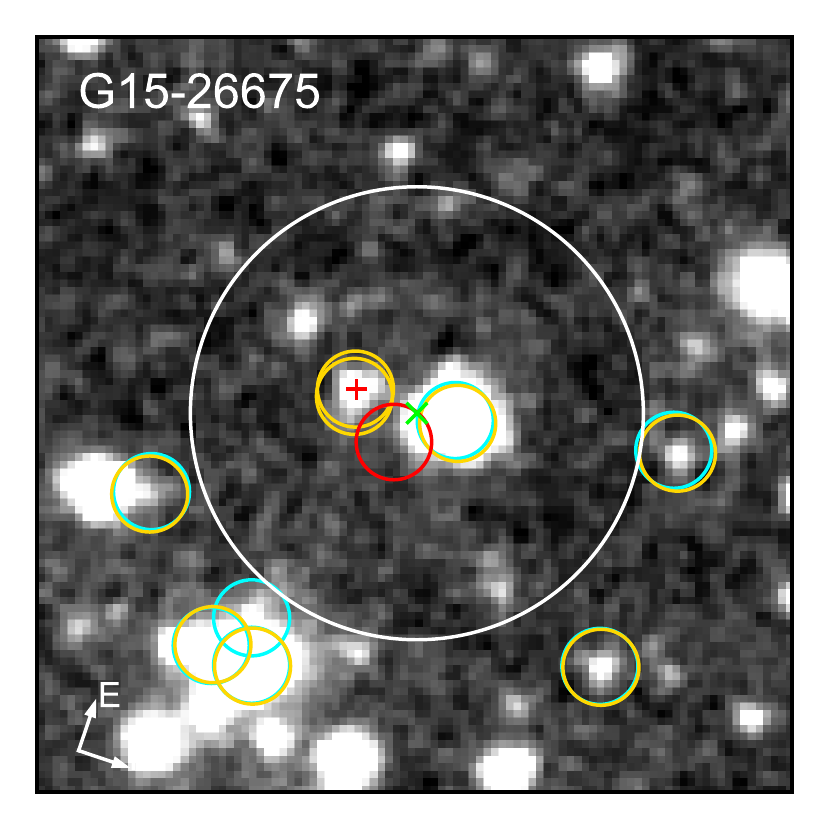}}
{\includegraphics[width=4.4cm, height=4.4cm]{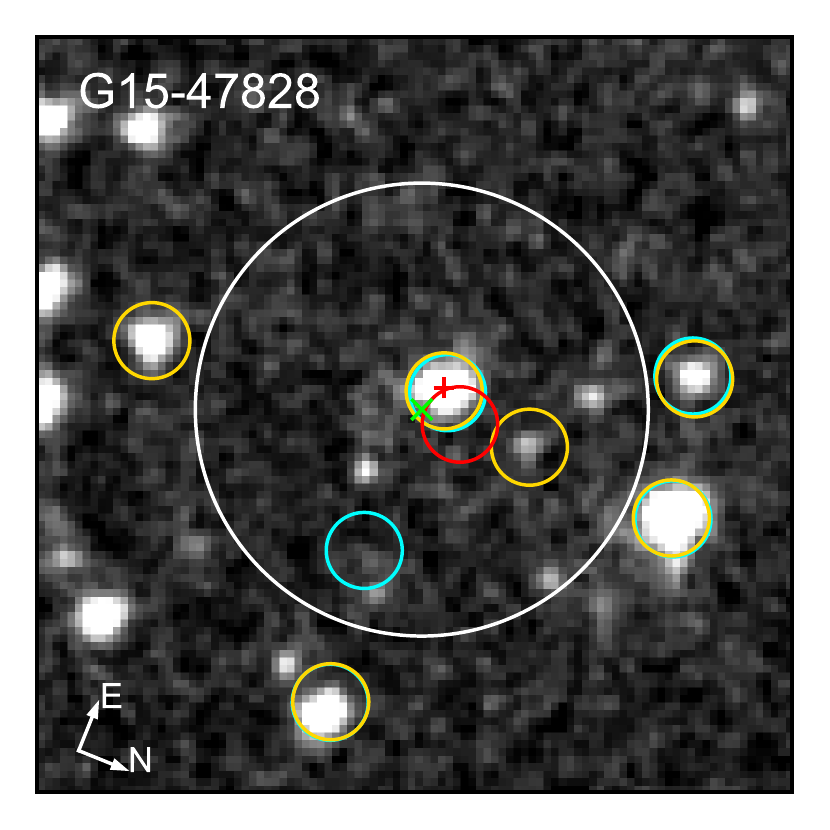}}
{\includegraphics[width=4.4cm, height=4.4cm]{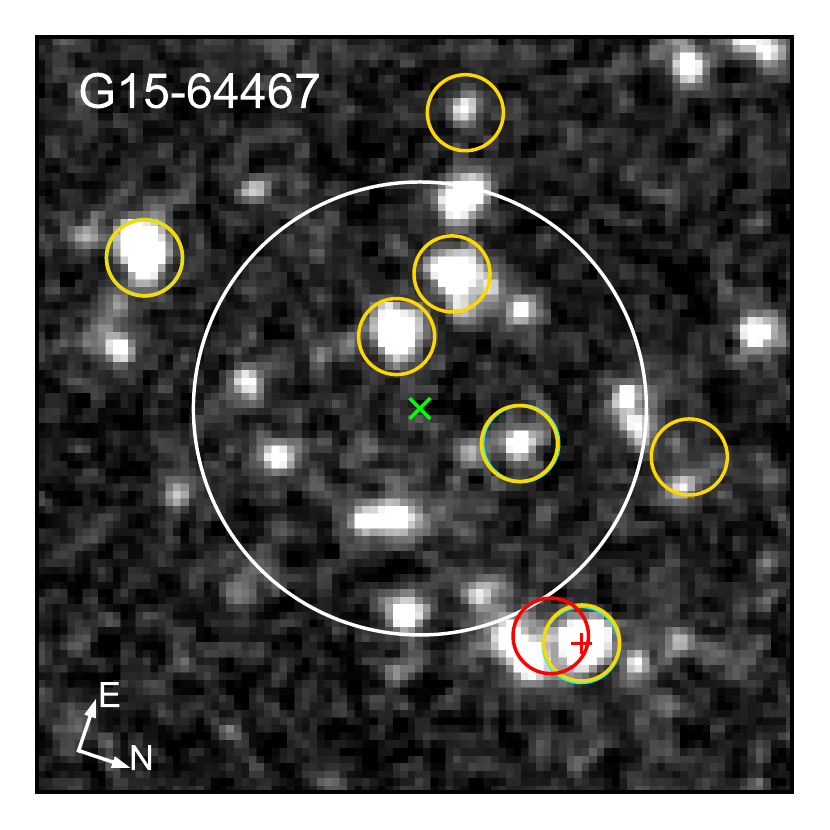}}
{\includegraphics[width=4.4cm, height=4.4cm]{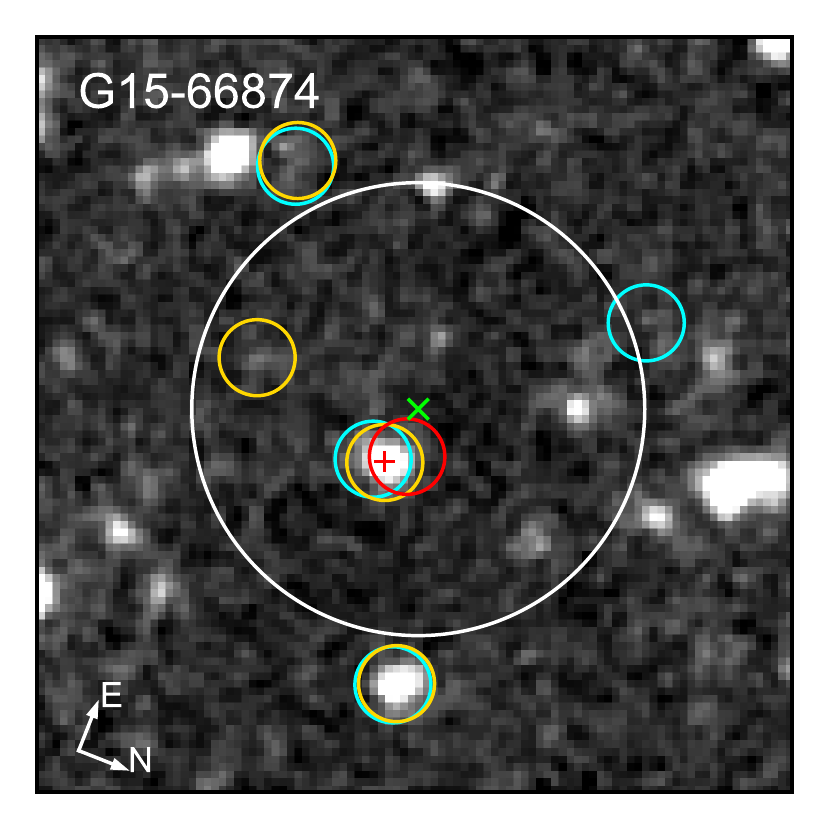}}
{\includegraphics[width=4.4cm, height=4.4cm]{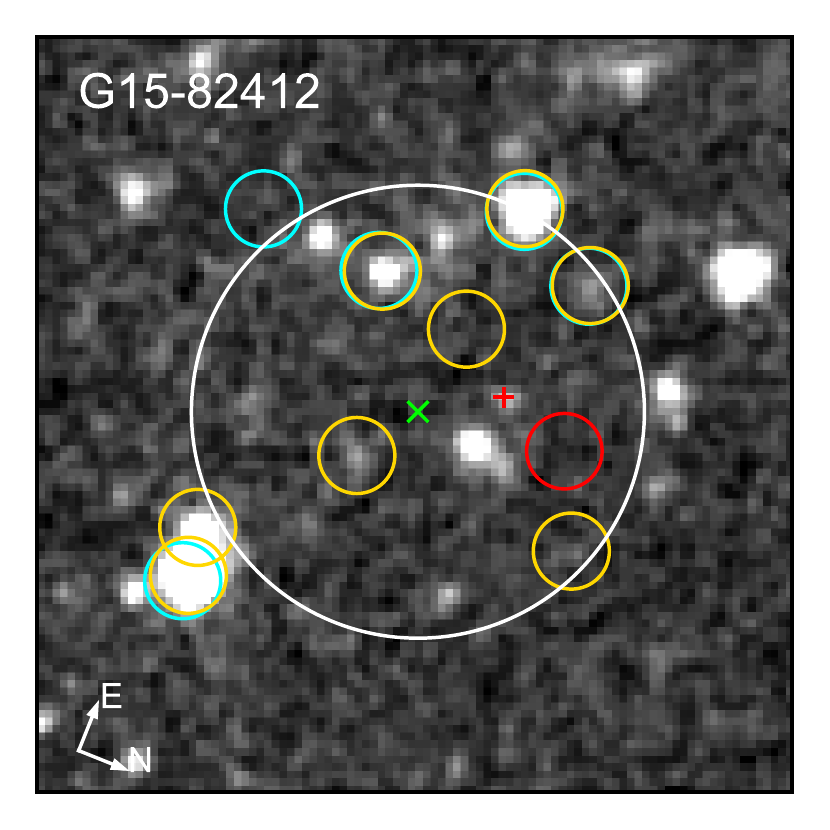}}
{\includegraphics[width=4.4cm, height=4.4cm]{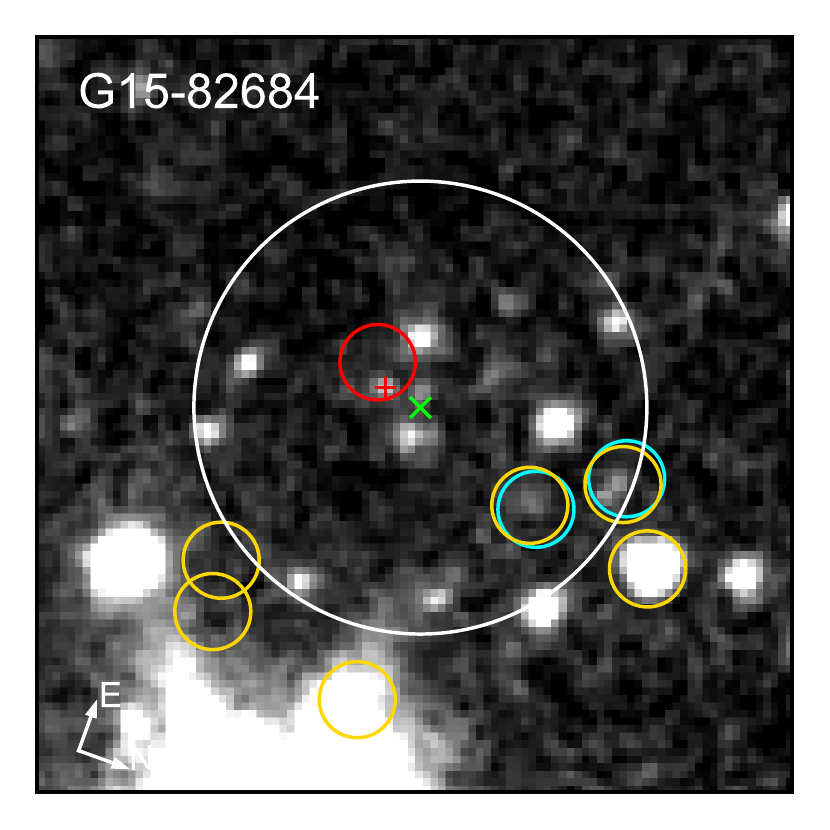}}
{\includegraphics[width=4.4cm, height=4.4cm]{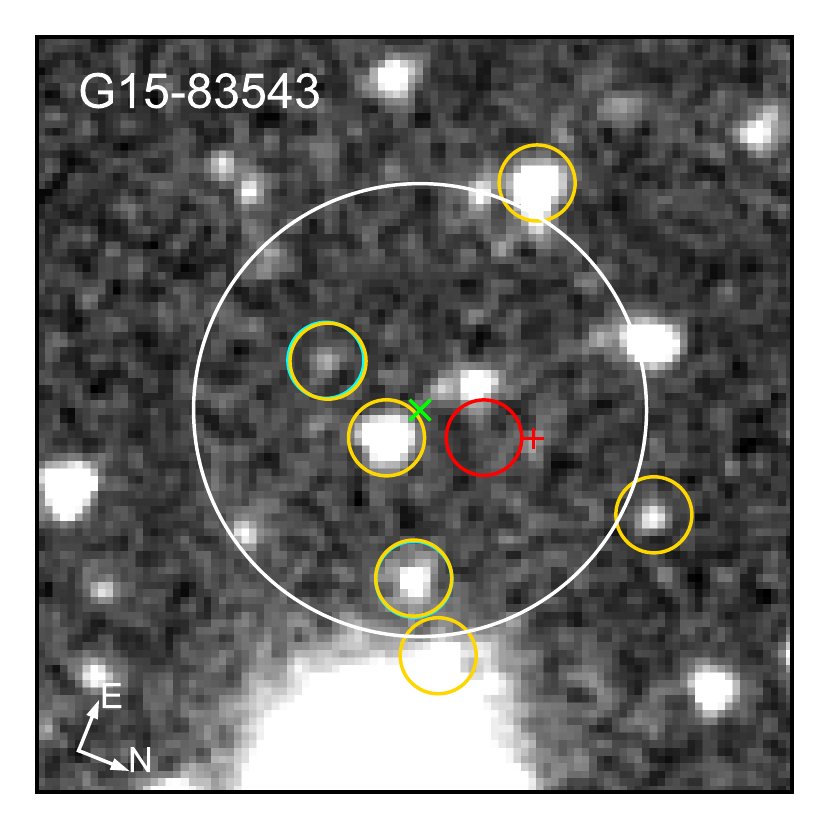}}
\caption{Continued 60$\arcsec$ $\times$ 60$\arcsec$ cutouts }
\label{fig:data}
\end{figure*}

\addtocounter{figure}{-1}

\begin{figure*}
\centering
{\includegraphics[width=4.4cm, height=4.4cm]{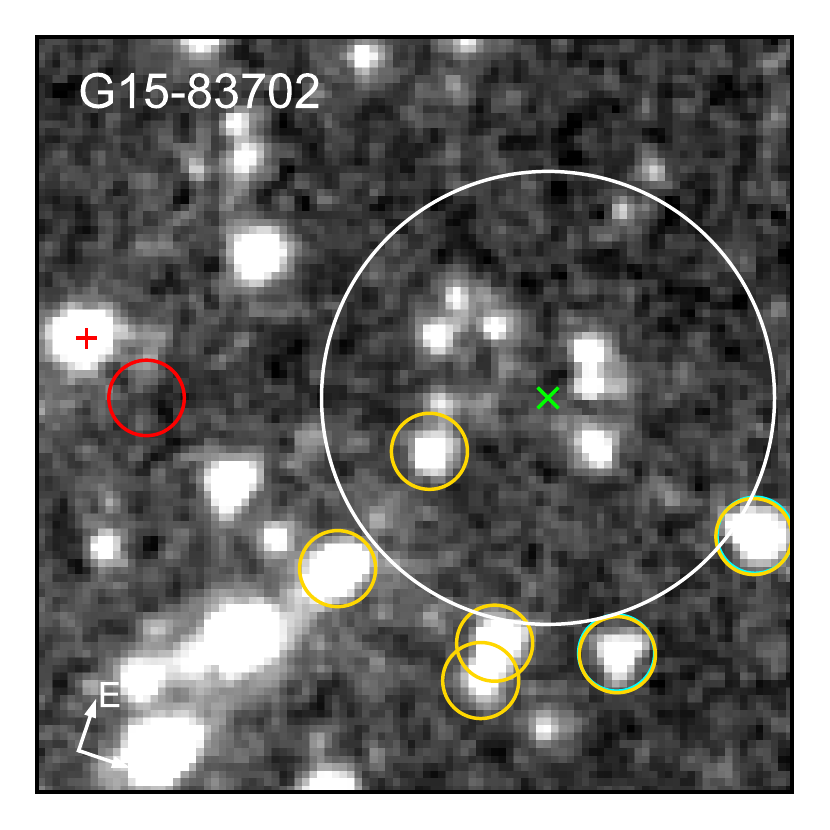}}
{\includegraphics[width=4.4cm, height=4.4cm]{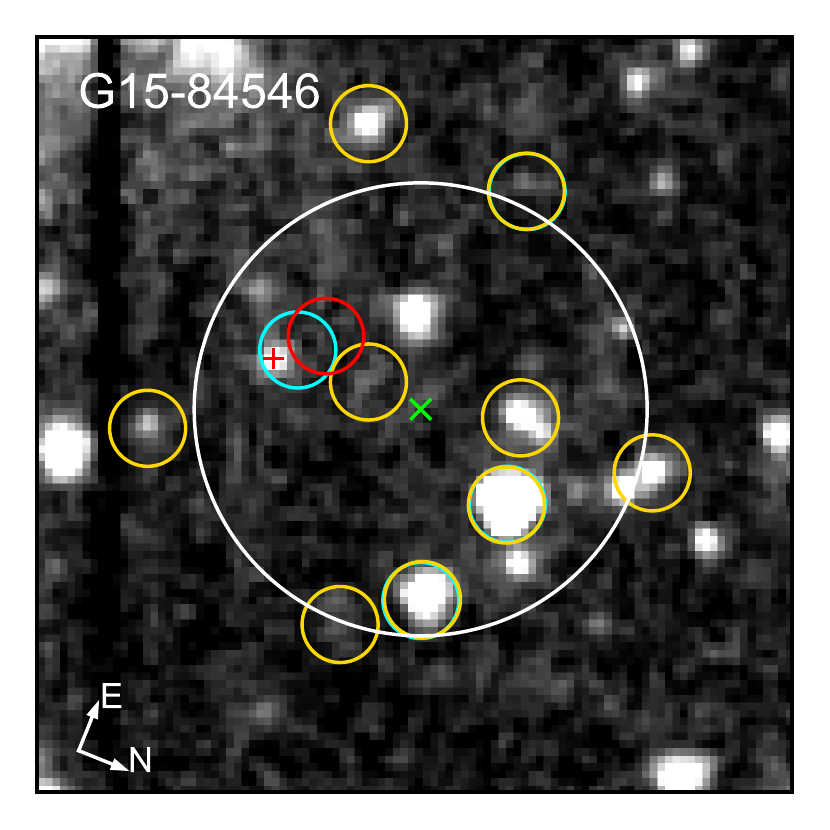}}
{\includegraphics[width=4.4cm, height=4.4cm]{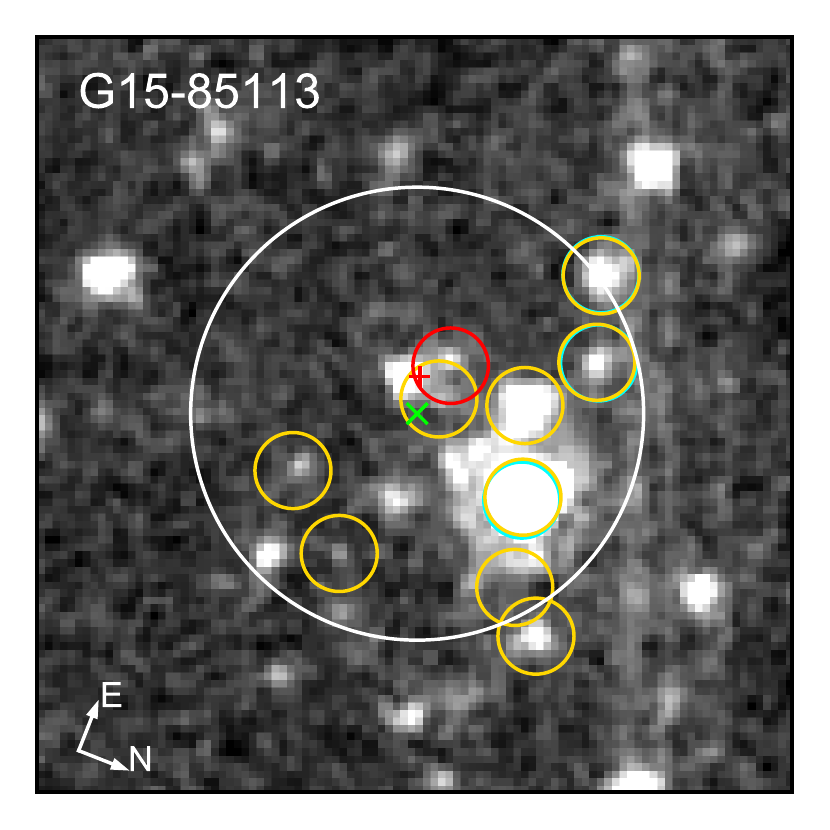}}
{\includegraphics[width=4.4cm, height=4.4cm]{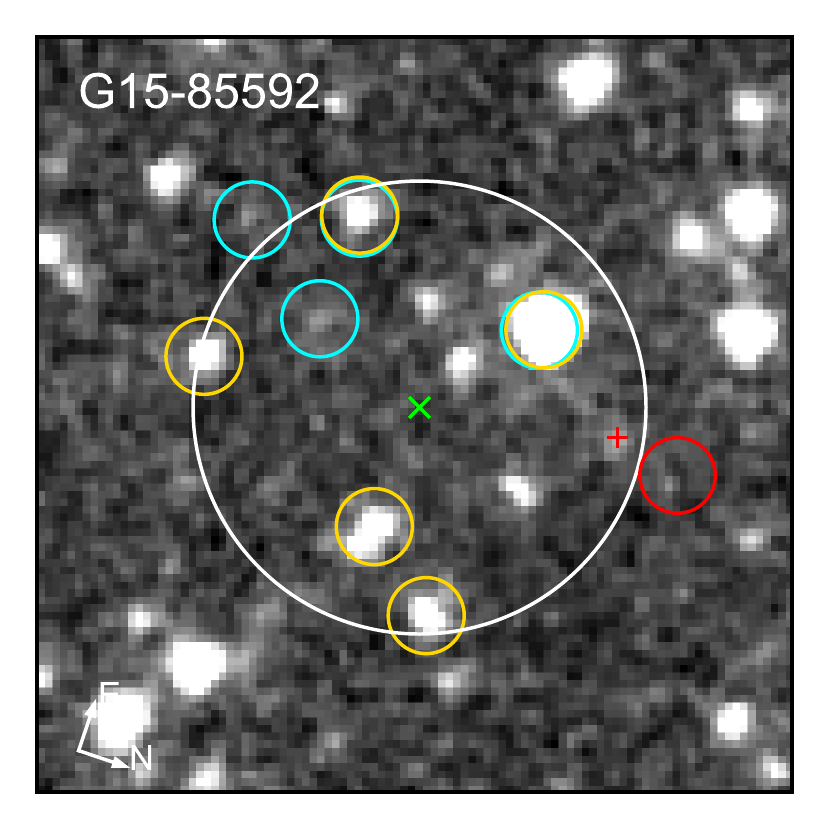}}
{\includegraphics[width=4.4cm, height=4.4cm]{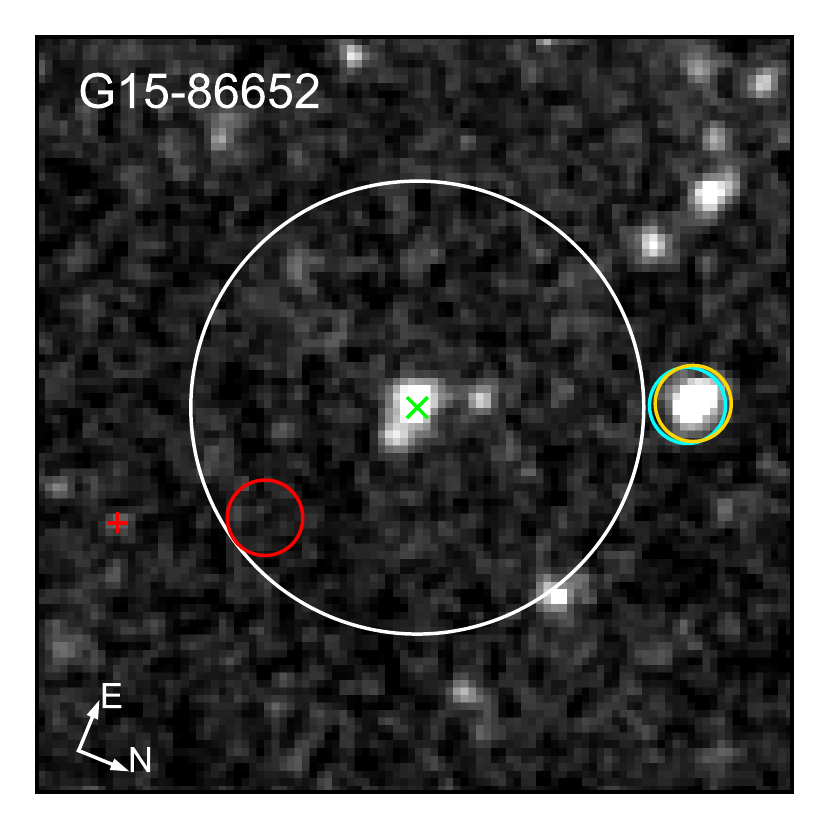}}
{\includegraphics[width=4.4cm, height=4.4cm]{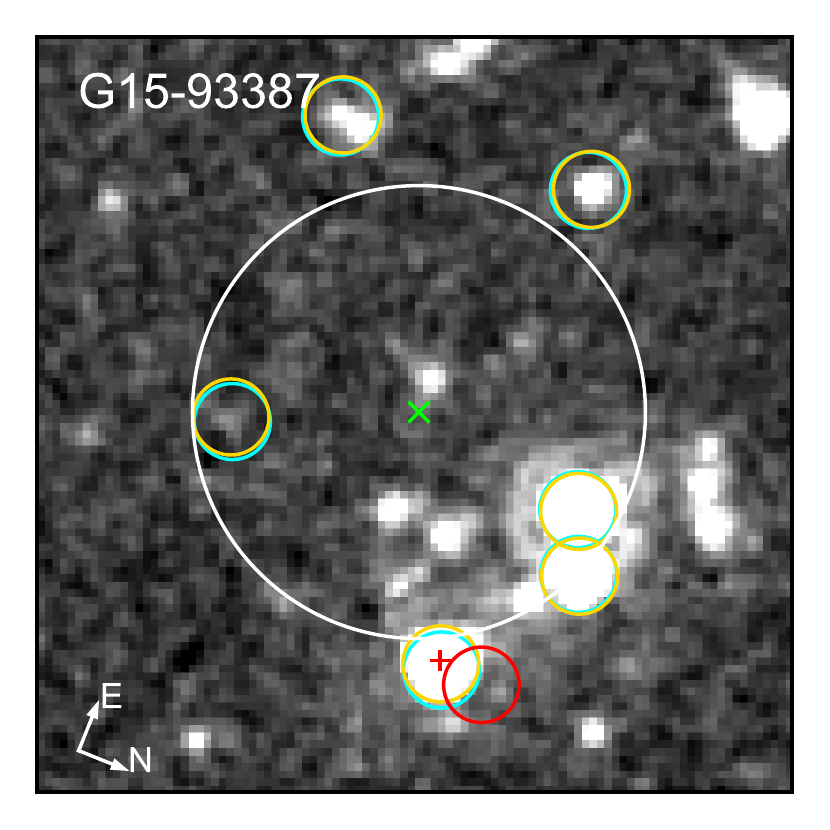}}
{\includegraphics[width=4.4cm, height=4.4cm]{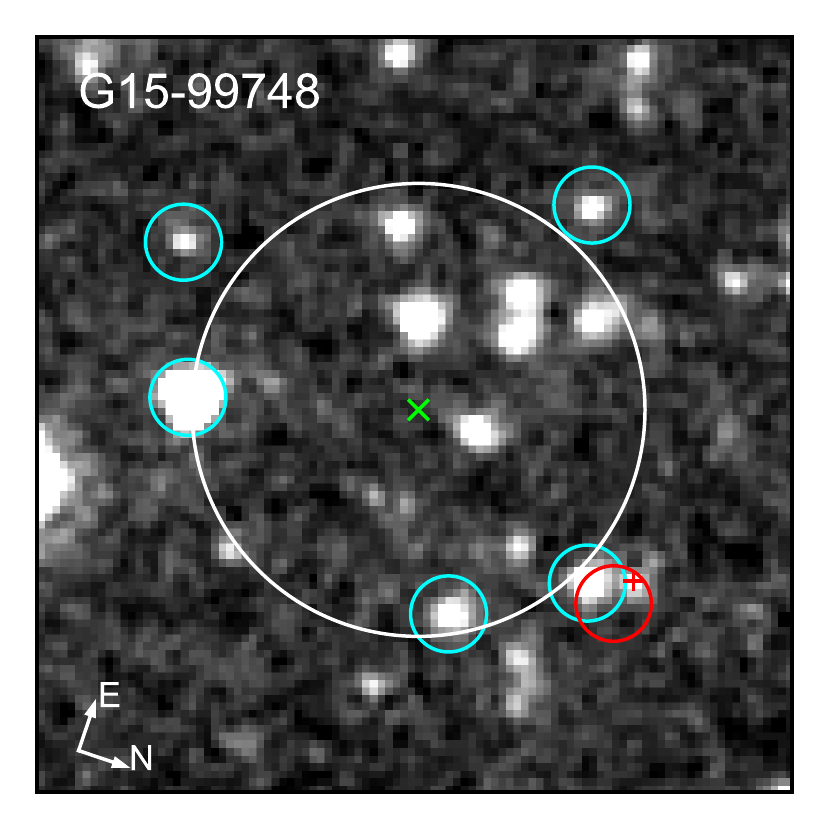}}
{\includegraphics[width=4.4cm, height=4.4cm]{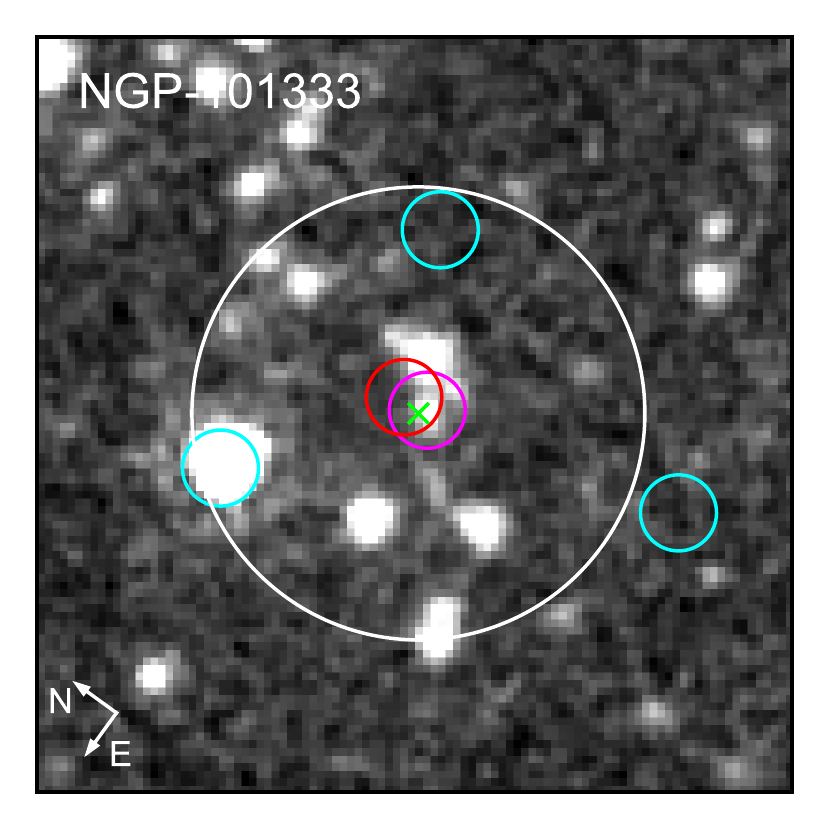}}
{\includegraphics[width=4.4cm, height=4.4cm]{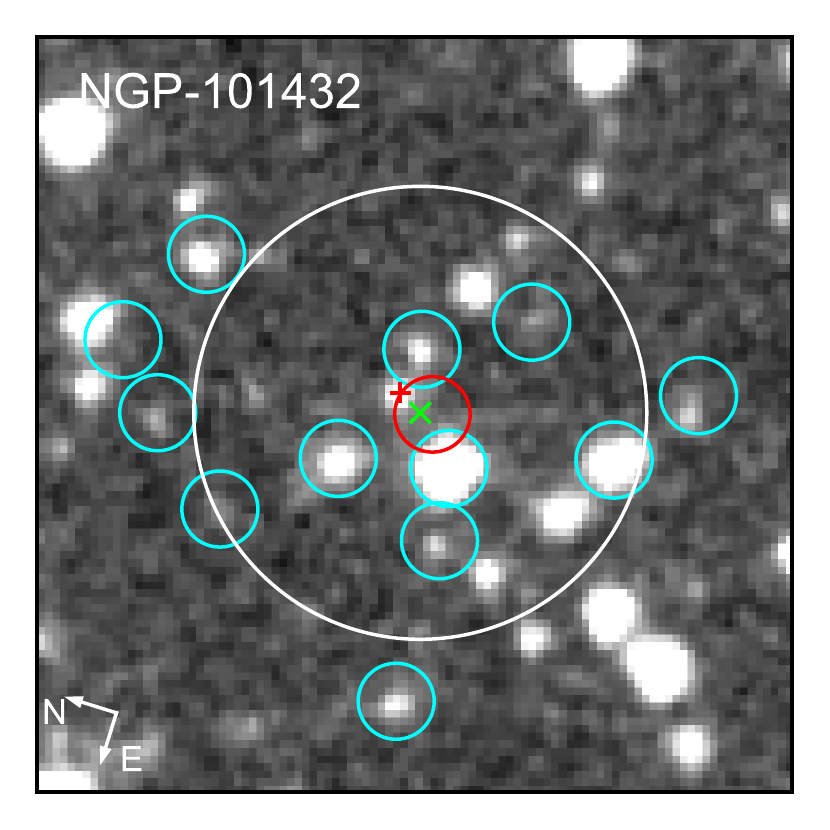}}
{\includegraphics[width=4.4cm, height=4.4cm]{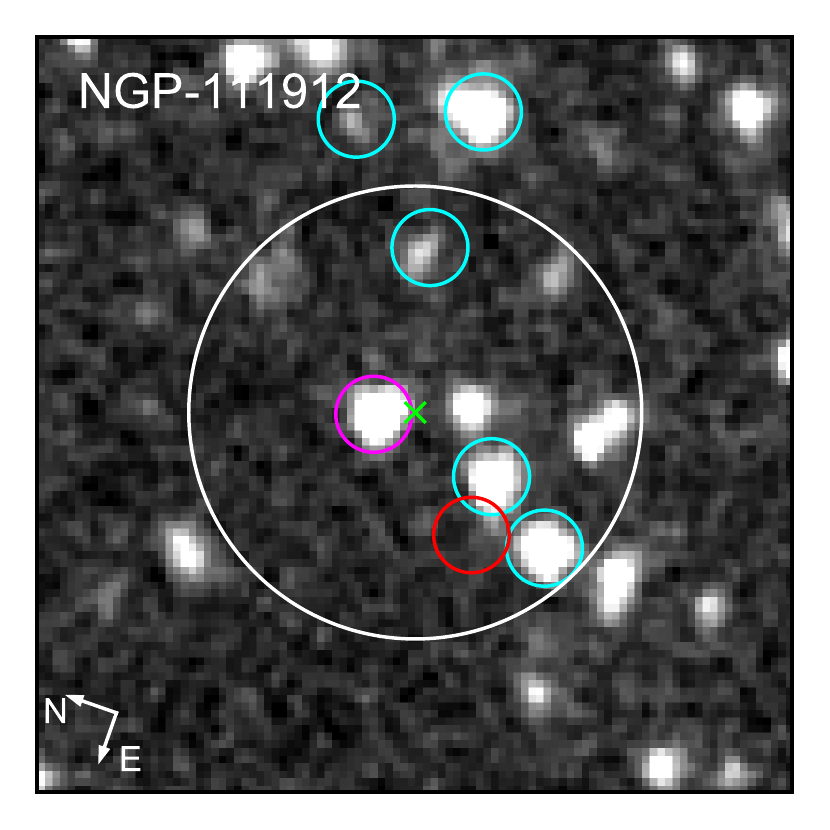}}
{\includegraphics[width=4.4cm, height=4.4cm]{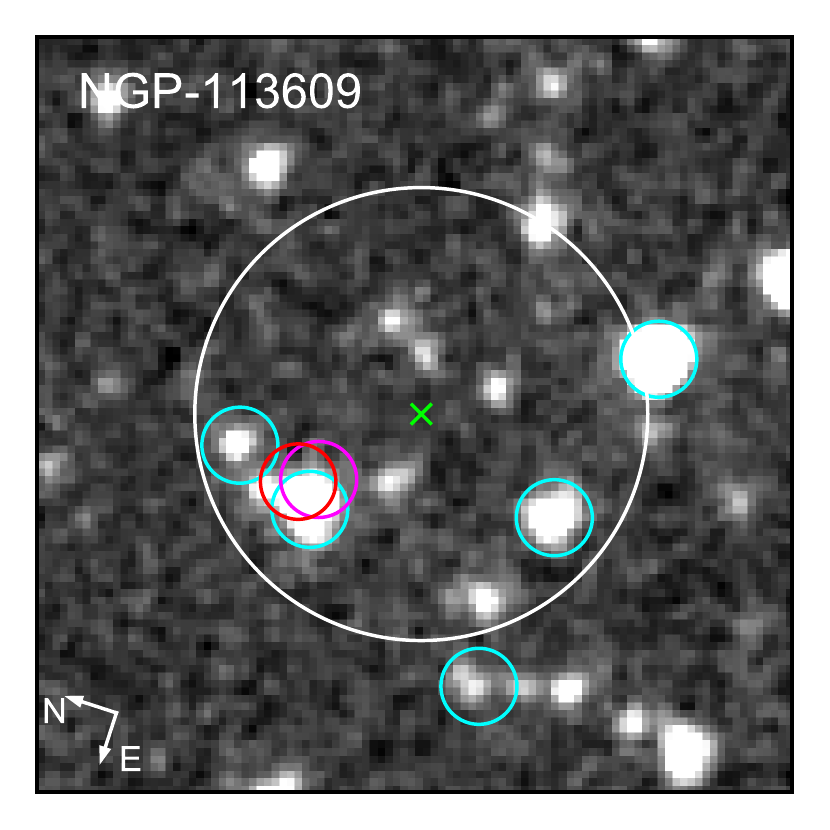}}
{\includegraphics[width=4.4cm, height=4.4cm]{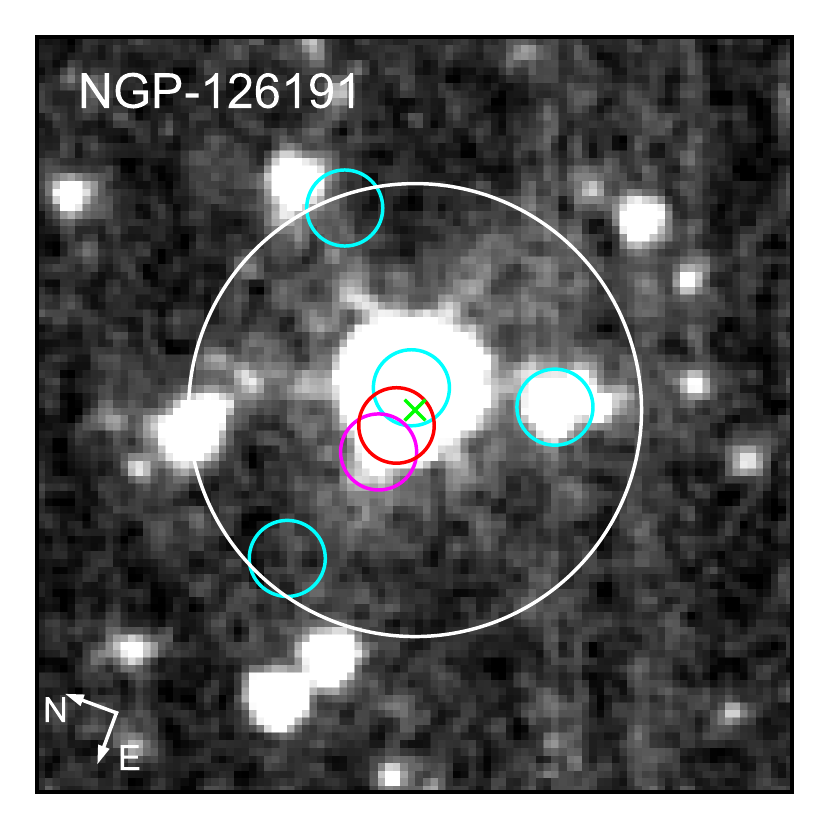}}
{\includegraphics[width=4.4cm, height=4.4cm]{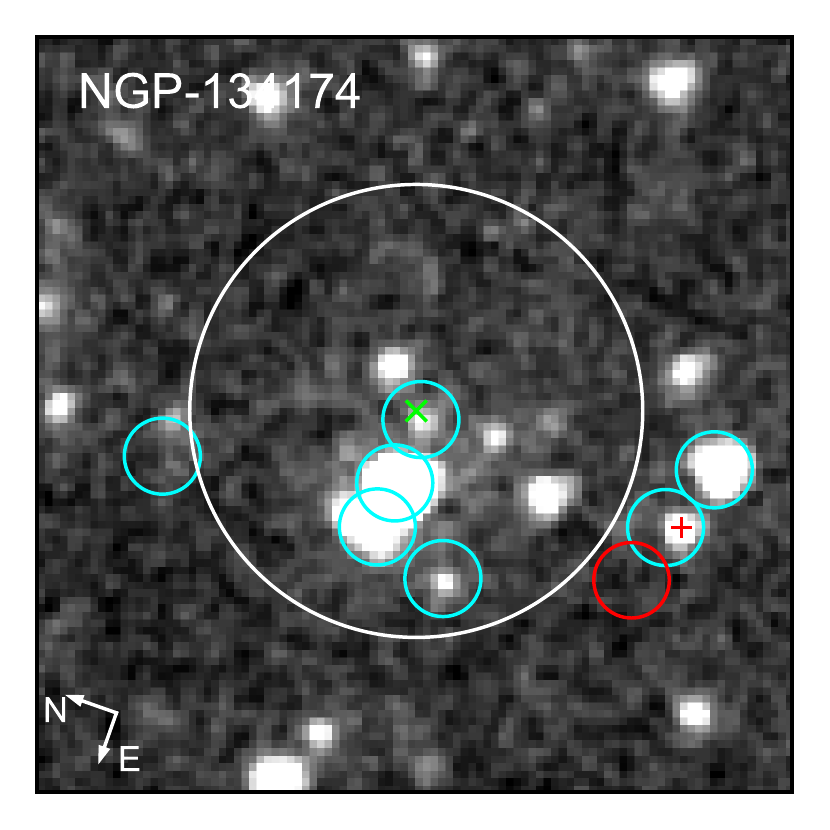}}
{\includegraphics[width=4.4cm, height=4.4cm]{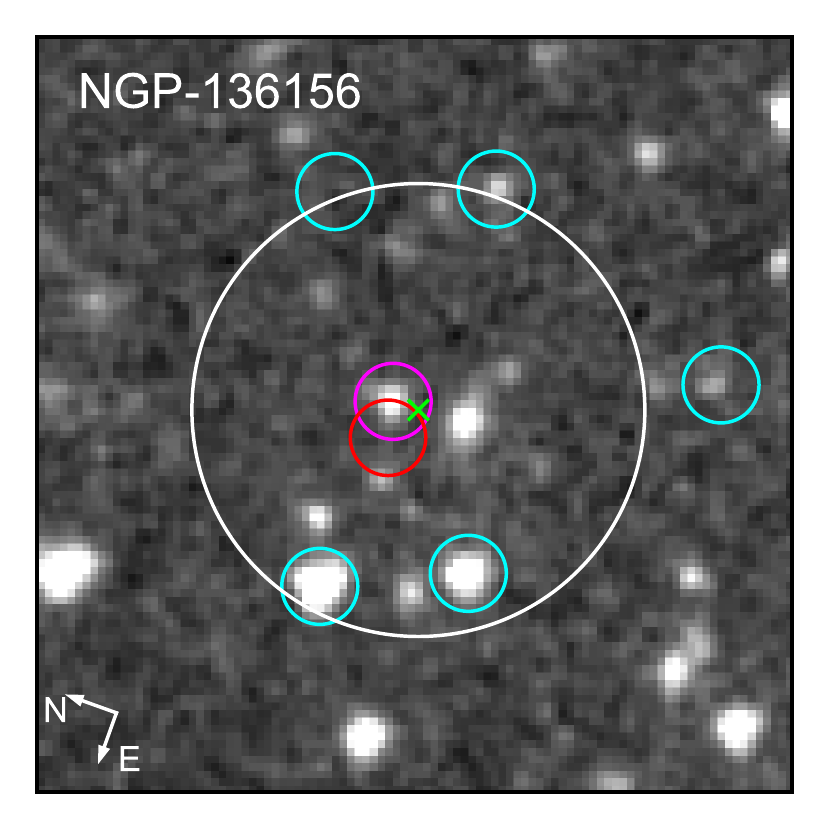}}
{\includegraphics[width=4.4cm, height=4.4cm]{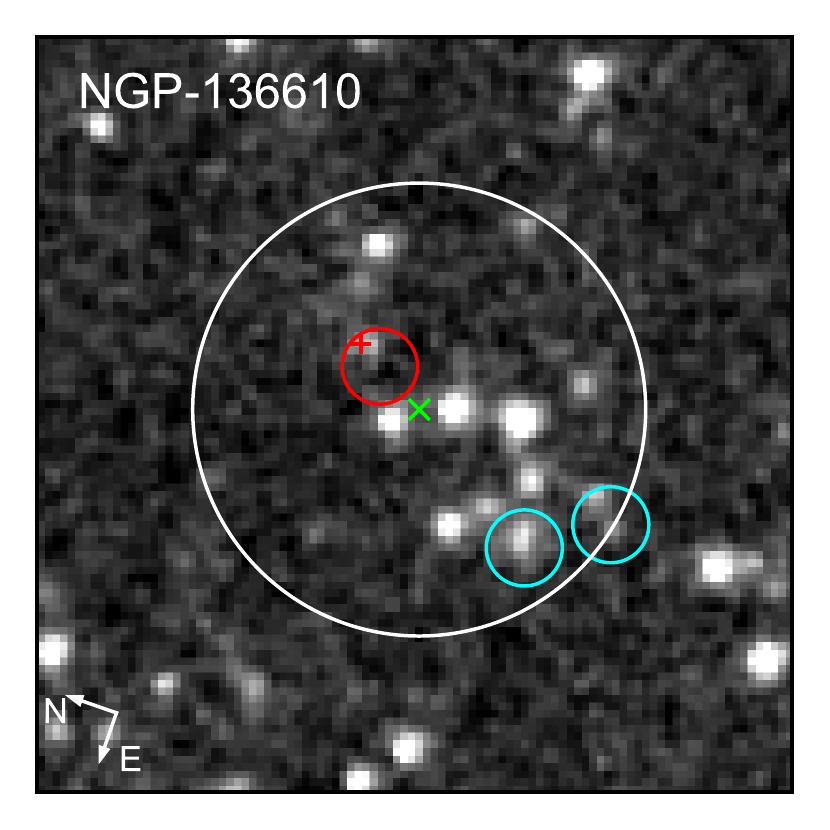}}
{\includegraphics[width=4.4cm, height=4.4cm]{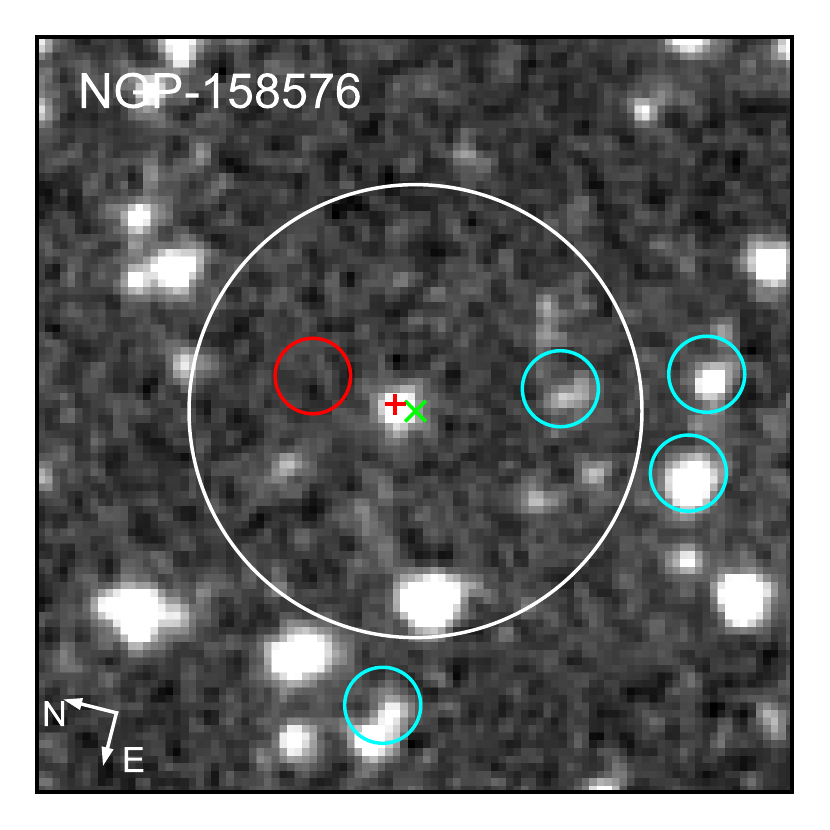}}
{\includegraphics[width=4.4cm, height=4.4cm]{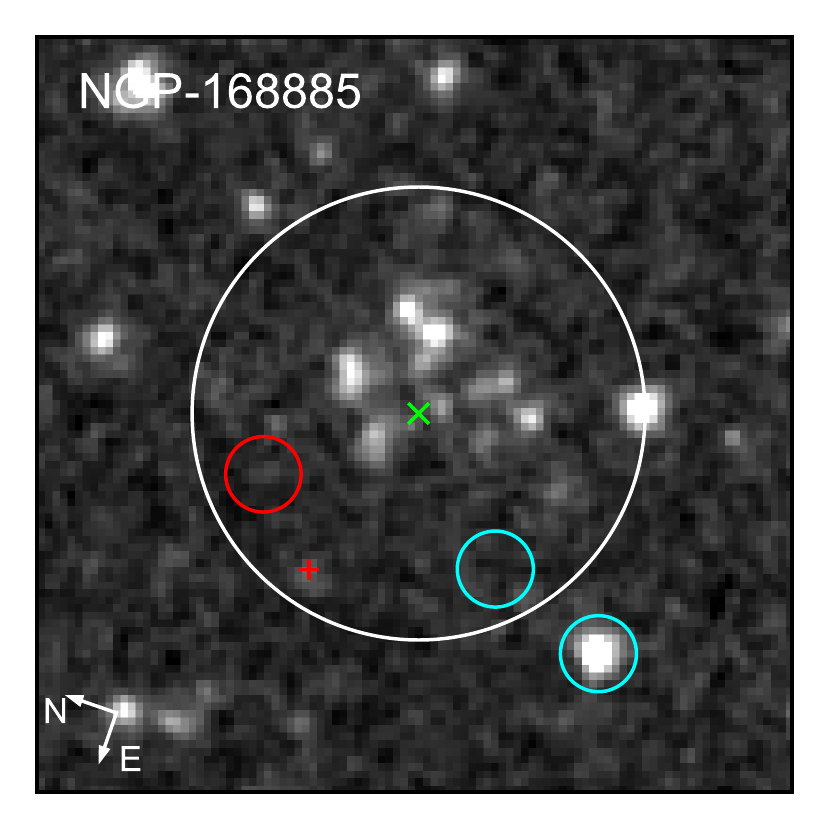}}
{\includegraphics[width=4.4cm, height=4.4cm]{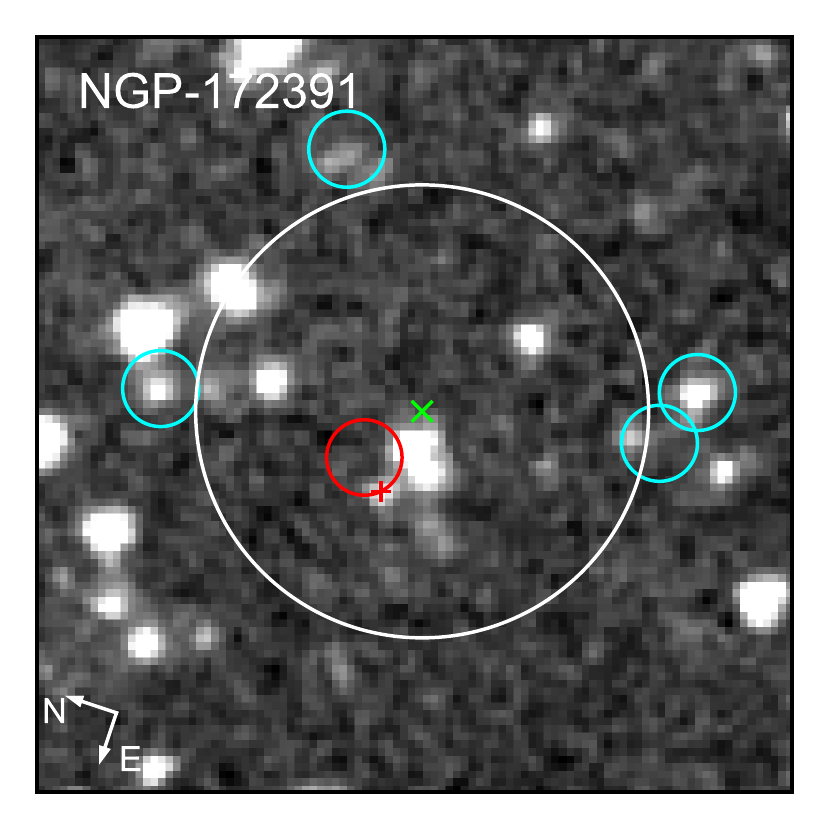}}
{\includegraphics[width=4.4cm, height=4.4cm]{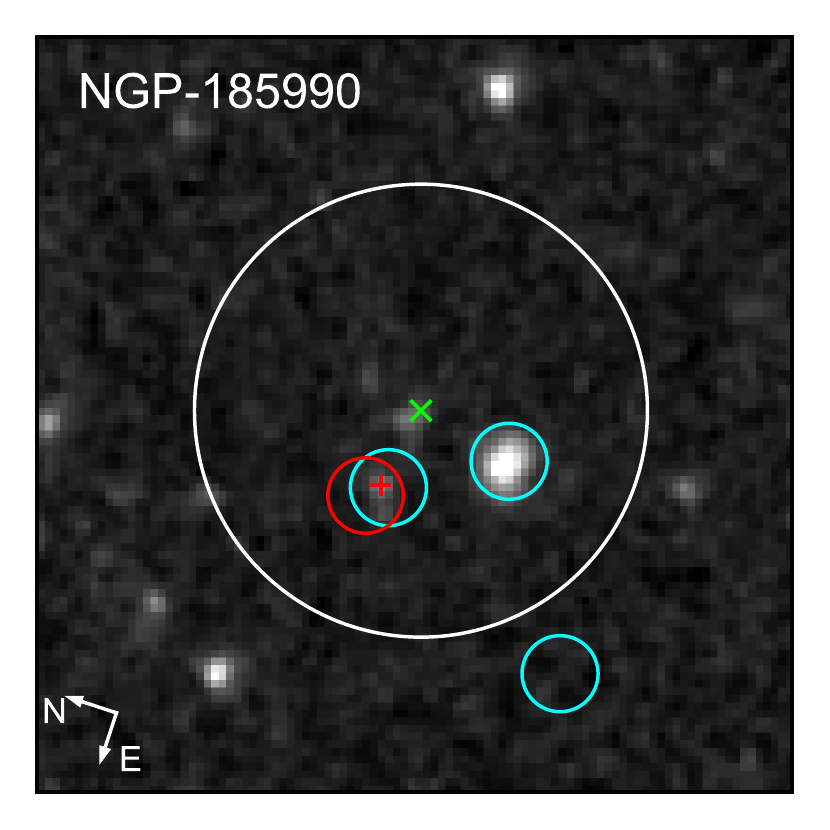}}
{\includegraphics[width=4.4cm, height=4.4cm]{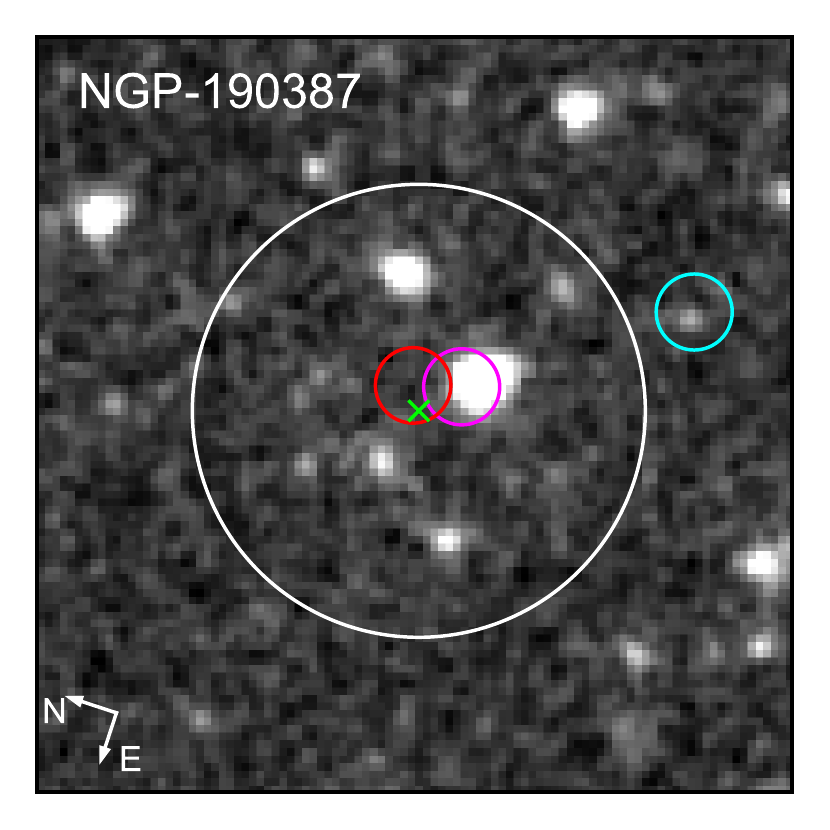}}
\caption{Continued 60$\arcsec$ $\times$ 60$\arcsec$ cutouts }
\label{fig:data}
\end{figure*}

\addtocounter{figure}{-1}
\begin{figure*}
\centering
{\includegraphics[width=4.4cm, height=4.4cm]{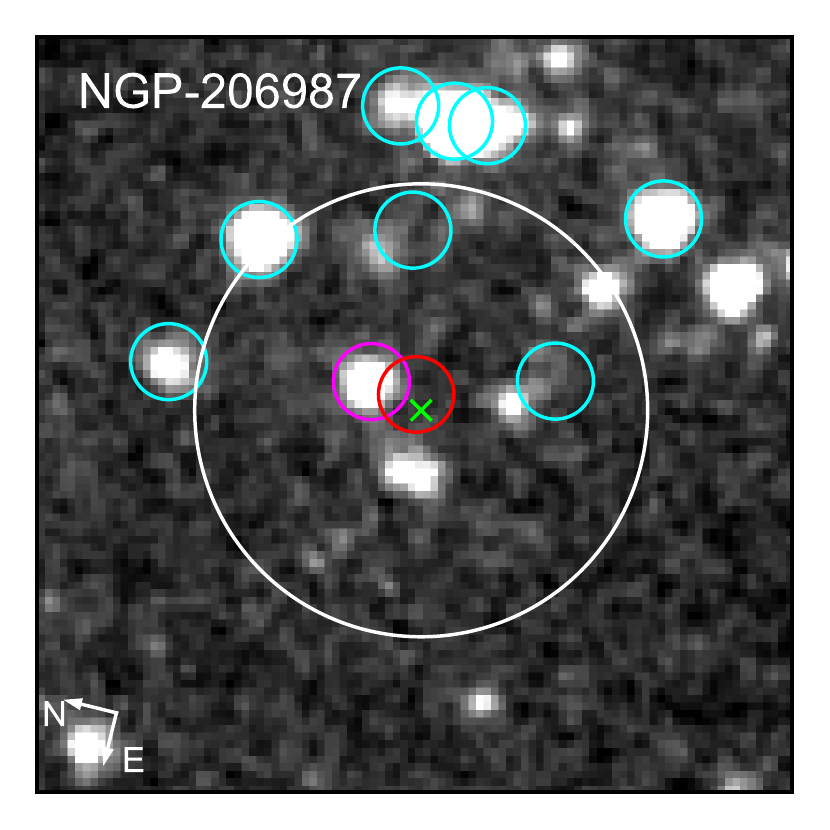}}
{\includegraphics[width=4.4cm, height=4.4cm]{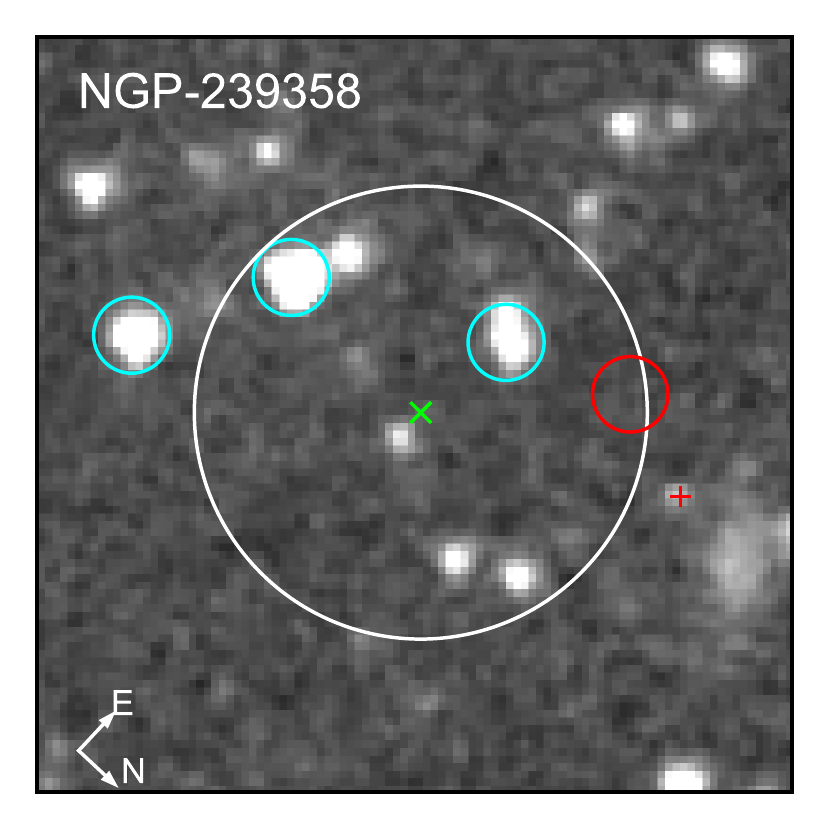}}
{\includegraphics[width=4.4cm, height=4.4cm]{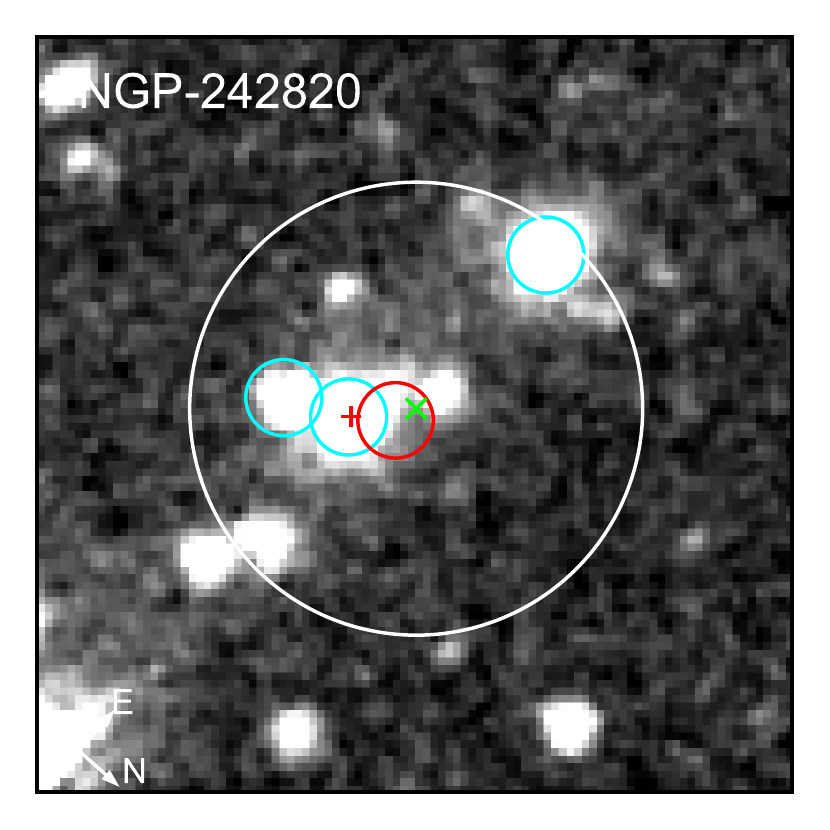}}
{\includegraphics[width=4.4cm, height=4.4cm]{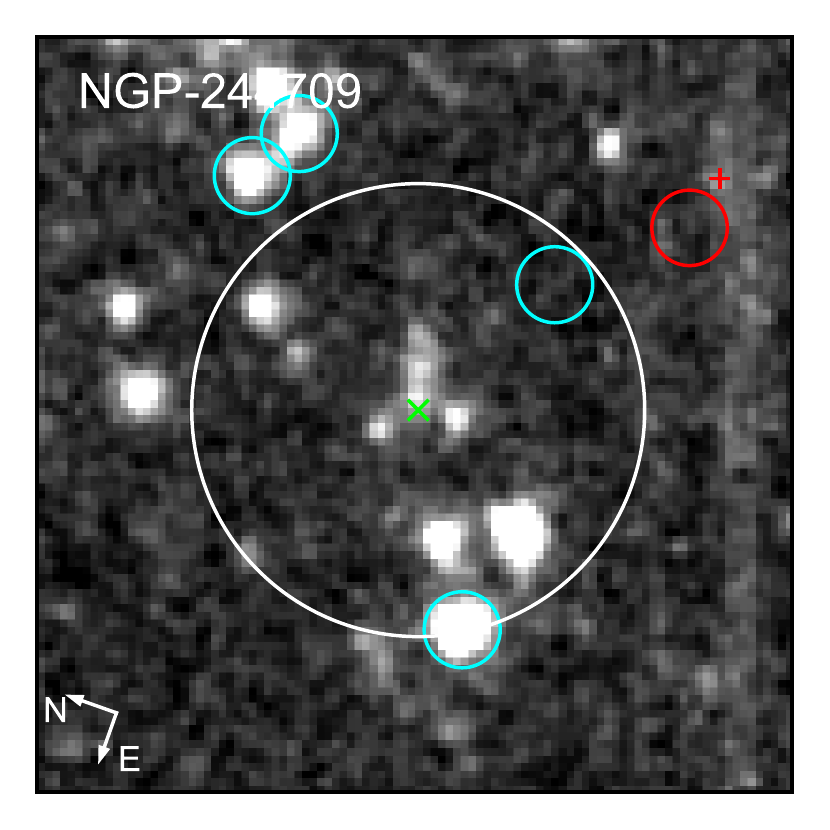}}
{\includegraphics[width=4.4cm, height=4.4cm]{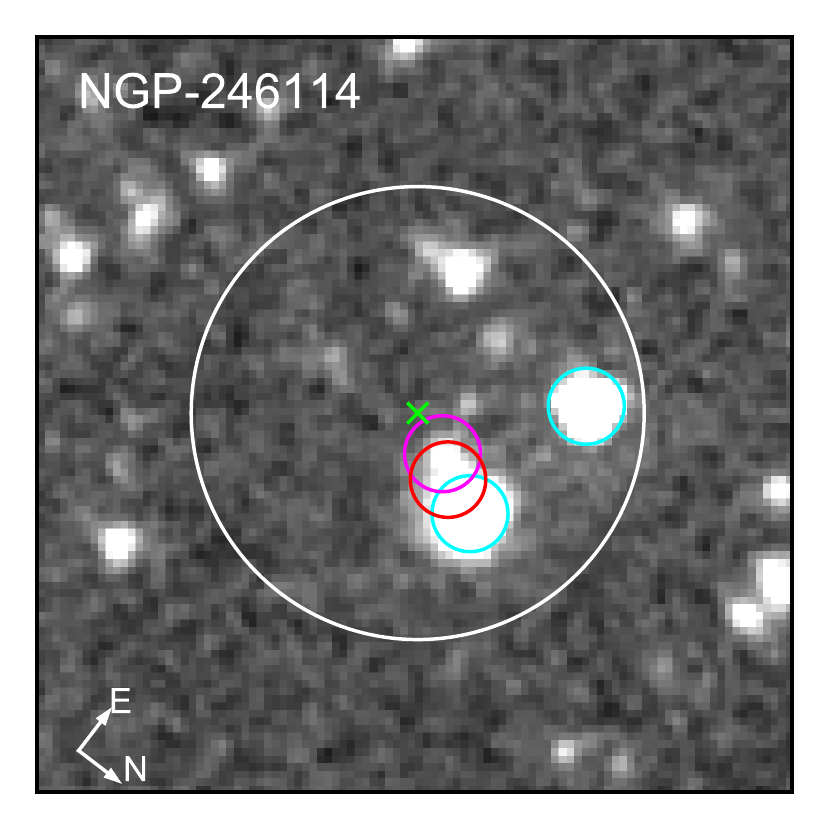}}
{\includegraphics[width=4.4cm, height=4.4cm]{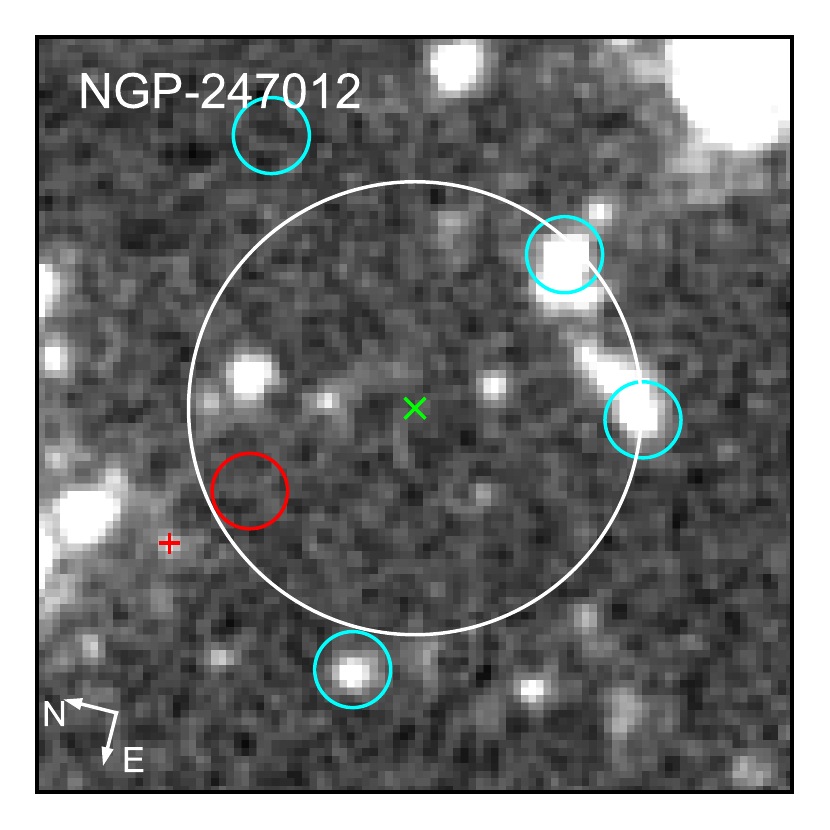}}
{\includegraphics[width=4.4cm, height=4.4cm]{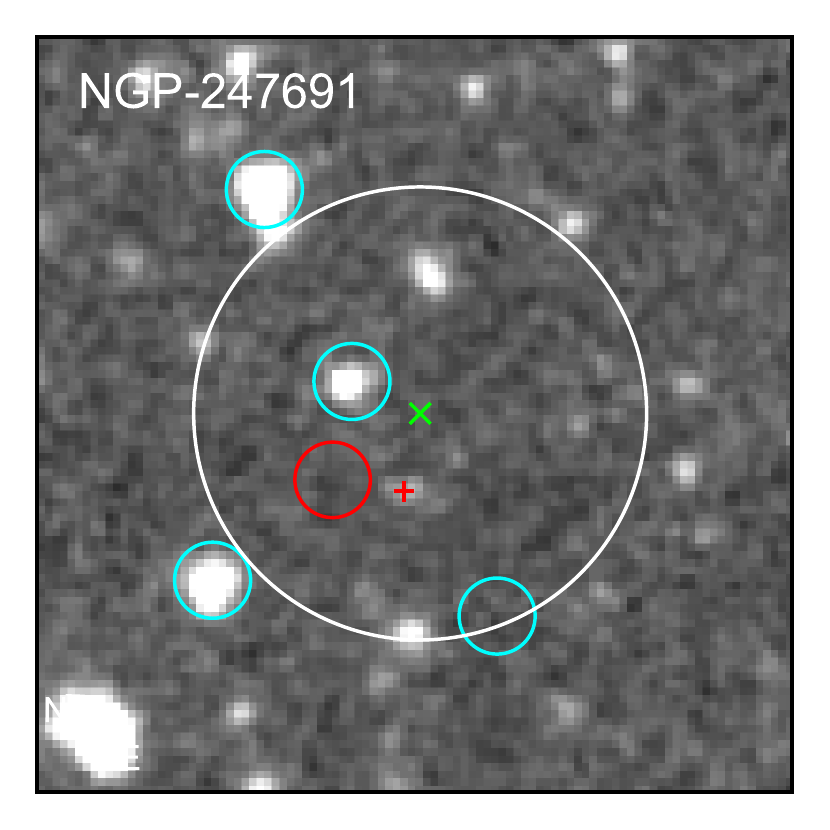}}
{\includegraphics[width=4.4cm, height=4.4cm]{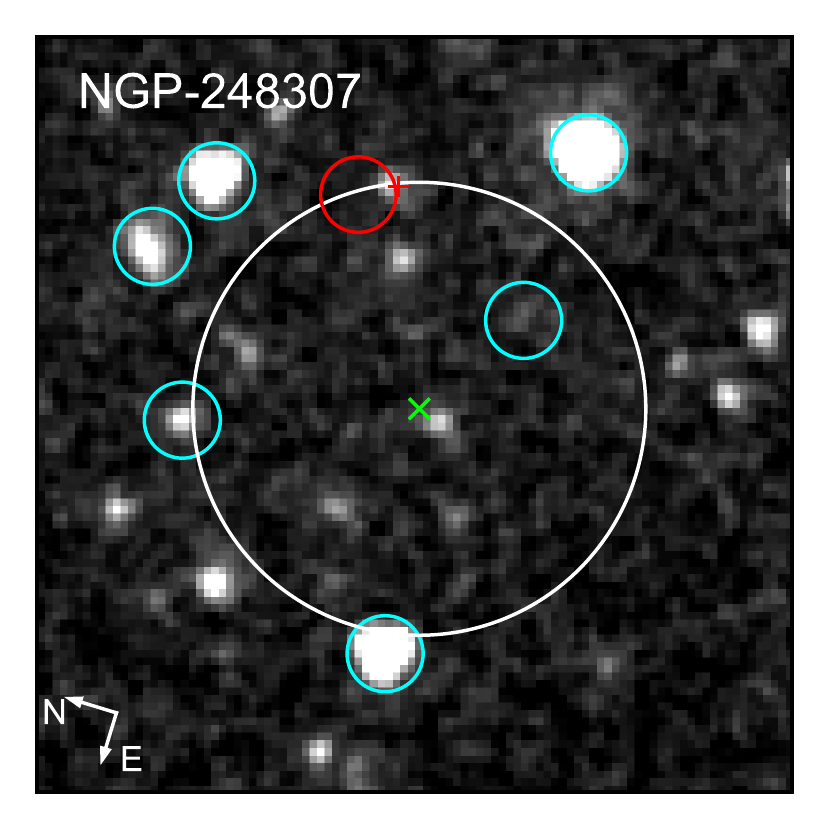}}
{\includegraphics[width=4.4cm, height=4.4cm]{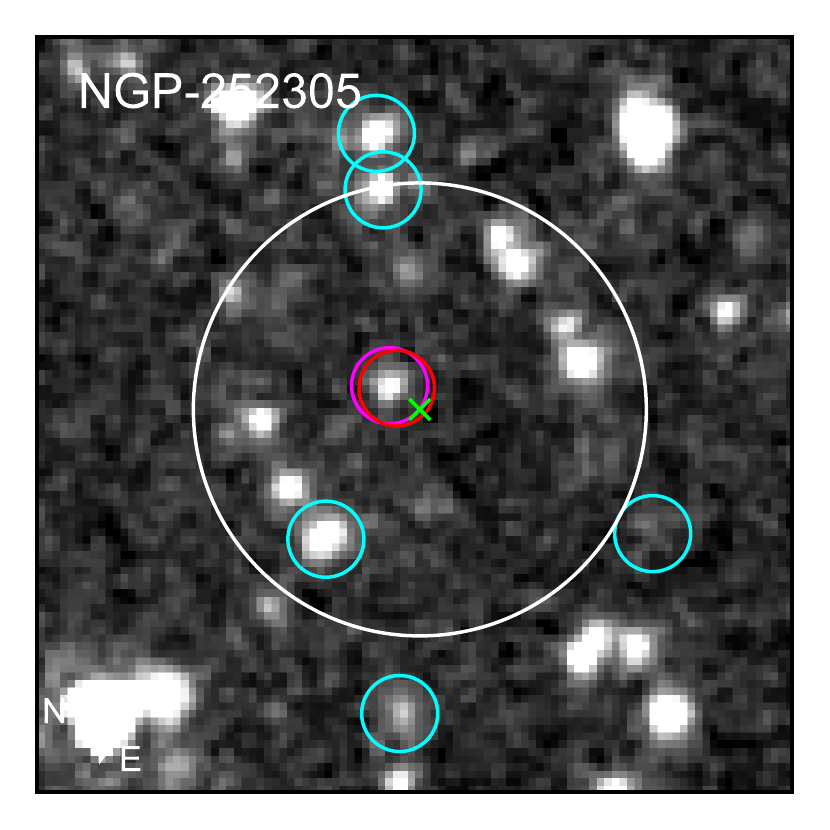}}
{\includegraphics[width=4.4cm, height=4.4cm]{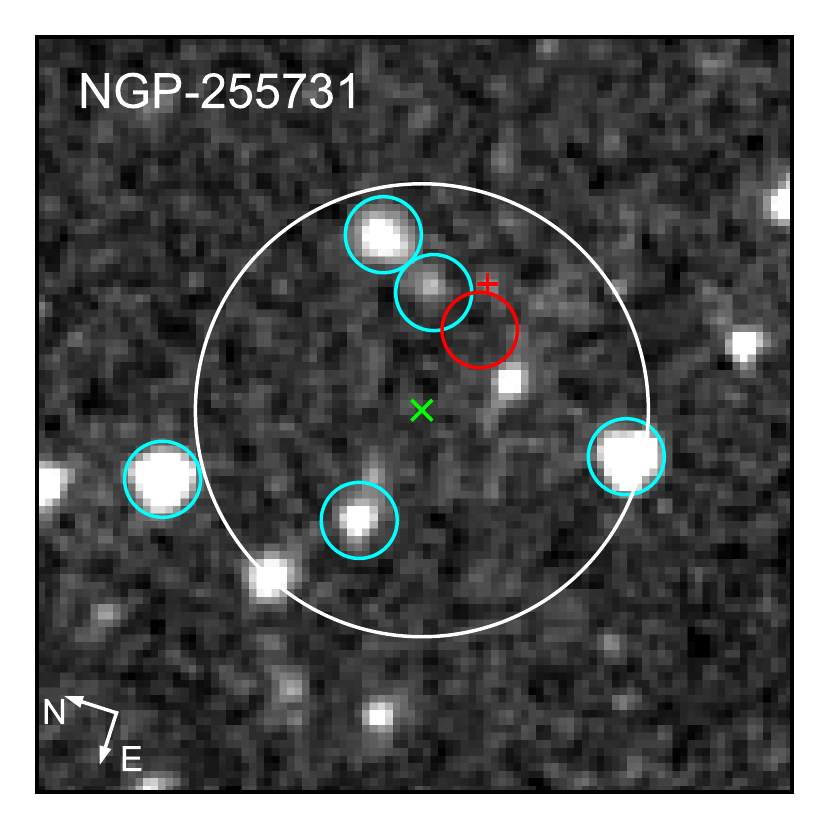}}
{\includegraphics[width=4.4cm, height=4.4cm]{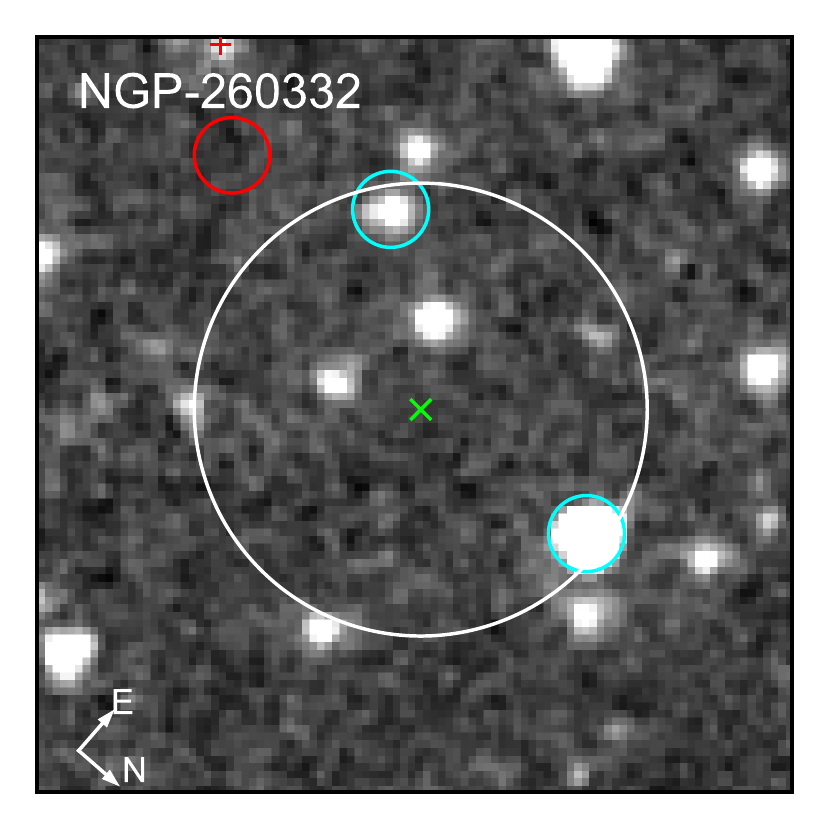}}
{\includegraphics[width=4.4cm, height=4.4cm]{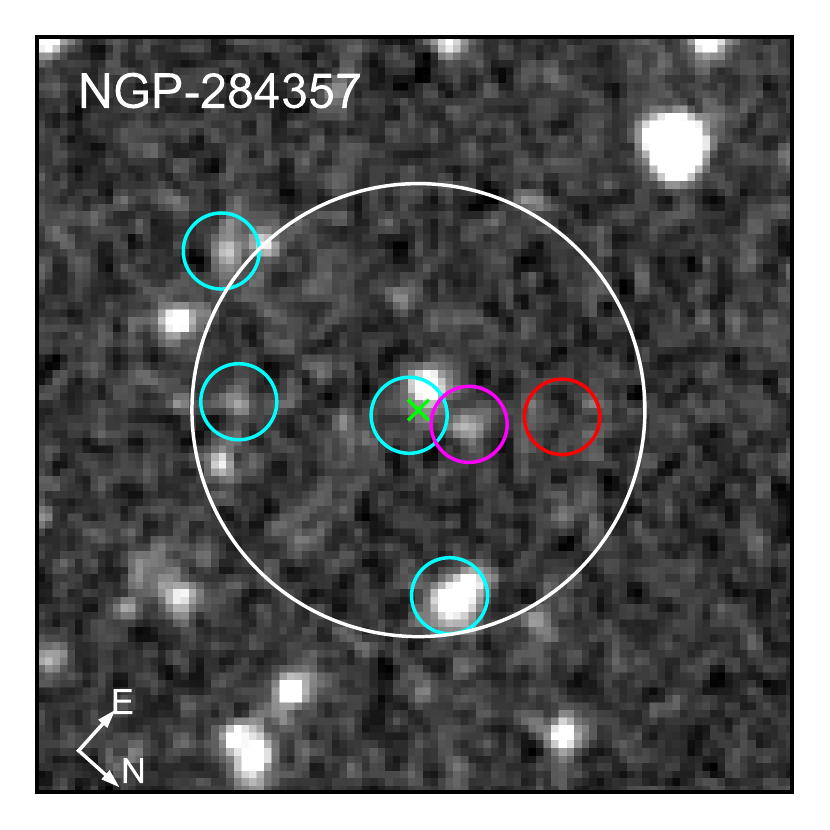}}
{\includegraphics[width=4.4cm, height=4.4cm]{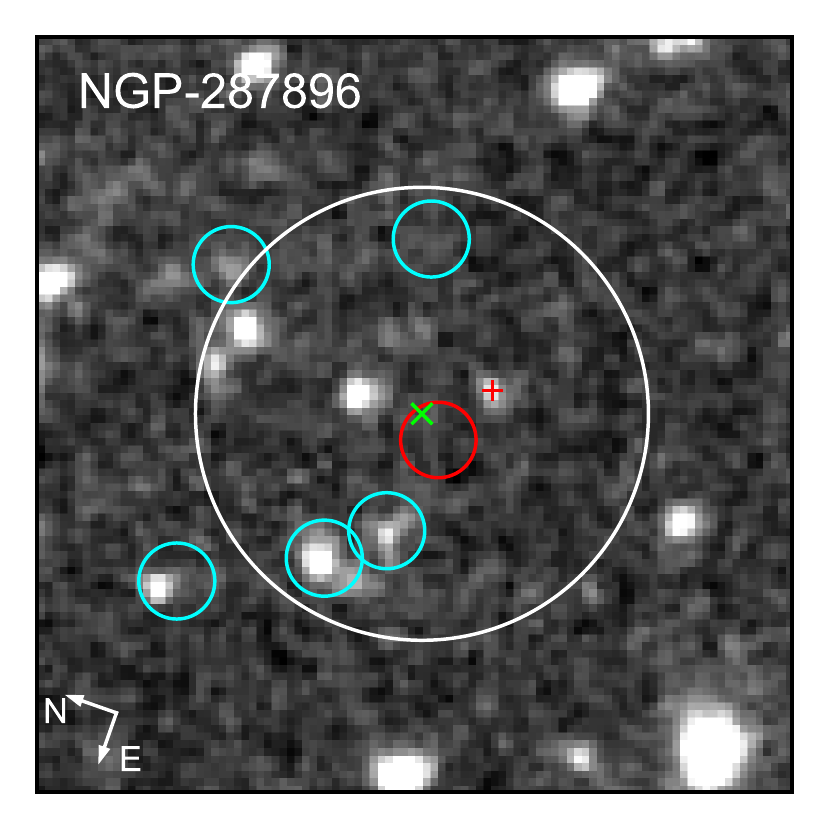}}
{\includegraphics[width=4.4cm, height=4.4cm]{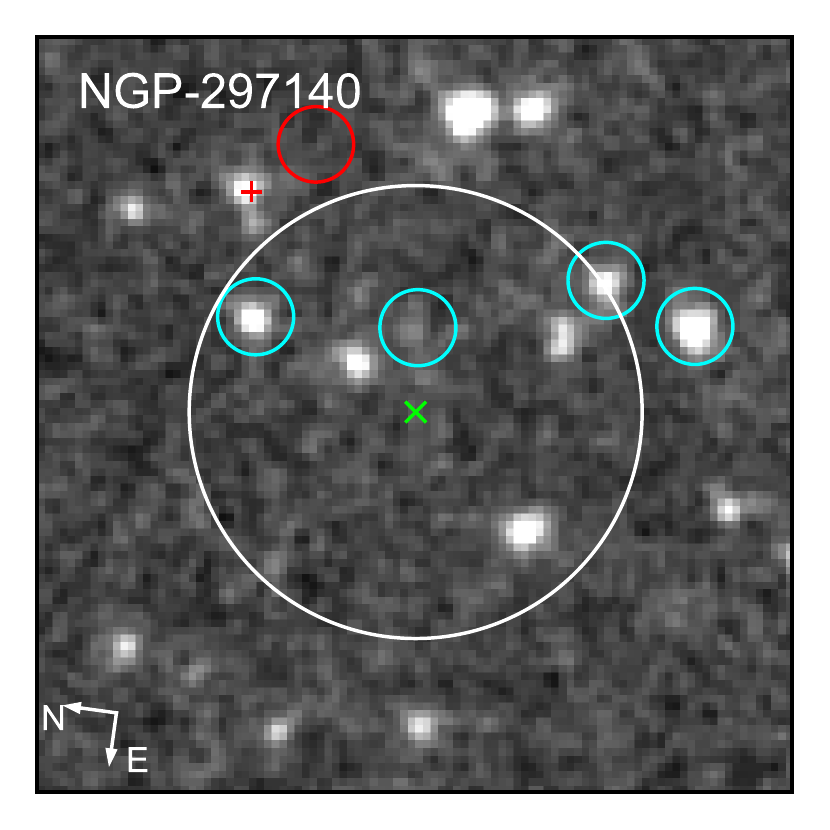}}
{\includegraphics[width=4.4cm, height=4.4cm]{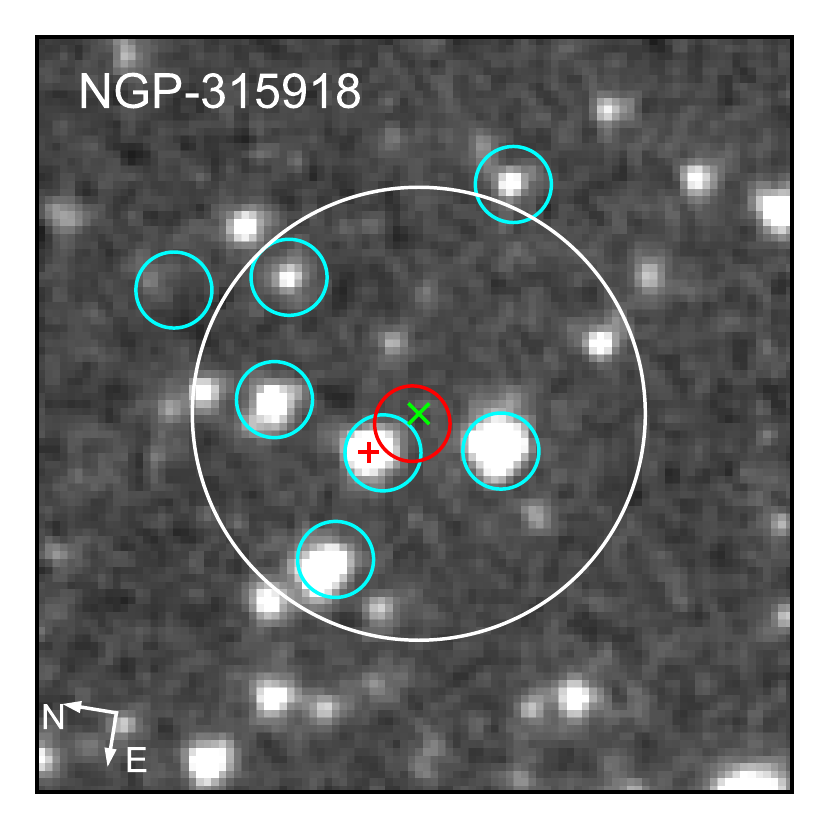}}
{\includegraphics[width=4.4cm, height=4.4cm]{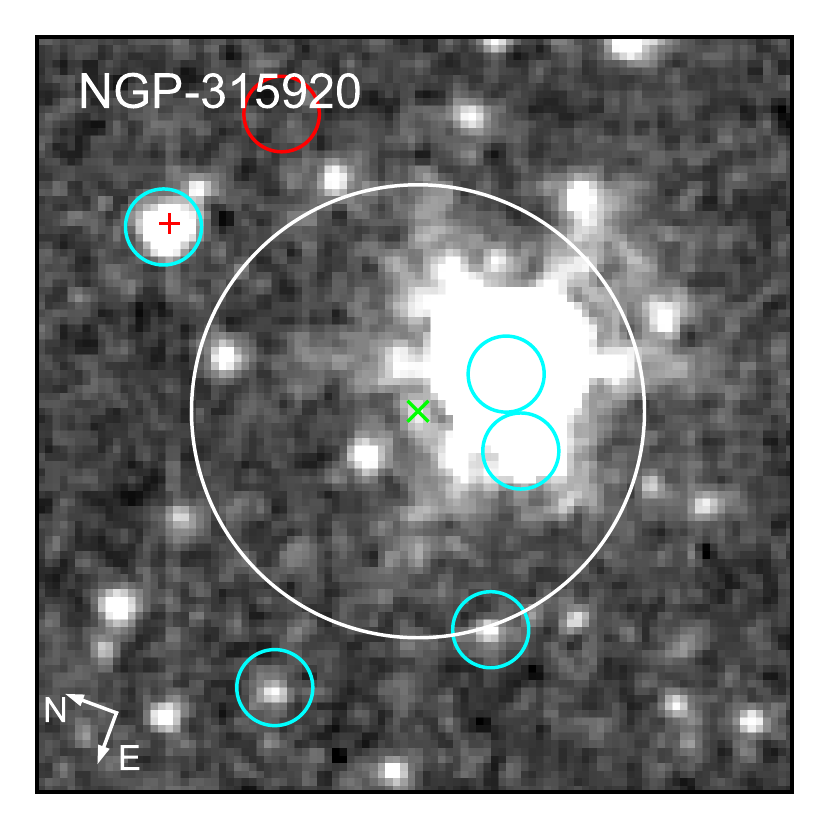}}
{\includegraphics[width=4.4cm, height=4.4cm]{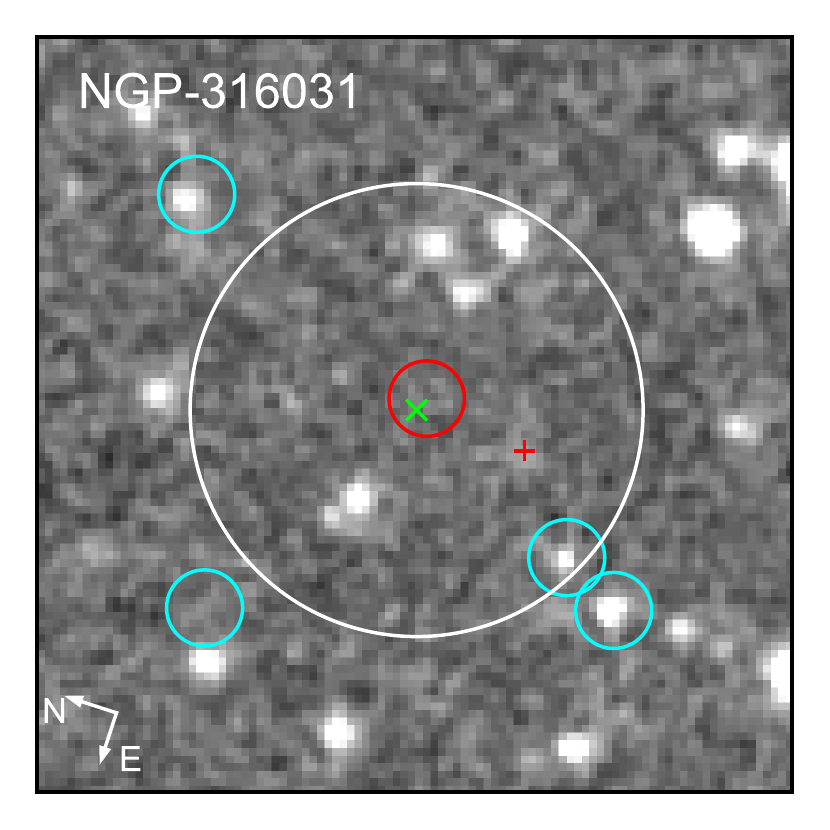}}
{\includegraphics[width=4.4cm, height=4.4cm]{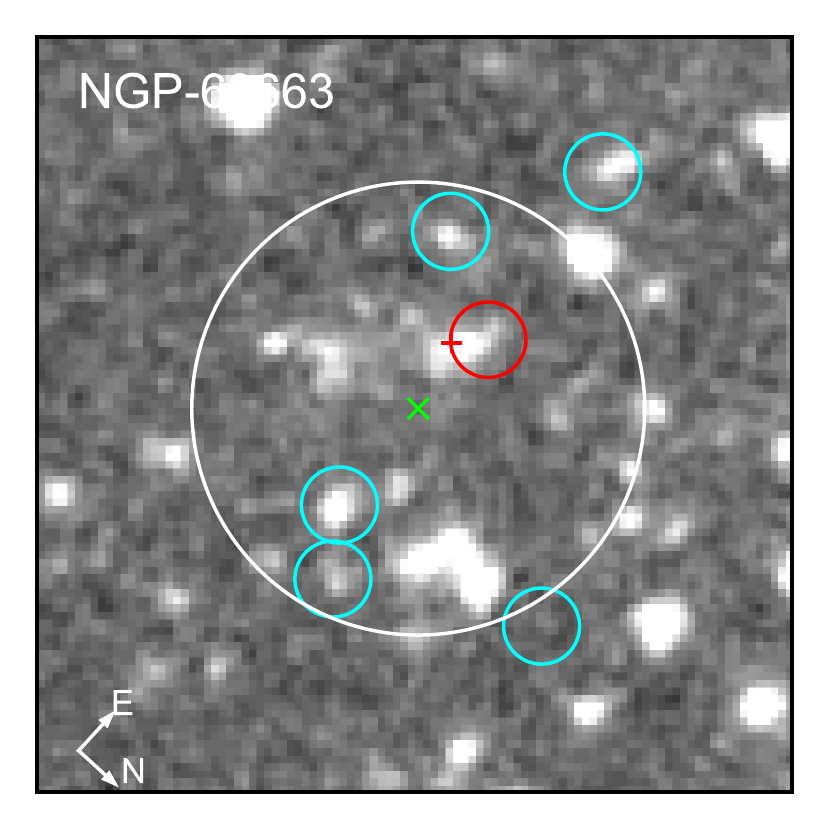}}
{\includegraphics[width=4.4cm, height=4.4cm]{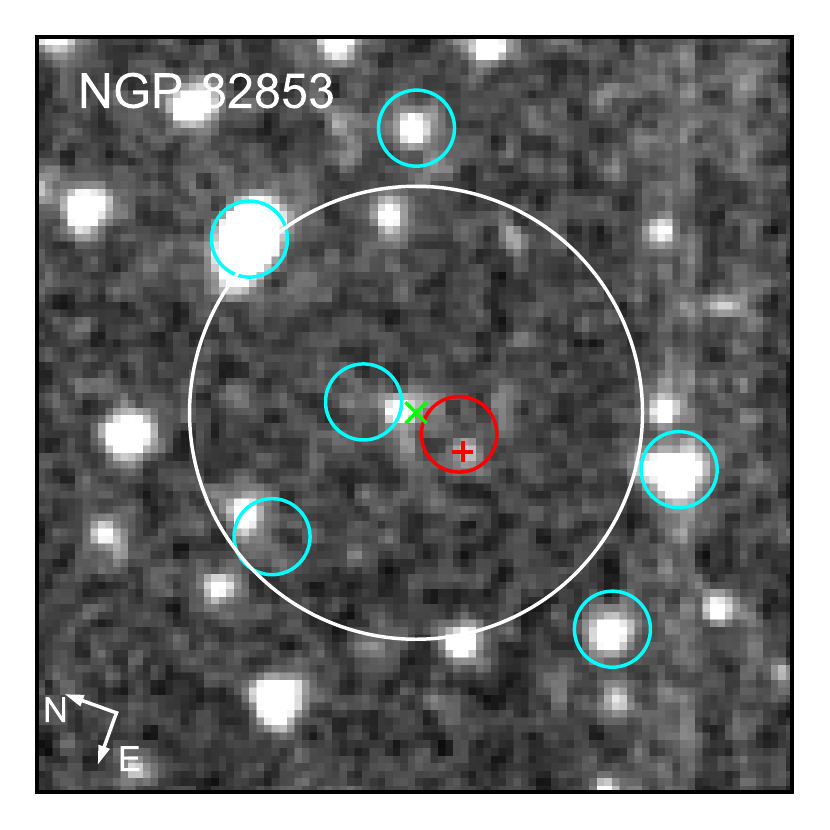}}
{\includegraphics[width=4.4cm, height=4.4cm]{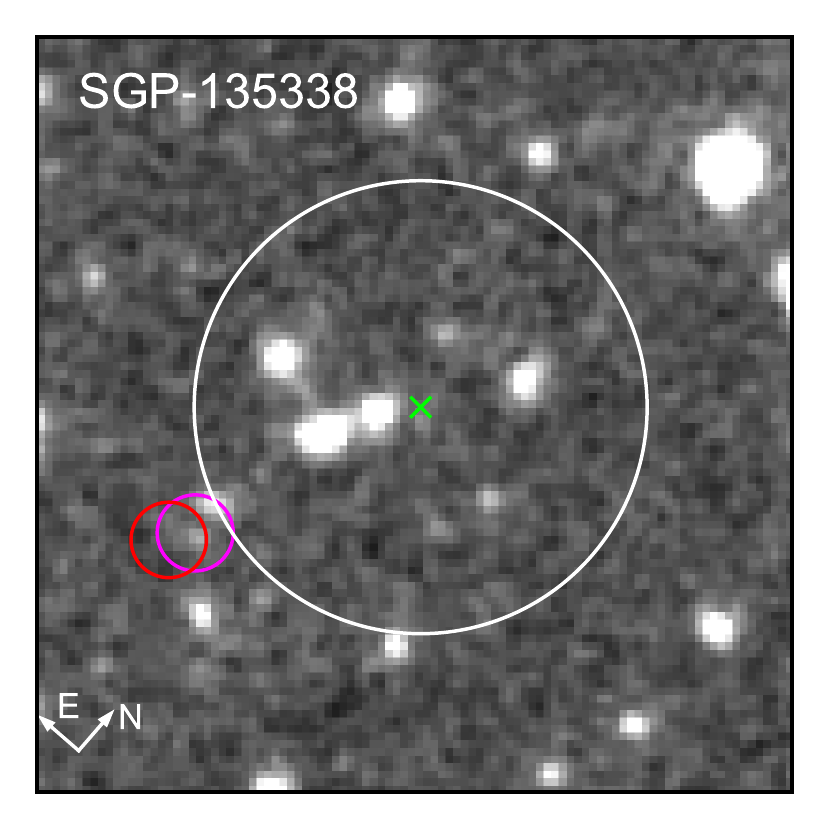}}
\caption{Continued 60$\arcsec$ $\times$ 60$\arcsec$ cutouts }
\label{fig:data}
\end{figure*}

\addtocounter{figure}{-1}
\begin{figure*}
\centering
{\includegraphics[width=4.4cm, height=4.4cm]{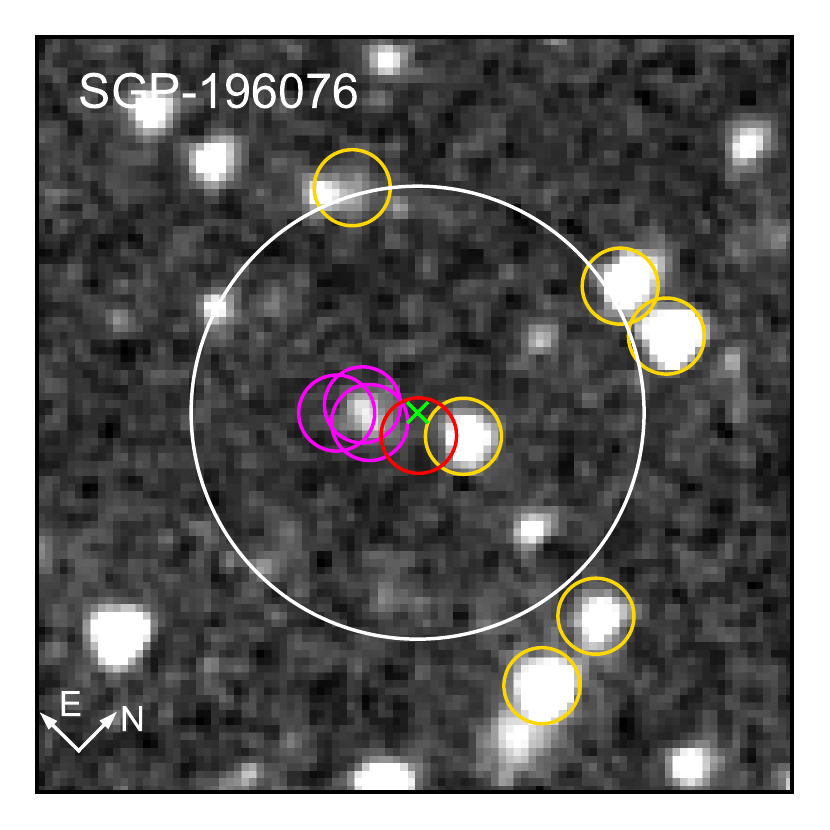}}
{\includegraphics[width=4.4cm, height=4.4cm]{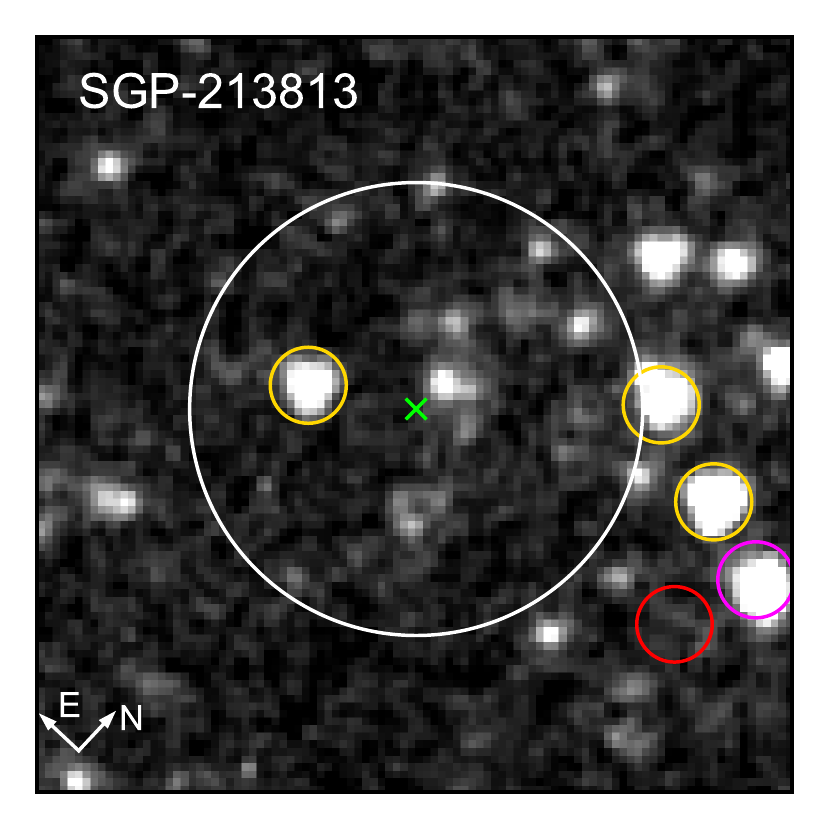}}
{\includegraphics[width=4.4cm, height=4.4cm]{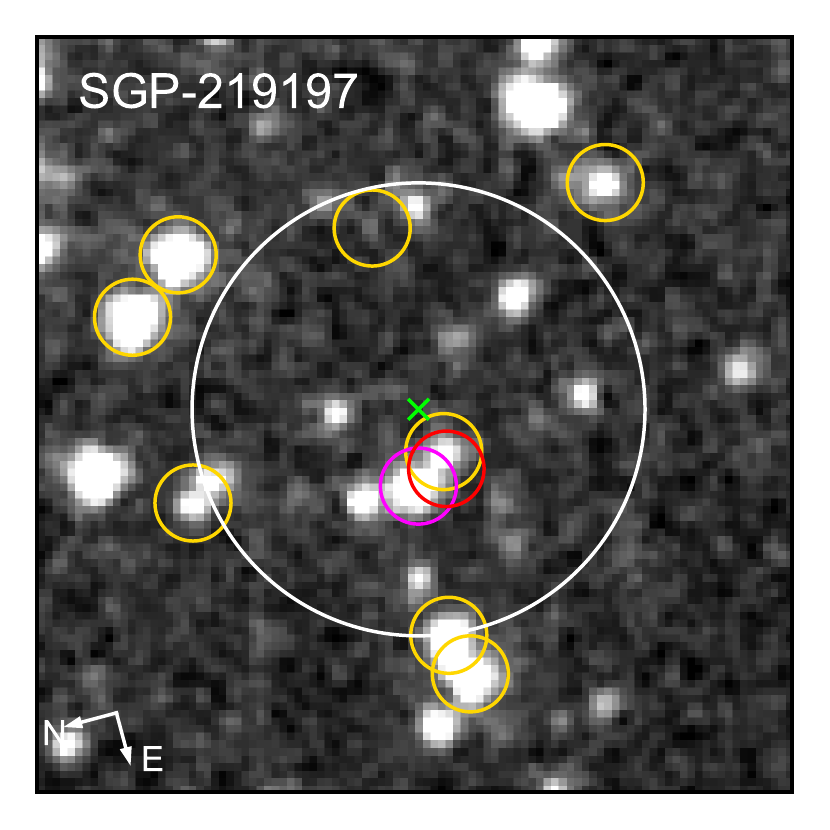}}
{\includegraphics[width=4.4cm, height=4.4cm]{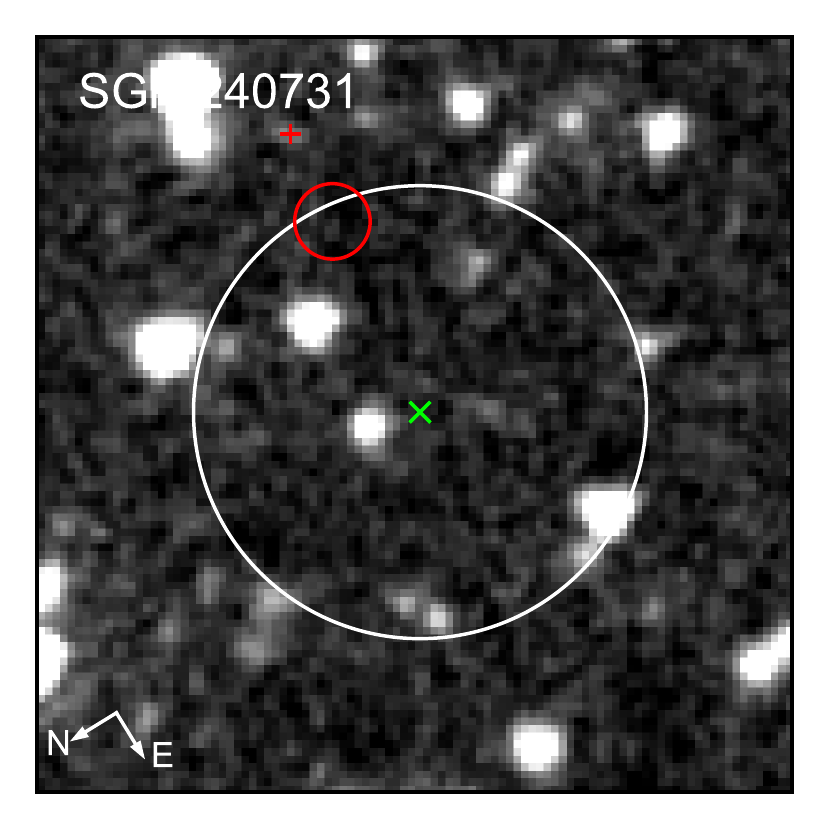}}
{\includegraphics[width=4.4cm, height=4.4cm]{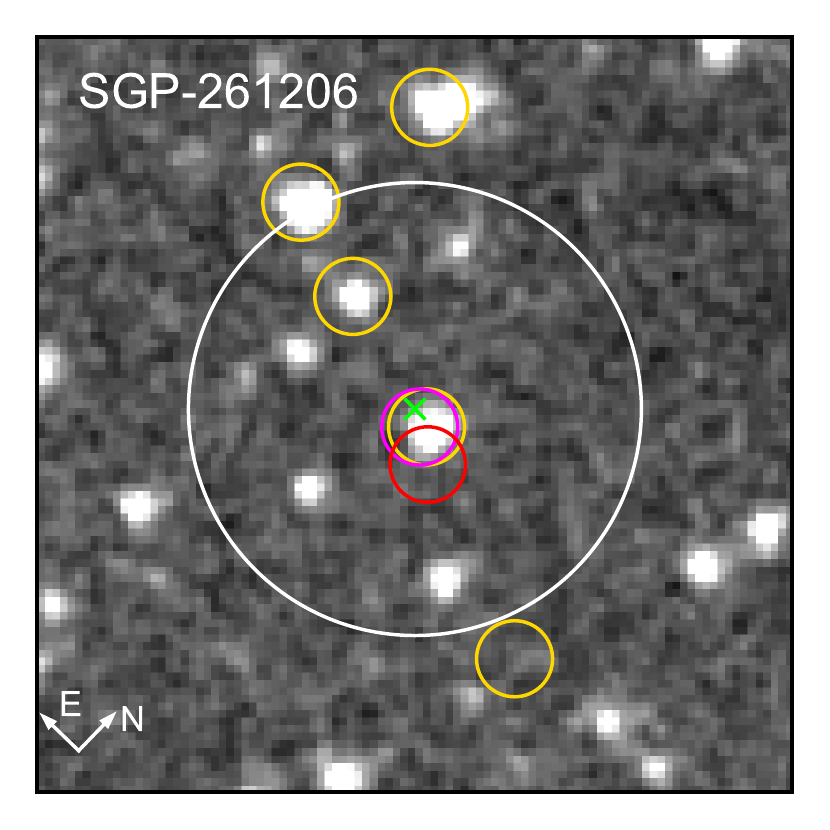}}
{\includegraphics[width=4.4cm, height=4.4cm]{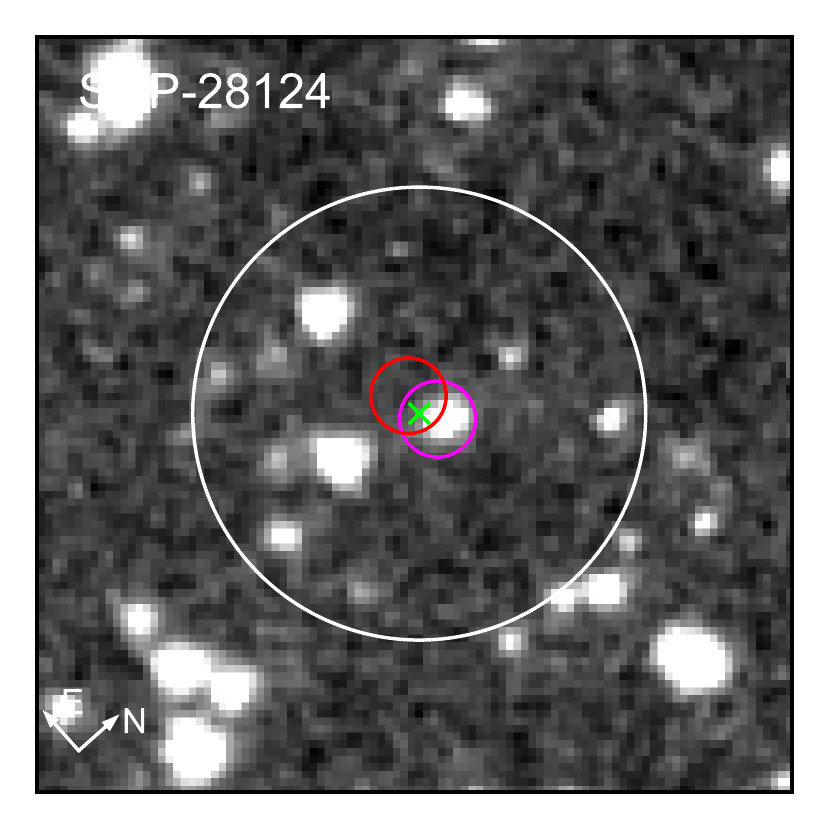}}
{\includegraphics[width=4.4cm, height=4.4cm]{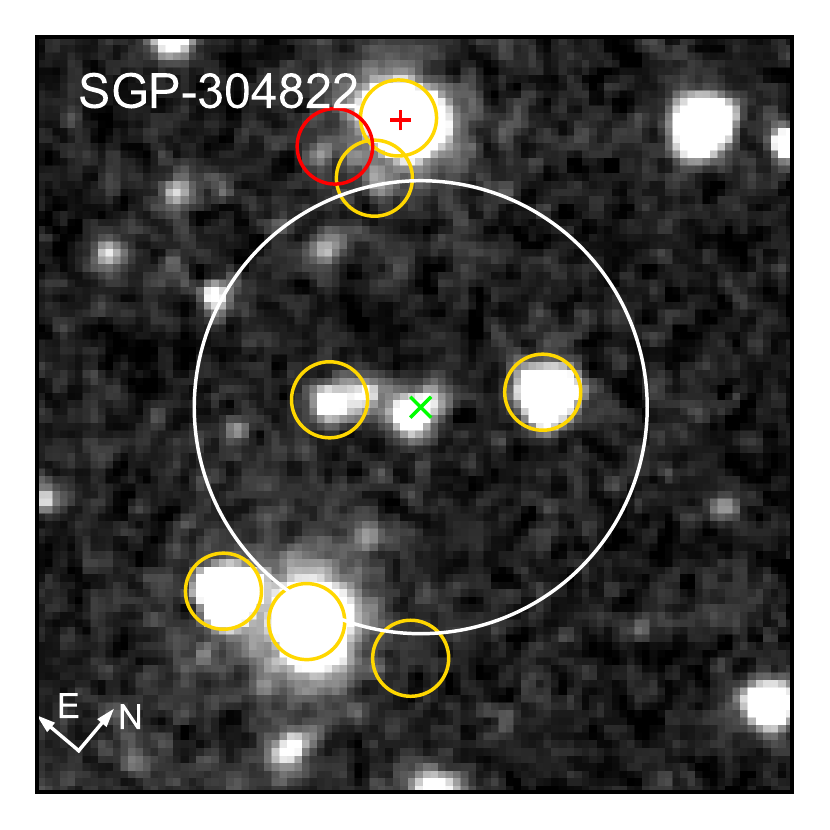}}
{\includegraphics[width=4.4cm, height=4.4cm]{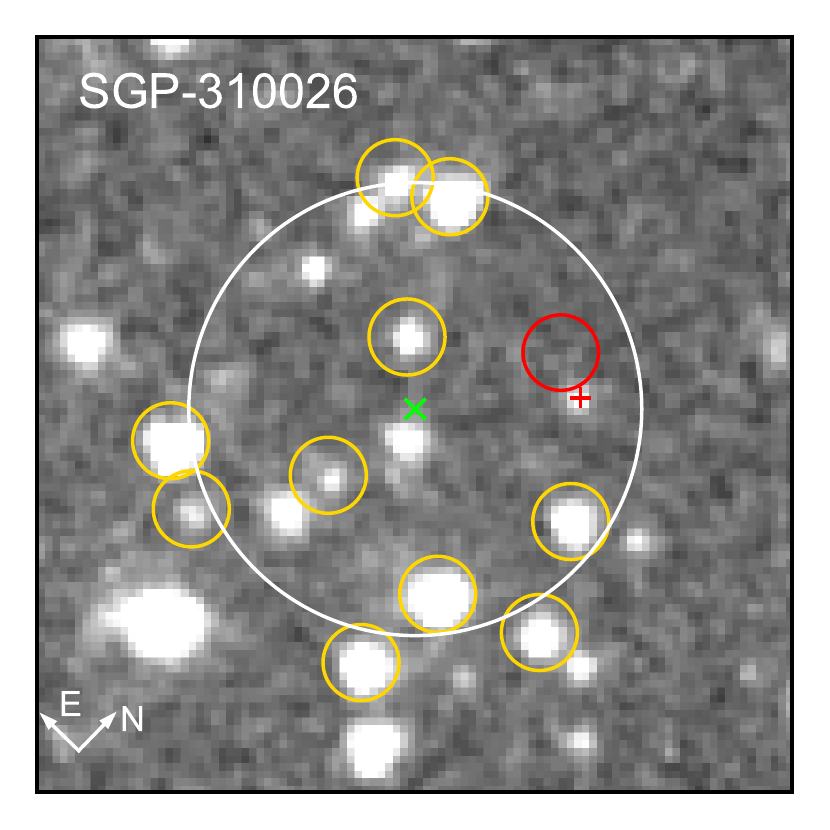}}
{\includegraphics[width=4.4cm, height=4.4cm]{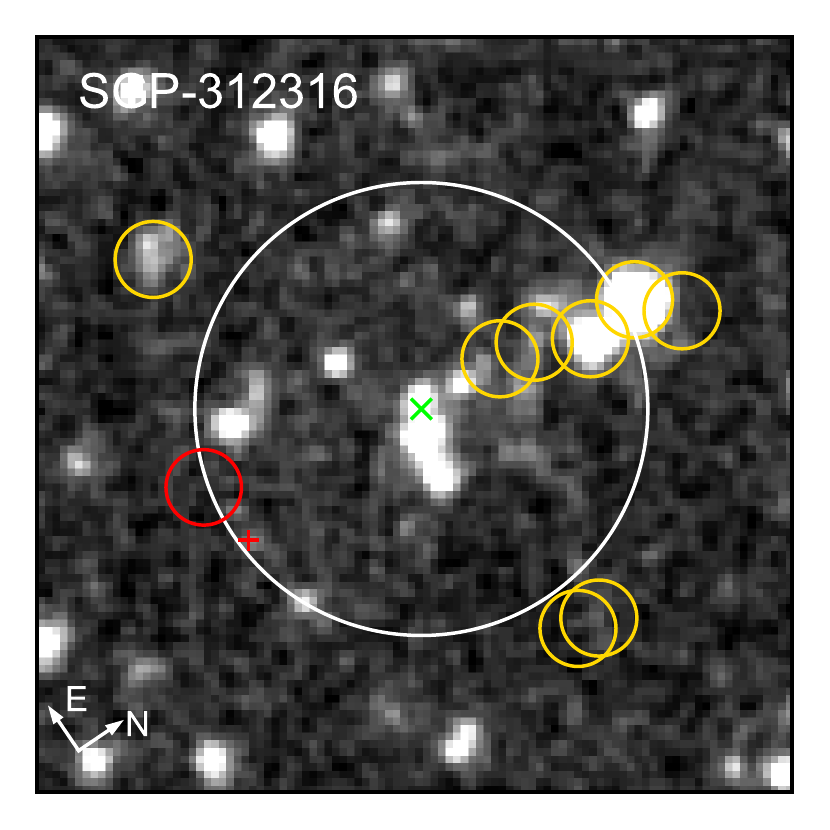}}
{\includegraphics[width=4.4cm, height=4.4cm]{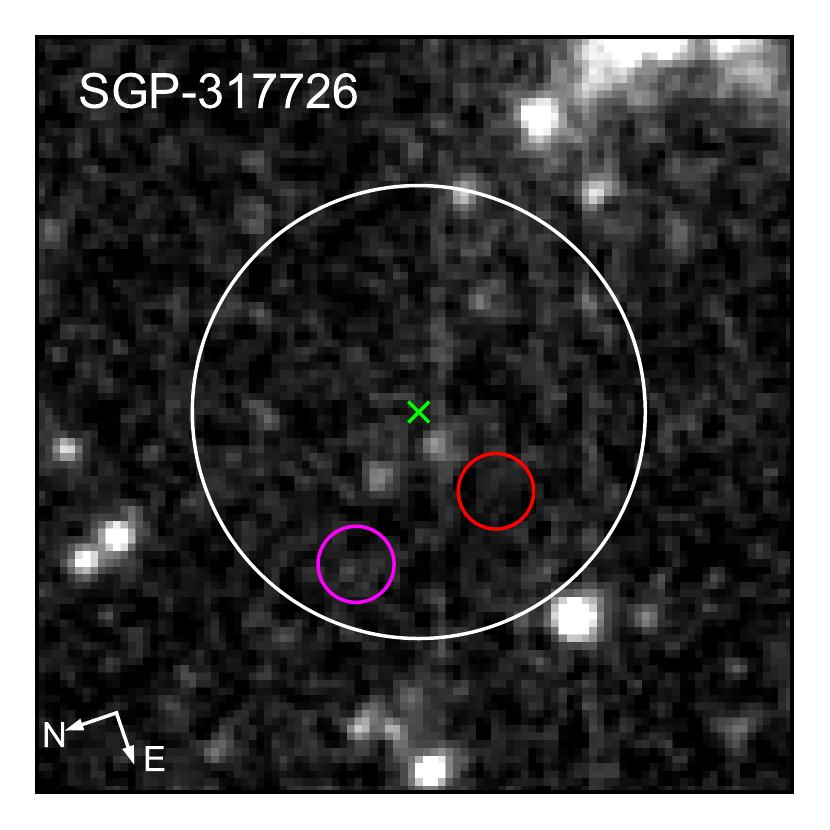}}
{\includegraphics[width=4.4cm, height=4.4cm]{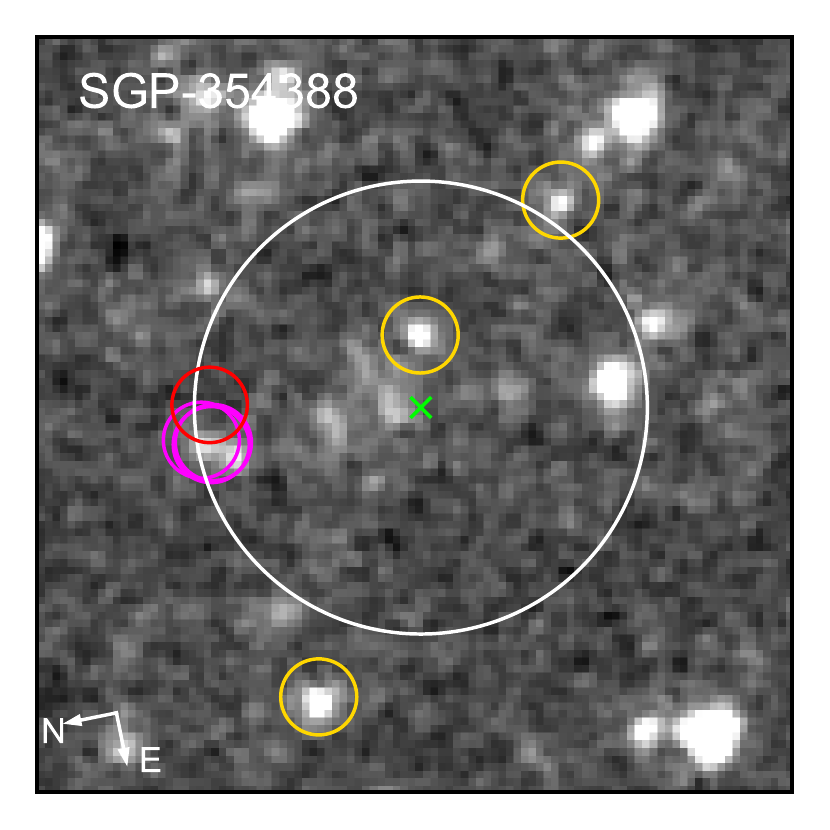}}
{\includegraphics[width=4.4cm, height=4.4cm]{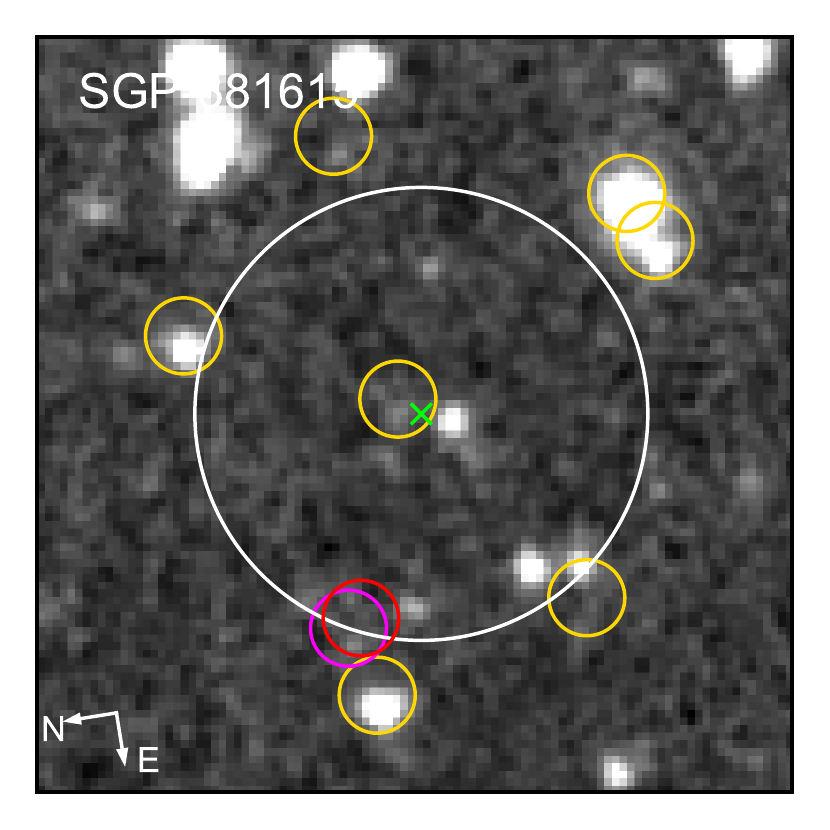}}
{\includegraphics[width=4.4cm, height=4.4cm]{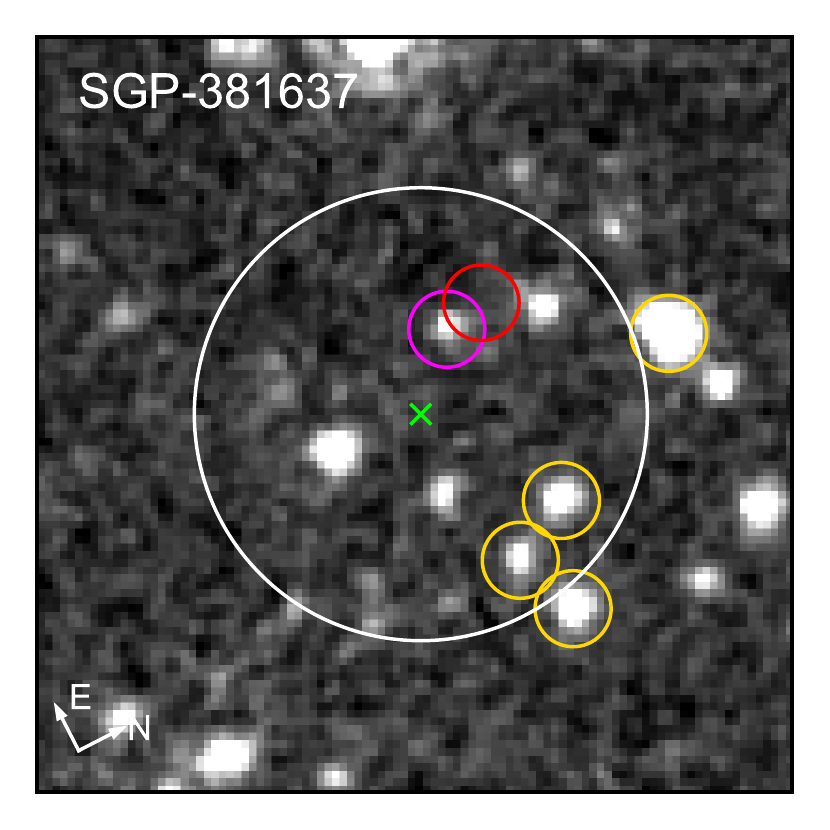}}
{\includegraphics[width=4.4cm, height=4.4cm]{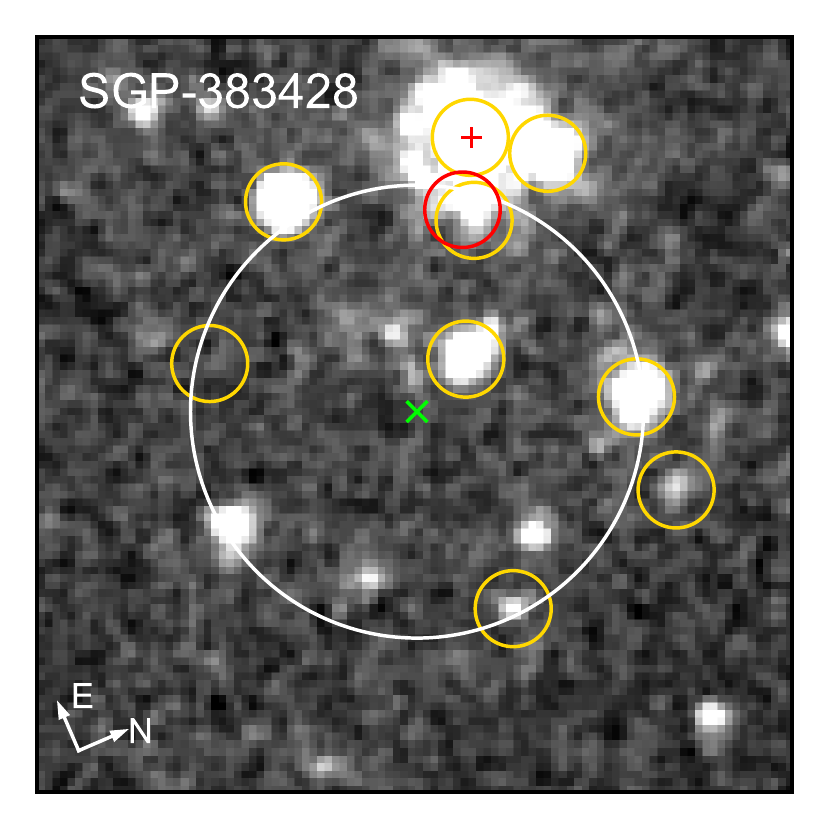}}
{\includegraphics[width=4.4cm, height=4.4cm]{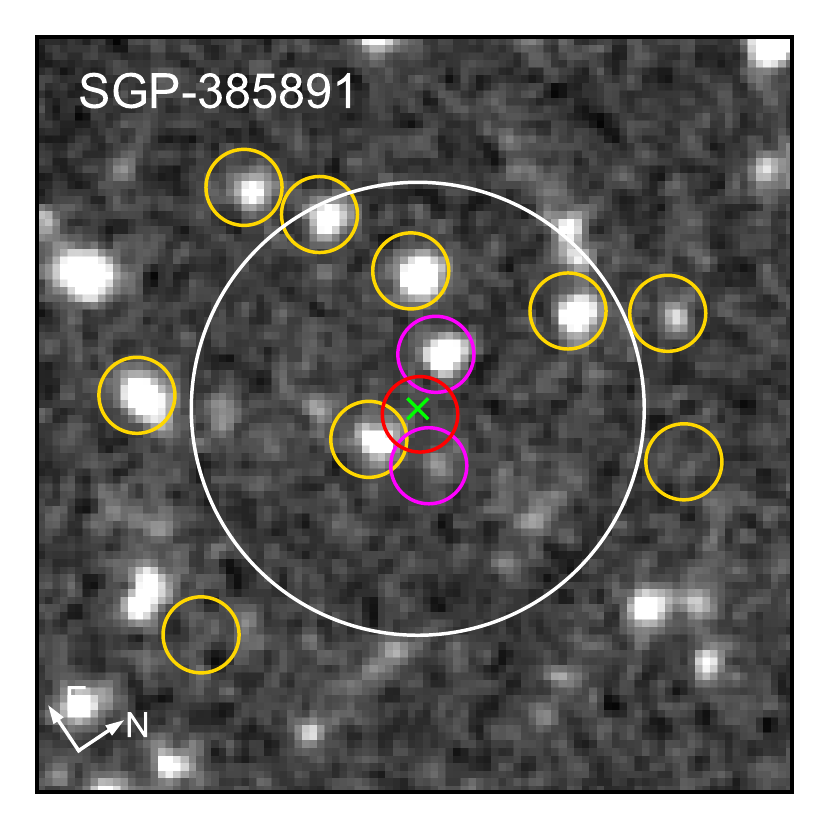}}
{\includegraphics[width=4.4cm, height=4.4cm]{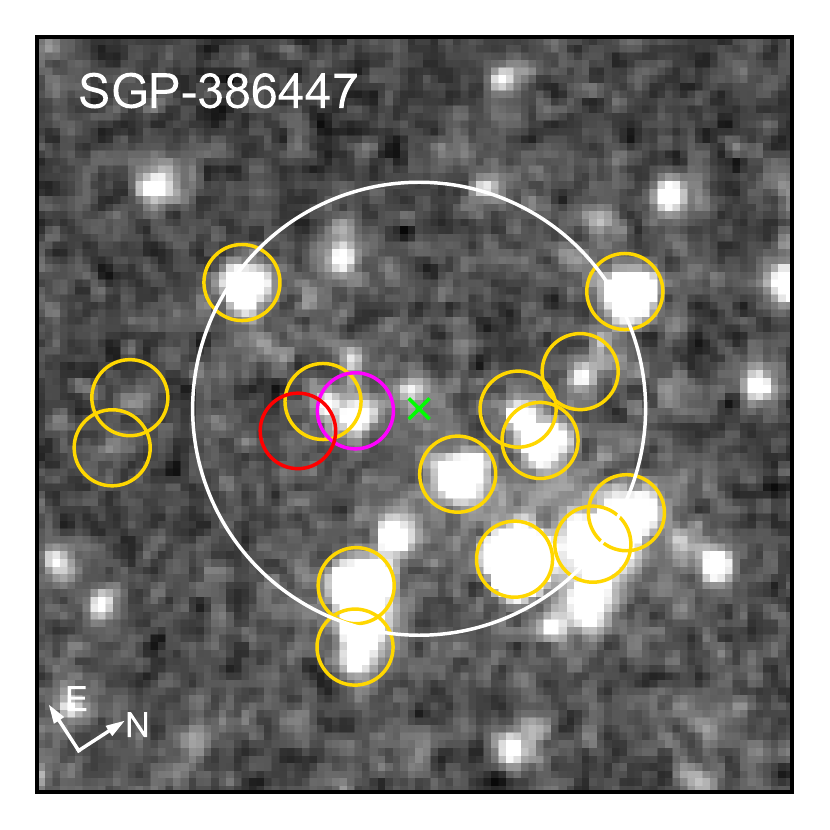}}
{\includegraphics[width=4.4cm, height=4.4cm]{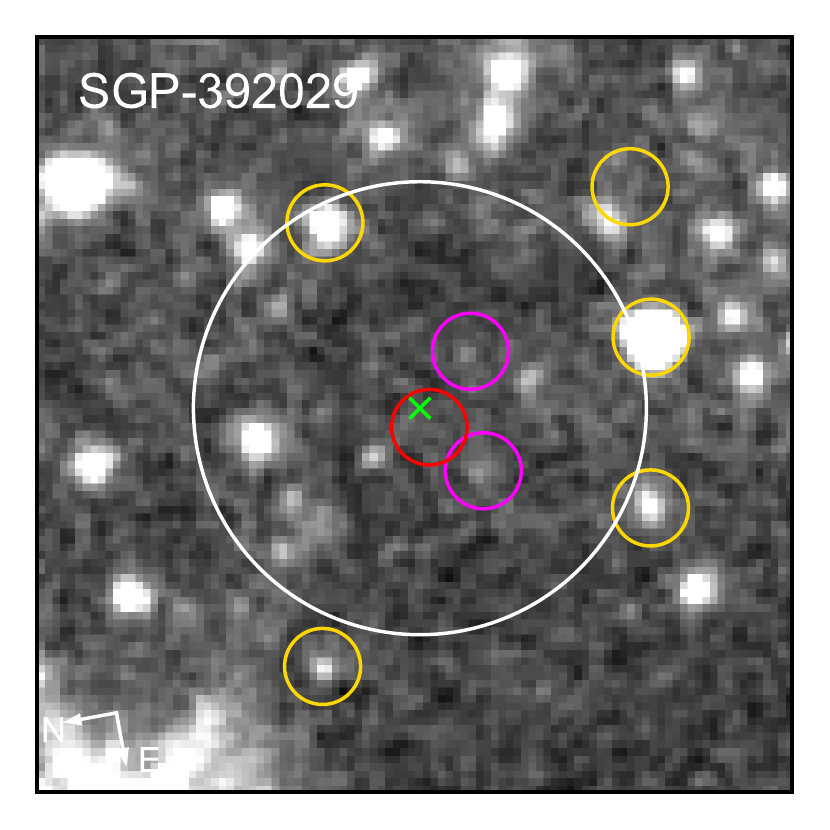}}
{\includegraphics[width=4.4cm, height=4.4cm]{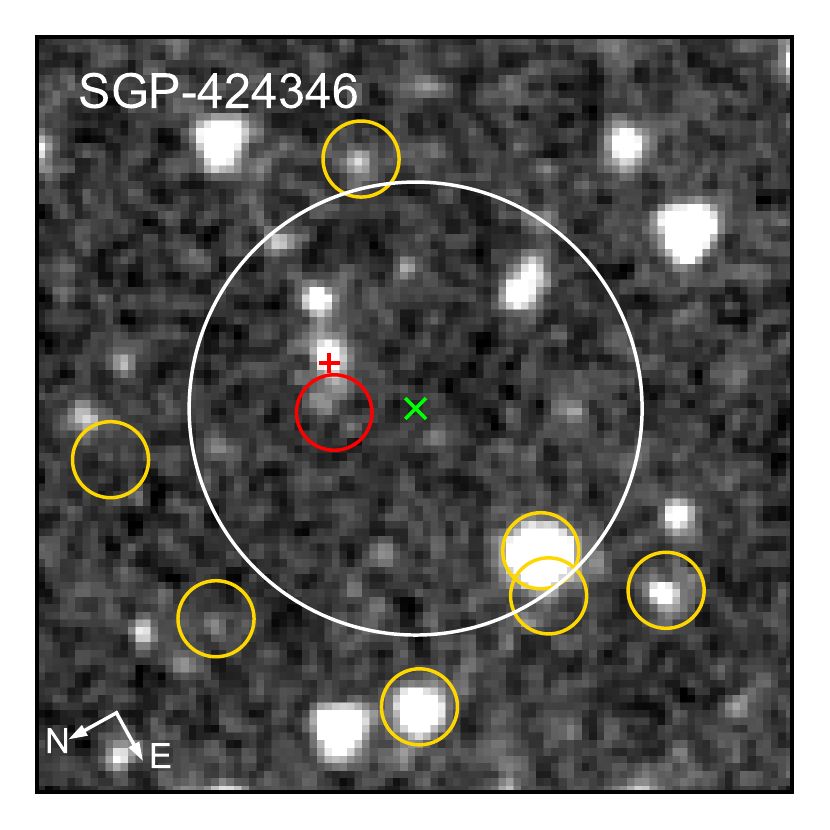}}
{\includegraphics[width=4.4cm, height=4.4cm]{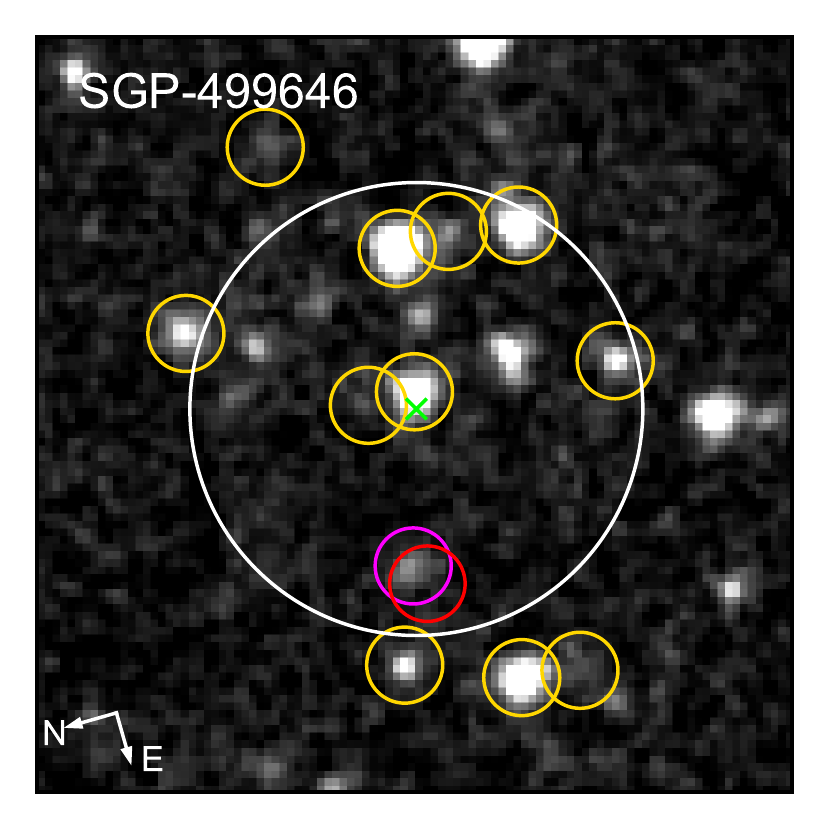}}
{\includegraphics[width=4.4cm, height=4.4cm]{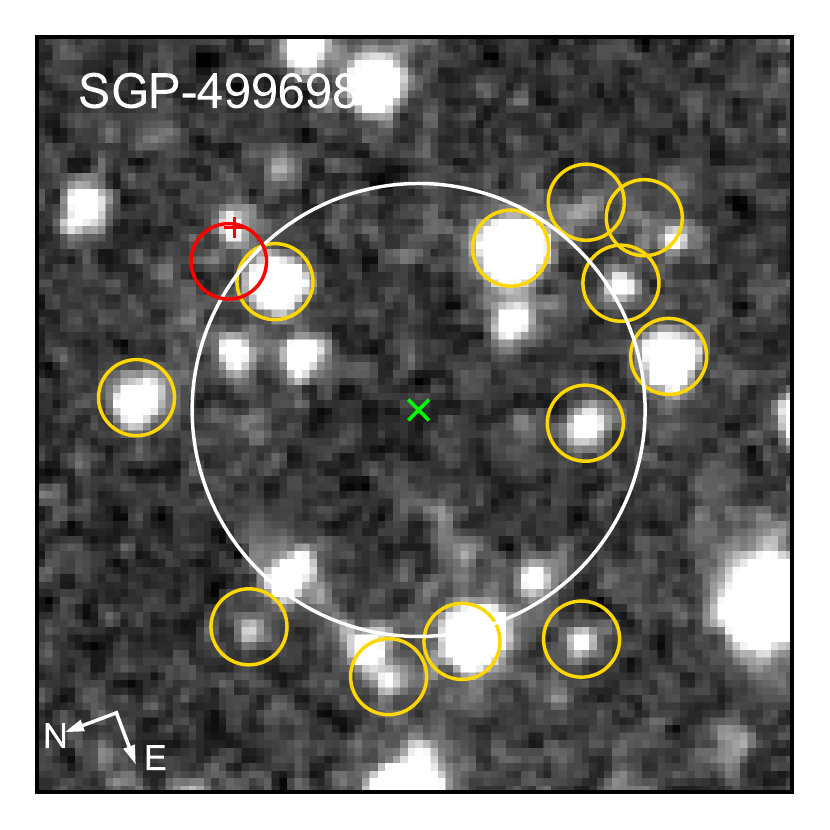}}
\caption{Continued 60$\arcsec$ $\times$ 60$\arcsec$ cutouts }
\label{fig:data}
\end{figure*}

\addtocounter{figure}{-1}
\begin{figure*}
\centering
{\includegraphics[width=4.4cm, height=4.4cm]{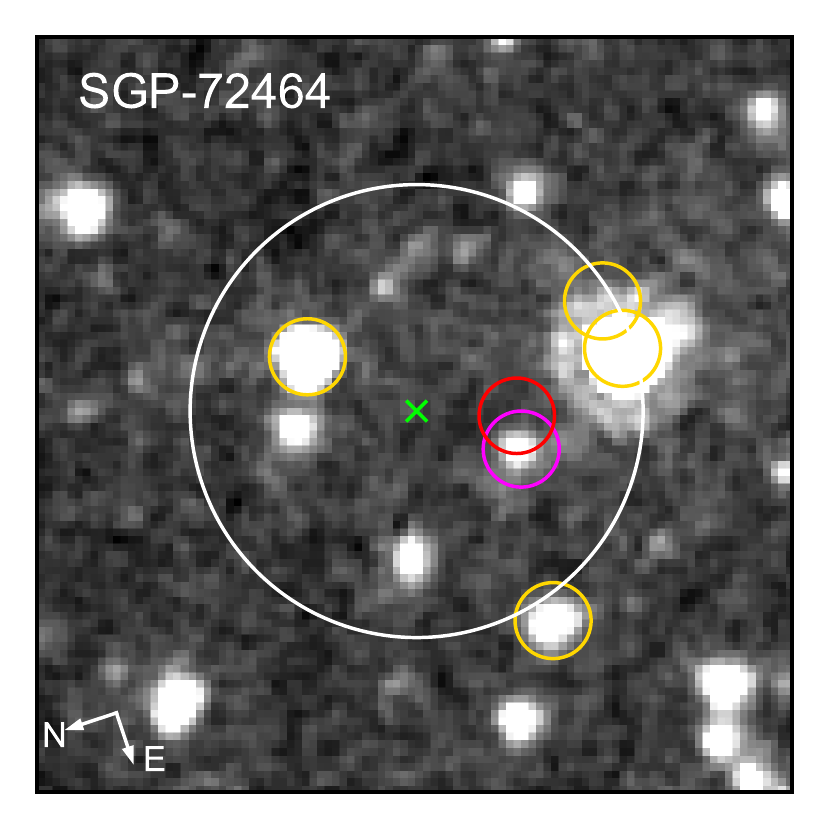}}
{\includegraphics[width=4.4cm, height=4.4cm]{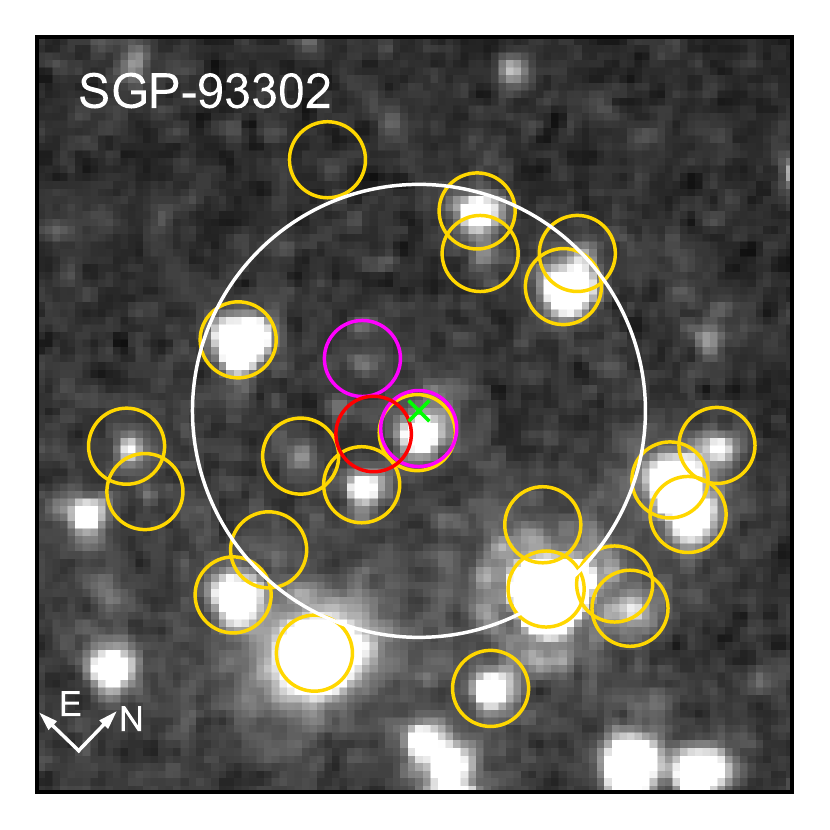}}
{\includegraphics[width=4.4cm, height=4.4cm]{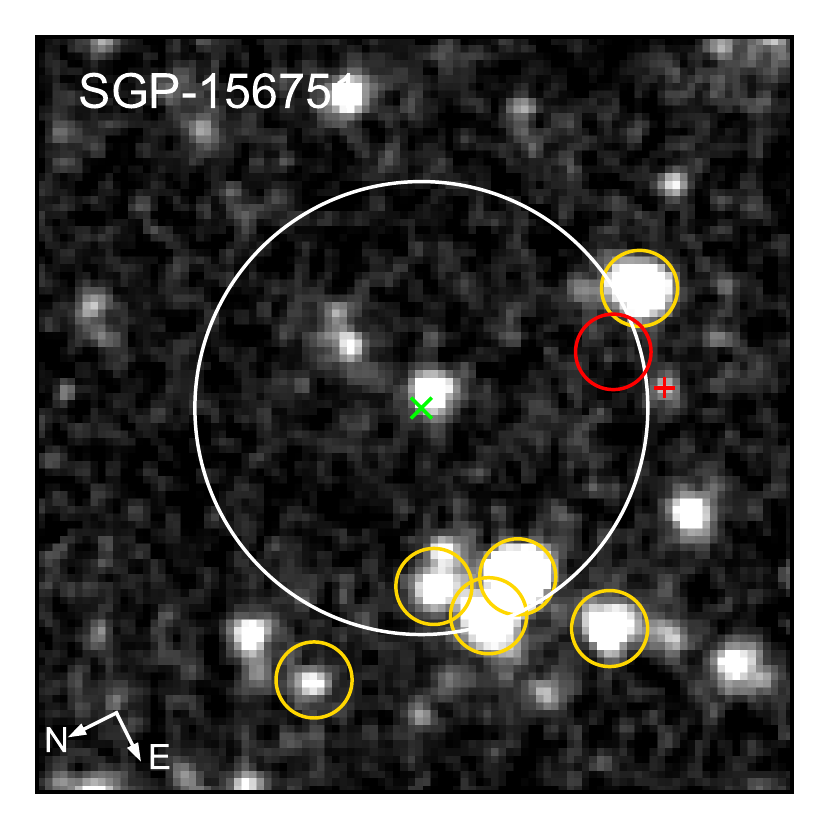}}
{\includegraphics[width=4.4cm, height=4.4cm]{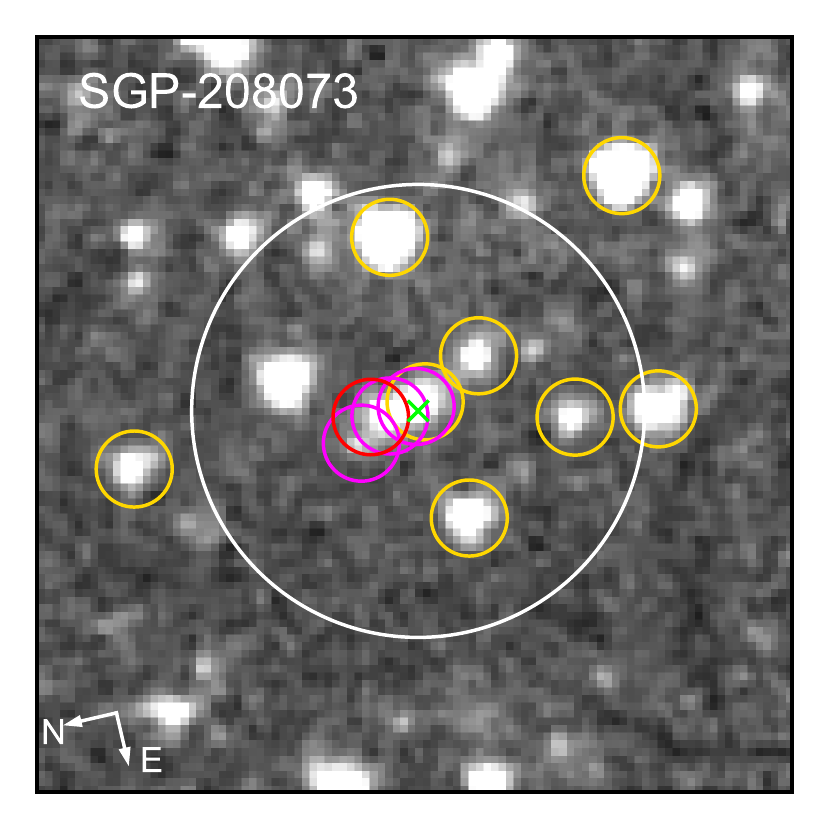}}
{\includegraphics[width=4.4cm, height=4.4cm]{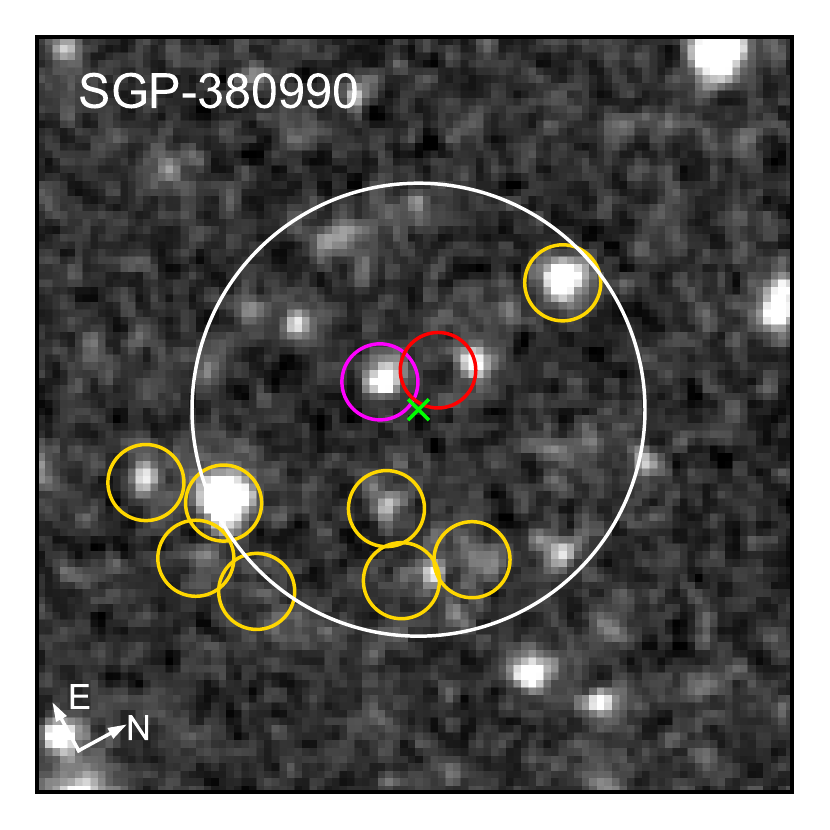}}
{\includegraphics[width=4.4cm, height=4.4cm]{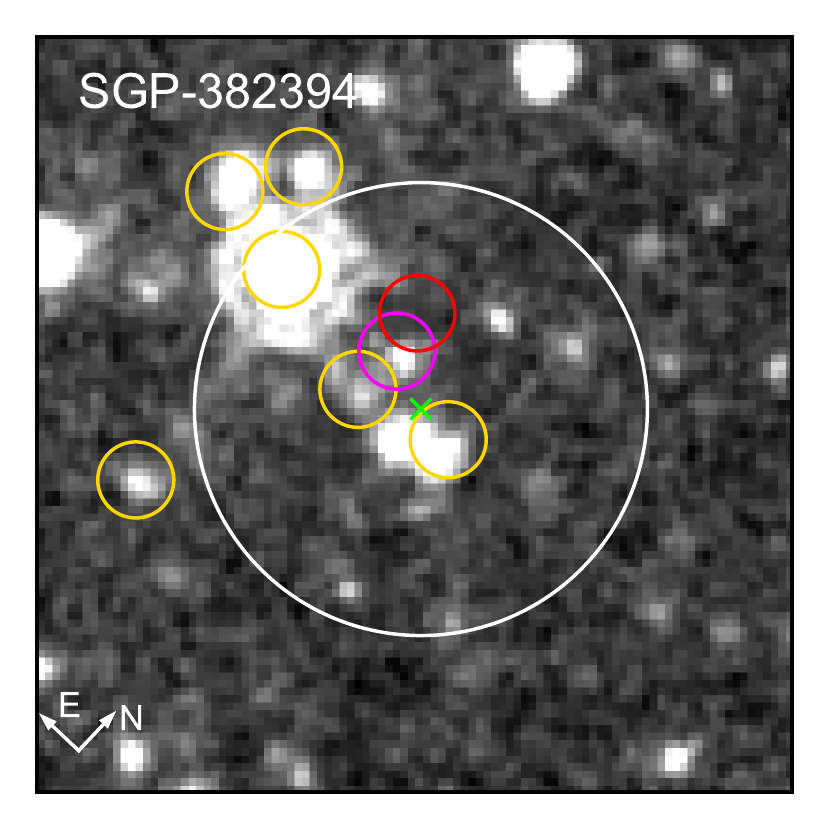}}
{\includegraphics[width=4.4cm, height=4.4cm]{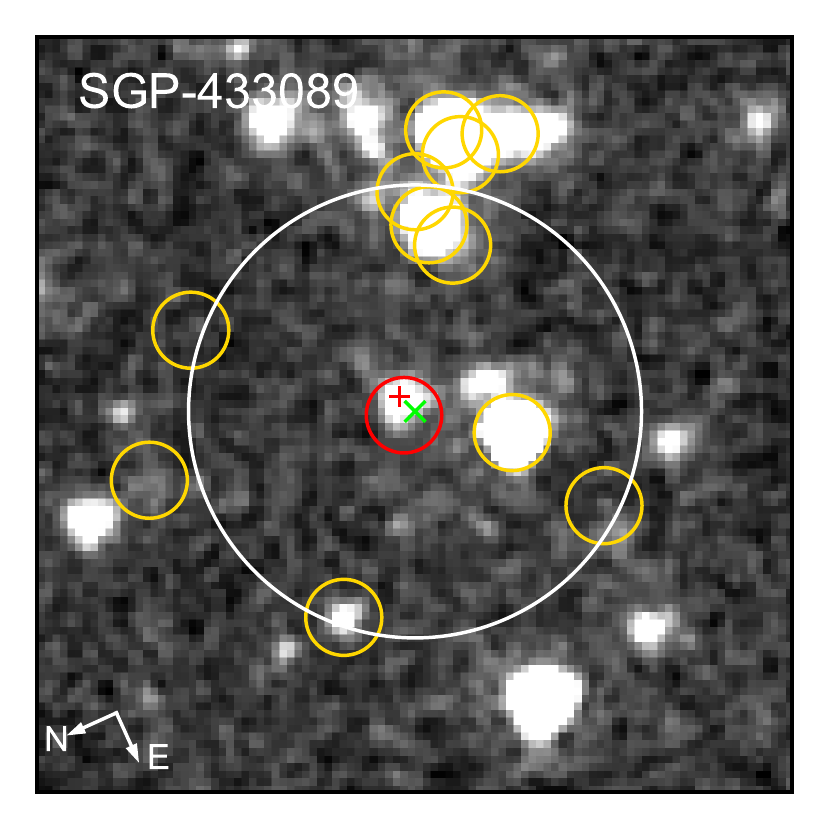}}
{\includegraphics[width=4.4cm, height=4.4cm]{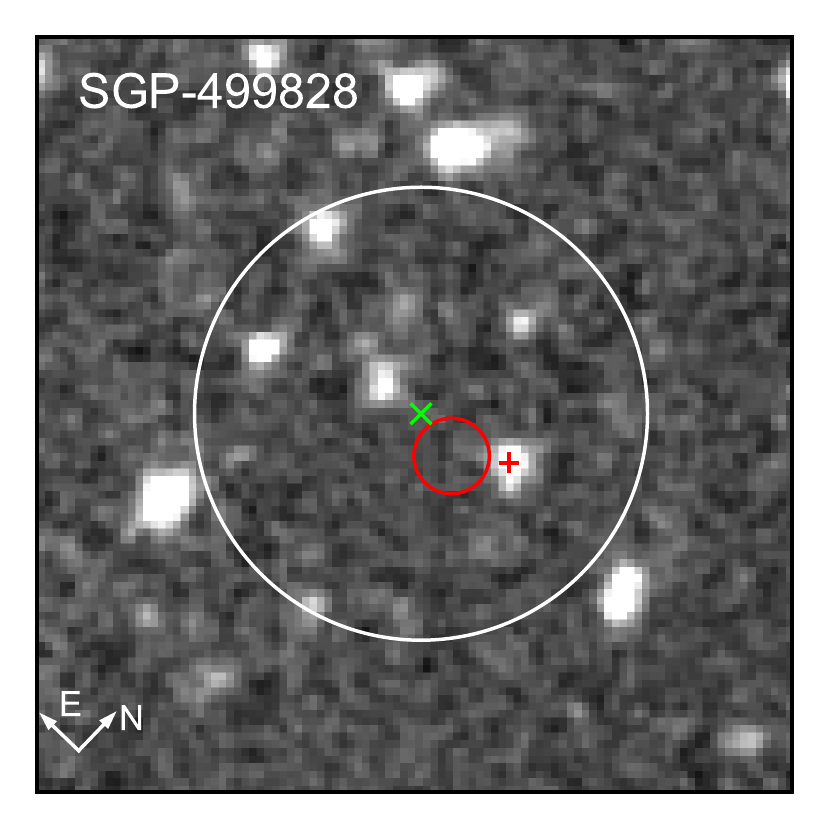}}
{\includegraphics[width=4.4cm, height=4.4cm]{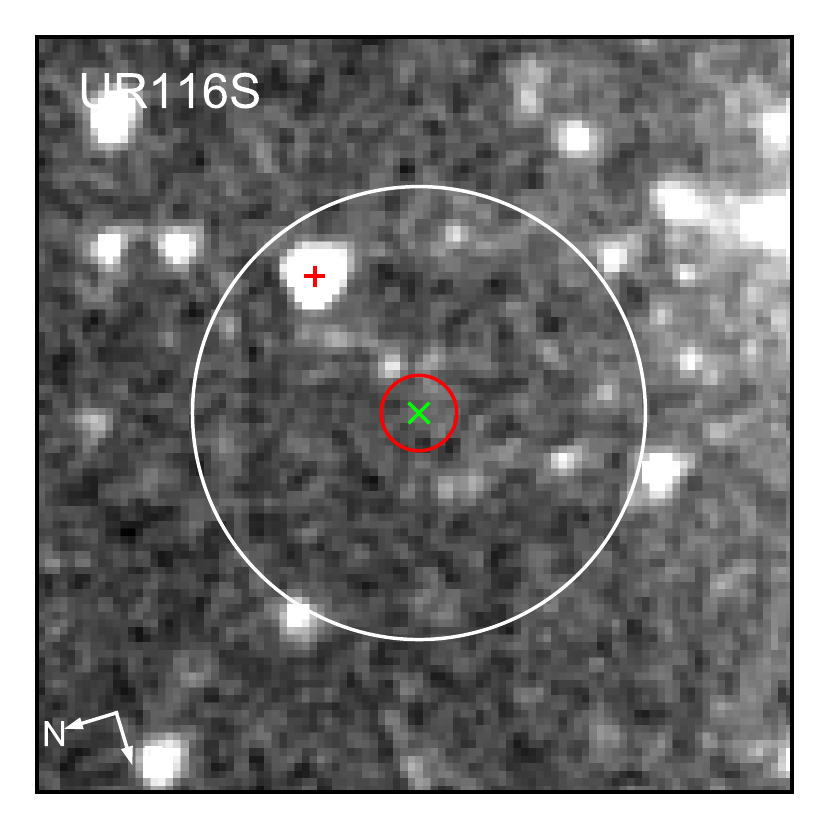}}
{\includegraphics[width=4.4cm, height=4.4cm]{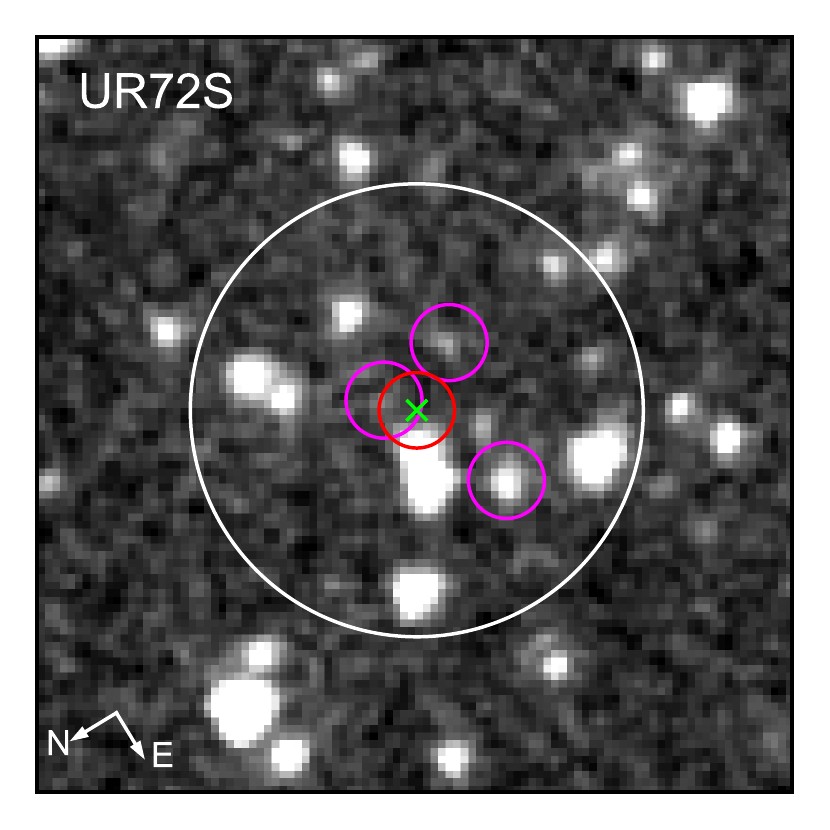}}
{\includegraphics[width=4.4cm, height=4.4cm]{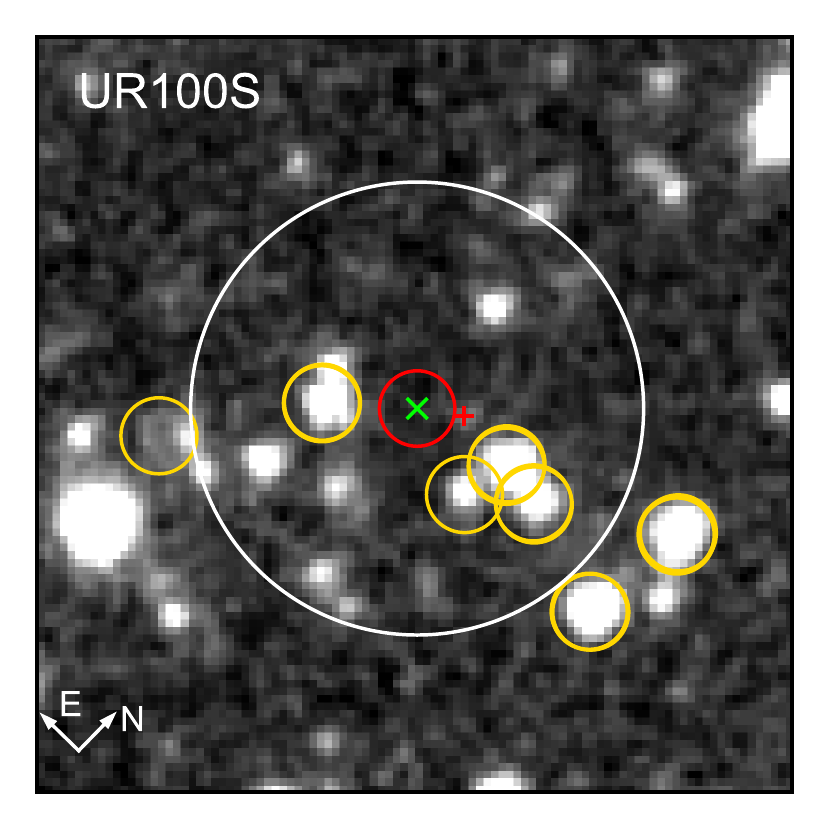}}
{\includegraphics[width=4.4cm, height=4.4cm]{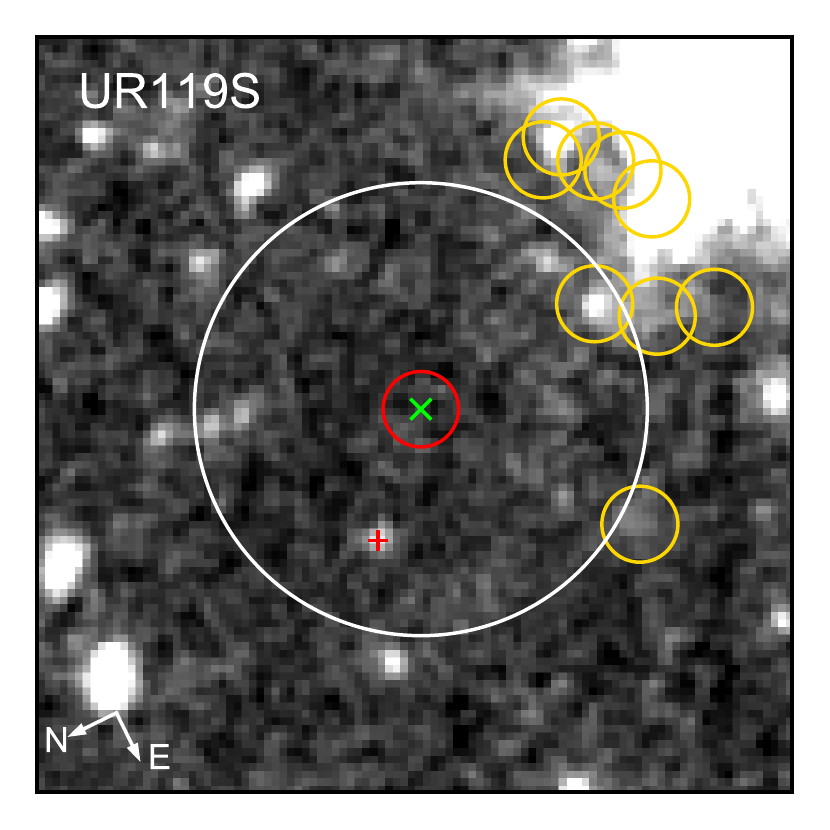}}
{\includegraphics[width=4.4cm, height=4.4cm]{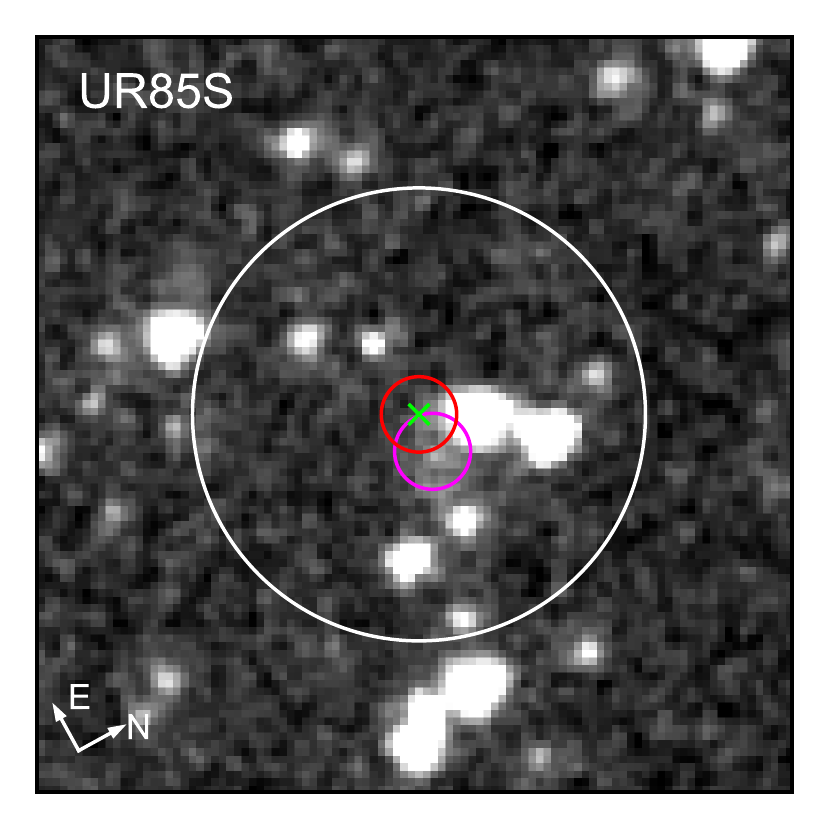}}
{\includegraphics[width=4.4cm, height=4.4cm]{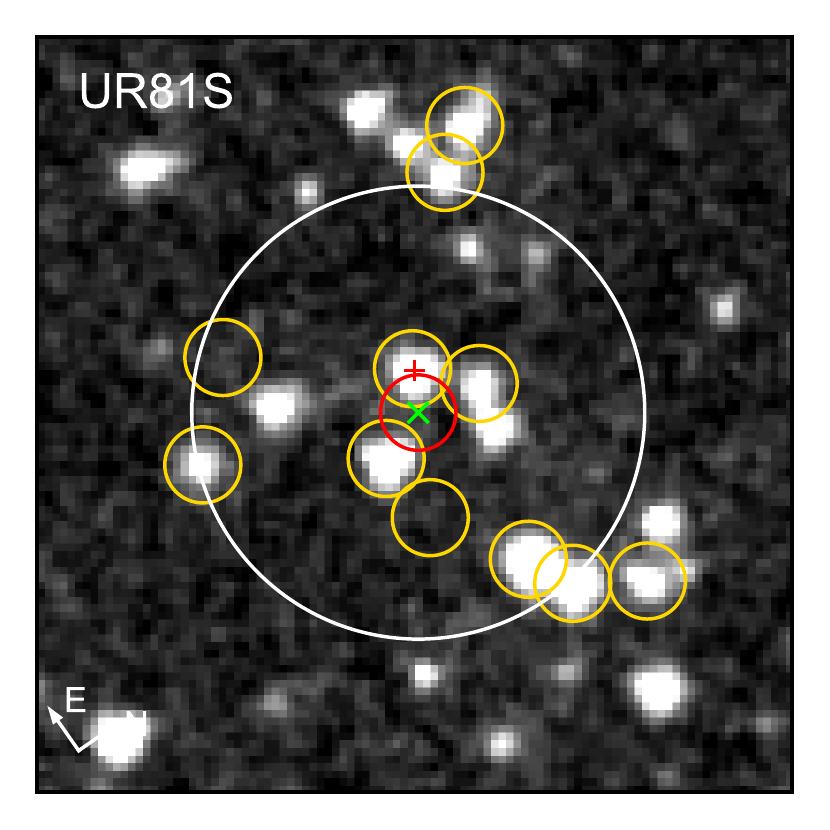}}
{\includegraphics[width=4.4cm, height=4.4cm]{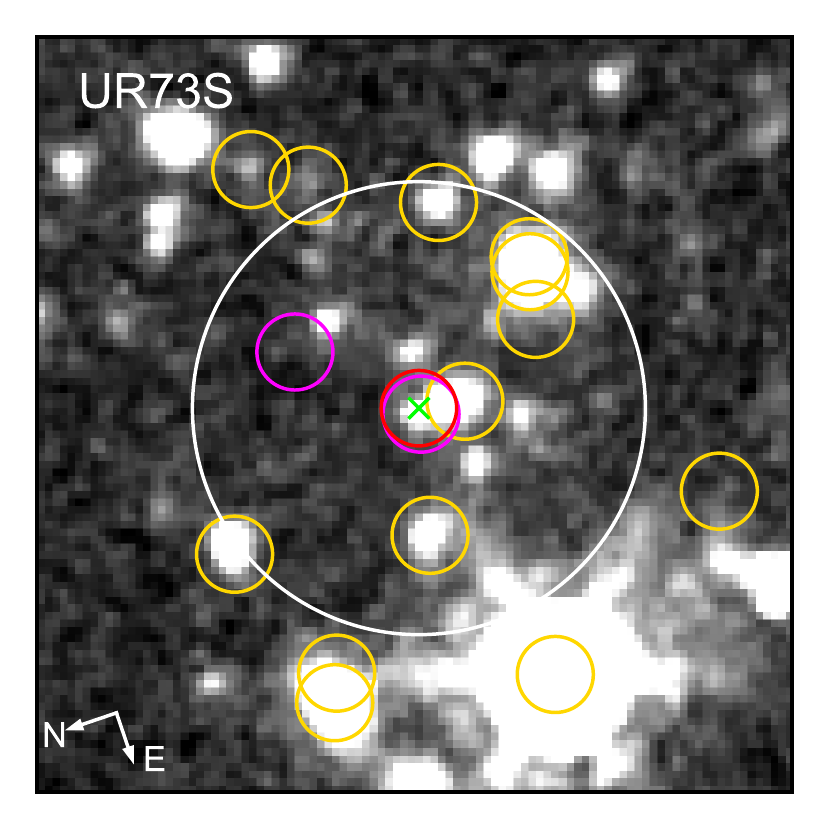}}
{\includegraphics[width=4.4cm, height=4.4cm]{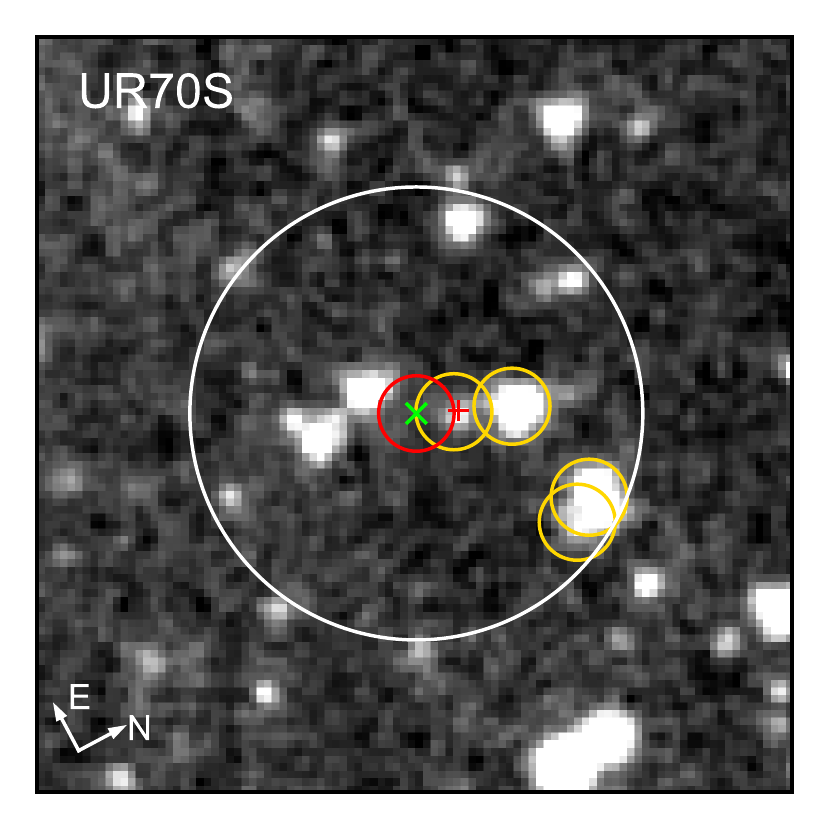}}
{\includegraphics[width=4.4cm, height=4.4cm]{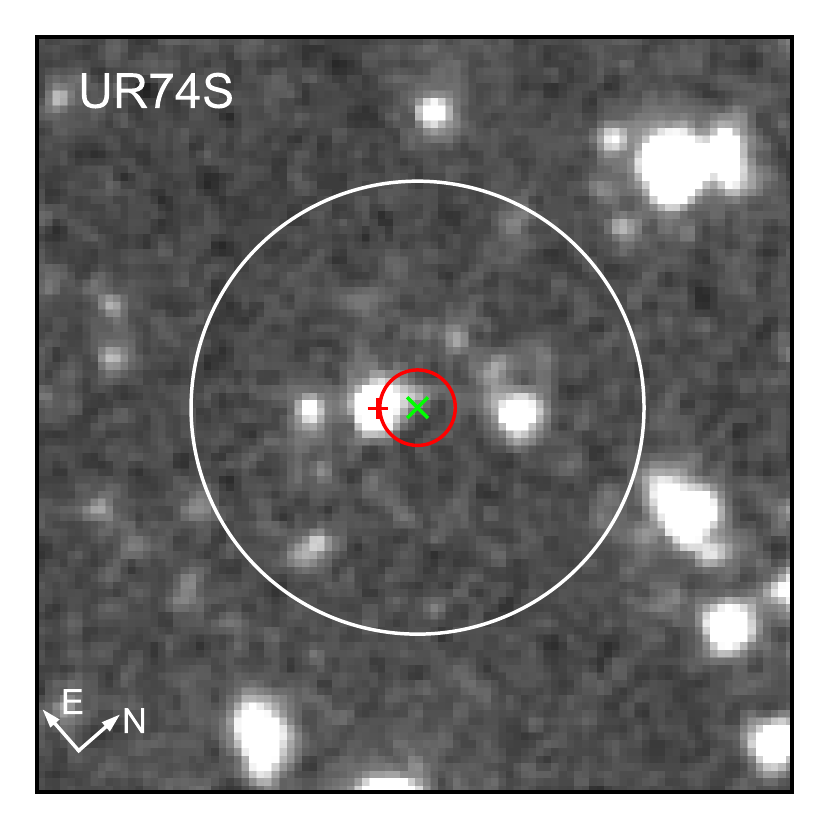}}
{\includegraphics[width=4.4cm, height=4.4cm]{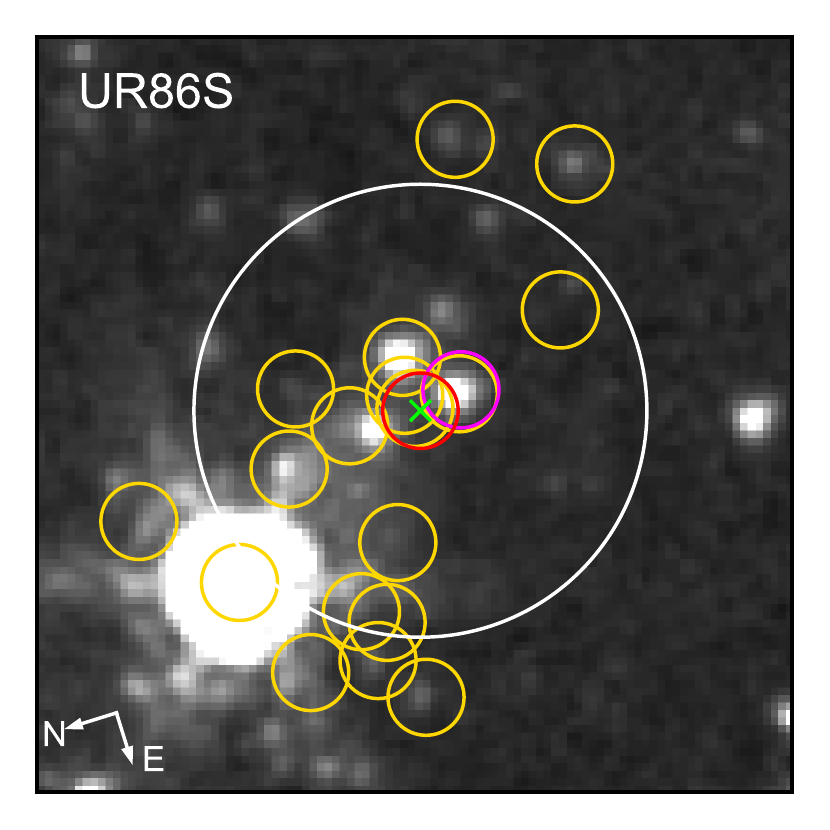}}
{\includegraphics[width=4.4cm, height=4.4cm]{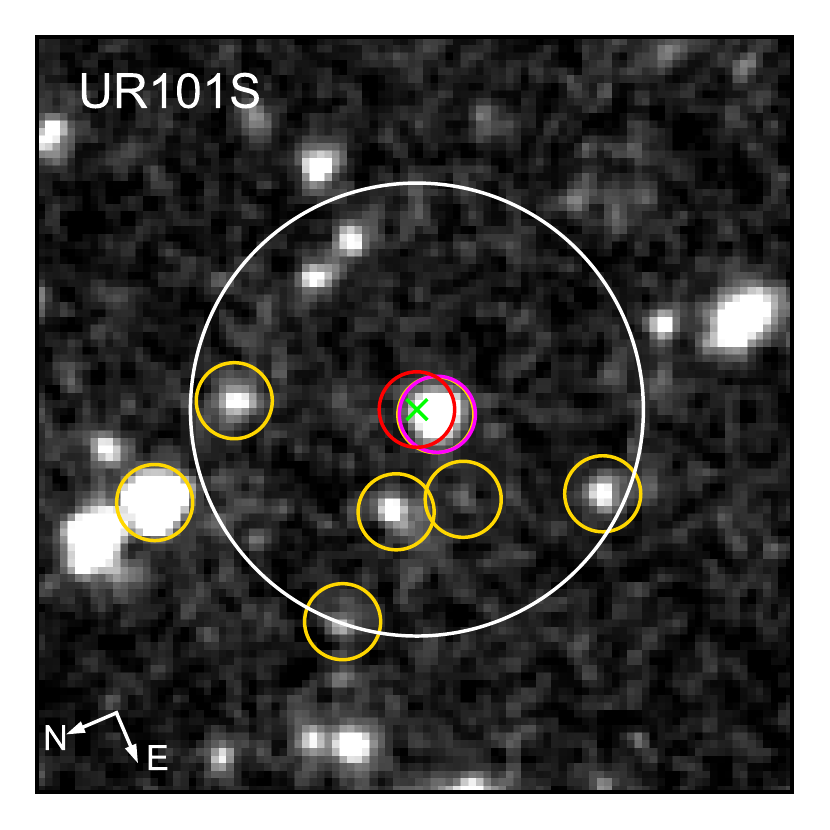}}
{\includegraphics[width=4.4cm, height=4.4cm]{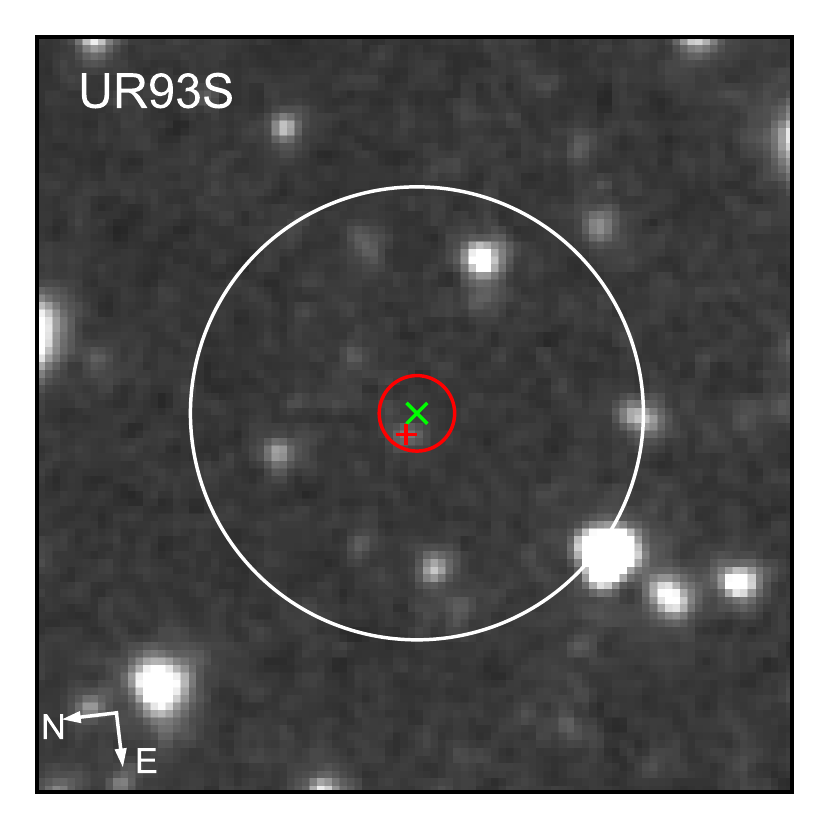}}
\caption{Continued 60$\arcsec$ $\times$ 60$\arcsec$ cutouts }
\label{fig:data}
\end{figure*}

\addtocounter{figure}{-1}
\begin{figure*}
\centering
{\includegraphics[width=4.4cm, height=4.4cm]{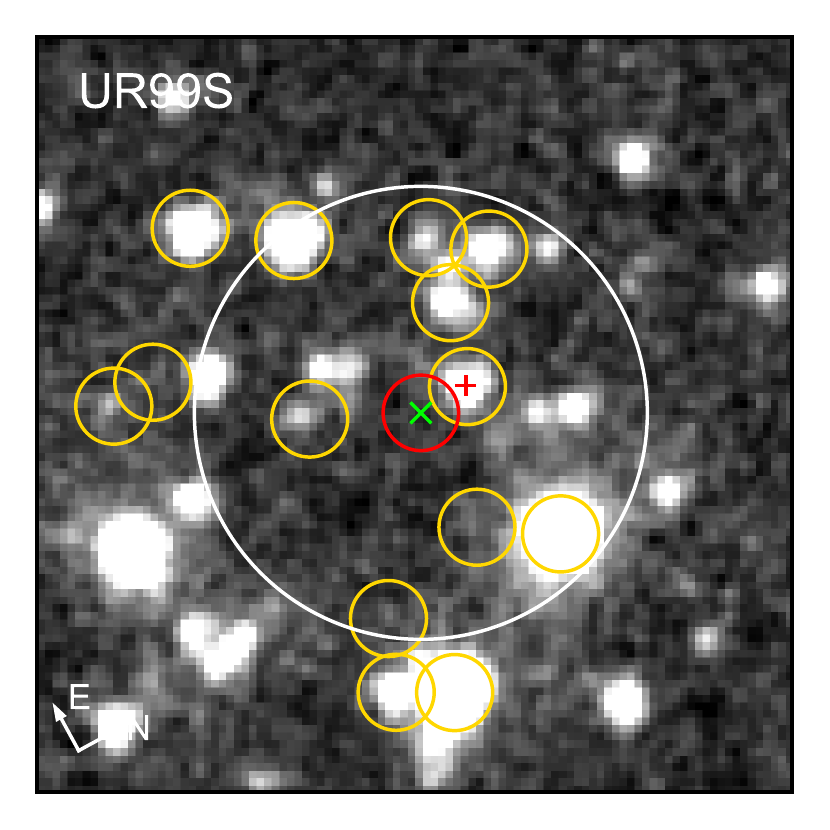}}
{\includegraphics[width=4.4cm, height=4.4cm]{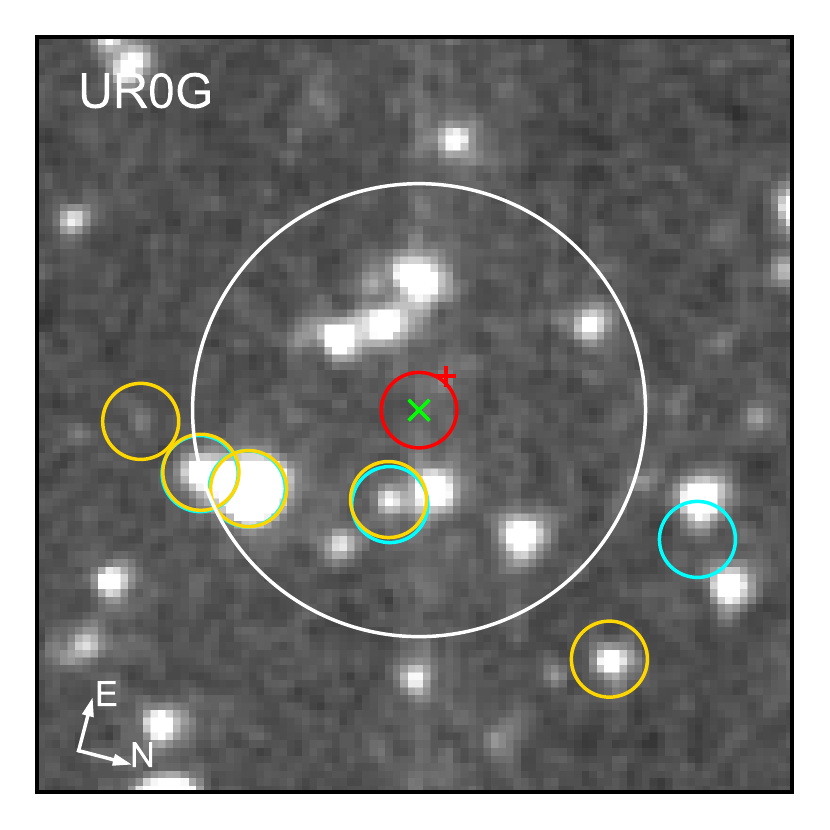}}
{\includegraphics[width=4.4cm, height=4.4cm]{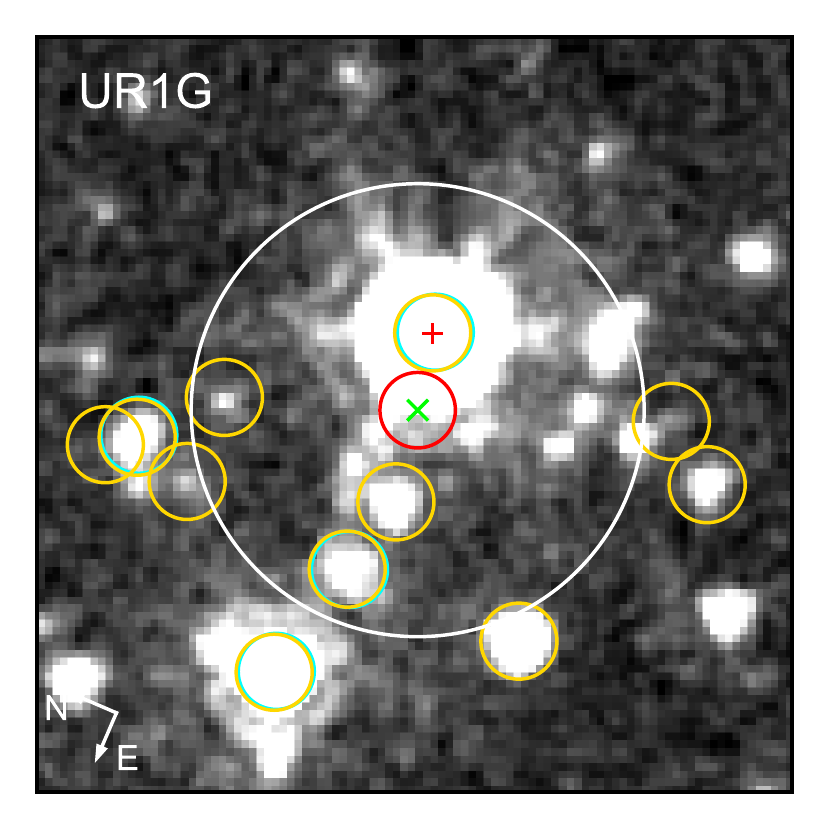}}
{\includegraphics[width=4.4cm, height=4.4cm]{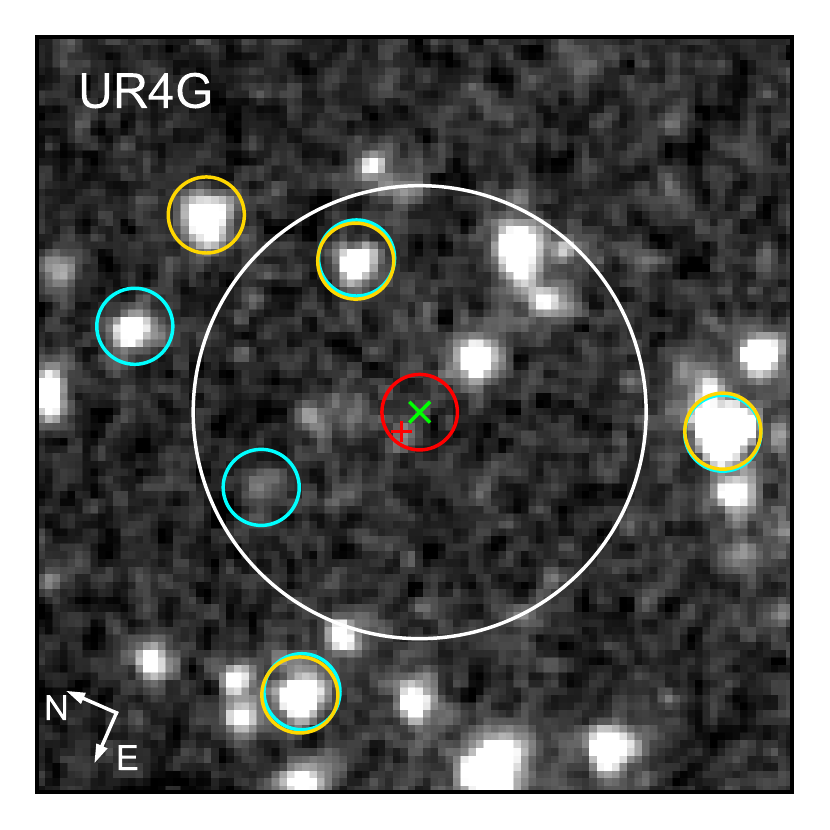}}
{\includegraphics[width=4.4cm, height=4.4cm]{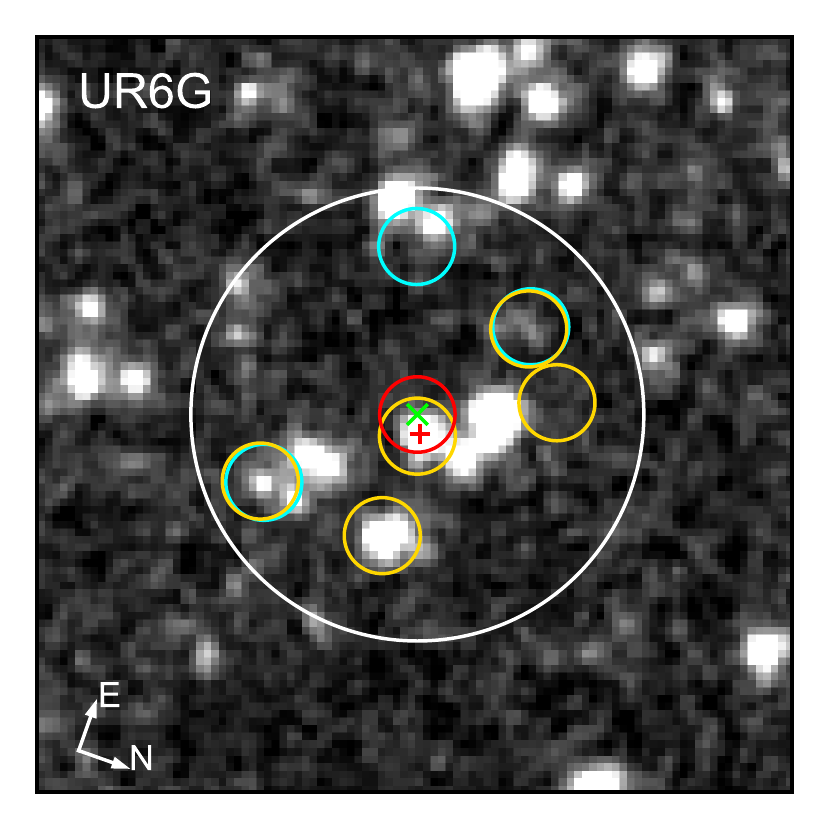}}
{\includegraphics[width=4.4cm, height=4.4cm]{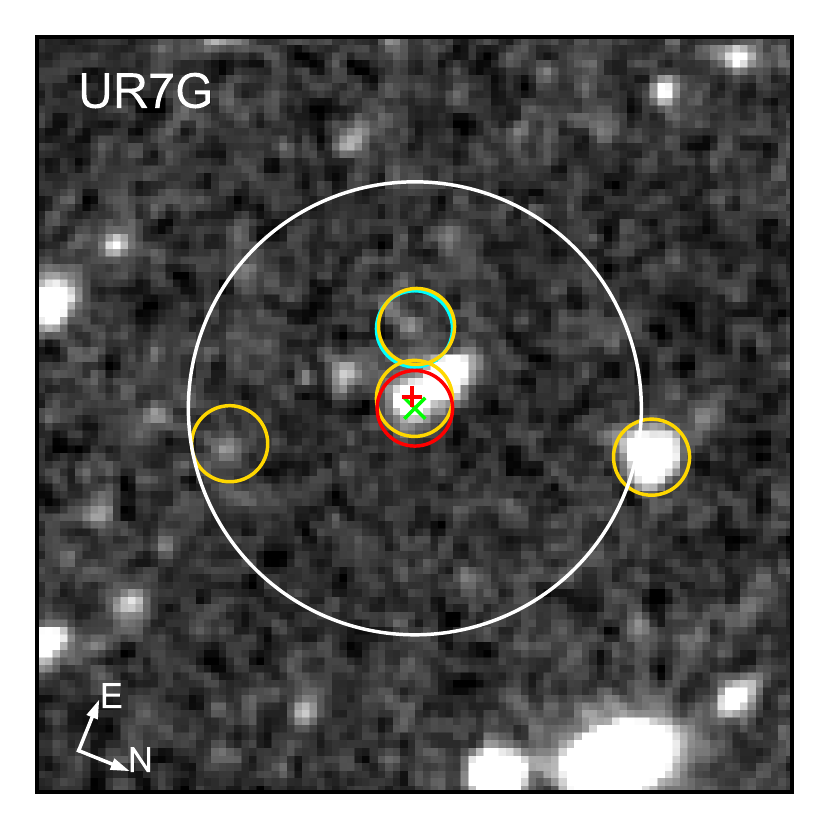}}
{\includegraphics[width=4.4cm, height=4.4cm]{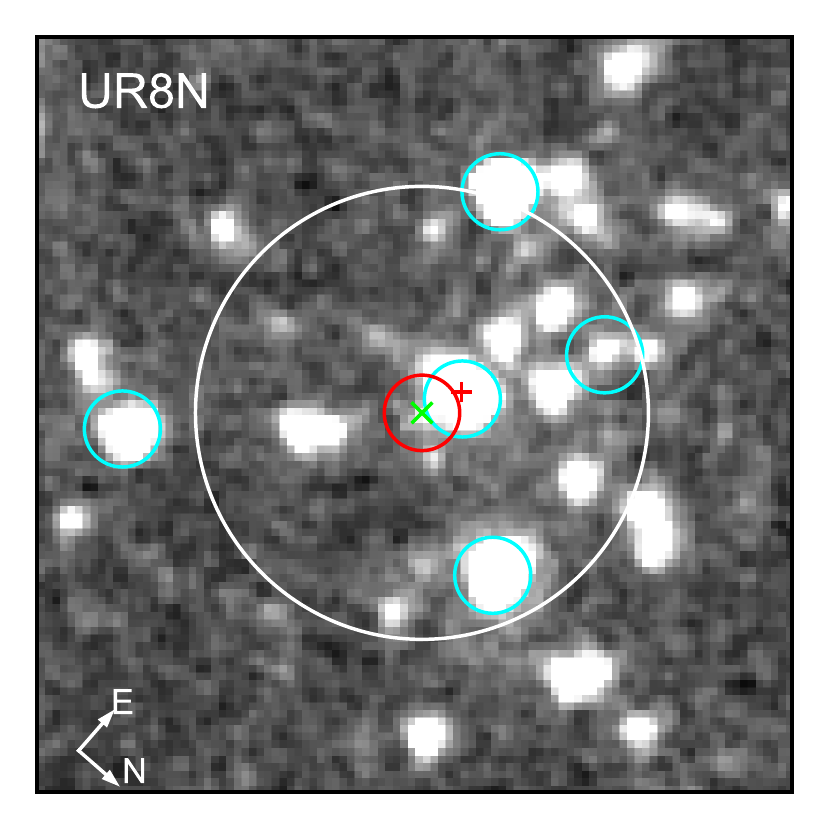}}
{\includegraphics[width=4.4cm, height=4.4cm]{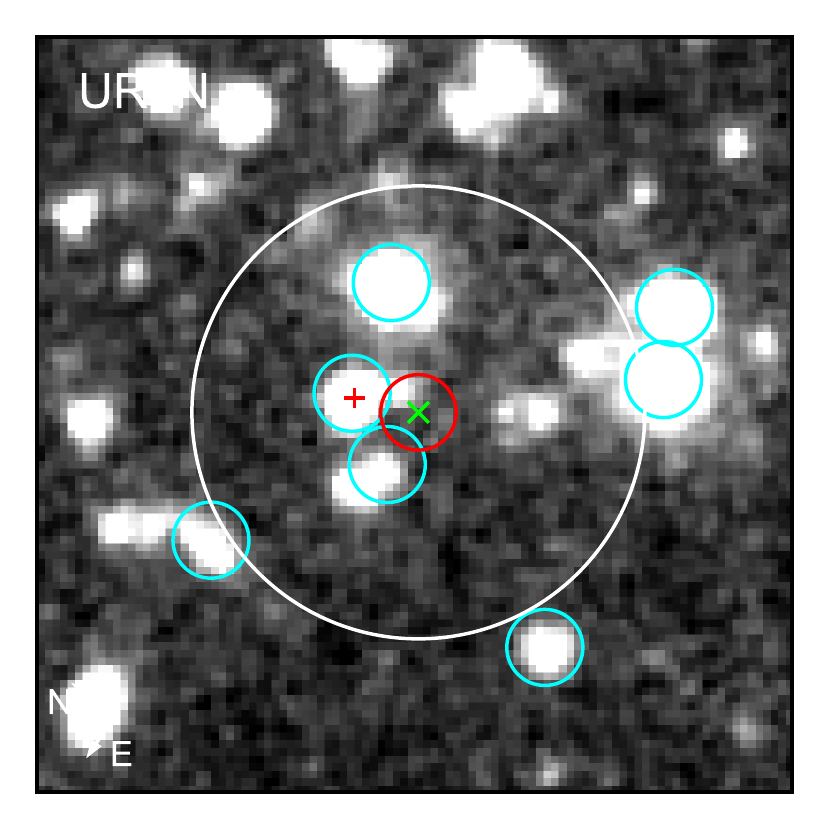}}
{\includegraphics[width=4.4cm, height=4.4cm]{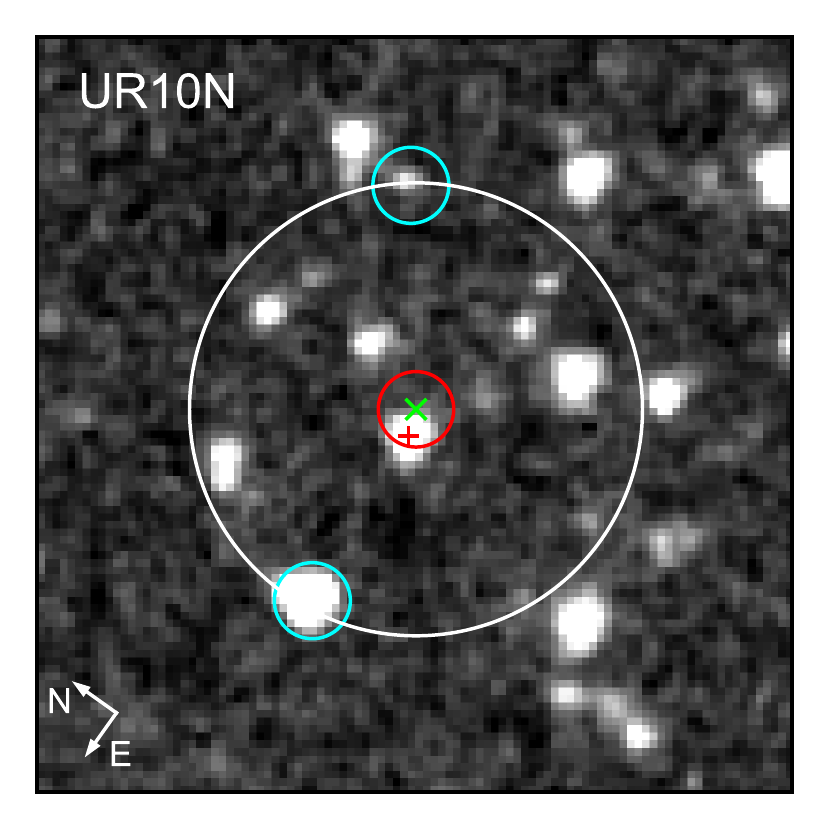}}
{\includegraphics[width=4.4cm, height=4.4cm]{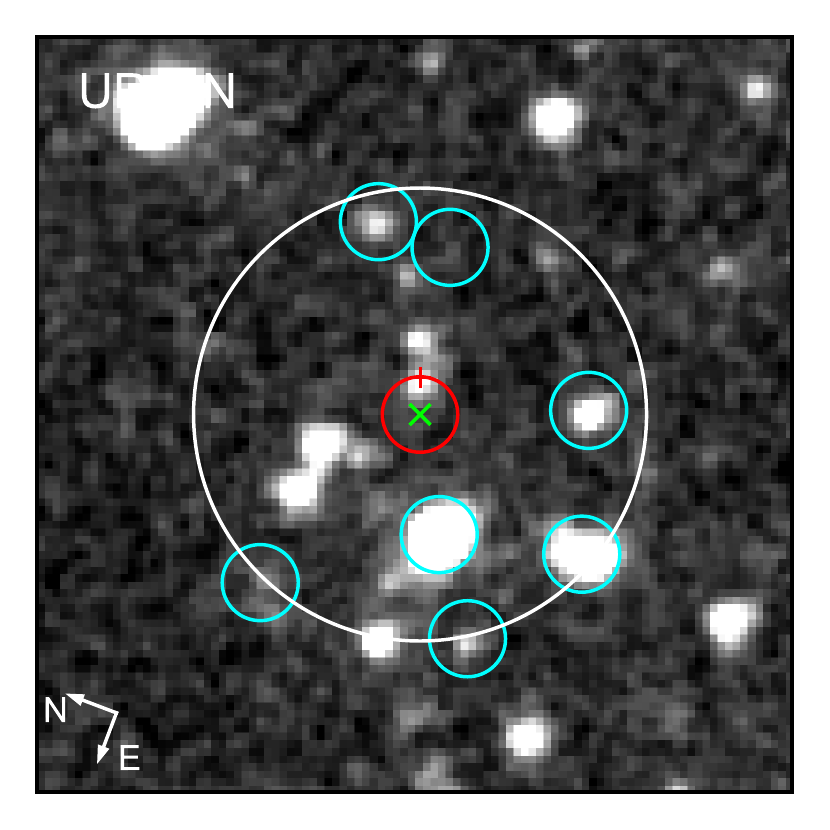}}
{\includegraphics[width=4.4cm, height=4.4cm]{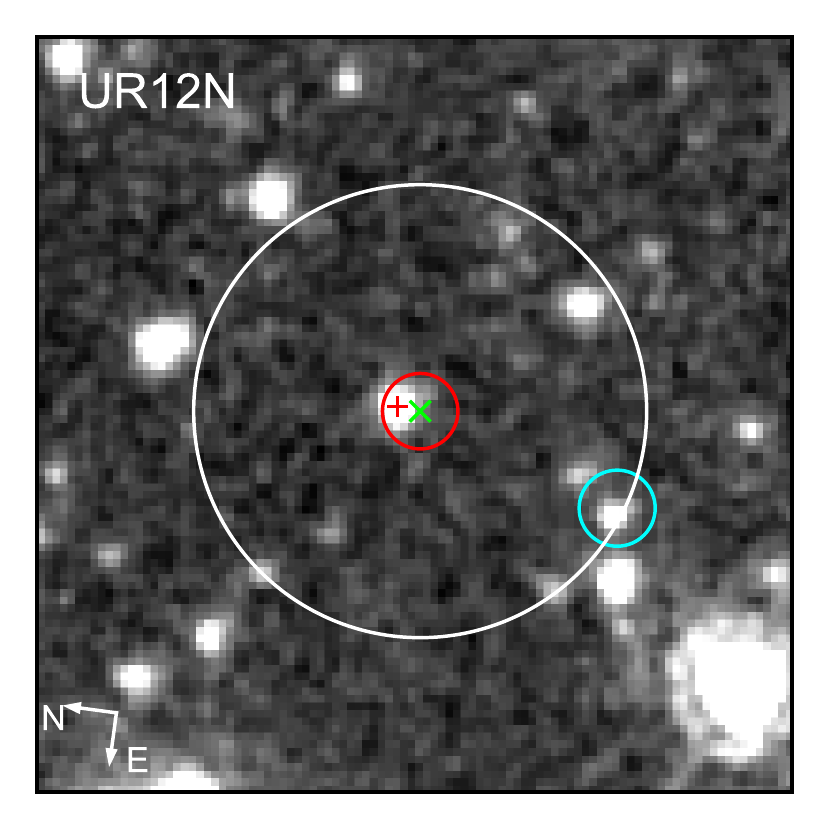}}
{\includegraphics[width=4.4cm, height=4.4cm]{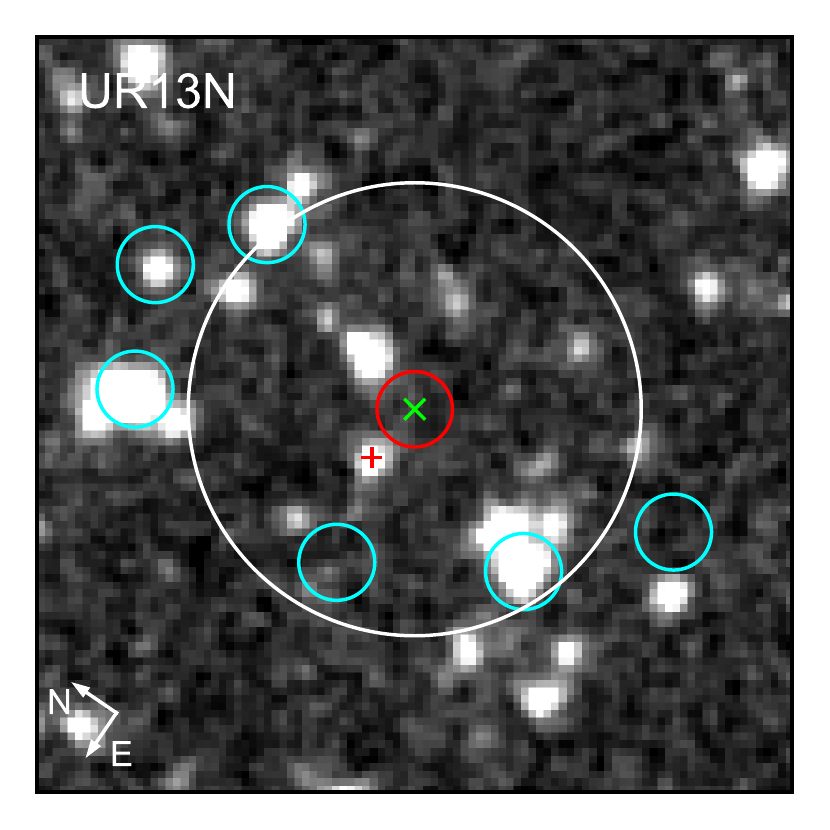}}
{\includegraphics[width=4.4cm, height=4.4cm]{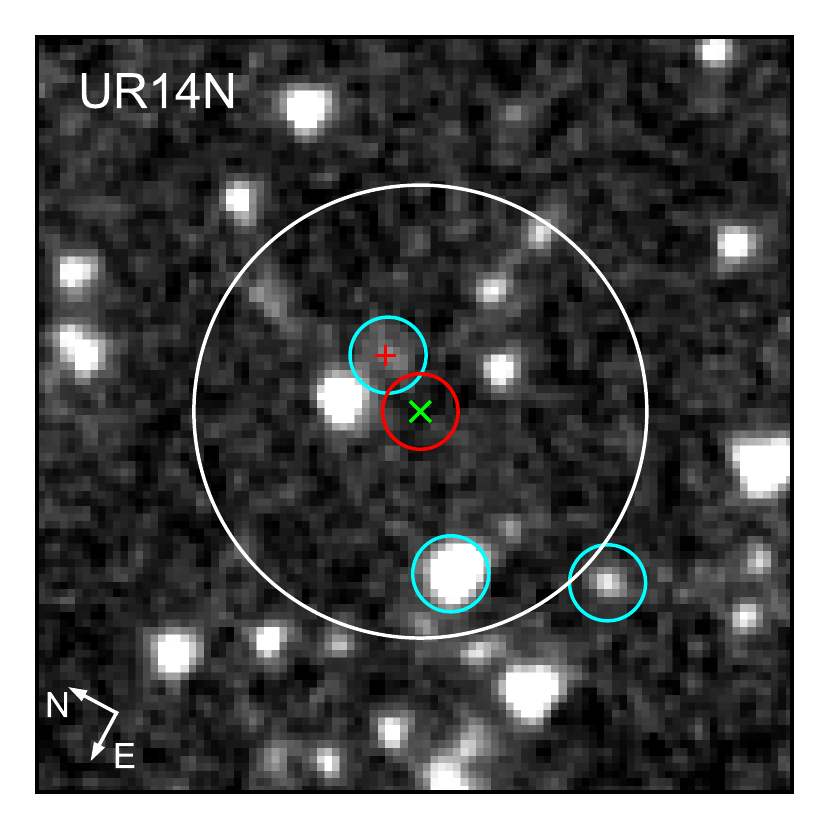}}
{\includegraphics[width=4.4cm, height=4.4cm]{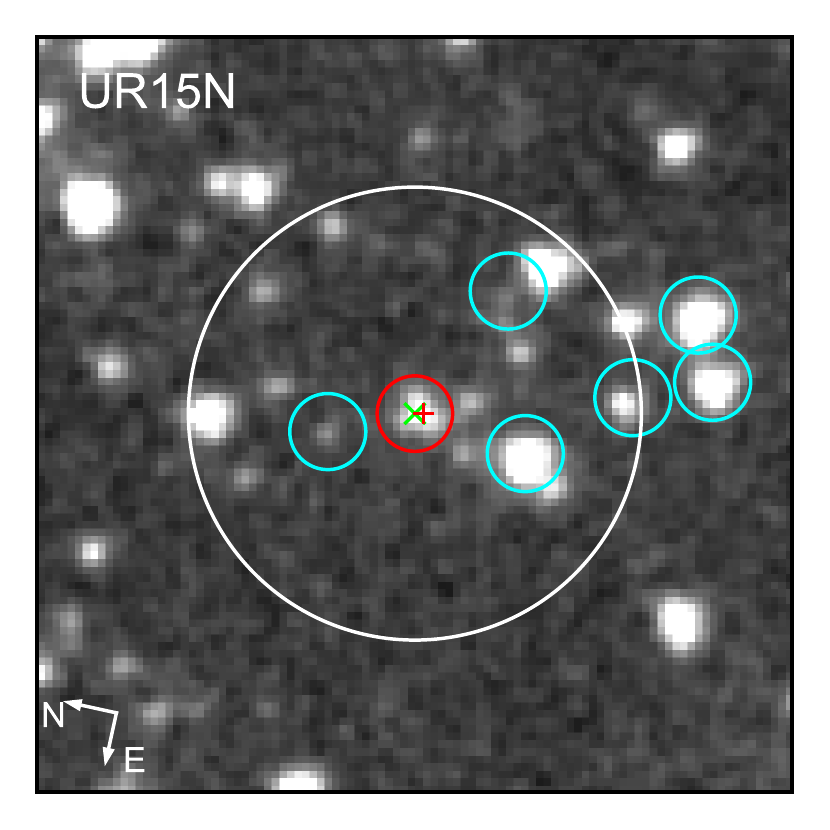}}
{\includegraphics[width=4.4cm, height=4.4cm]{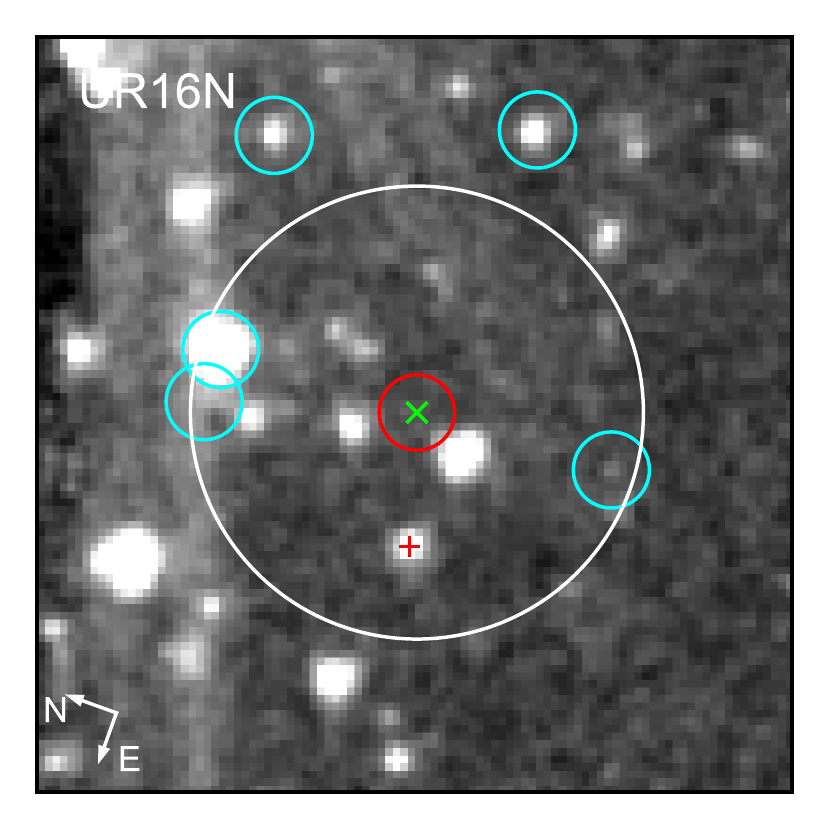}}
{\includegraphics[width=4.4cm, height=4.4cm]{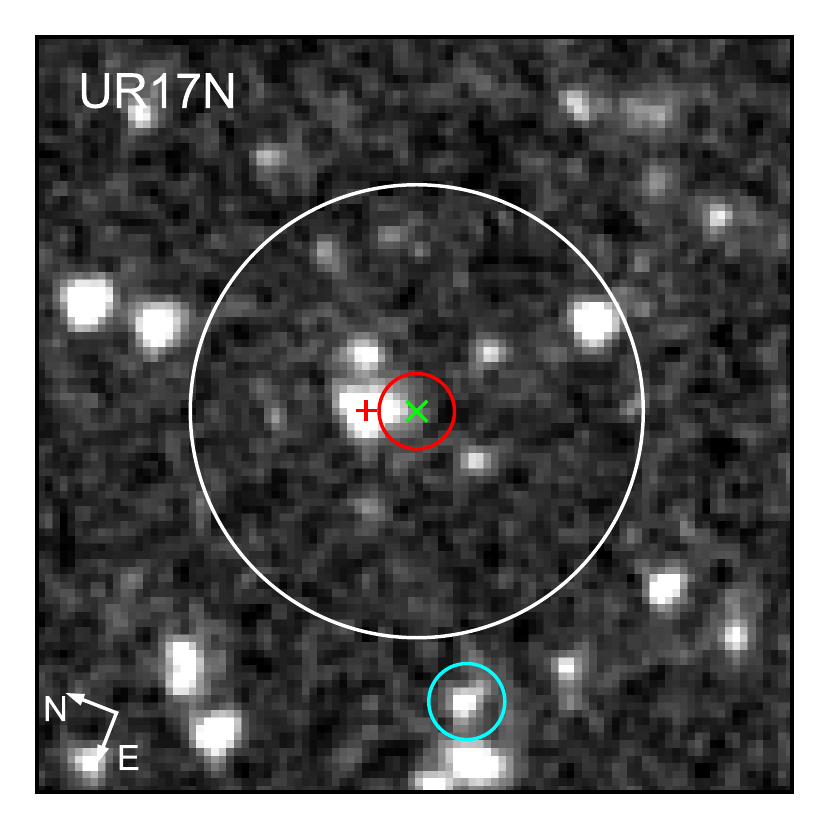}}
{\includegraphics[width=4.4cm, height=4.4cm]{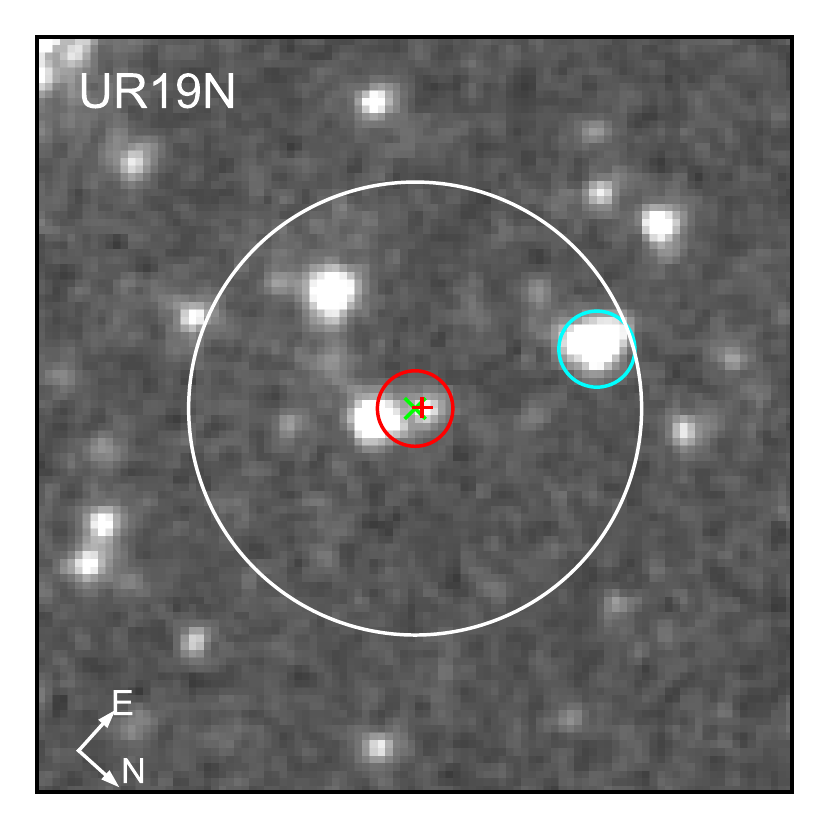}}
{\includegraphics[width=4.4cm, height=4.4cm]{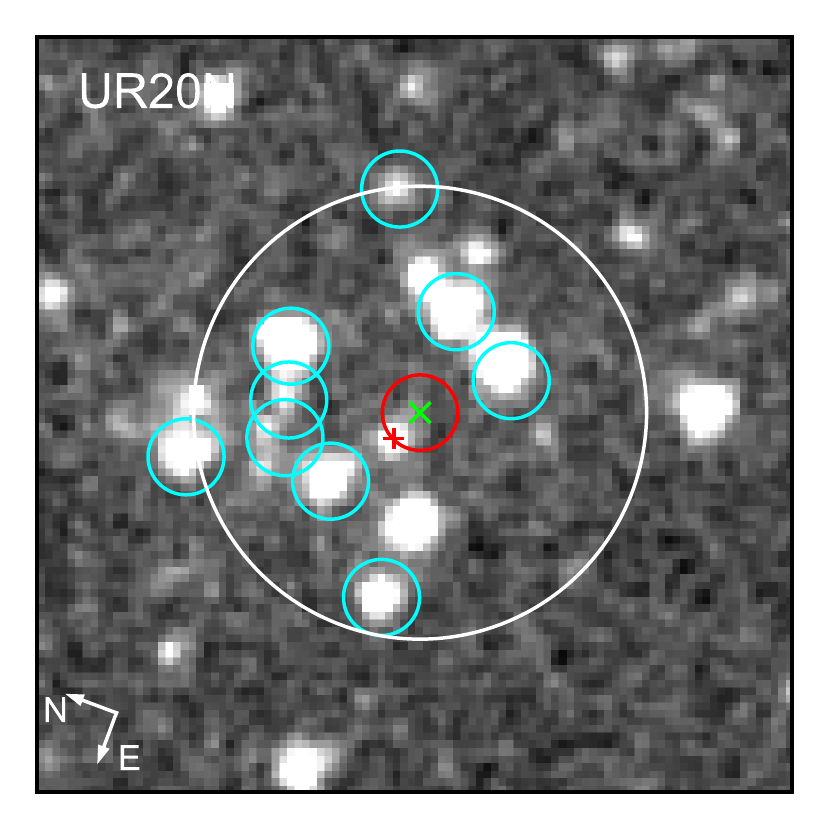}}
{\includegraphics[width=4.4cm, height=4.4cm]{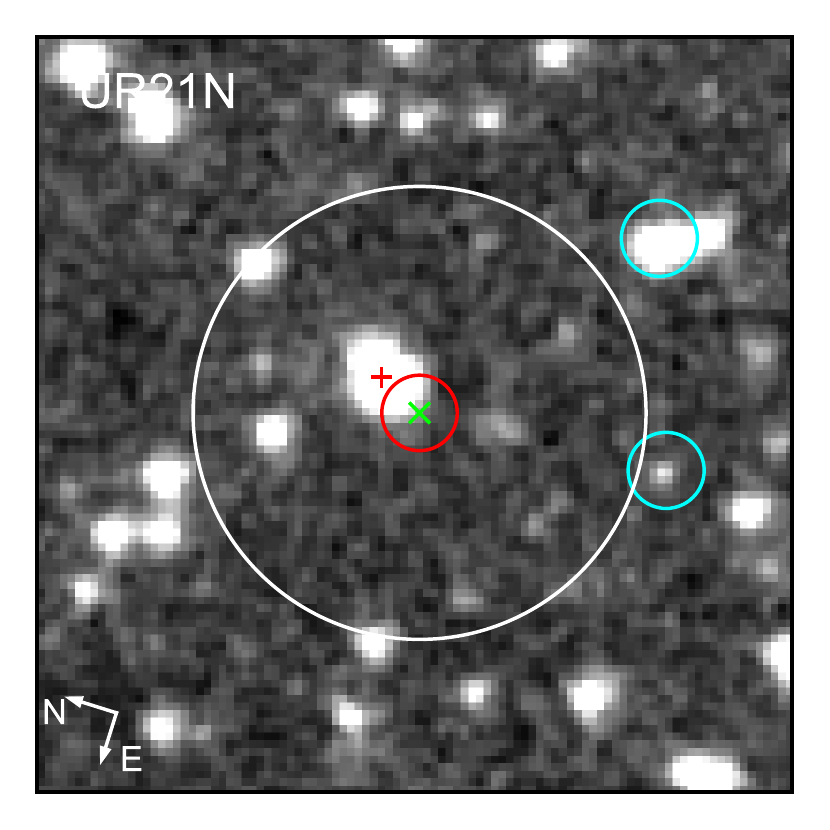}}
{\includegraphics[width=4.4cm, height=4.4cm]{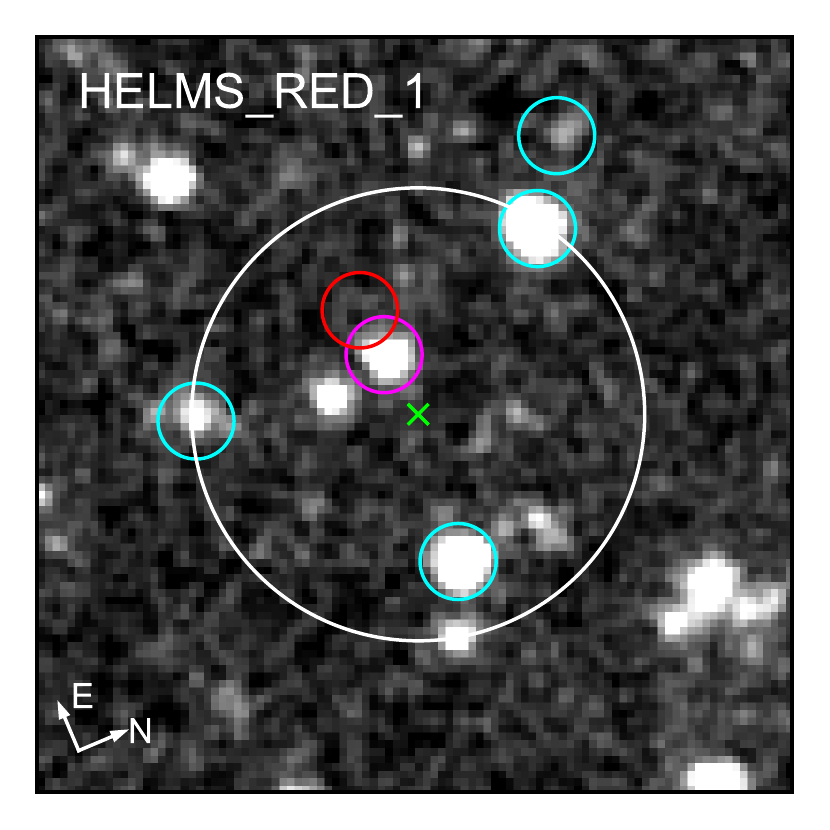}}
\caption{Continued 60$\arcsec$ $\times$ 60$\arcsec$ cutouts }
\label{fig:data}
\end{figure*}

\addtocounter{figure}{-1}
\begin{figure*}
\centering
{\includegraphics[width=4.4cm, height=4.4cm]{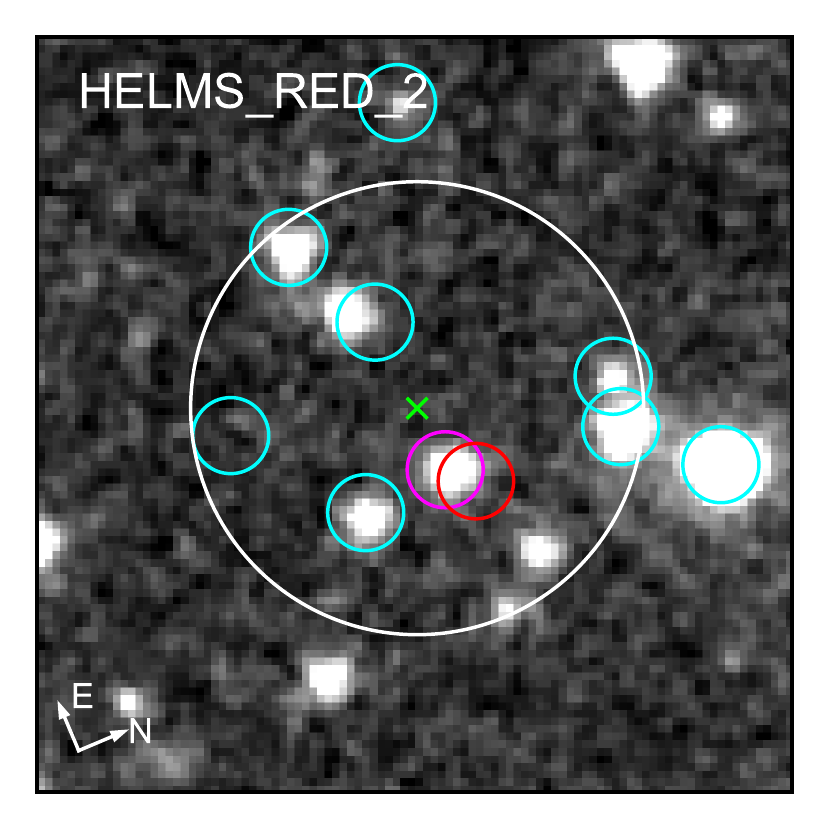}}
{\includegraphics[width=4.4cm, height=4.4cm]{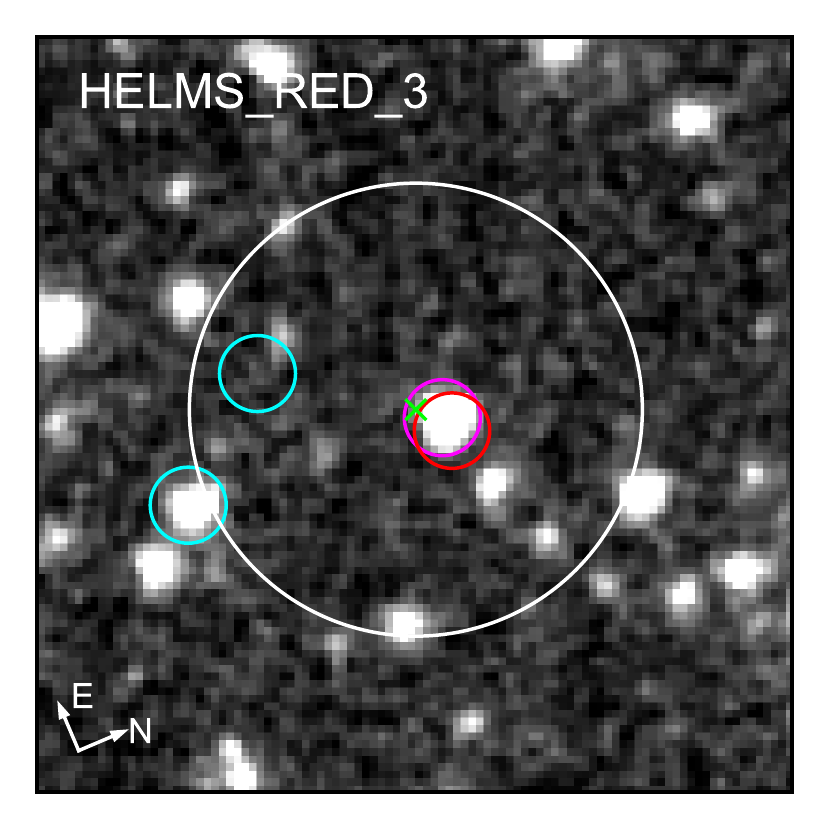}}
{\includegraphics[width=4.4cm, height=4.4cm]{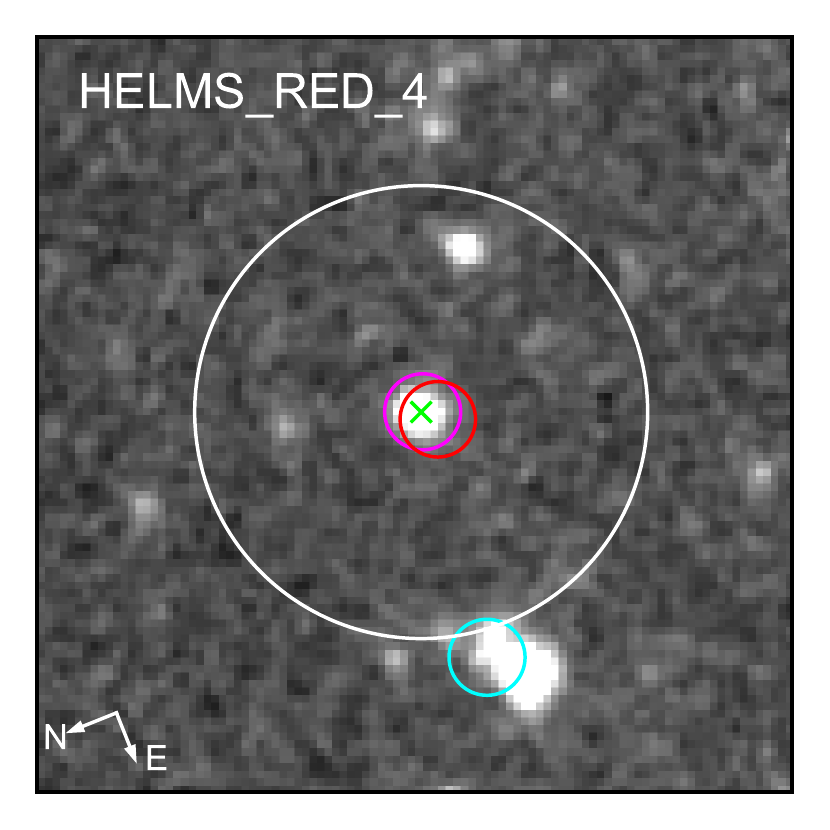}}
{\includegraphics[width=4.4cm, height=4.4cm]{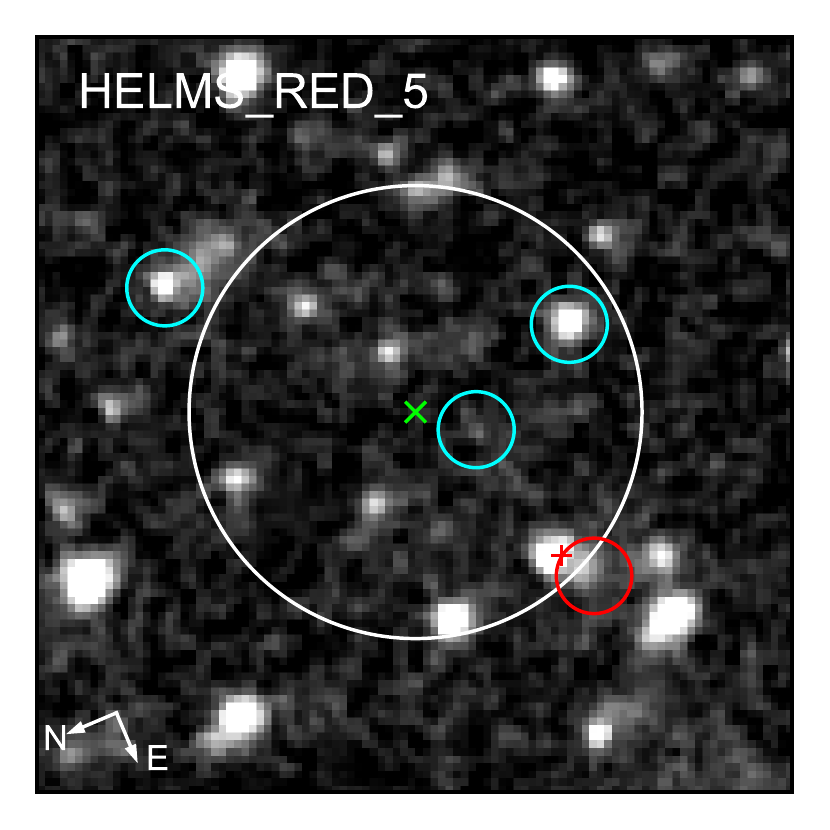}}
{\includegraphics[width=4.4cm, height=4.4cm]{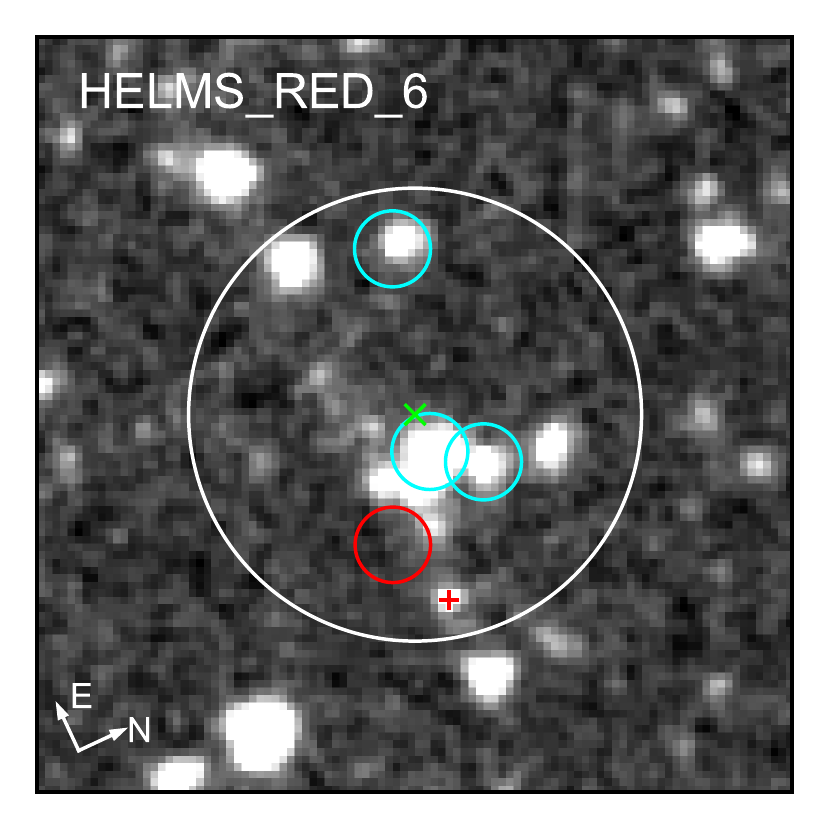}}
{\includegraphics[width=4.4cm, height=4.4cm]{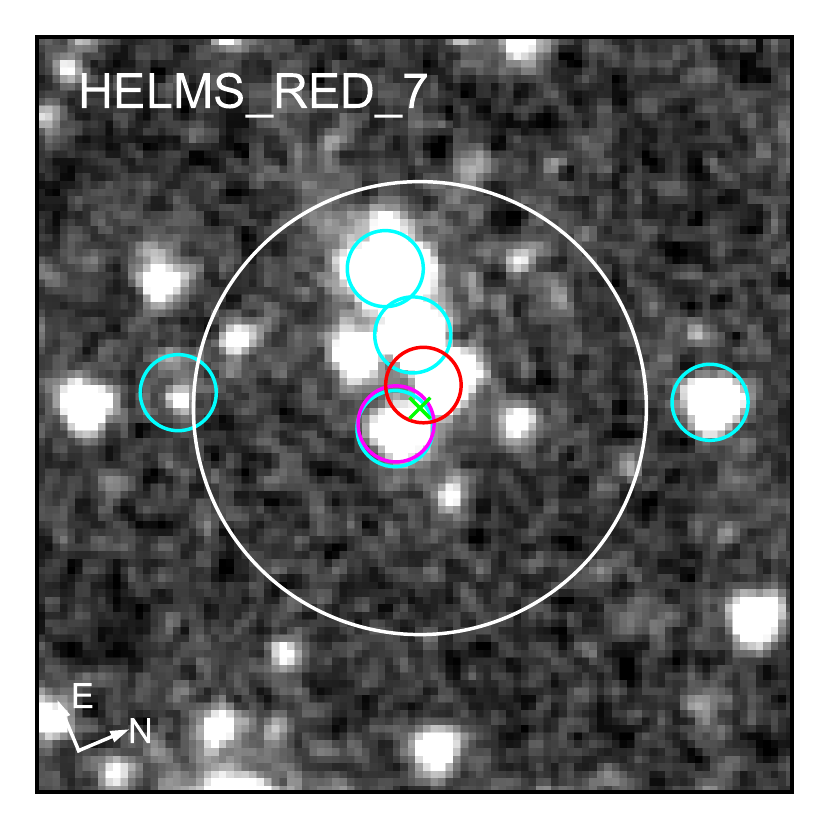}}
{\includegraphics[width=4.4cm, height=4.4cm]{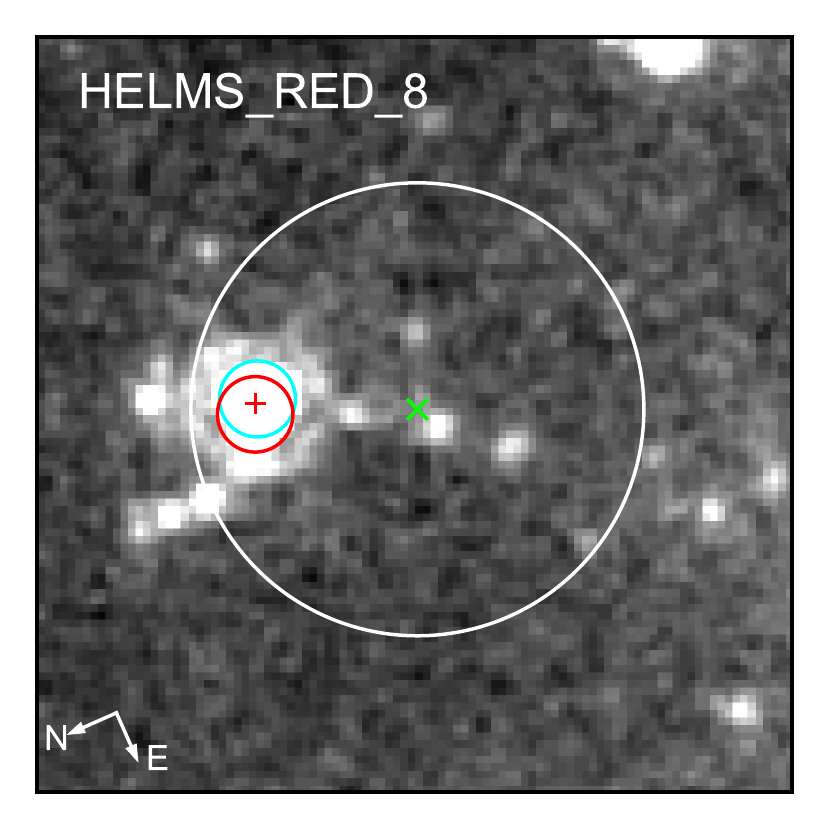}}
{\includegraphics[width=4.4cm, height=4.4cm]{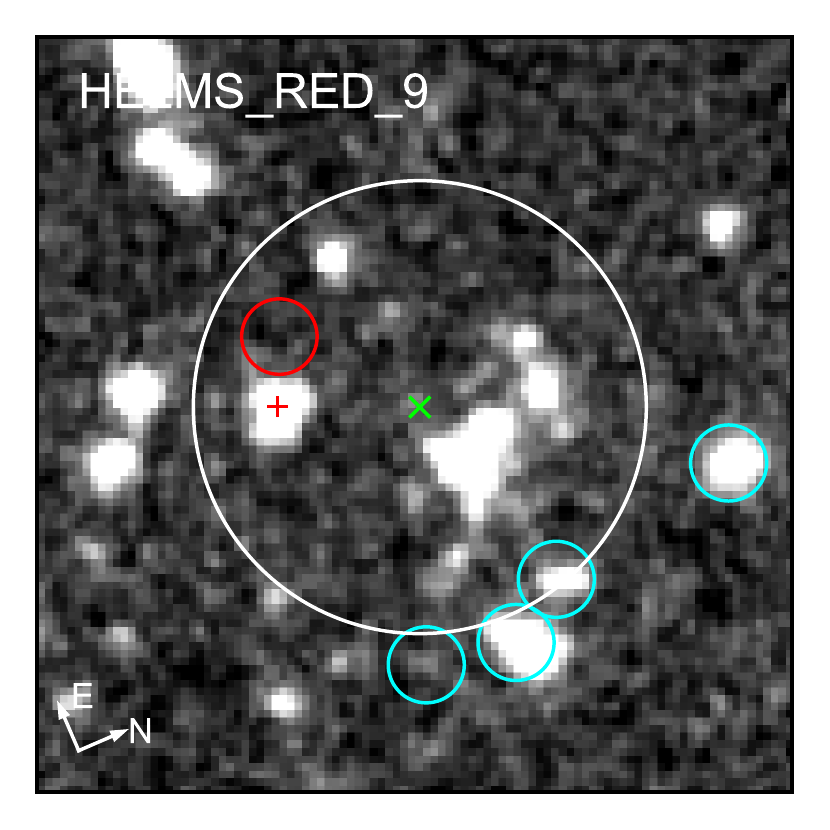}}
{\includegraphics[width=4.4cm, height=4.4cm]{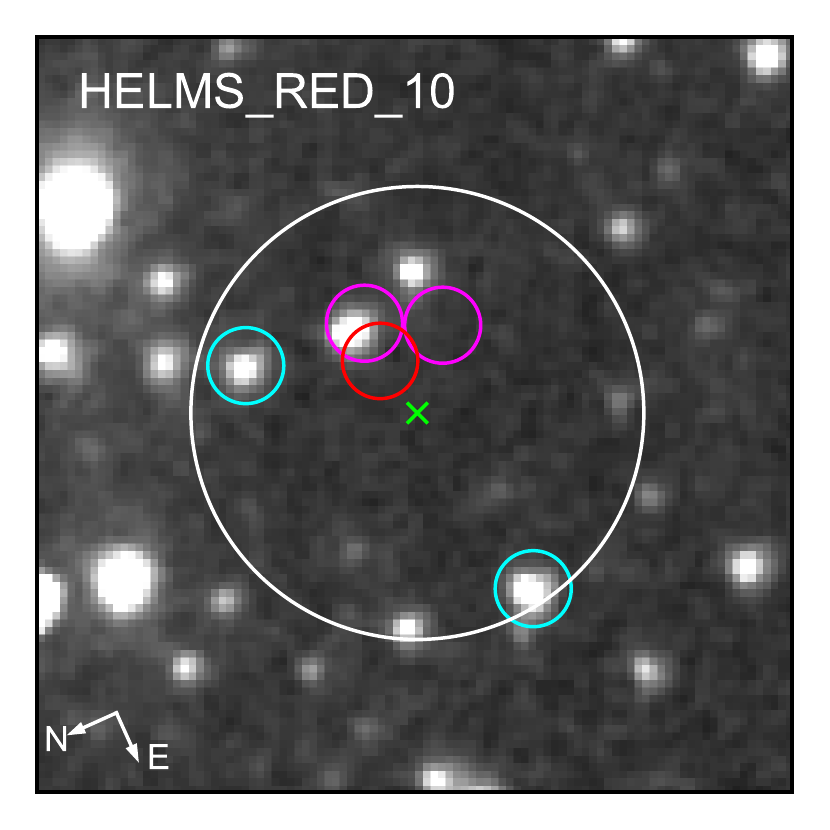}}
{\includegraphics[width=4.4cm, height=4.4cm]{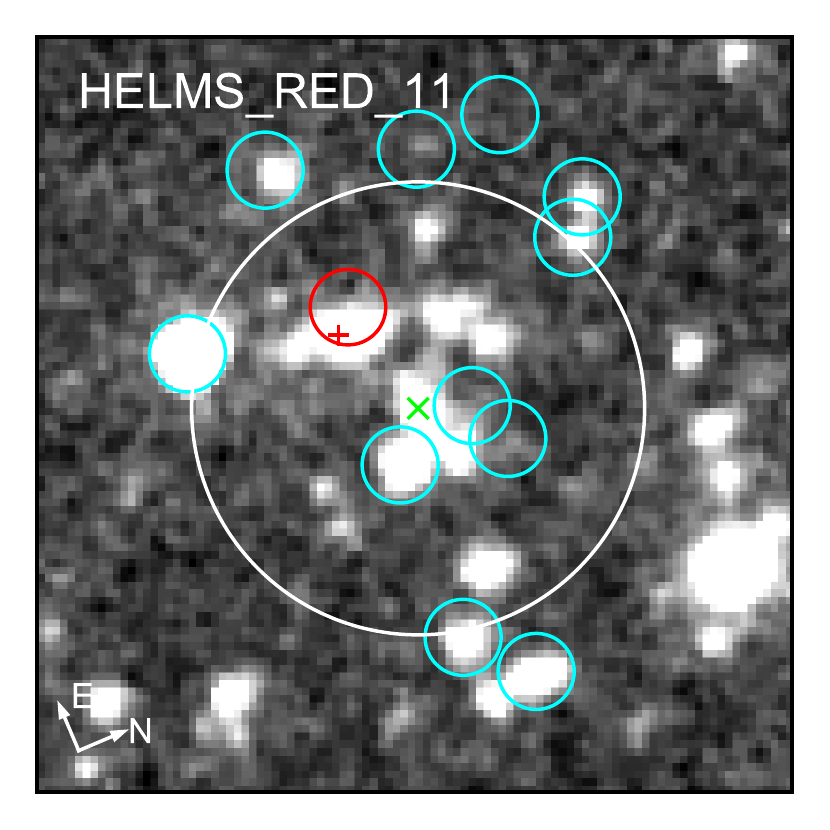}}
{\includegraphics[width=4.4cm, height=4.4cm]{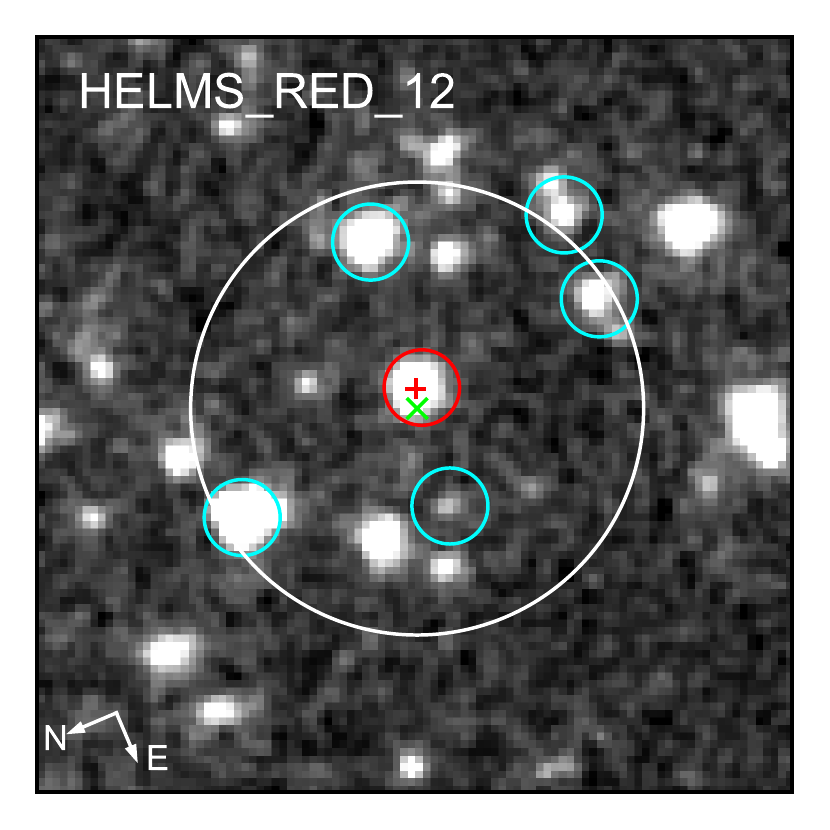}}
{\includegraphics[width=4.4cm, height=4.4cm]{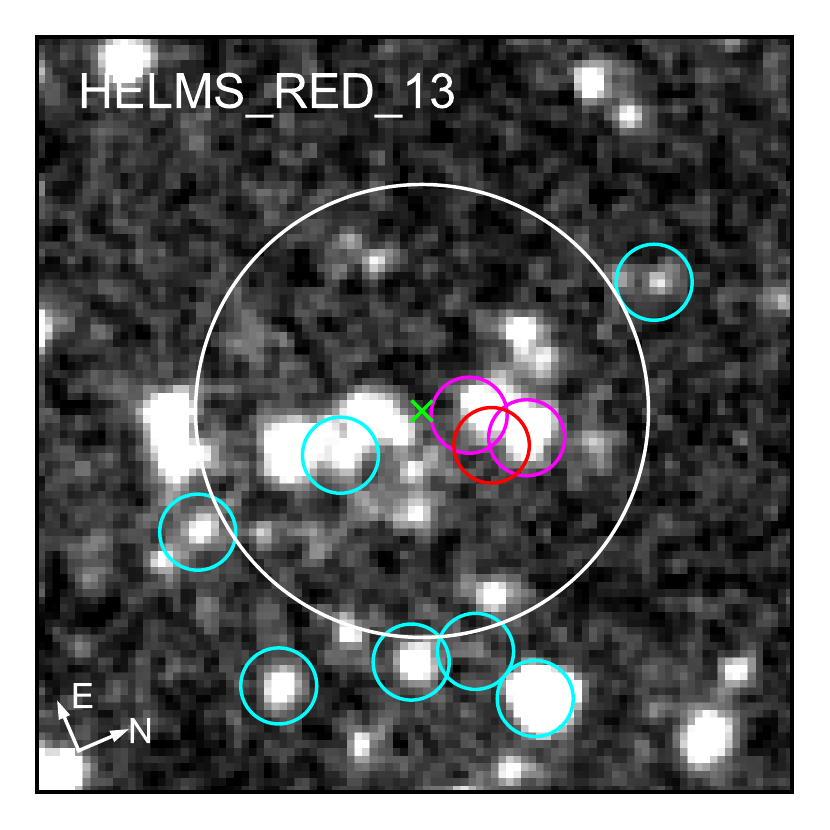}}
{\includegraphics[width=4.4cm, height=4.4cm]{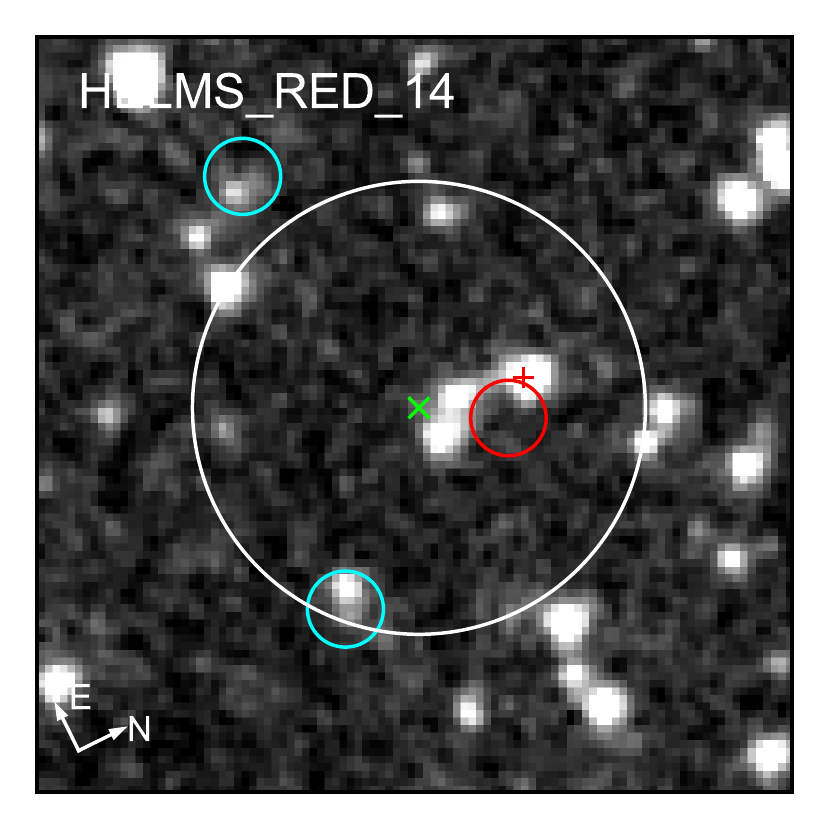}}
{\includegraphics[width=4.4cm, height=4.4cm]{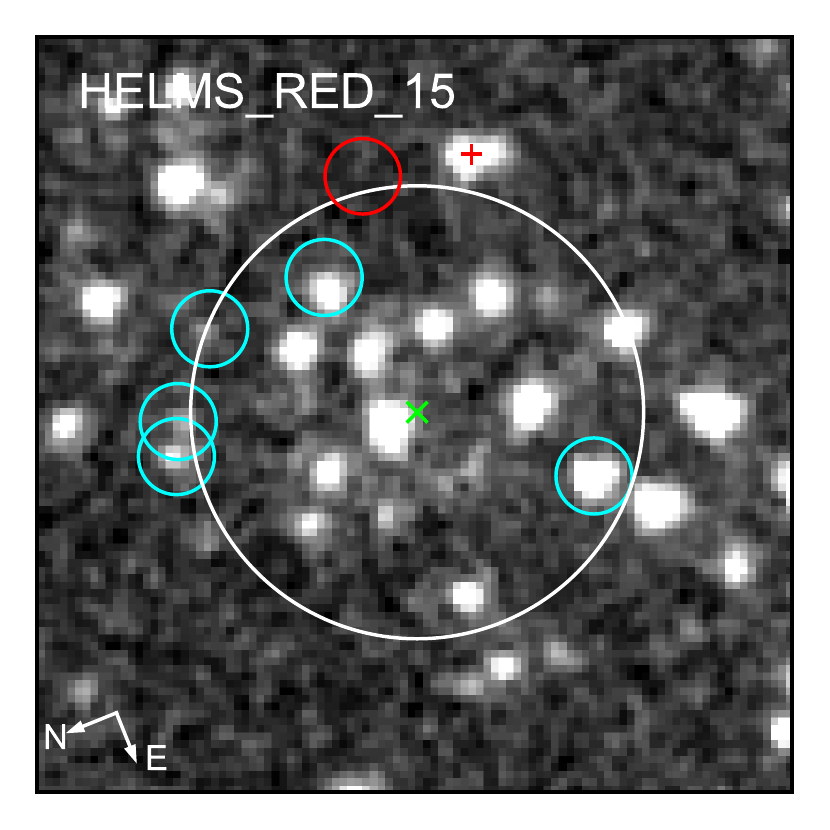}}
{\includegraphics[width=4.4cm, height=4.4cm]{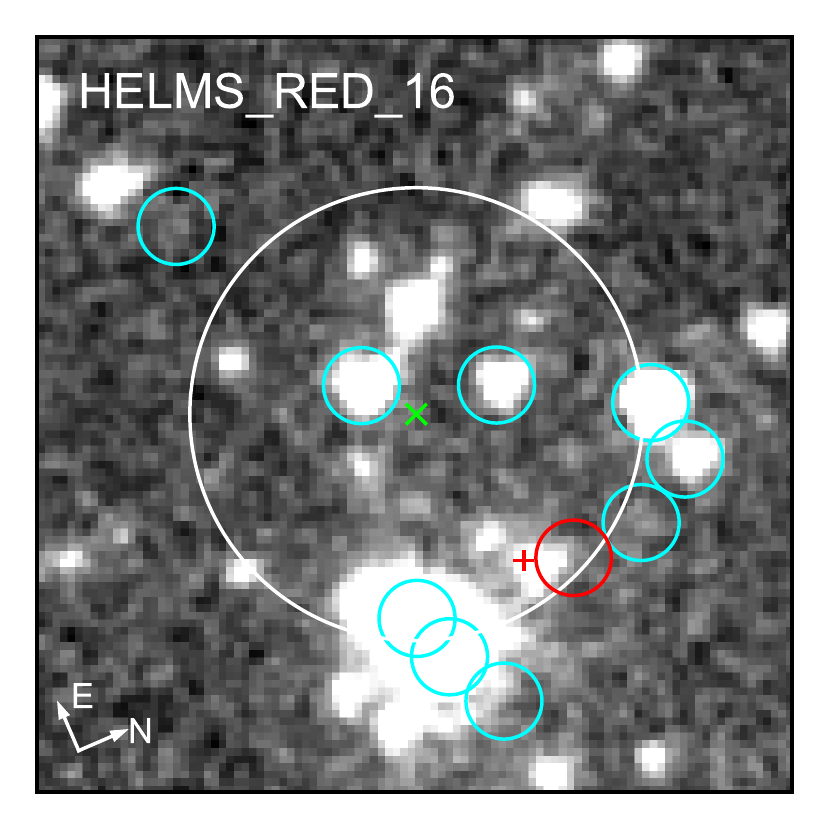}}
{\includegraphics[width=4.4cm, height=4.4cm]{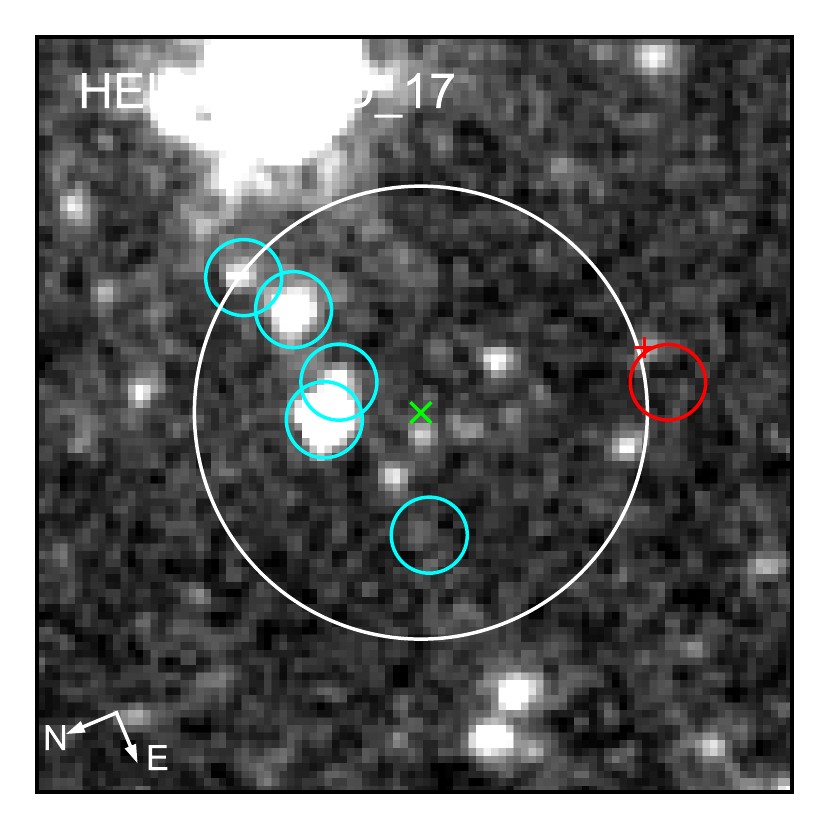}}
{\includegraphics[width=4.4cm, height=4.4cm]{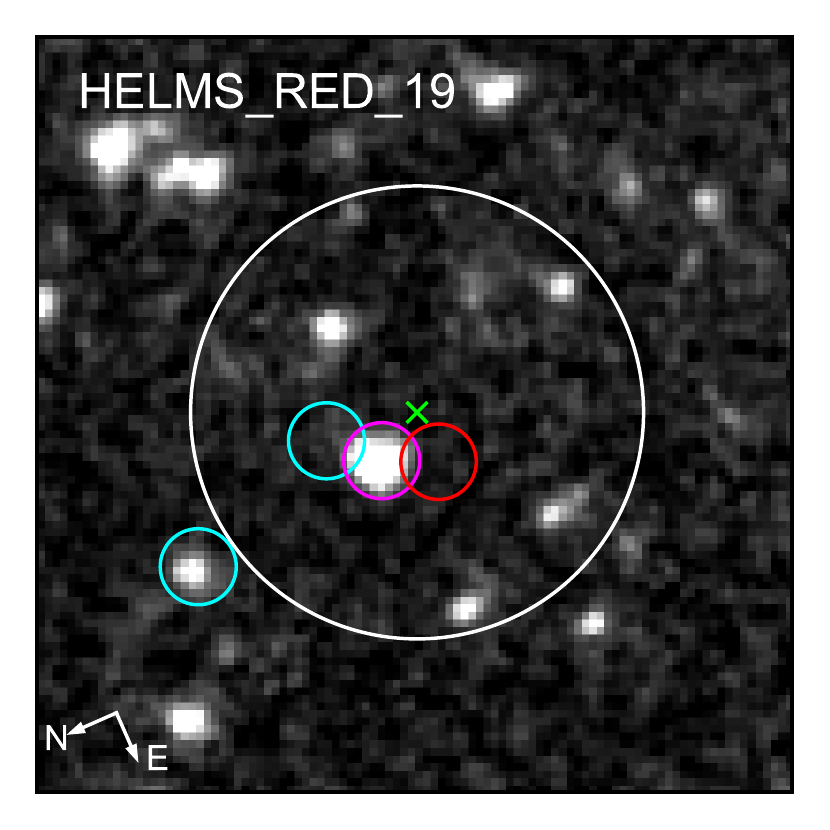}}
{\includegraphics[width=4.4cm, height=4.4cm]{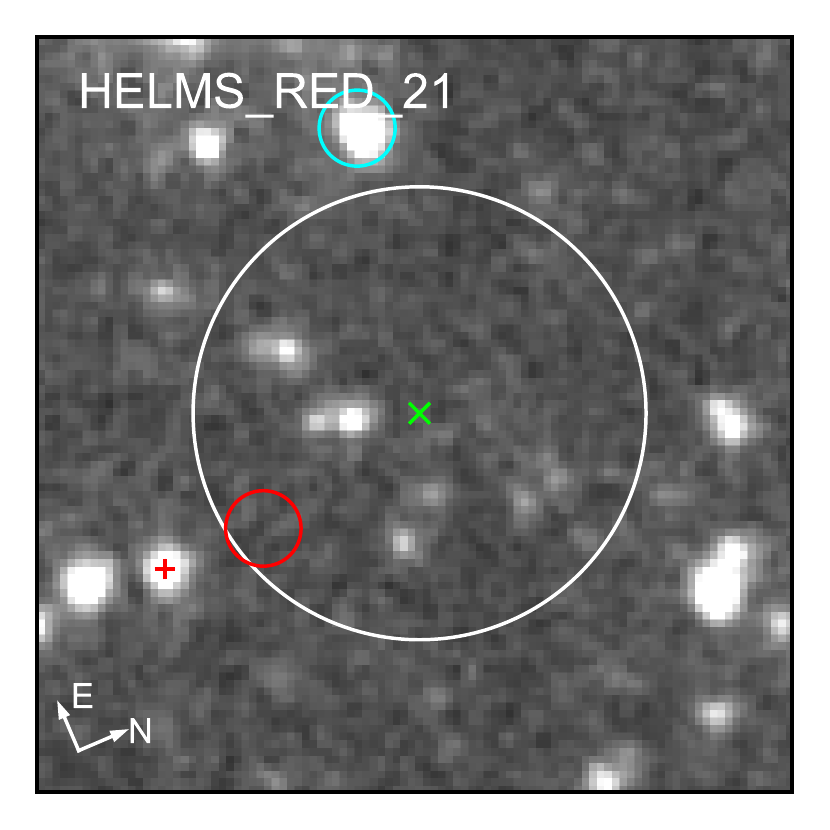}}
{\includegraphics[width=4.4cm, height=4.4cm]{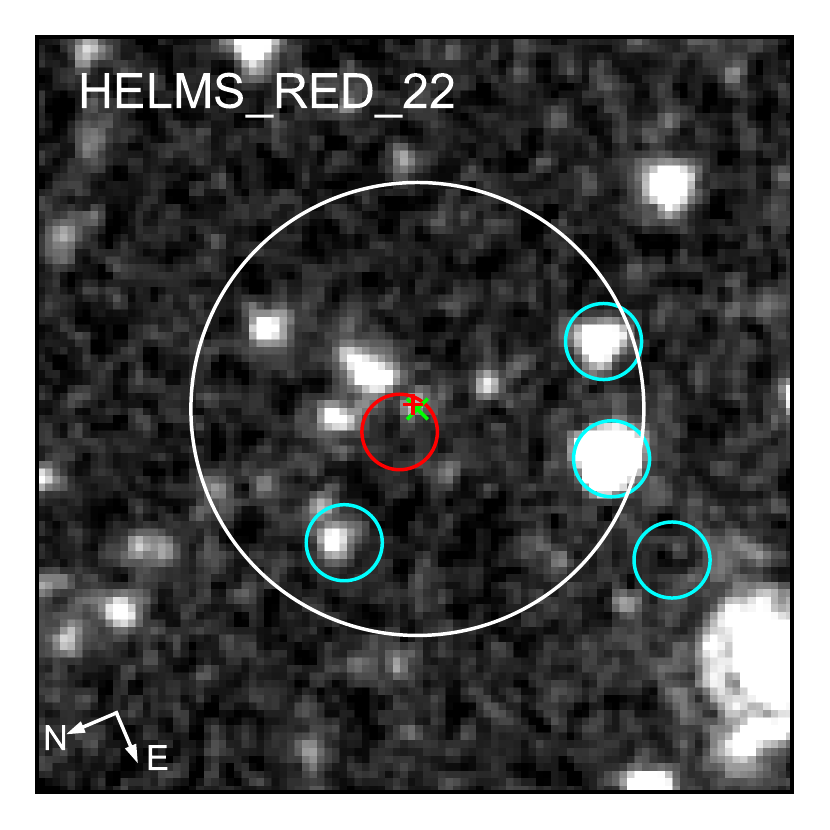}}
{\includegraphics[width=4.4cm, height=4.4cm]{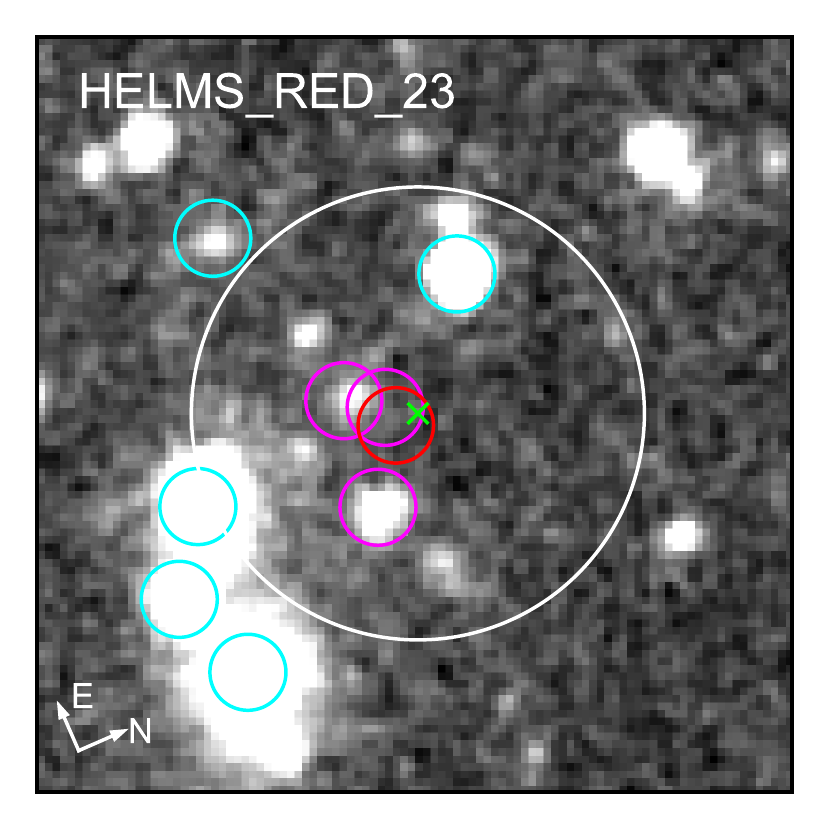}}
\caption{Continued 60$\arcsec$ $\times$ 60$\arcsec$ cutouts }
\label{fig:data}
\end{figure*}

\addtocounter{figure}{-1}
\begin{figure*}
\centering
{\includegraphics[width=4.4cm, height=4.4cm]{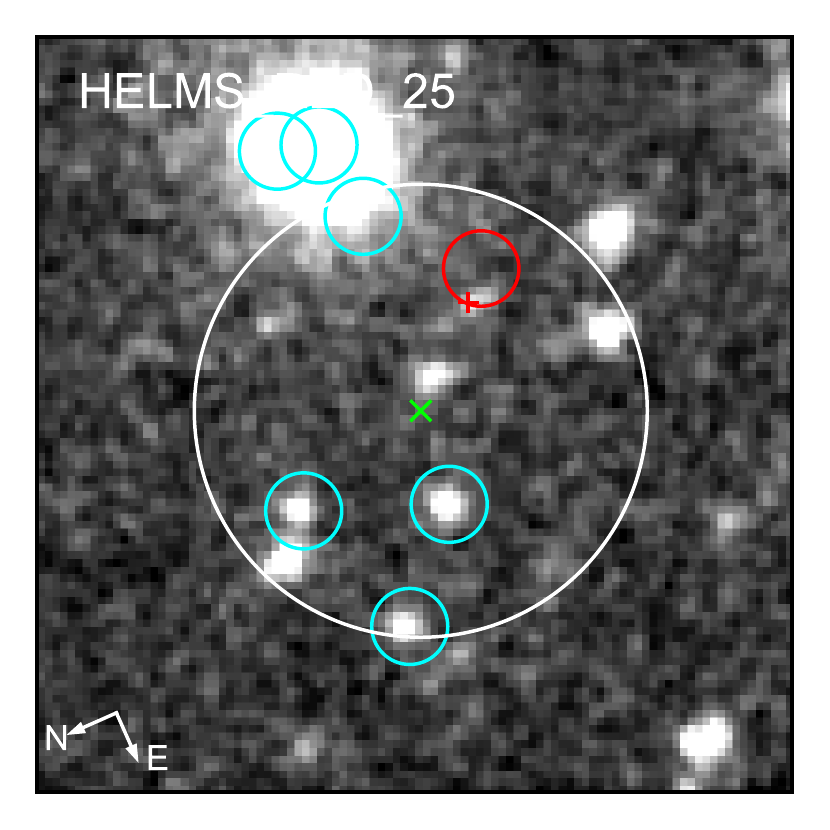}}
{\includegraphics[width=4.4cm, height=4.4cm]{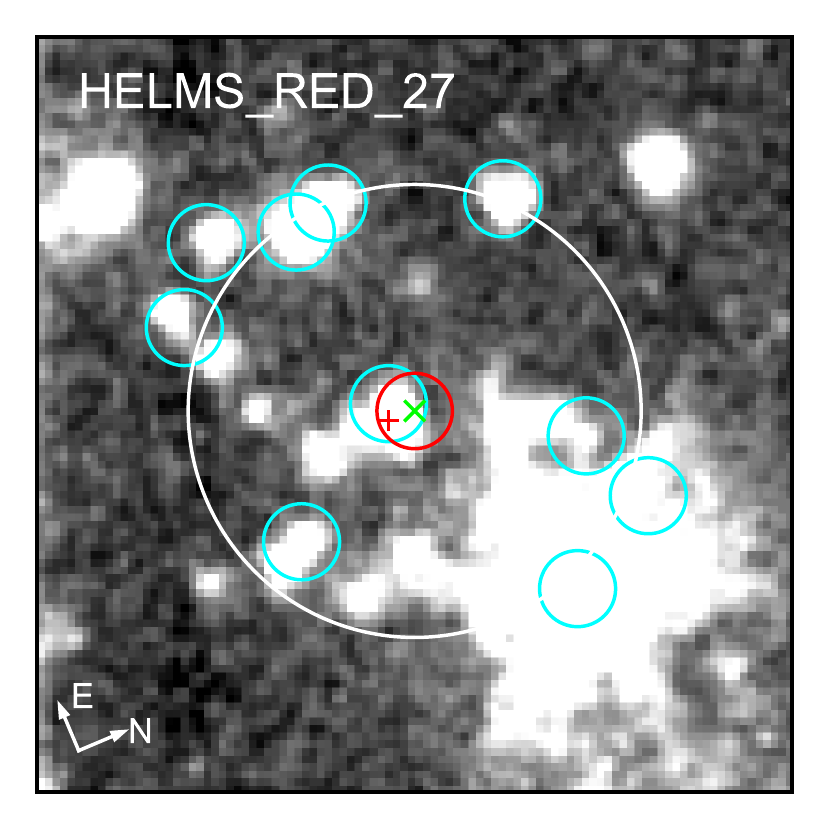}}
{\includegraphics[width=4.4cm, height=4.4cm]{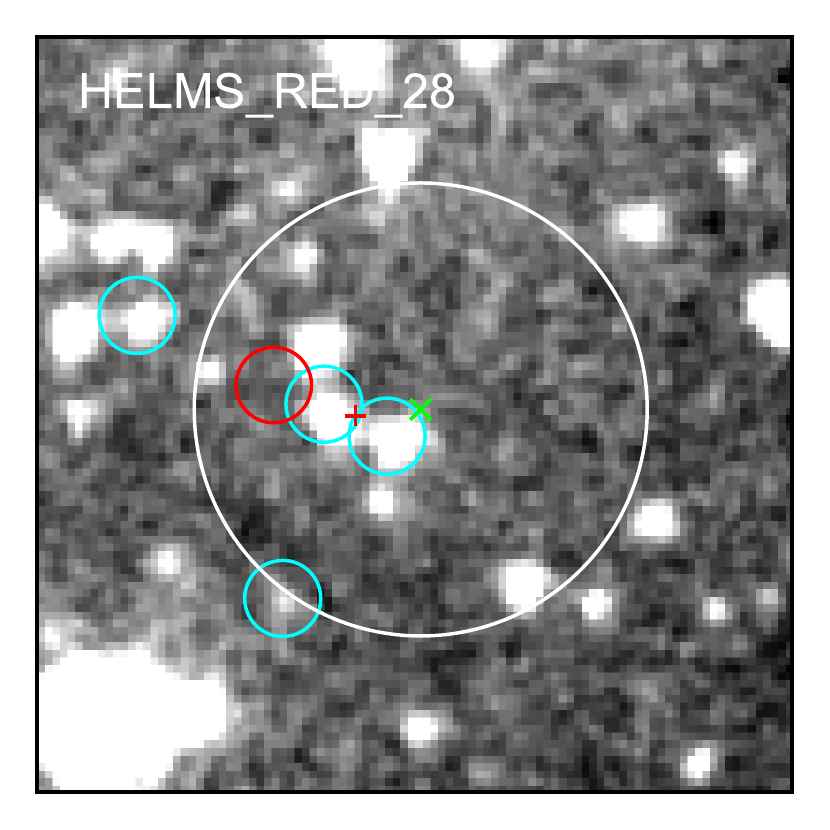}}
{\includegraphics[width=4.4cm, height=4.4cm]{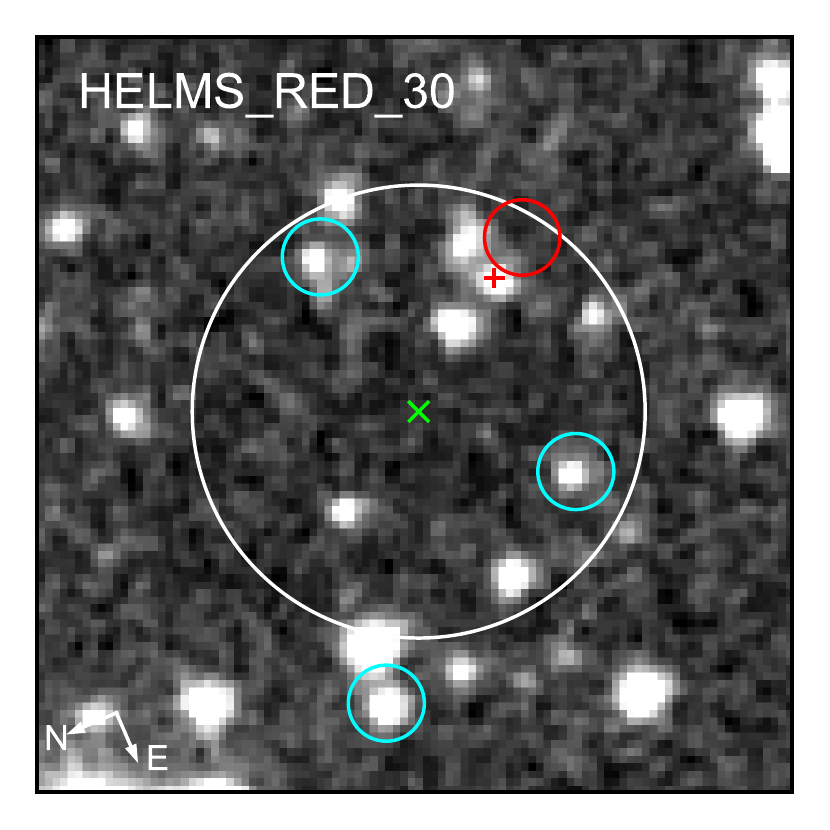}}
{\includegraphics[width=4.4cm, height=4.4cm]{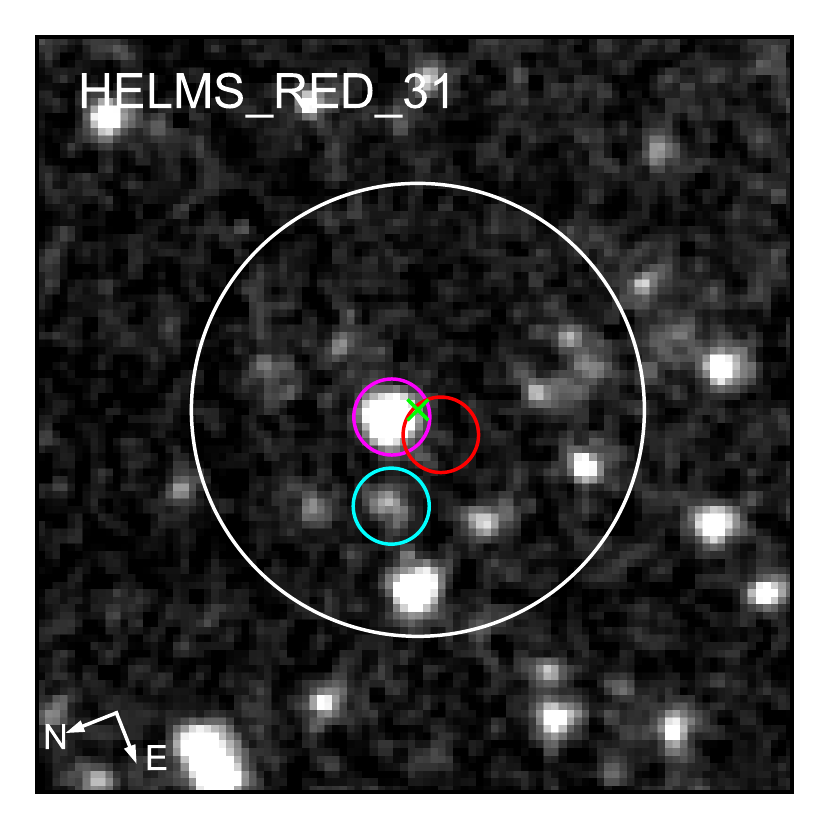}}
{\includegraphics[width=4.4cm, height=4.4cm]{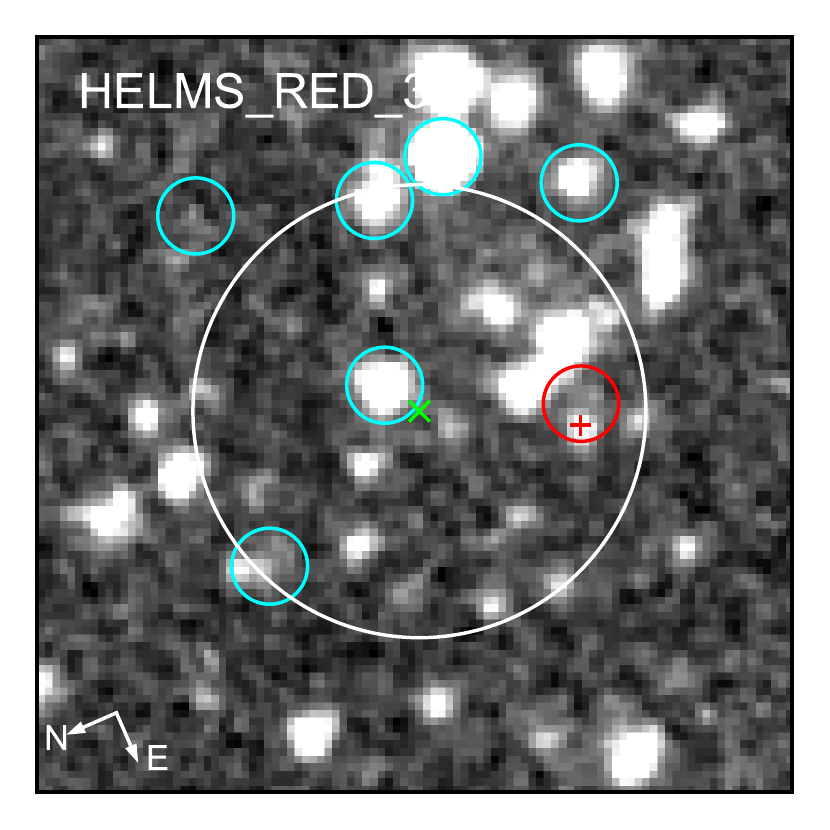}}
{\includegraphics[width=4.4cm, height=4.4cm]{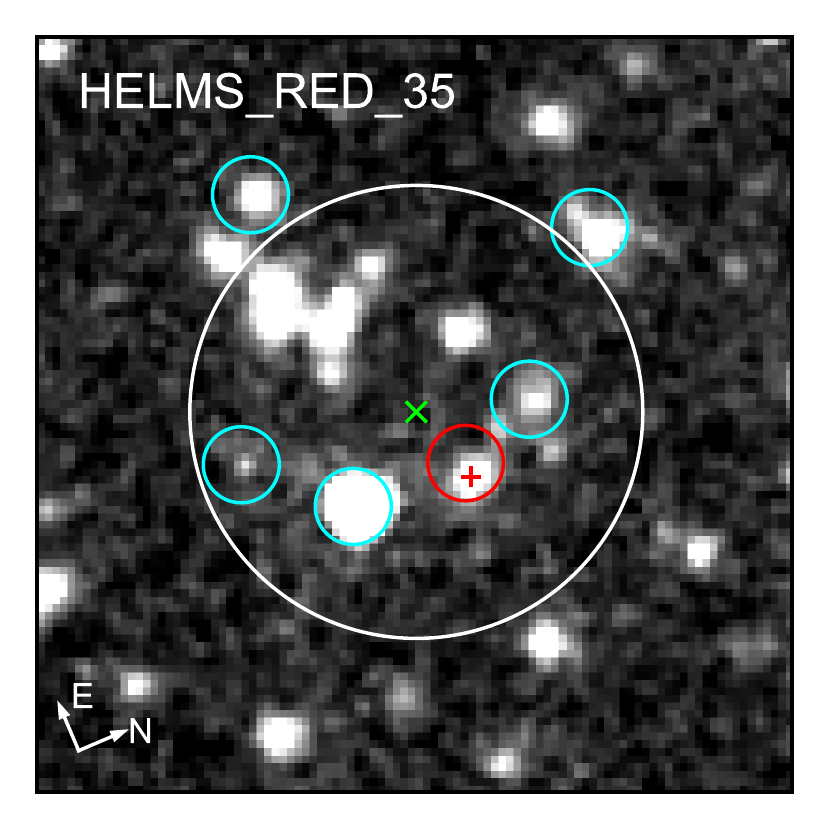}}
{\includegraphics[width=4.4cm, height=4.4cm]{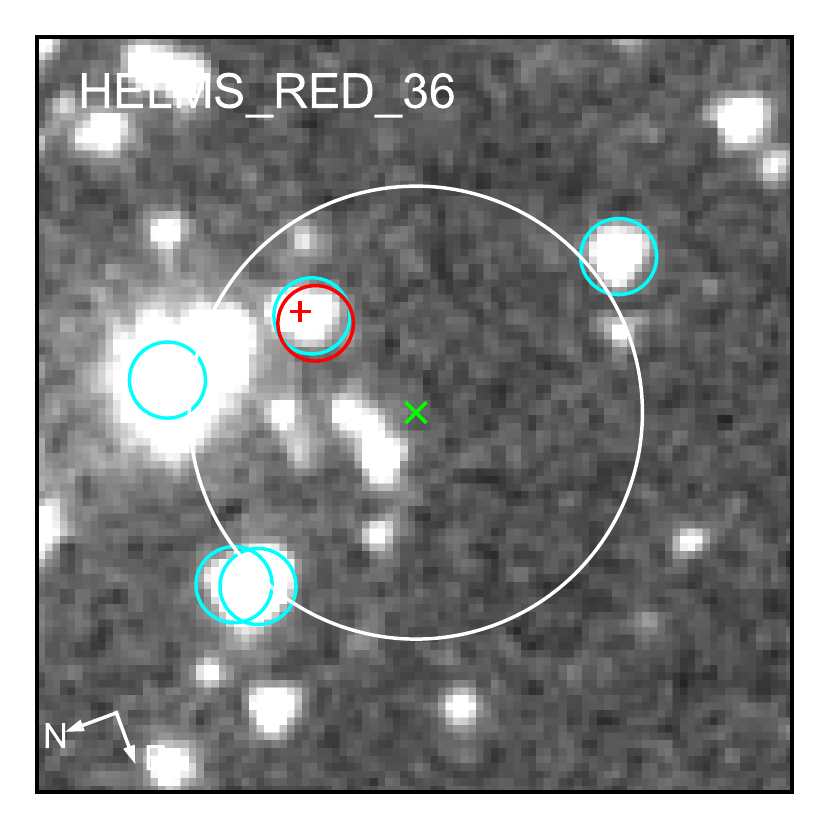}}
{\includegraphics[width=4.4cm, height=4.4cm]{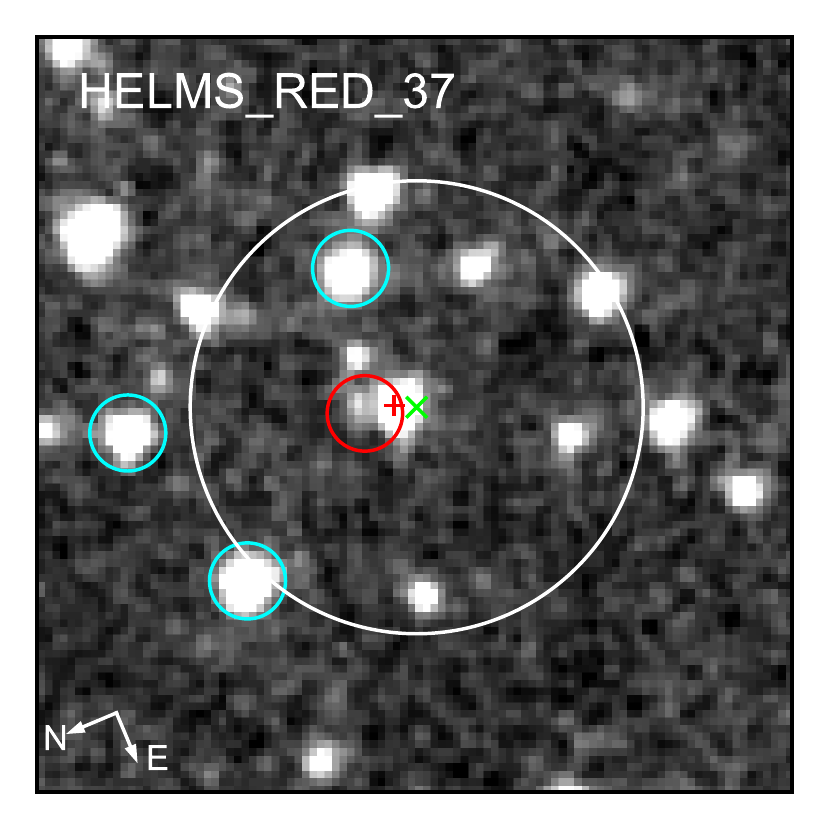}}
{\includegraphics[width=4.4cm, height=4.4cm]{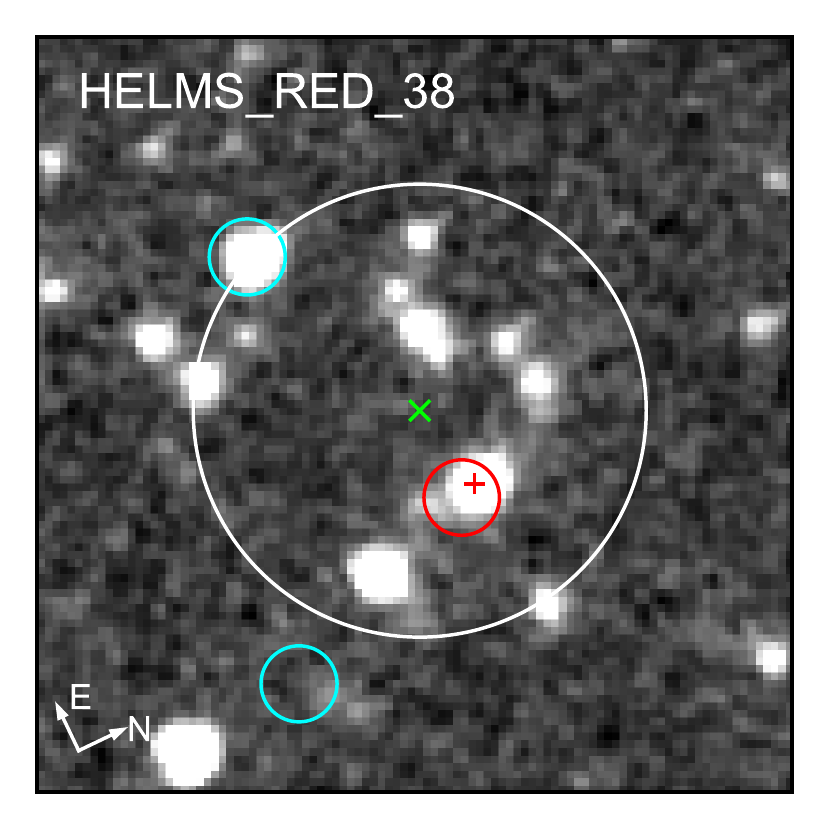}}
{\includegraphics[width=4.4cm, height=4.4cm]{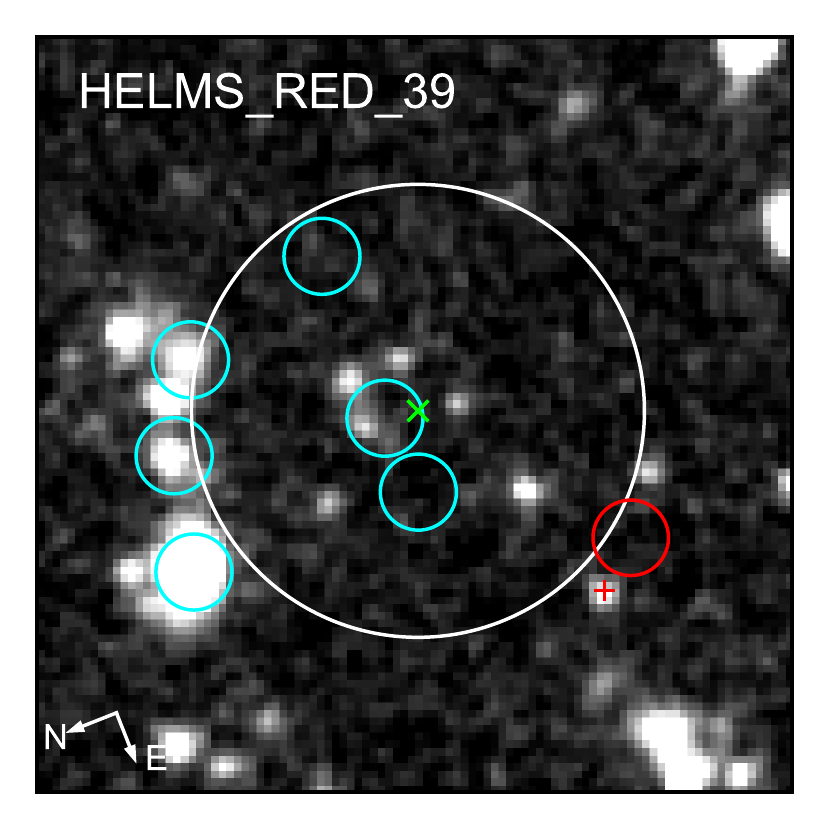}}
{\includegraphics[width=4.4cm, height=4.4cm]{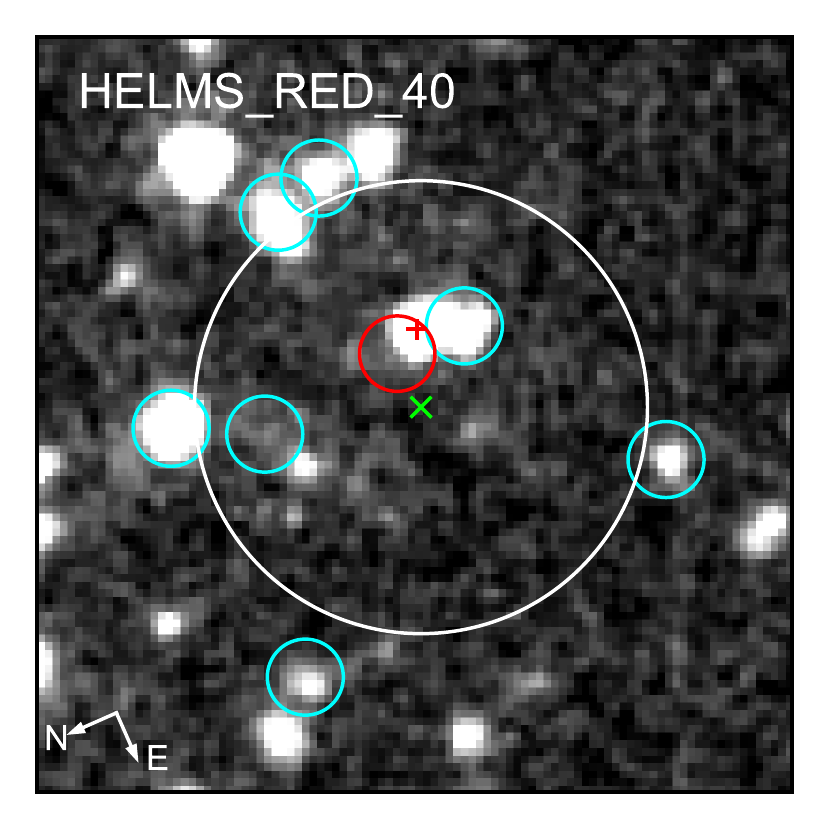}}
{\includegraphics[width=4.4cm, height=4.4cm]{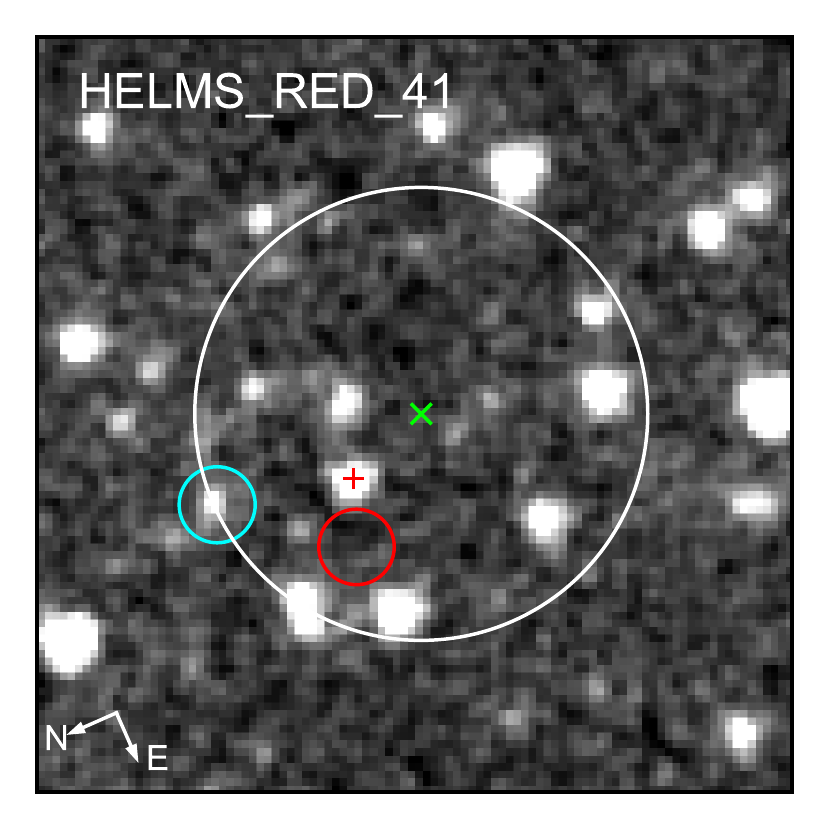}}
{\includegraphics[width=4.4cm, height=4.4cm]{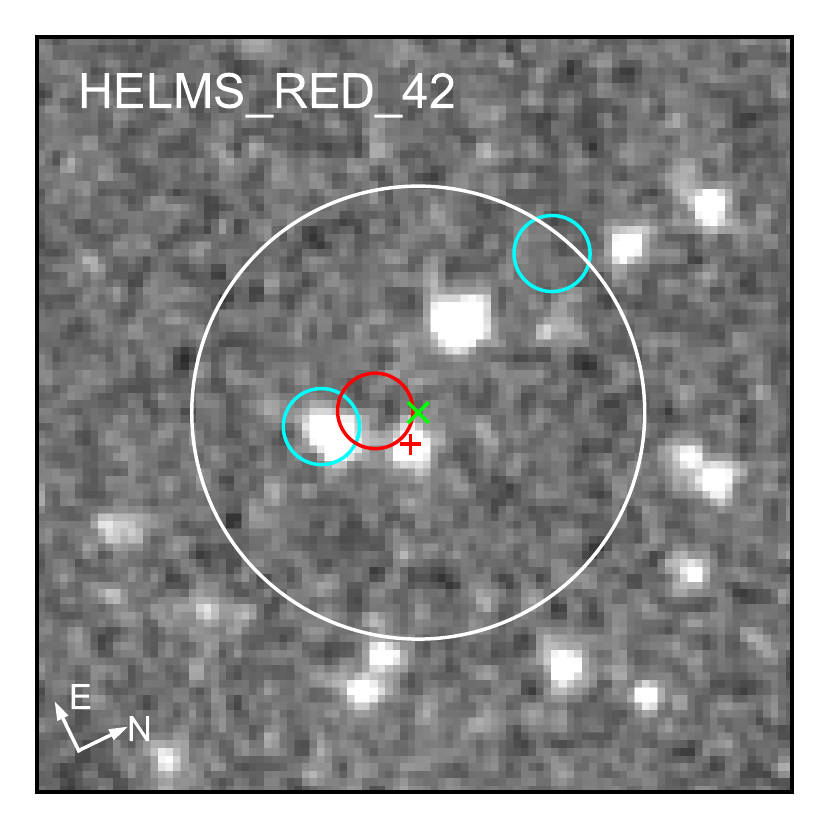}}
{\includegraphics[width=4.4cm, height=4.4cm]{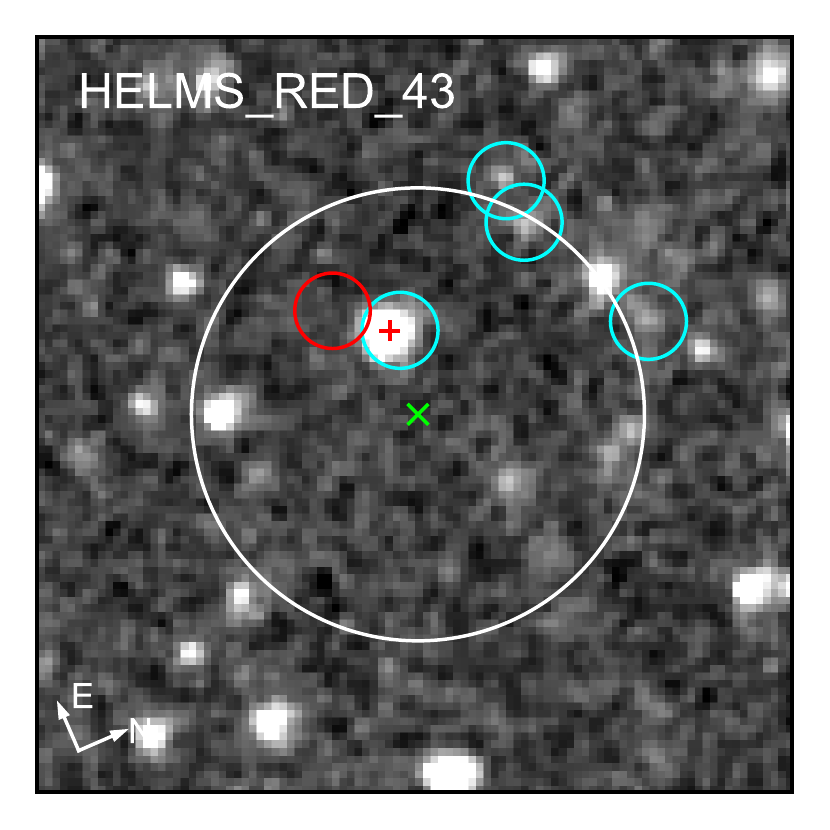}}
{\includegraphics[width=4.4cm, height=4.4cm]{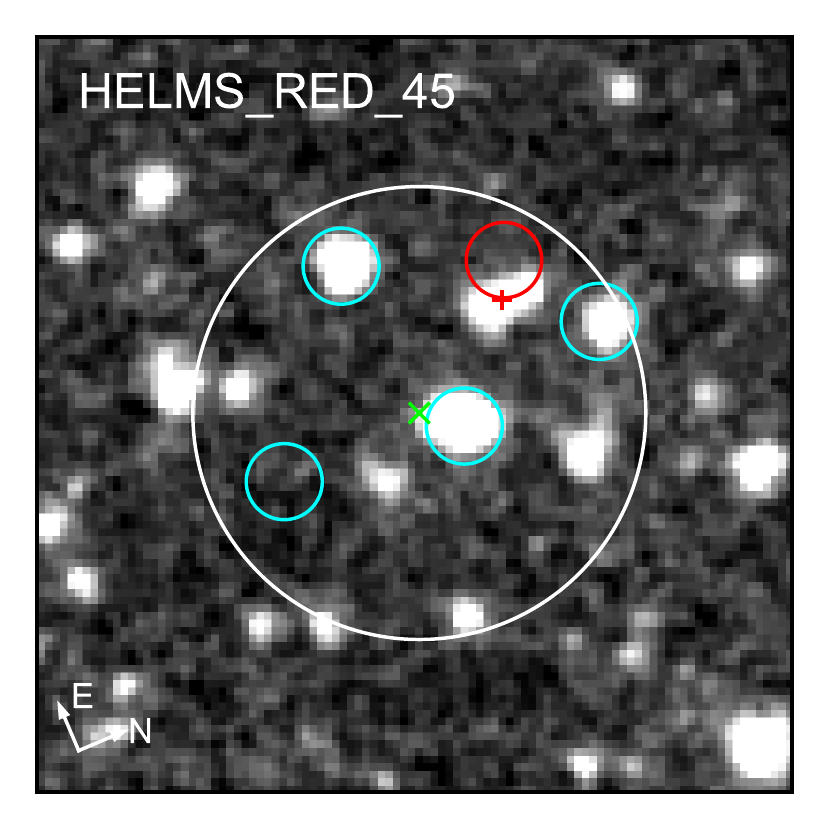}}
{\includegraphics[width=4.4cm, height=4.4cm]{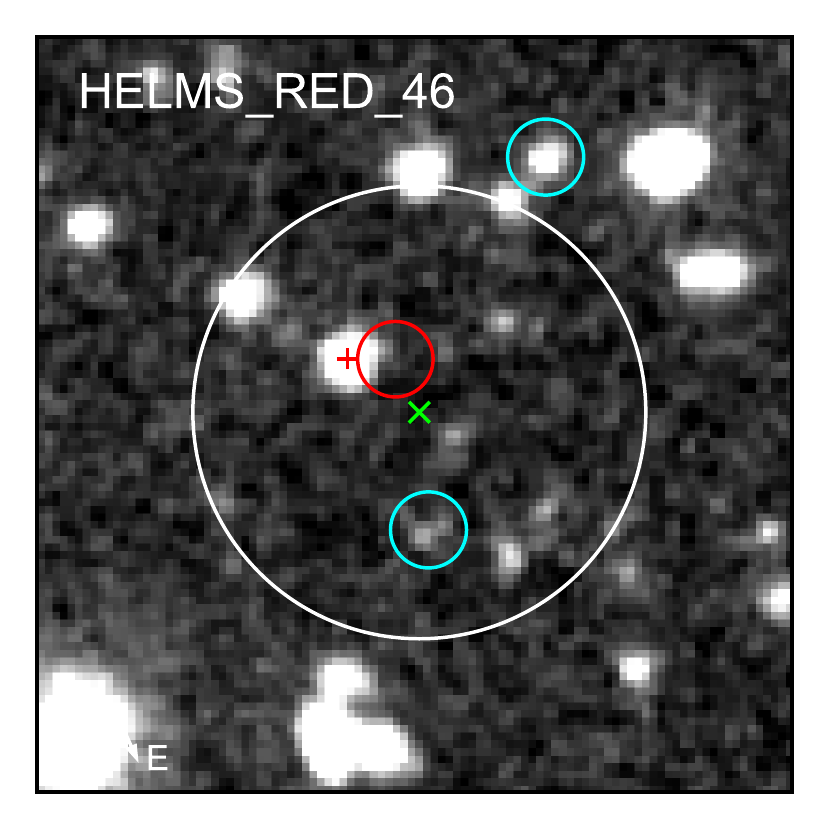}}
{\includegraphics[width=4.4cm, height=4.4cm]{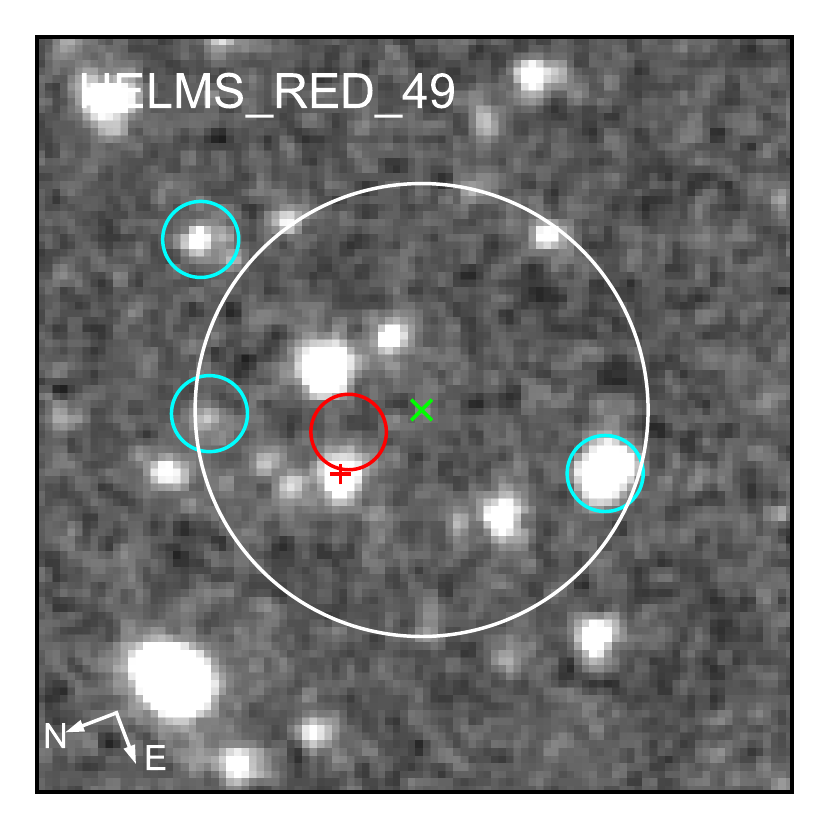}}
{\includegraphics[width=4.4cm, height=4.4cm]{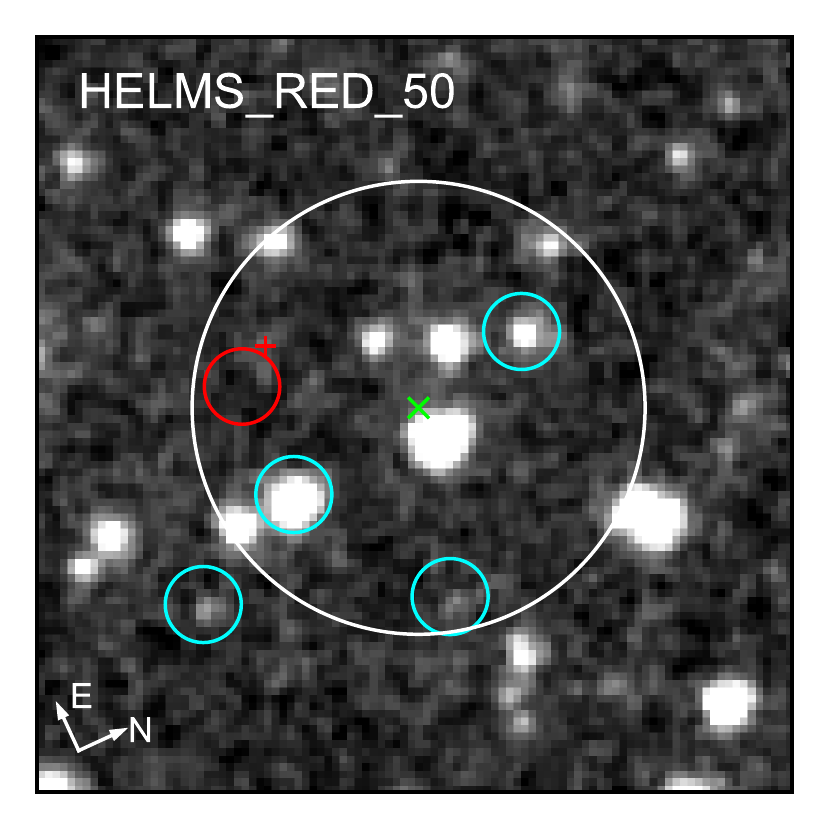}}
{\includegraphics[width=4.4cm, height=4.4cm]{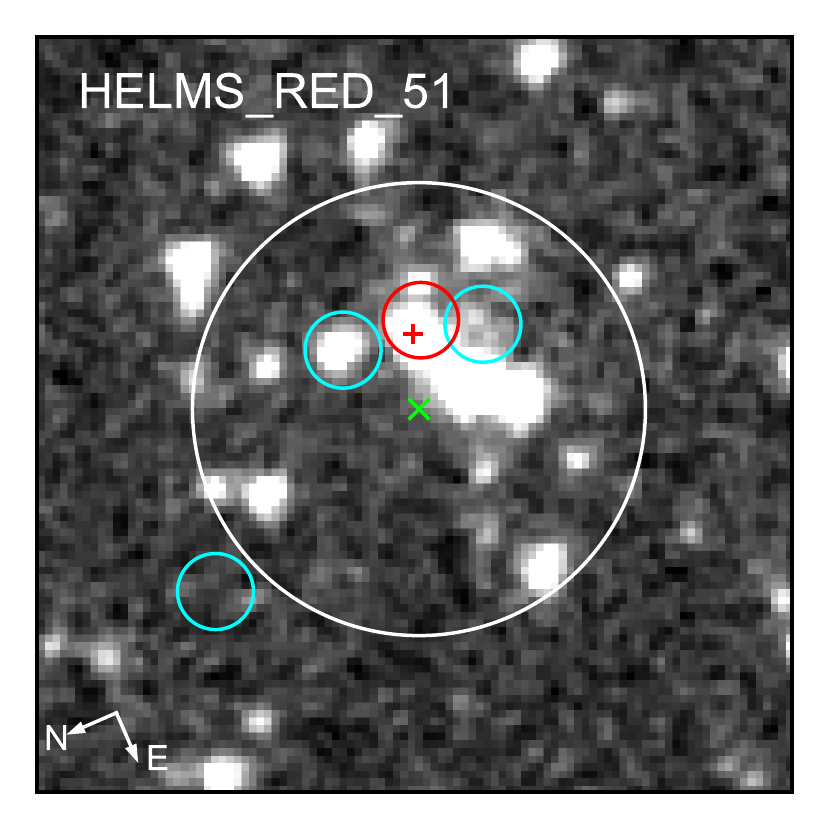}}
\caption{Continued 60$\arcsec$ $\times$ 60$\arcsec$ cutouts }
\label{fig:data}
\end{figure*}

\addtocounter{figure}{-1}
\begin{figure*}
\centering
{\includegraphics[width=4.4cm, height=4.4cm]{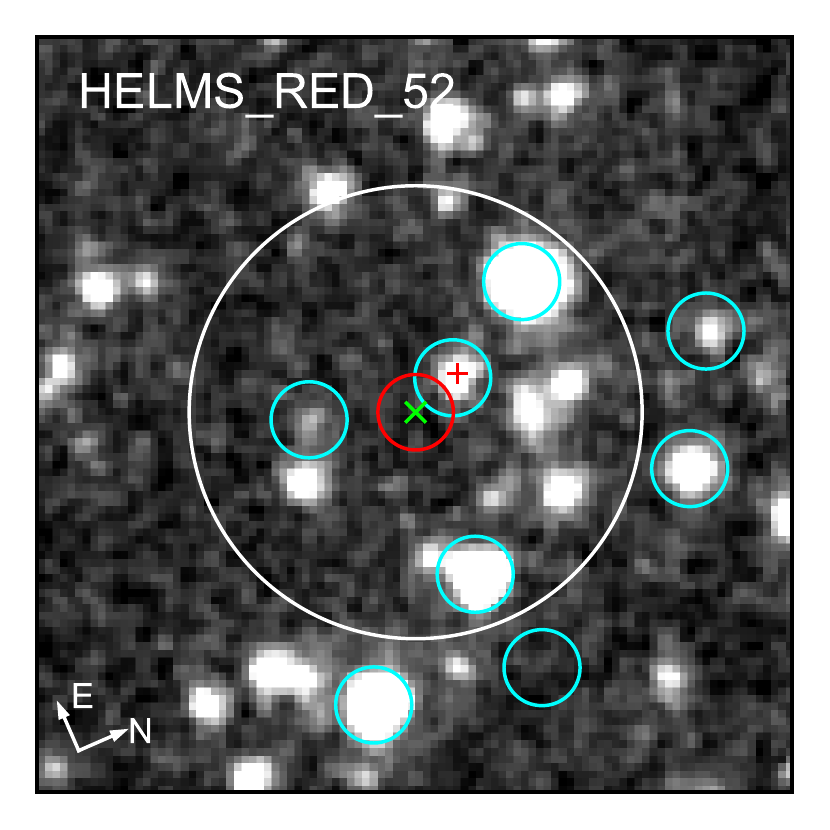}}
{\includegraphics[width=4.4cm, height=4.4cm]{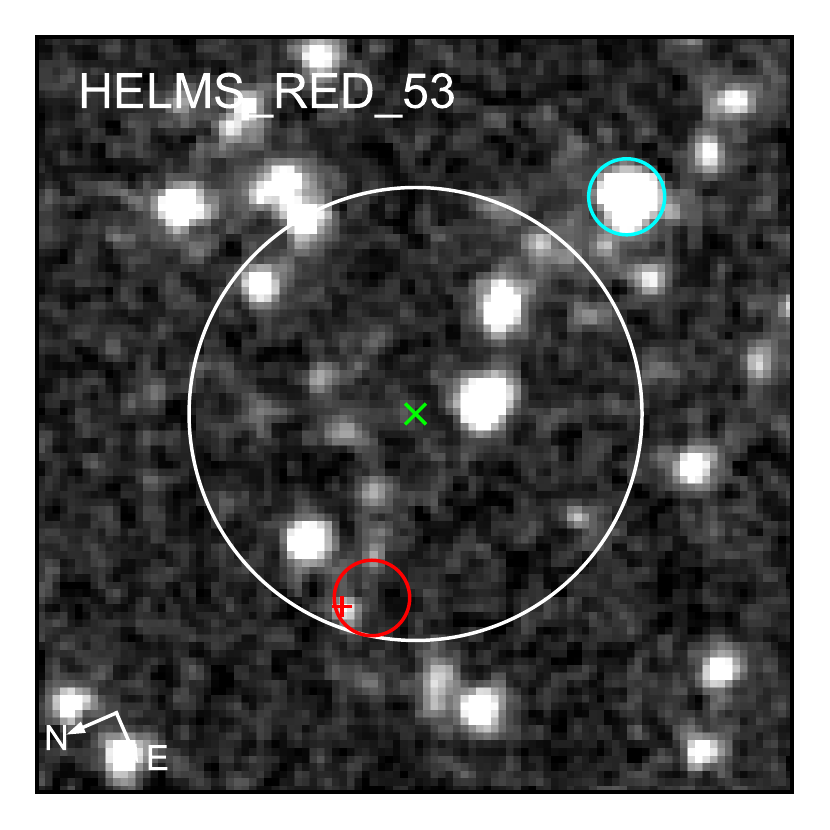}}
{\includegraphics[width=4.4cm, height=4.4cm]{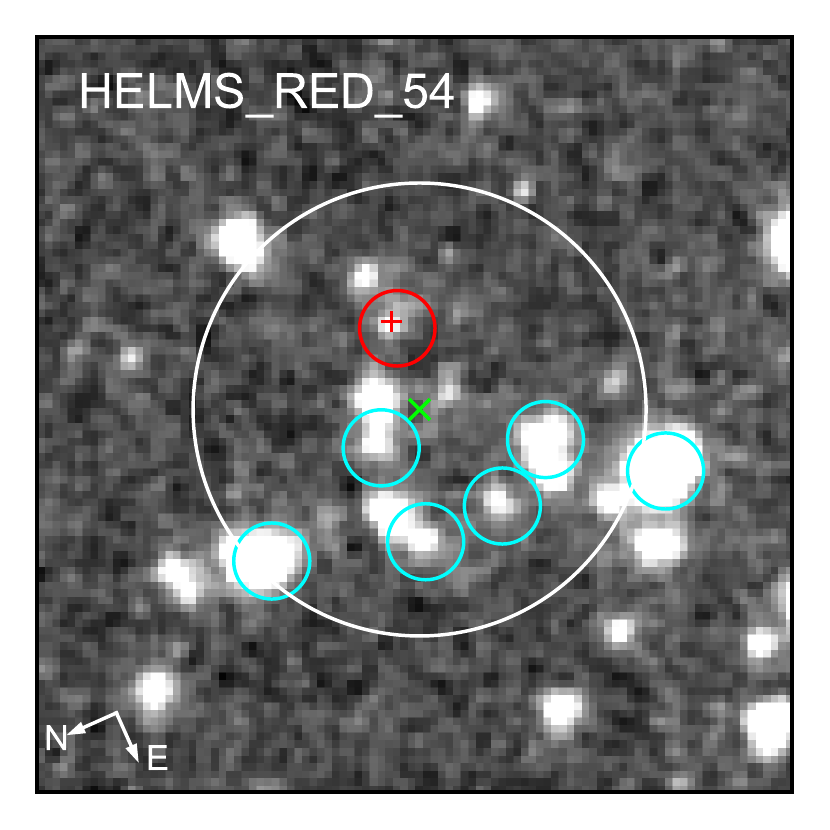}}
{\includegraphics[width=4.4cm, height=4.4cm]{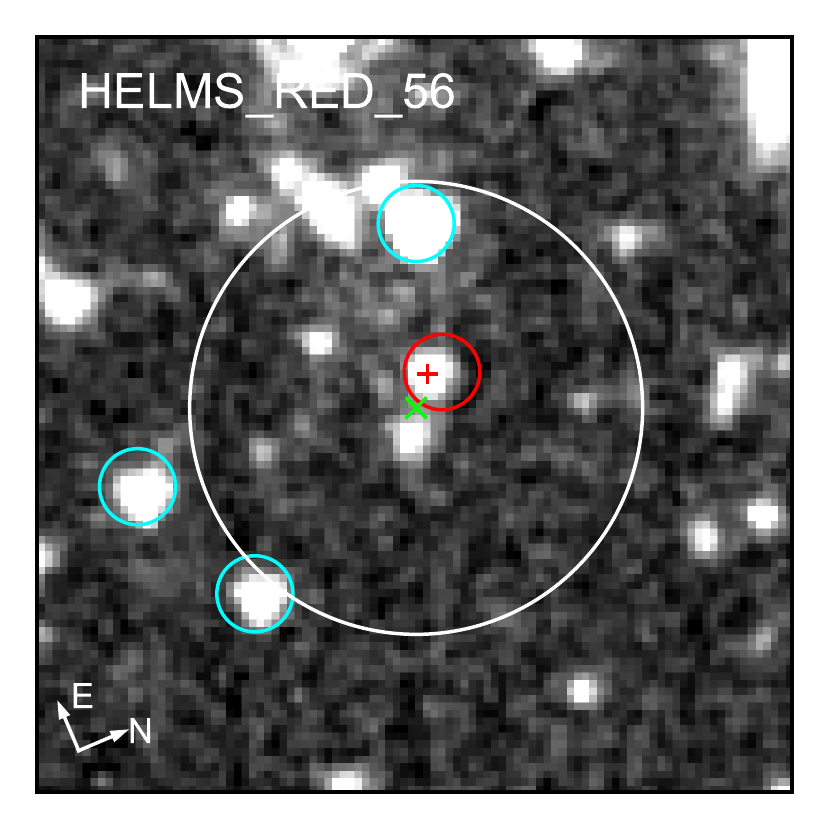}}
{\includegraphics[width=4.4cm, height=4.4cm]{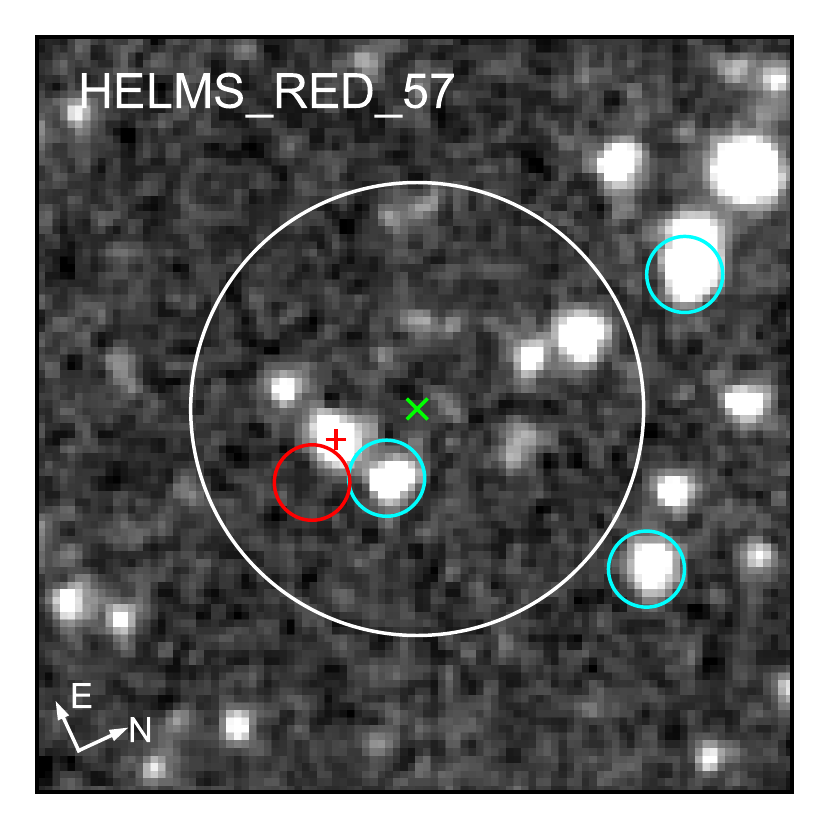}}
{\includegraphics[width=4.4cm, height=4.4cm]{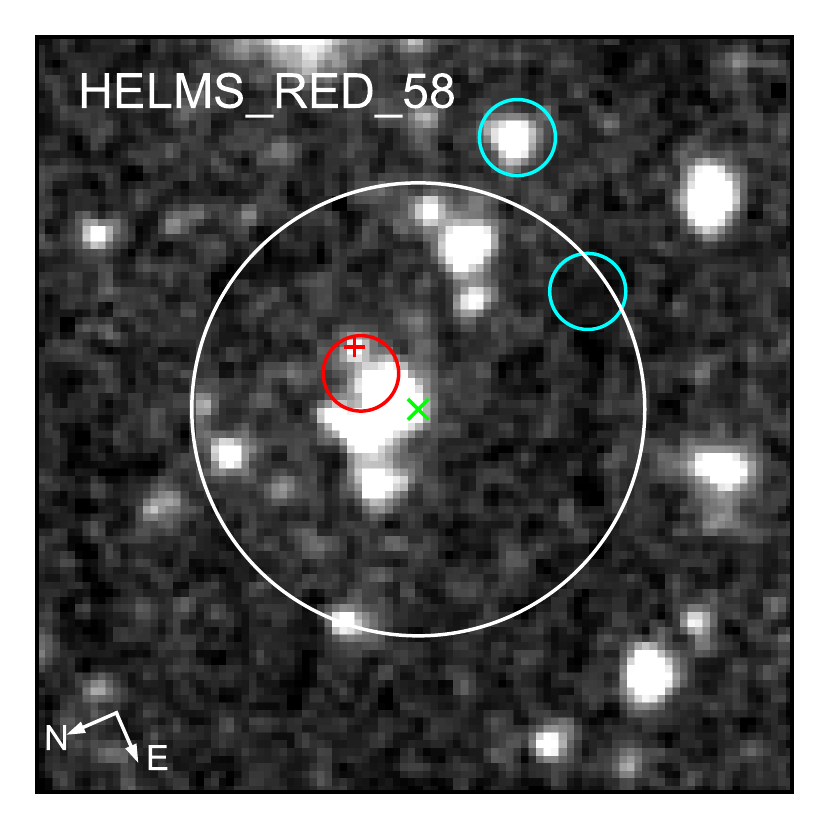}}
{\includegraphics[width=4.4cm, height=4.4cm]{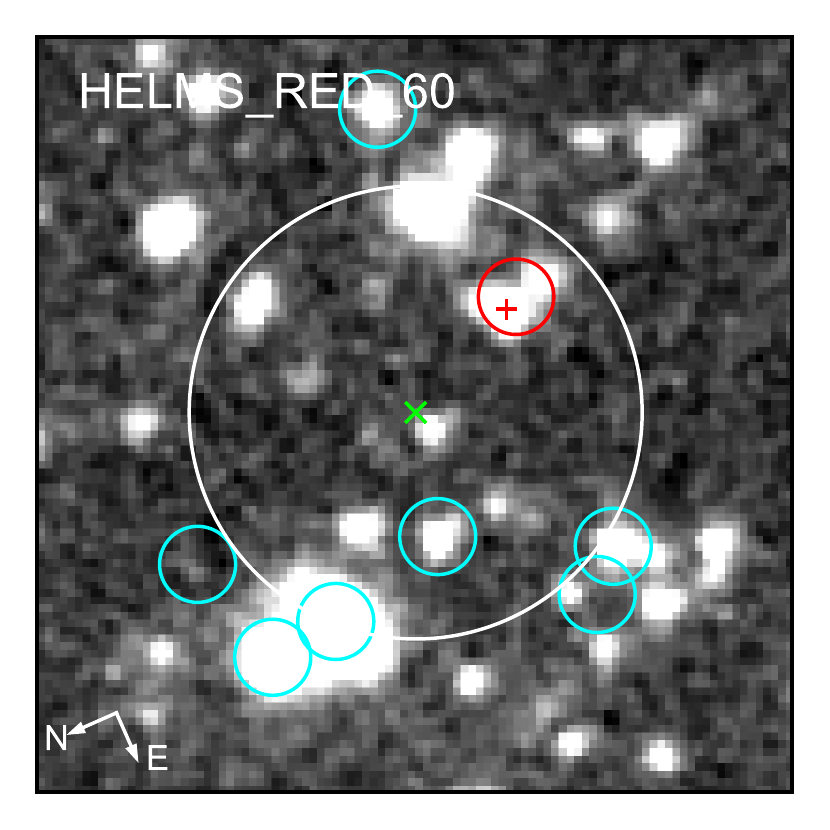}}
{\includegraphics[width=4.4cm, height=4.4cm]{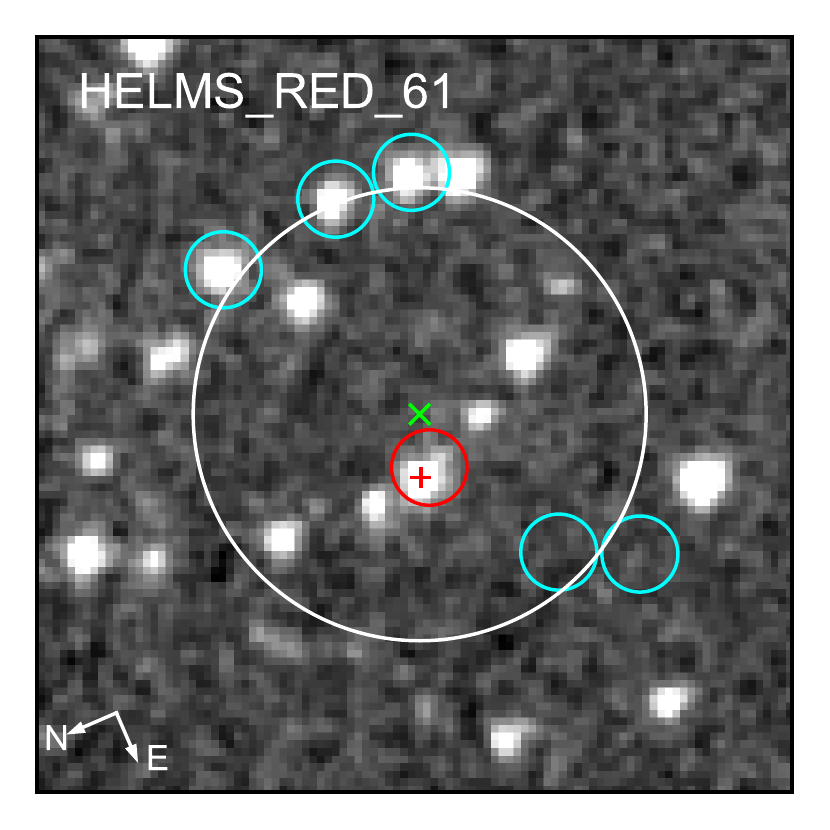}}
{\includegraphics[width=4.4cm, height=4.4cm]{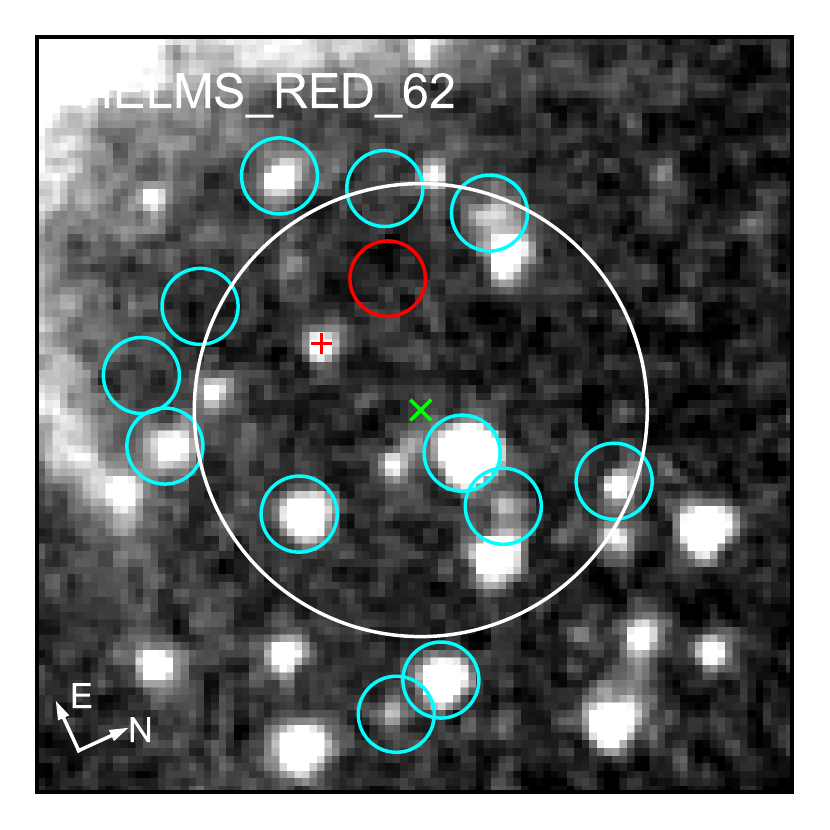}}
{\includegraphics[width=4.4cm, height=4.4cm]{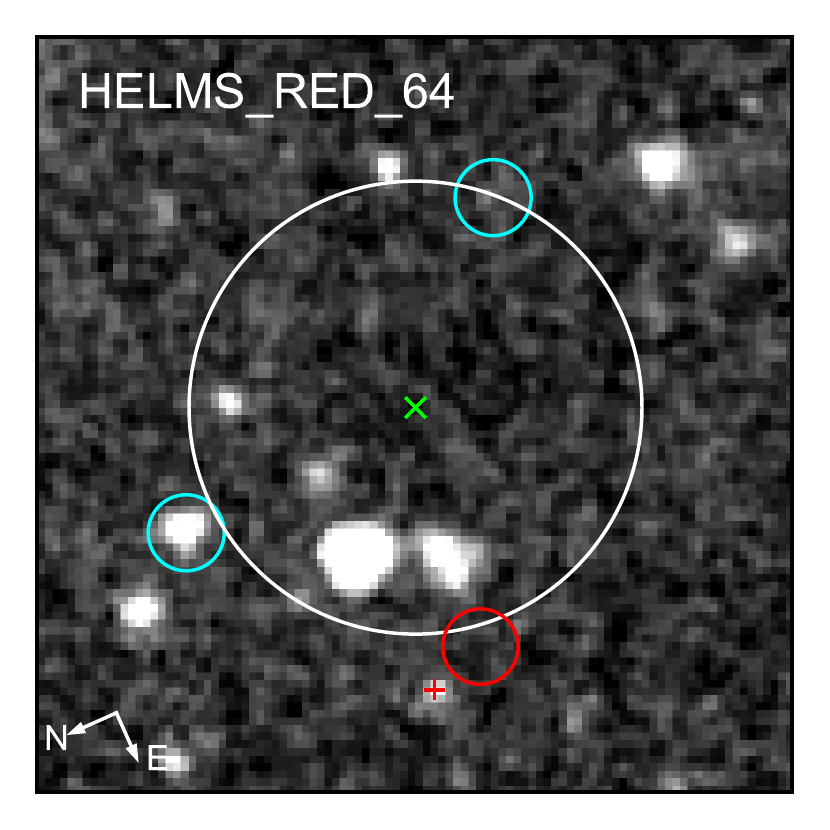}}
{\includegraphics[width=4.4cm, height=4.4cm]{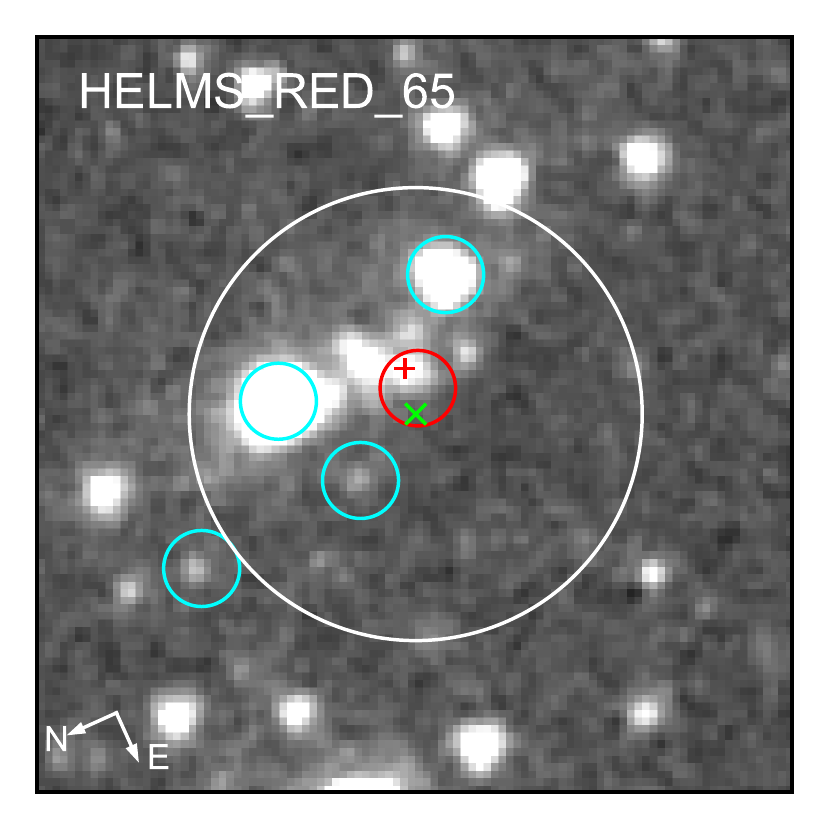}}
{\includegraphics[width=4.4cm, height=4.4cm]{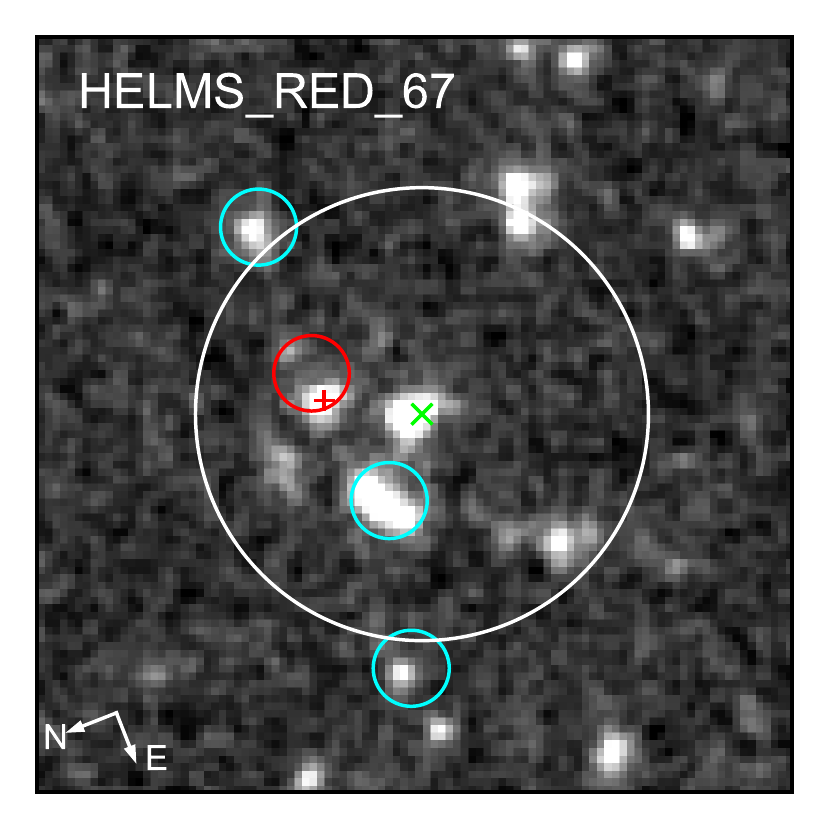}}
{\includegraphics[width=4.4cm, height=4.4cm]{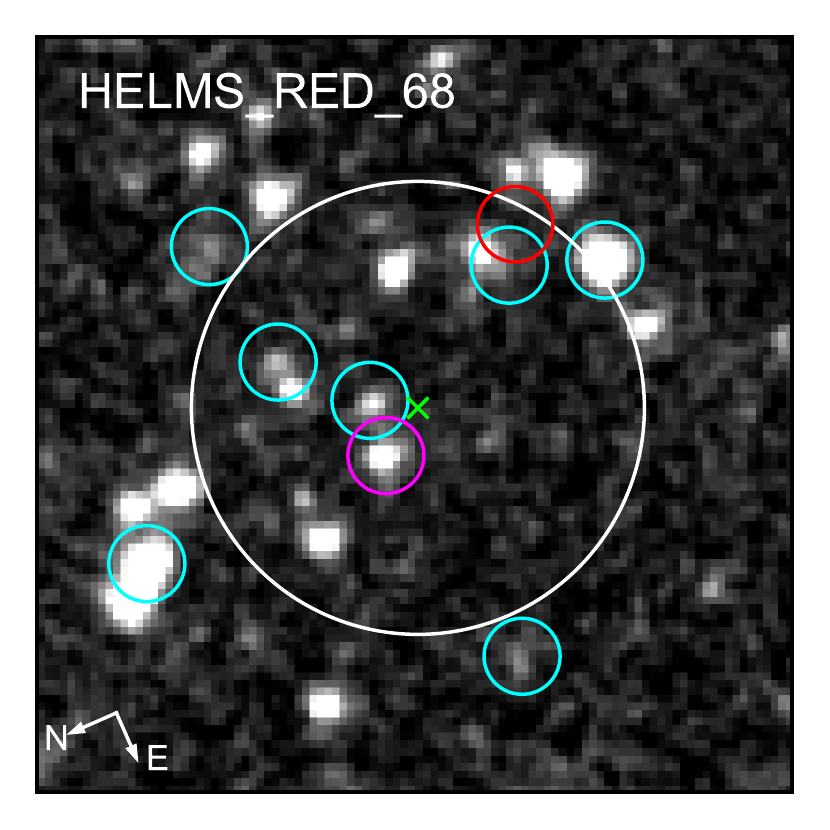}}
{\includegraphics[width=4.4cm, height=4.4cm]{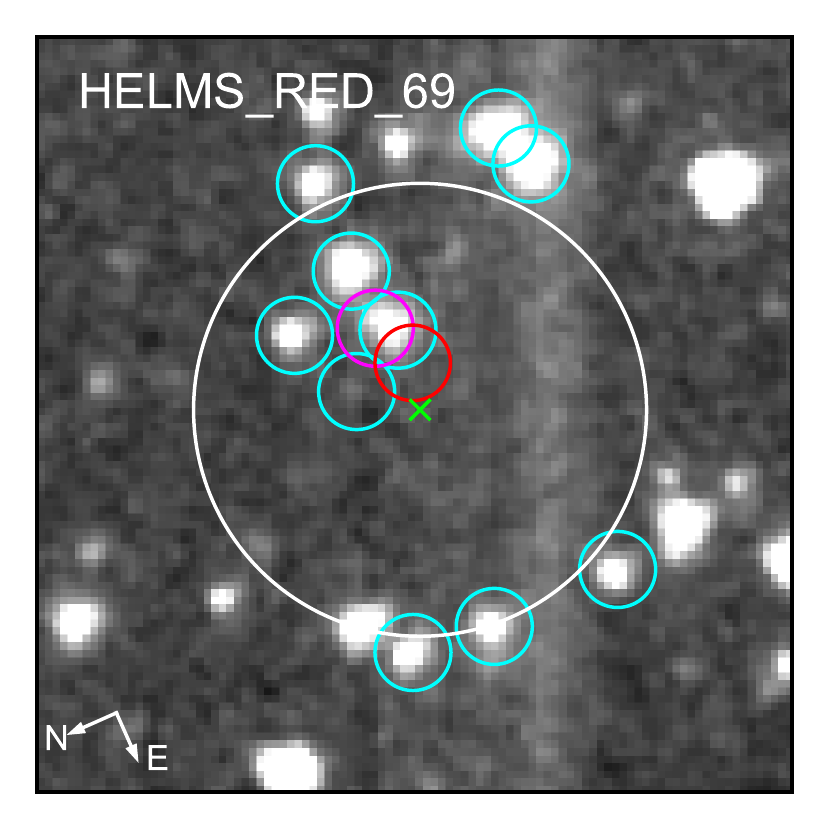}}
{\includegraphics[width=4.4cm, height=4.4cm]{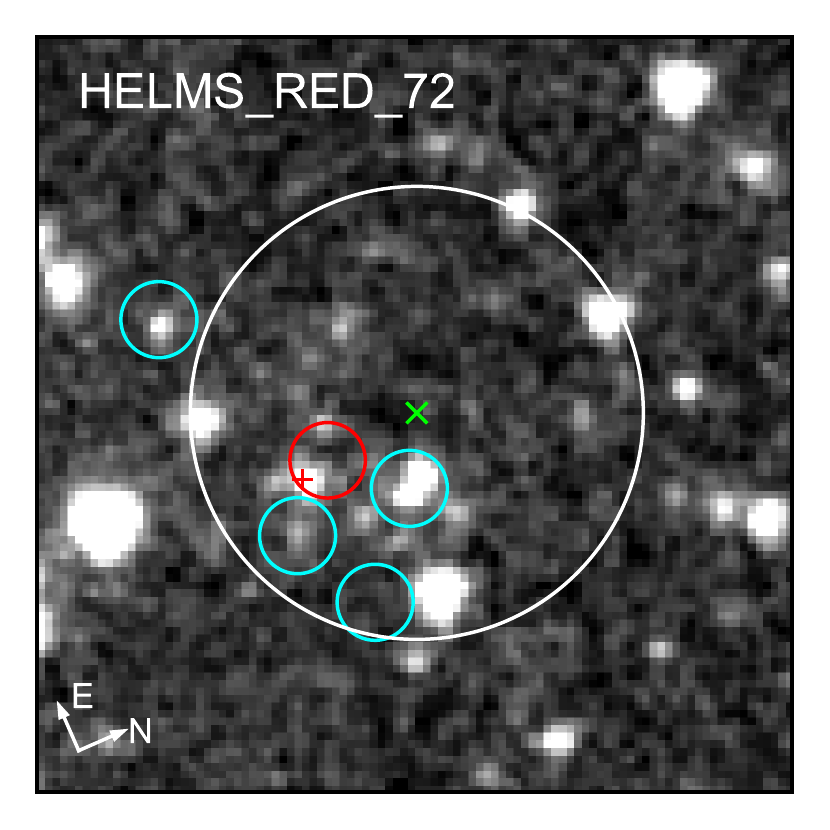}}
{\includegraphics[width=4.4cm, height=4.4cm]{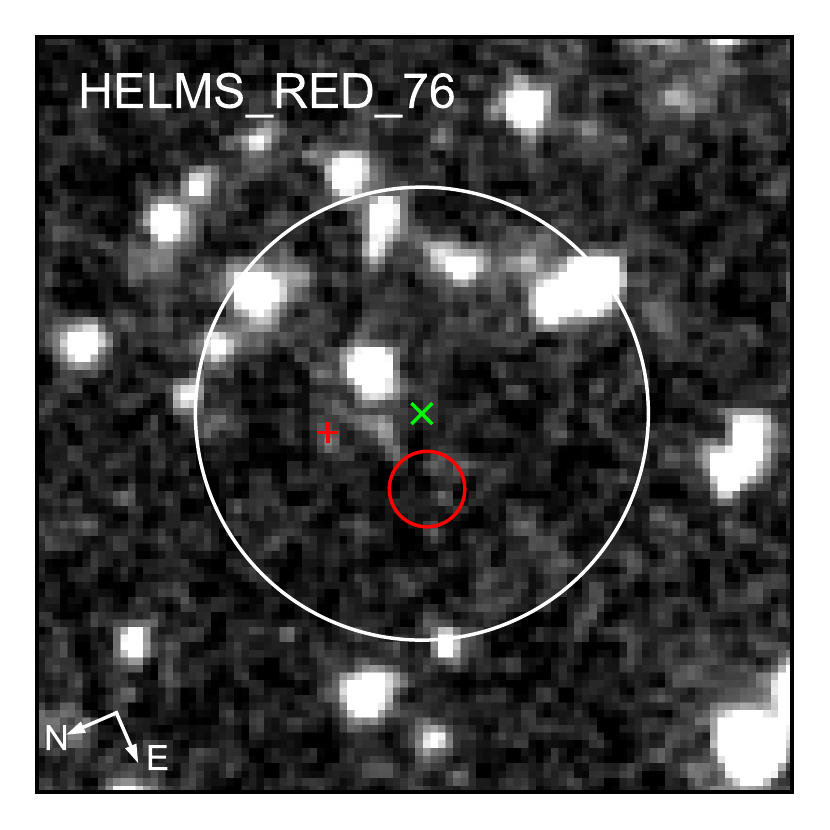}}
{\includegraphics[width=4.4cm, height=4.4cm]{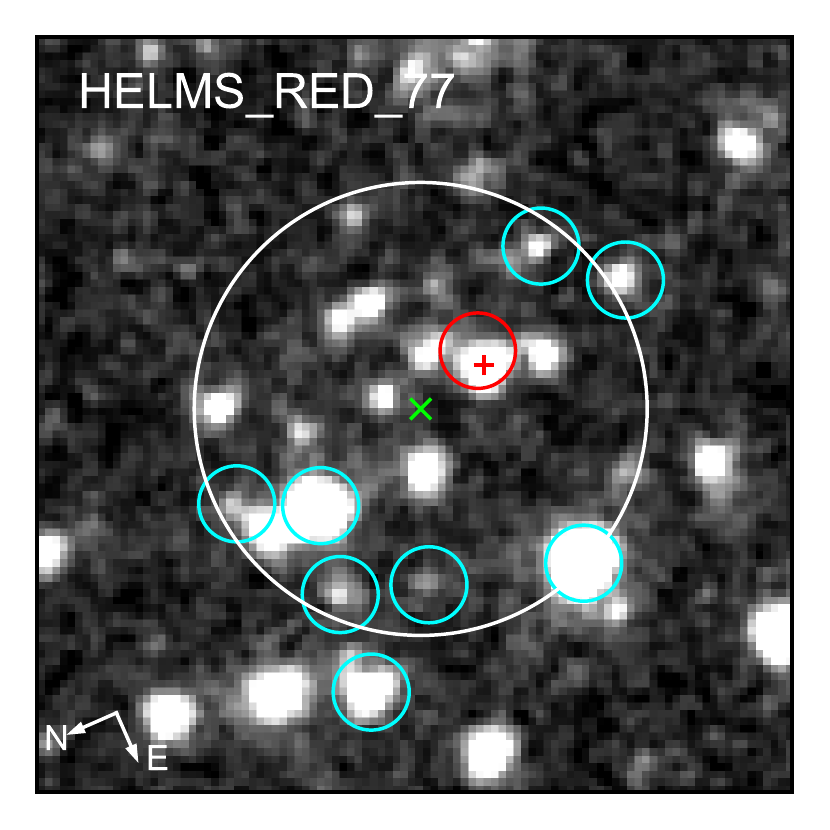}}
{\includegraphics[width=4.4cm, height=4.4cm]{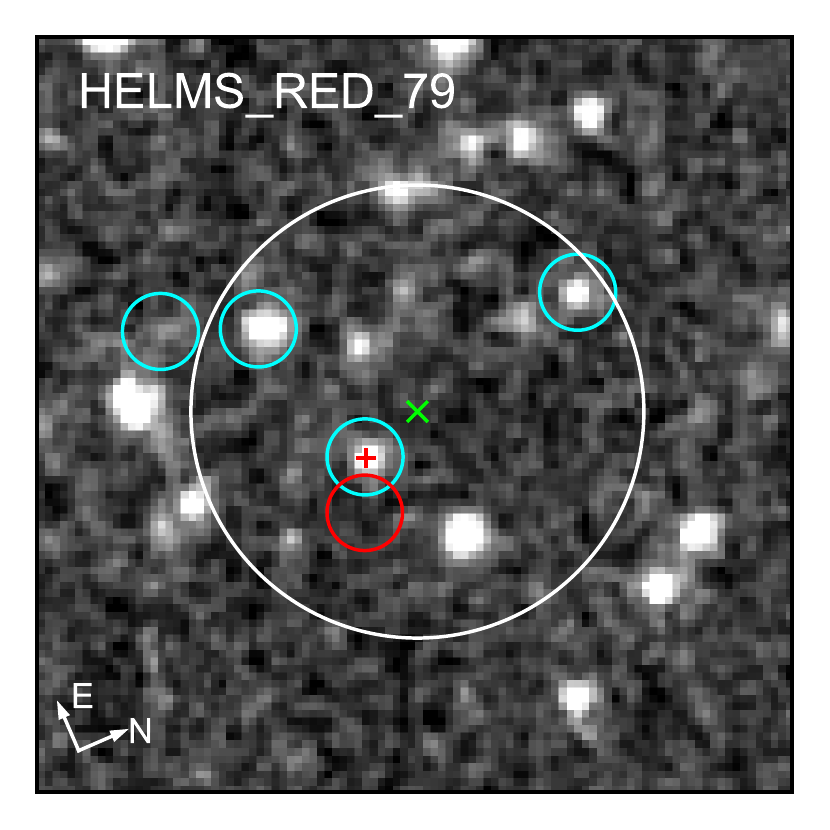}}
{\includegraphics[width=4.4cm, height=4.4cm]{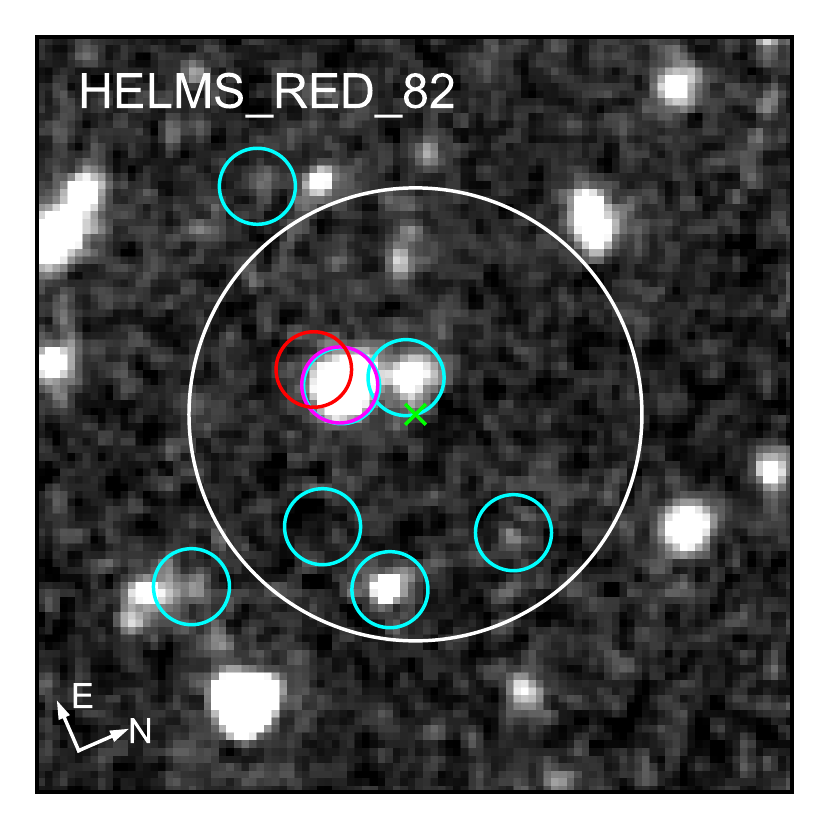}}
{\includegraphics[width=4.4cm, height=4.4cm]{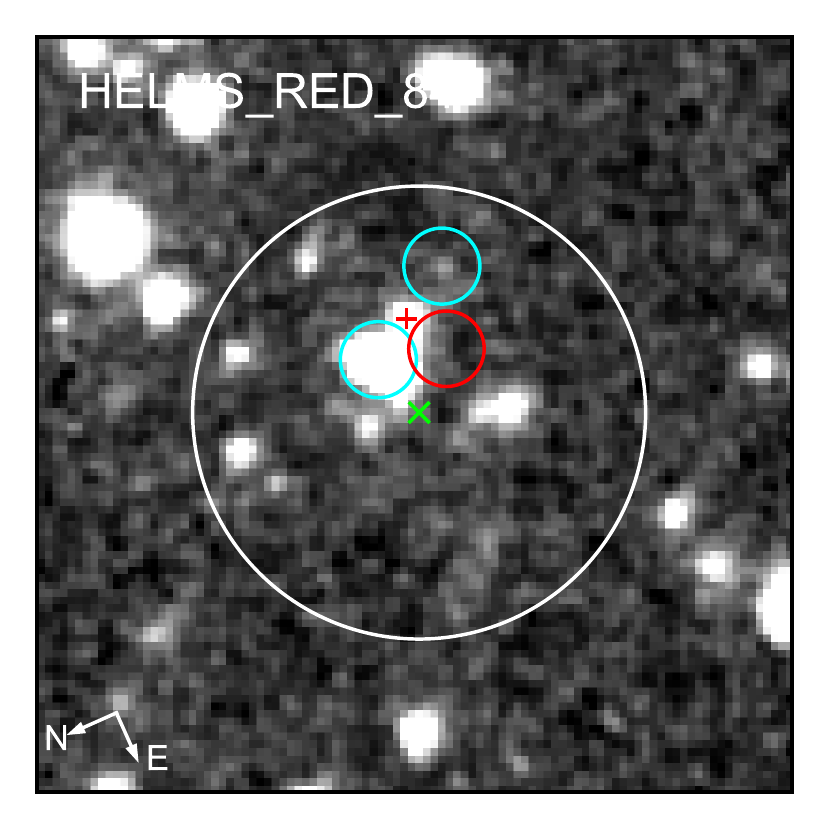}}
\caption{Continued 60$\arcsec$ $\times$ 60$\arcsec$ cutouts }
\label{fig:data}
\end{figure*}

\addtocounter{figure}{-1}
\begin{figure*}
\centering
{\includegraphics[width=4.4cm, height=4.4cm]{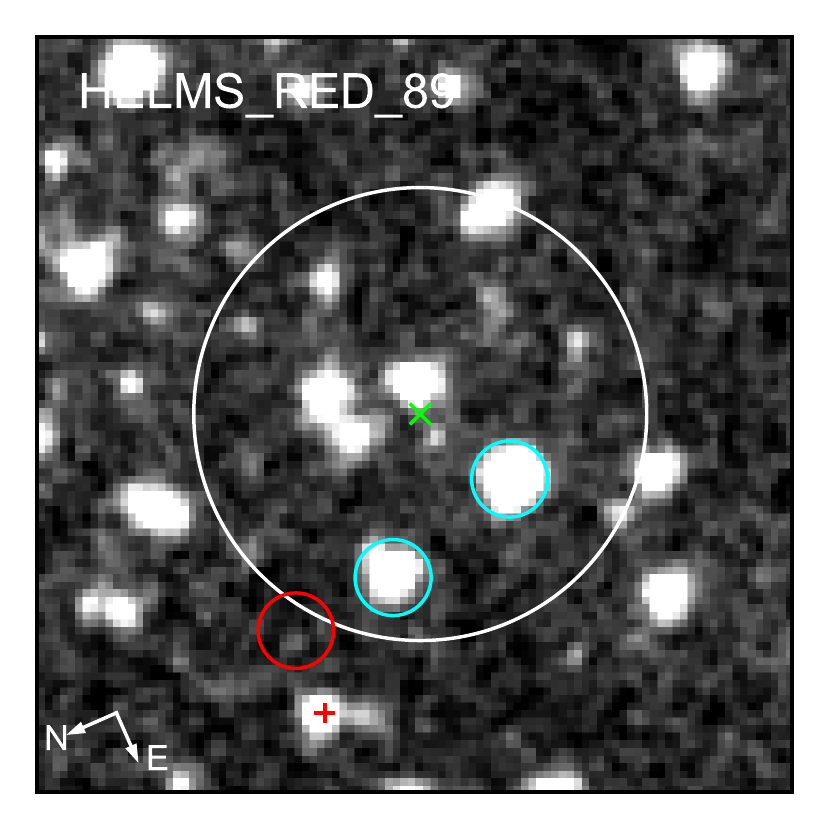}}
{\includegraphics[width=4.4cm, height=4.4cm]{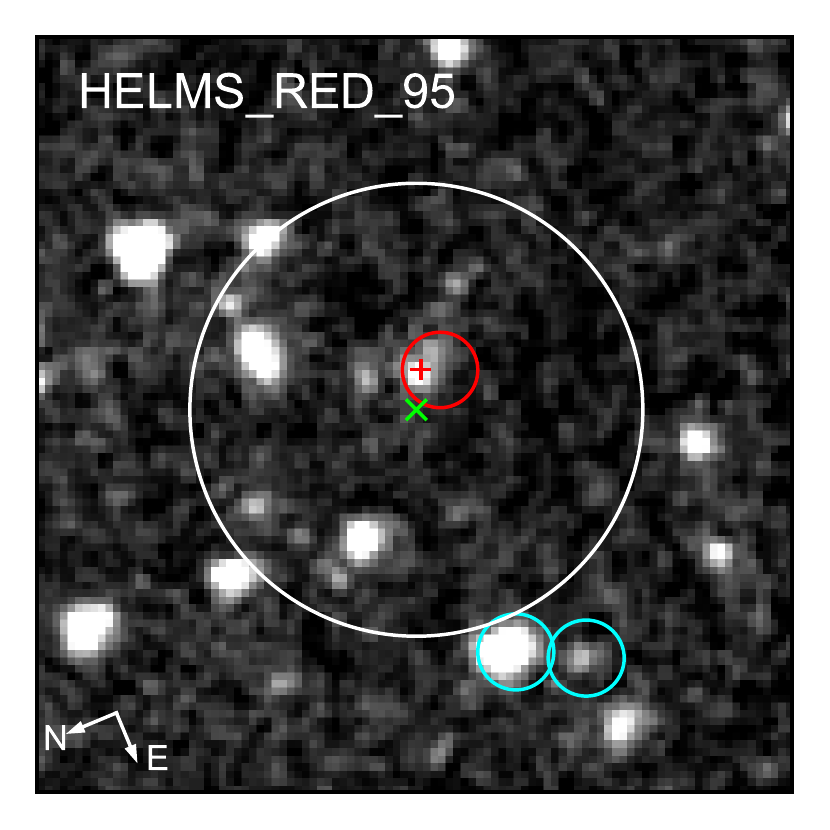}}
{\includegraphics[width=4.4cm, height=4.4cm]{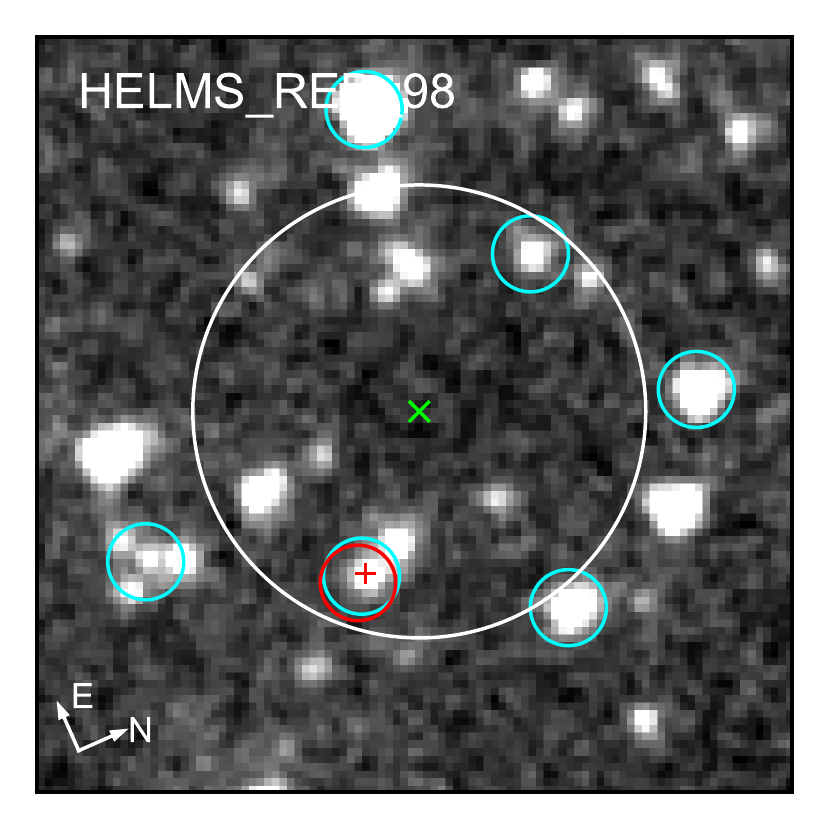}}
{\includegraphics[width=4.4cm, height=4.4cm]{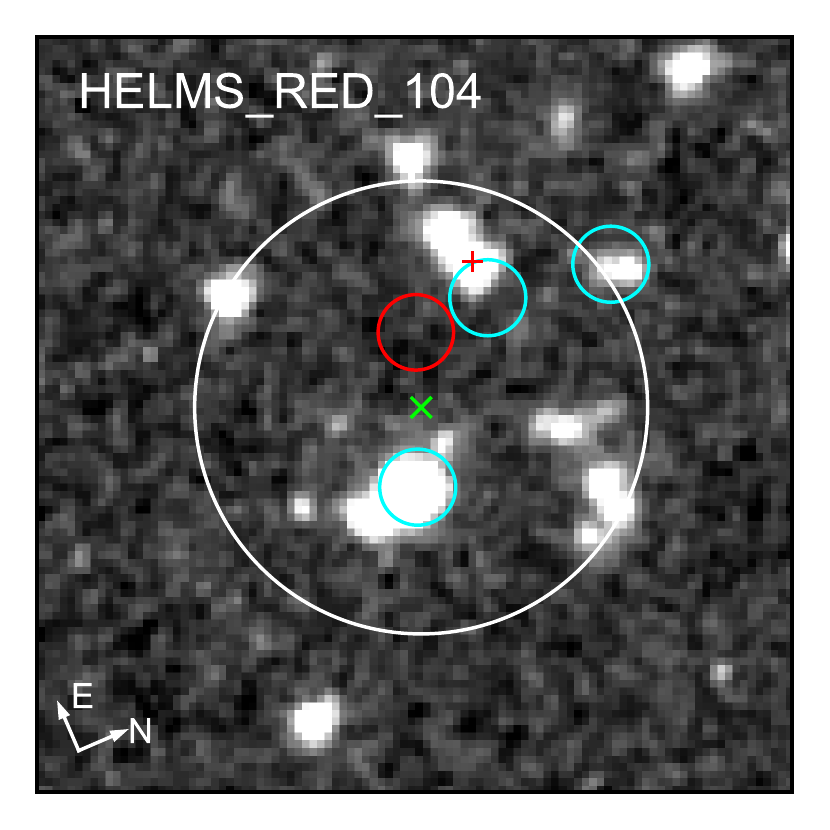}}
{\includegraphics[width=4.4cm, height=4.4cm]{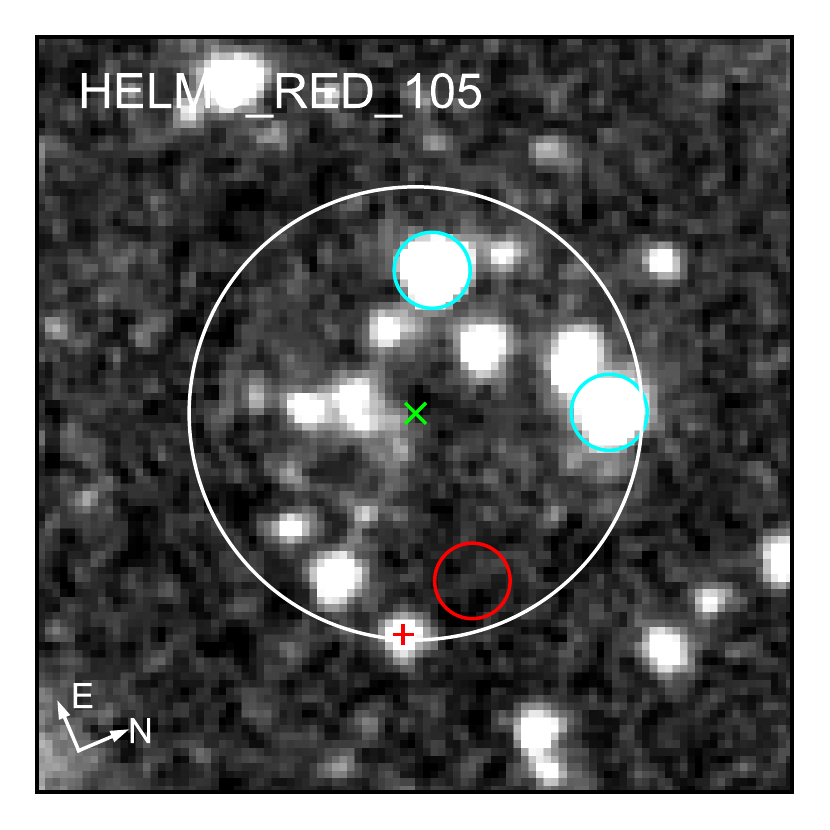}}
{\includegraphics[width=4.4cm, height=4.4cm]{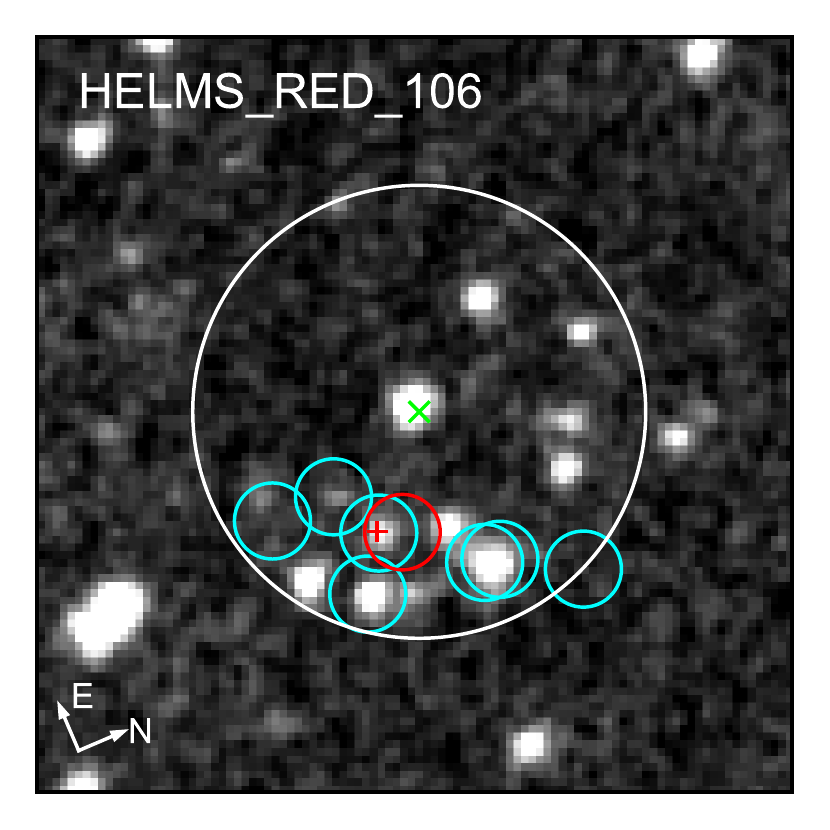}}
{\includegraphics[width=4.4cm, height=4.4cm]{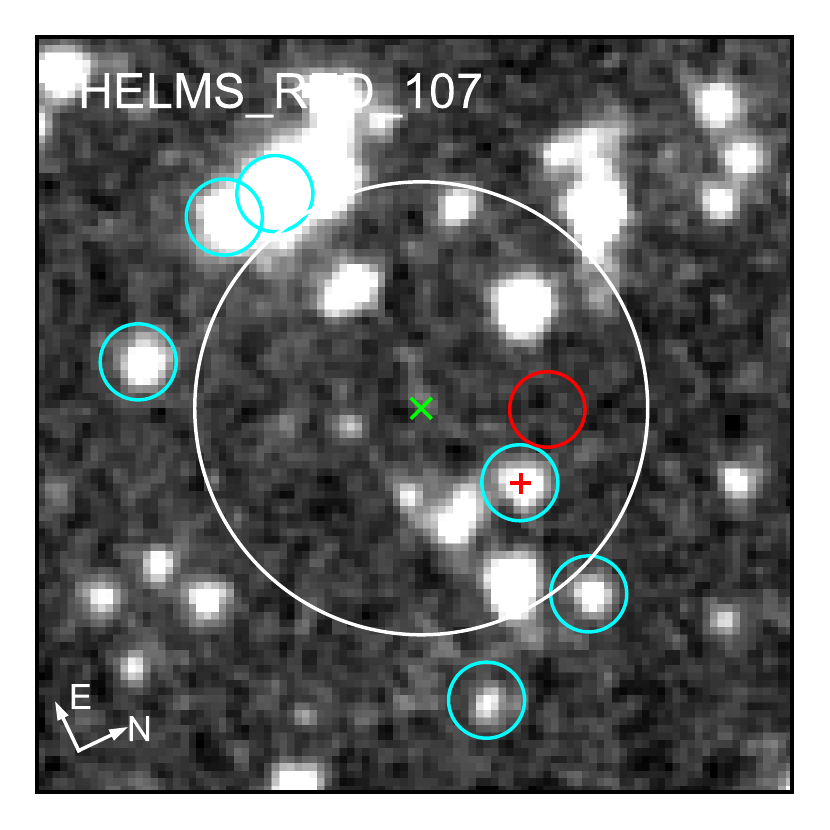}}
{\includegraphics[width=4.4cm, height=4.4cm]{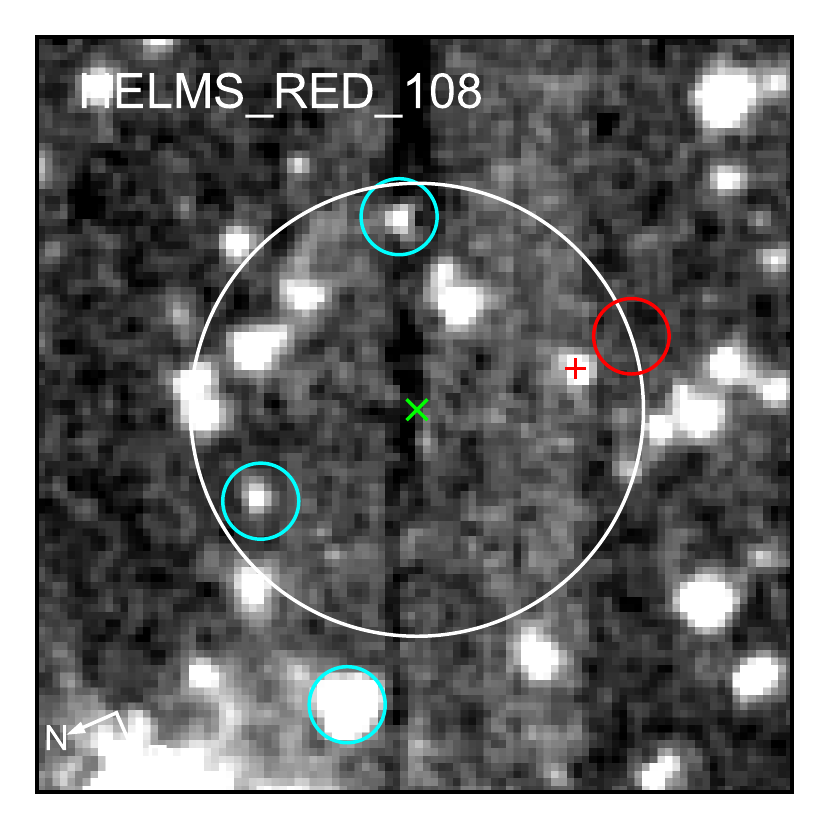}}
{\includegraphics[width=4.4cm, height=4.4cm]{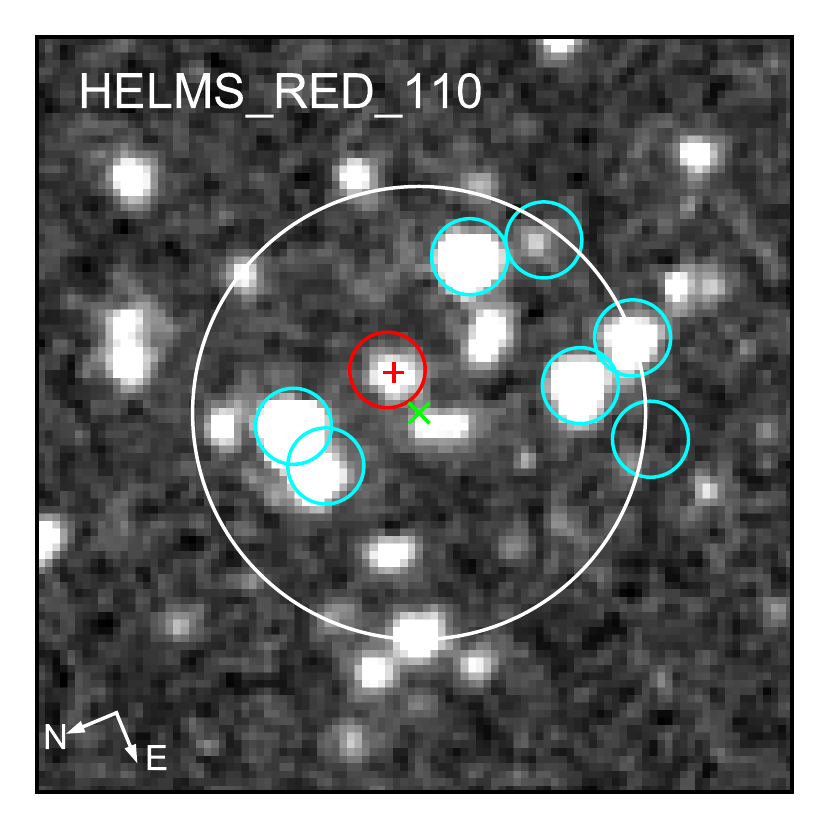}}
{\includegraphics[width=4.4cm, height=4.4cm]{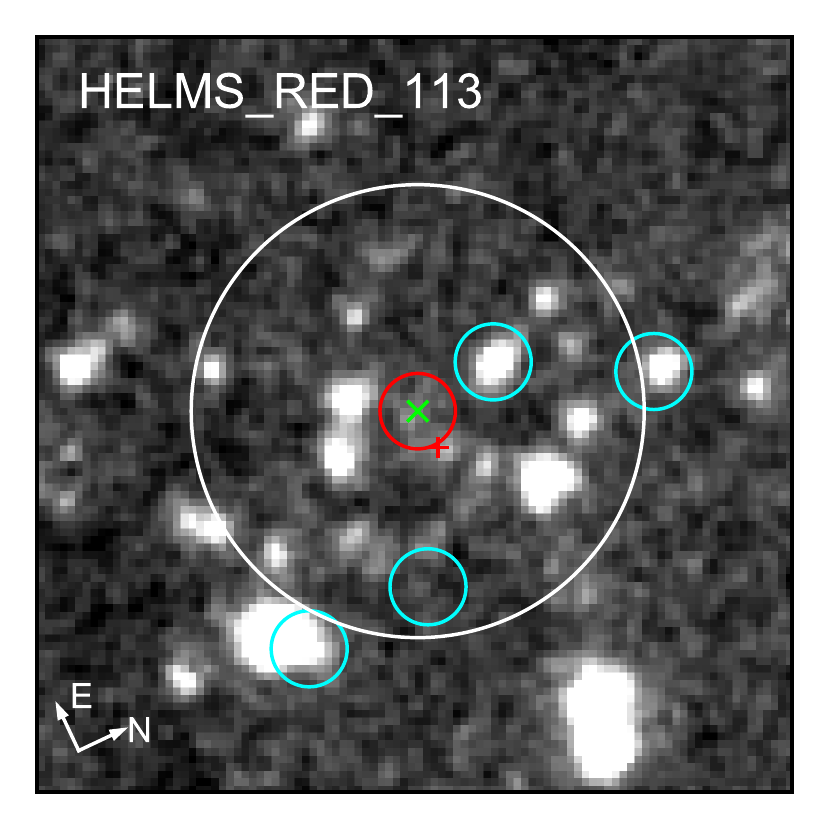}}
{\includegraphics[width=4.4cm, height=4.4cm]{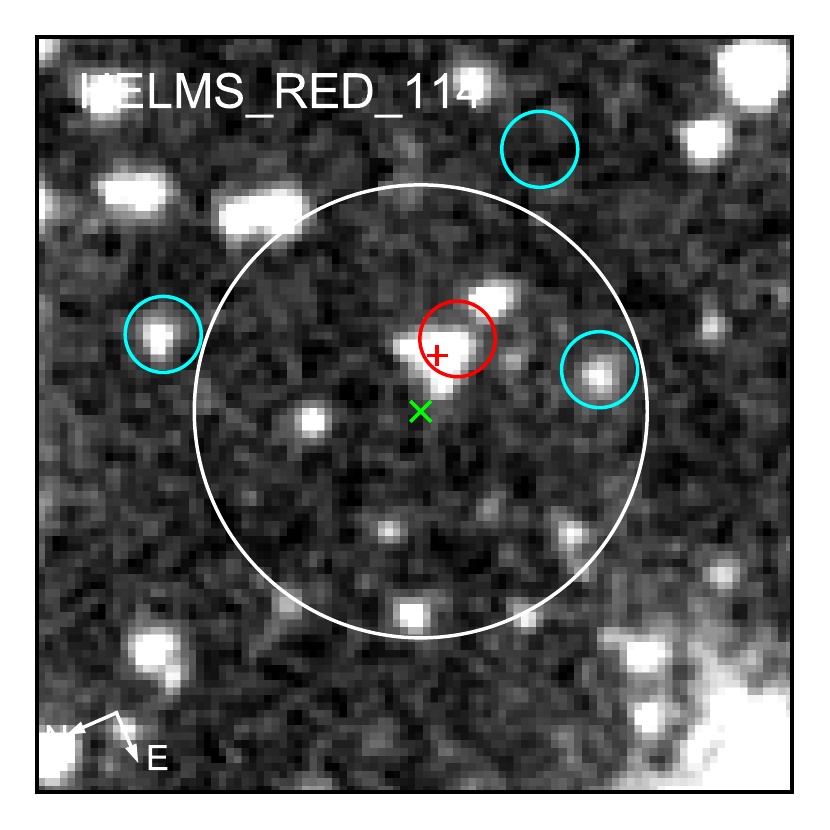}}
{\includegraphics[width=4.4cm, height=4.4cm]{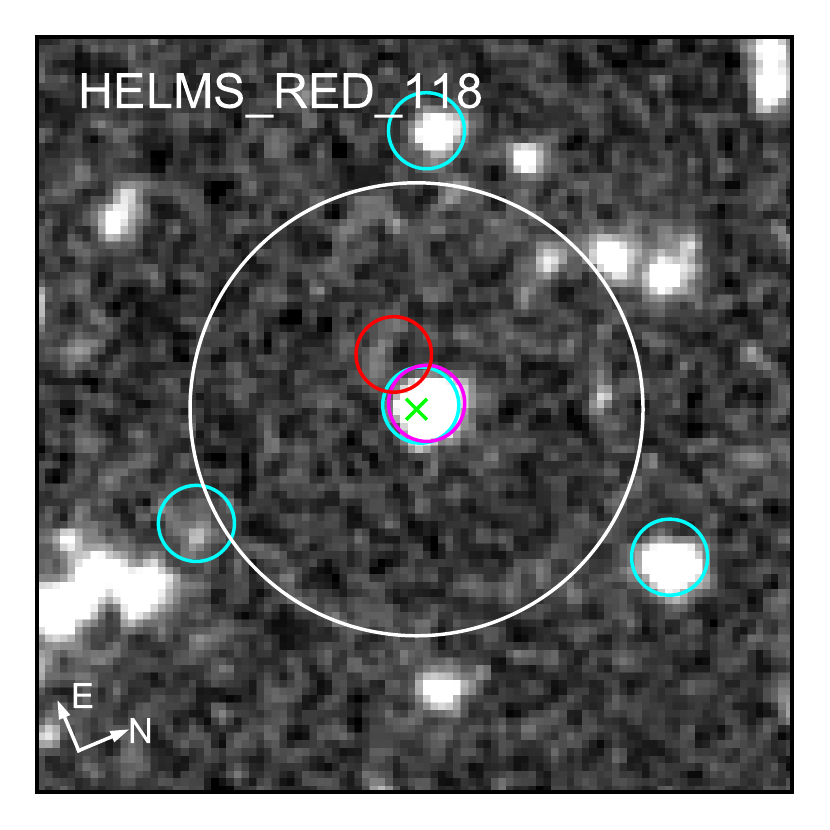}}
{\includegraphics[width=4.4cm, height=4.4cm]{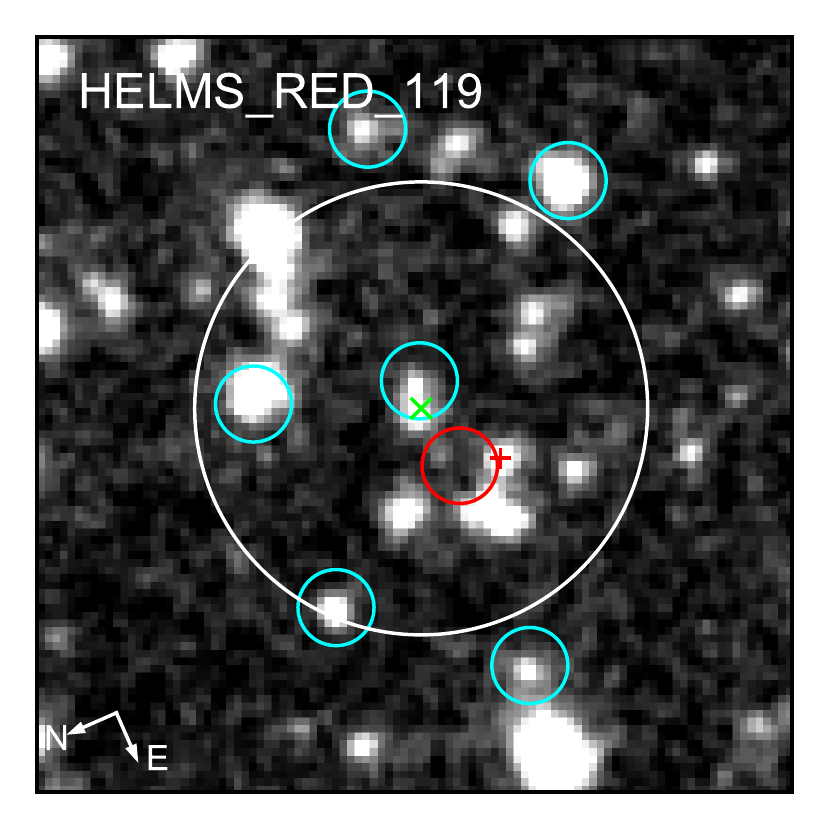}}
{\includegraphics[width=4.4cm, height=4.4cm]{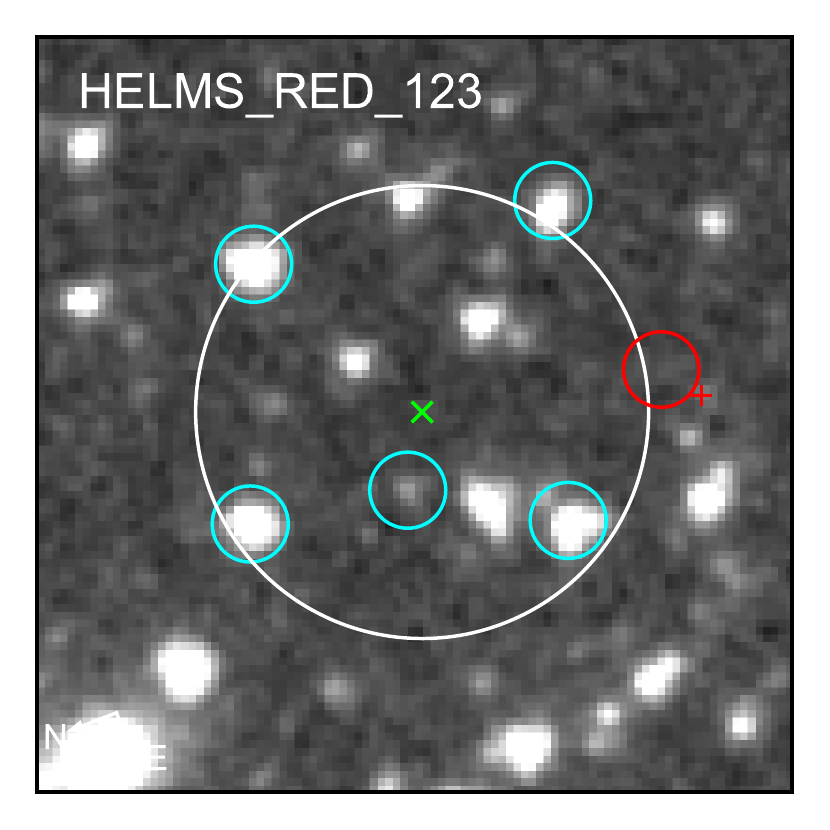}}
{\includegraphics[width=4.4cm, height=4.4cm]{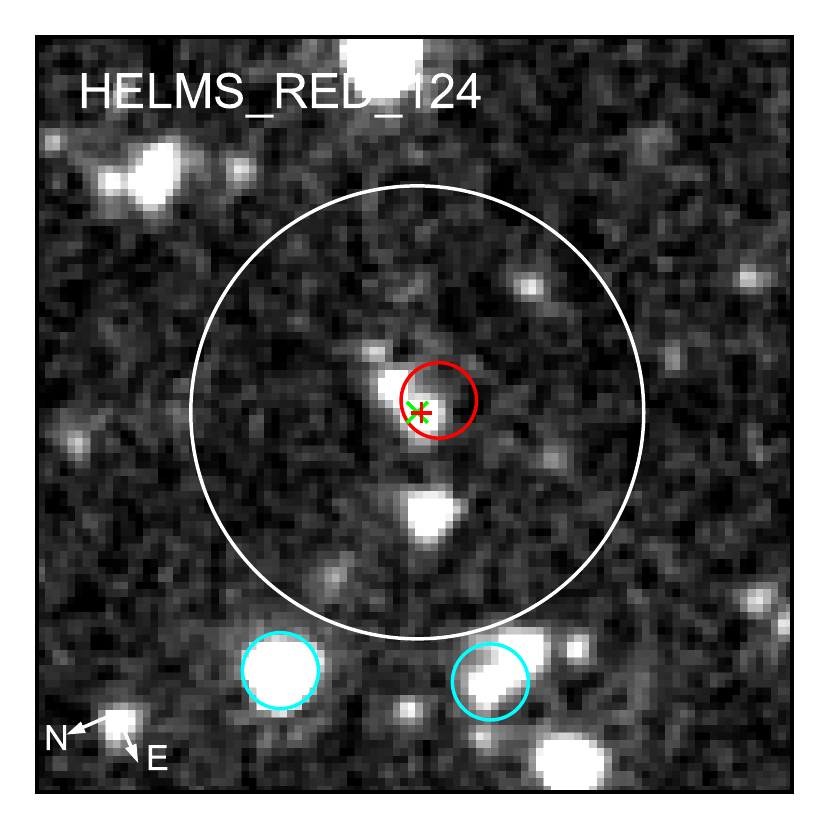}}
{\includegraphics[width=4.4cm, height=4.4cm]{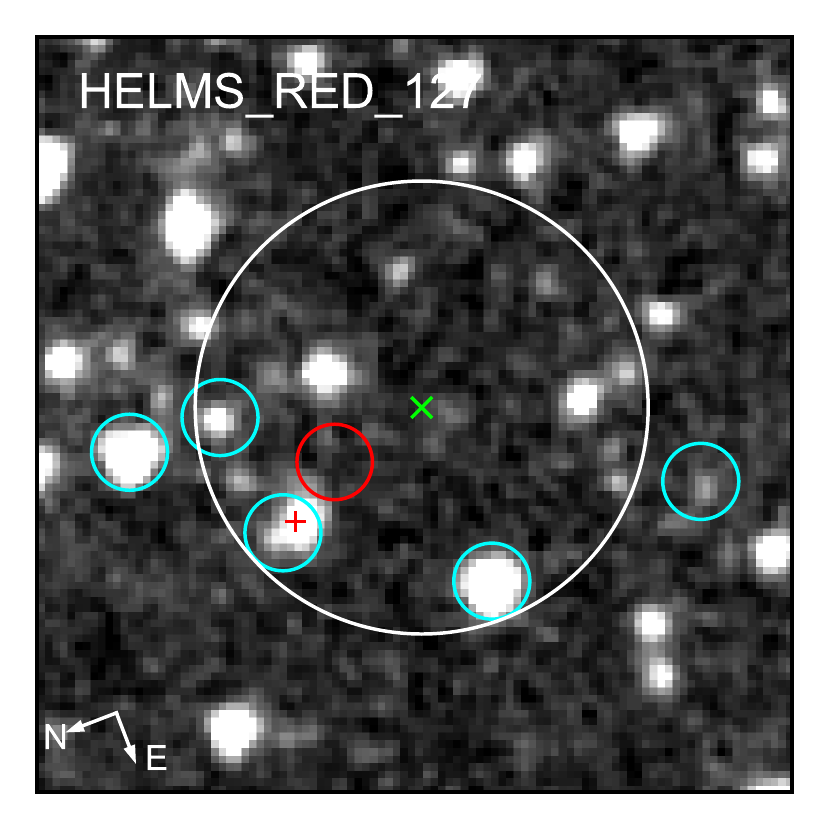}}
{\includegraphics[width=4.4cm, height=4.4cm]{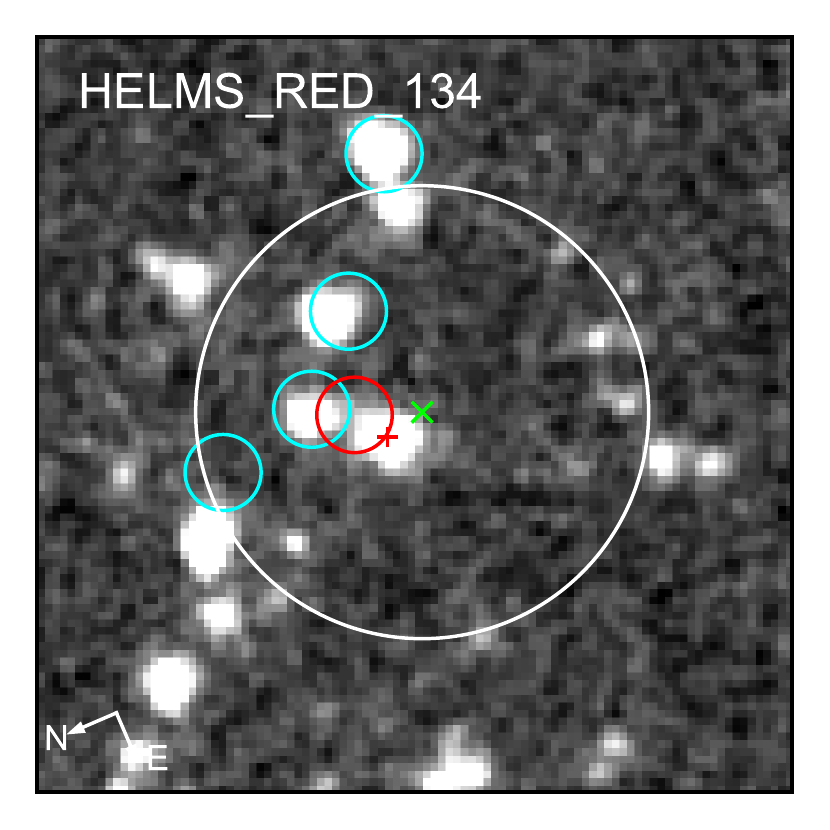}}
{\includegraphics[width=4.4cm, height=4.4cm]{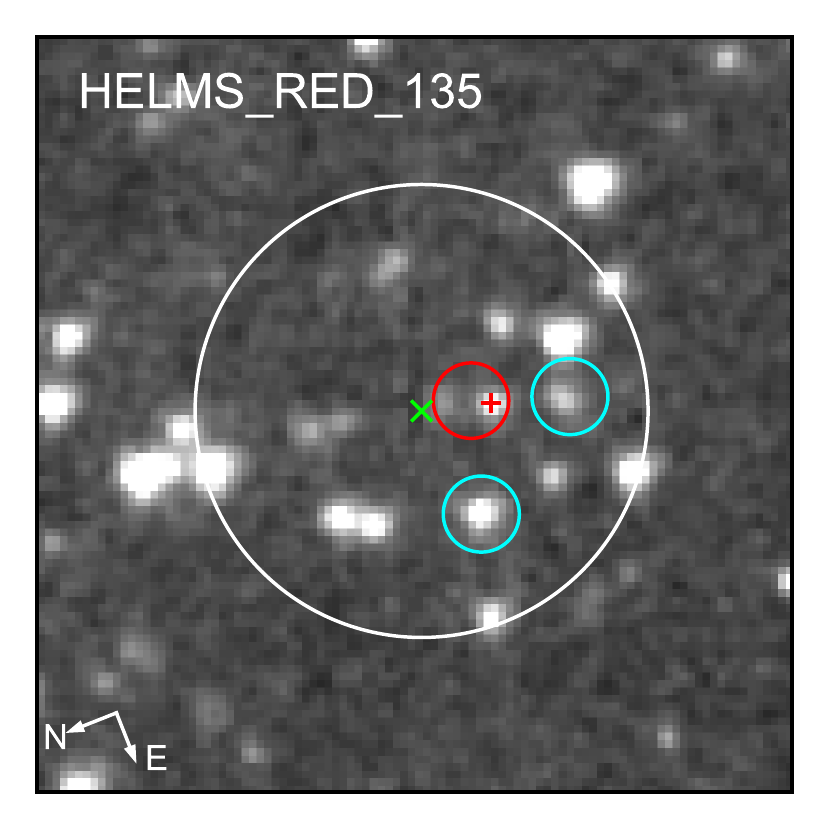}}
{\includegraphics[width=4.4cm, height=4.4cm]{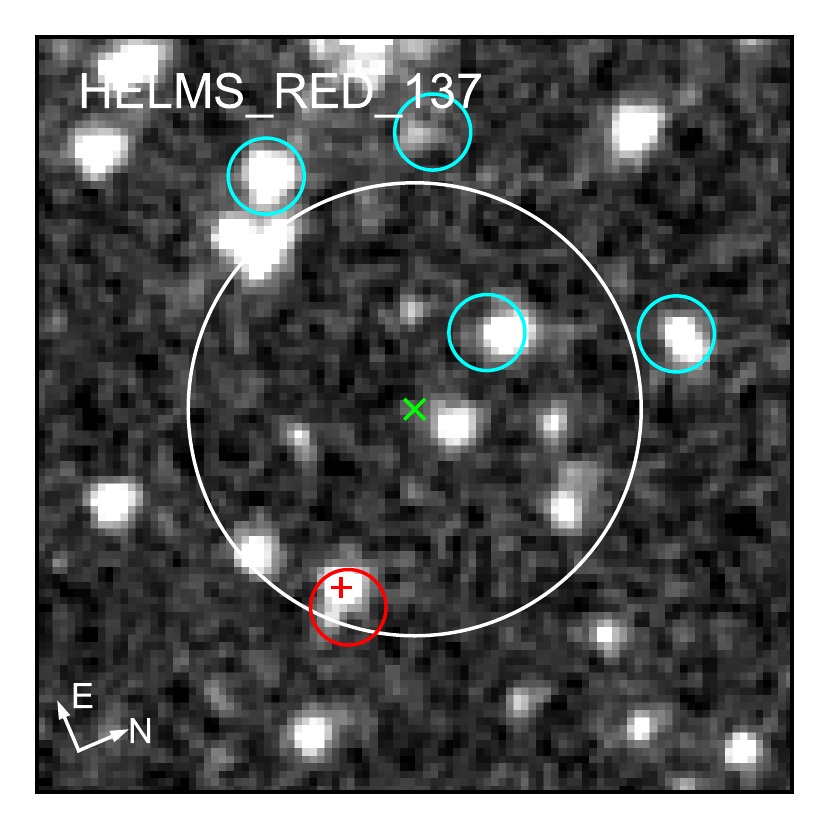}}
{\includegraphics[width=4.4cm, height=4.4cm]{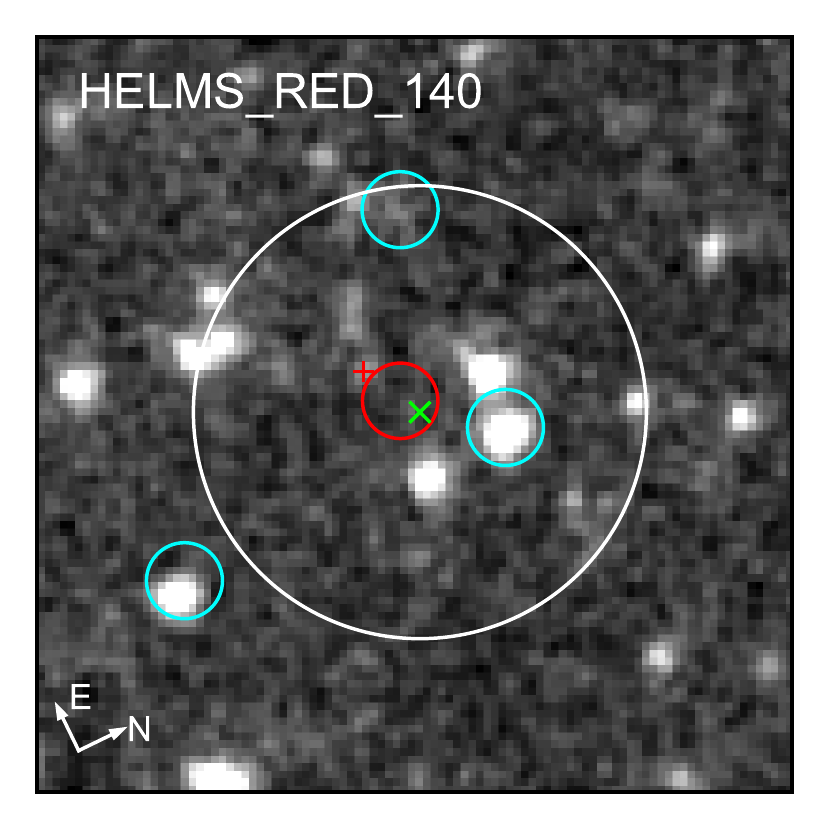}}
\caption{Continued 60$\arcsec$ $\times$ 60$\arcsec$ cutouts }
\label{fig:data}
\end{figure*}

\addtocounter{figure}{-1}
\begin{figure*}
\centering
{\includegraphics[width=4.4cm, height=4.4cm]{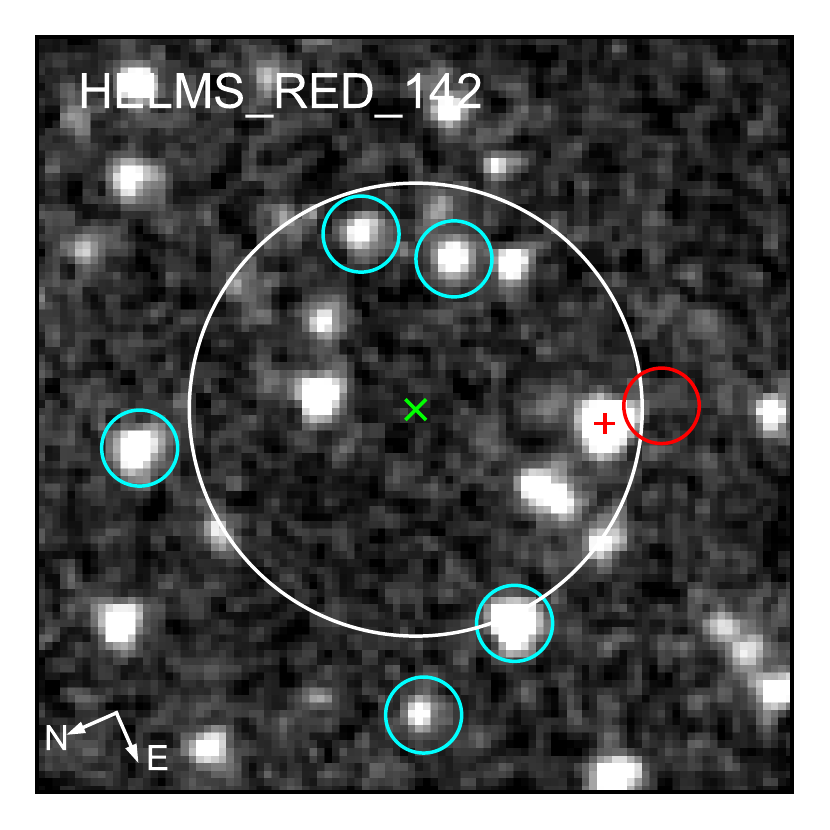}}
{\includegraphics[width=4.4cm, height=4.4cm]{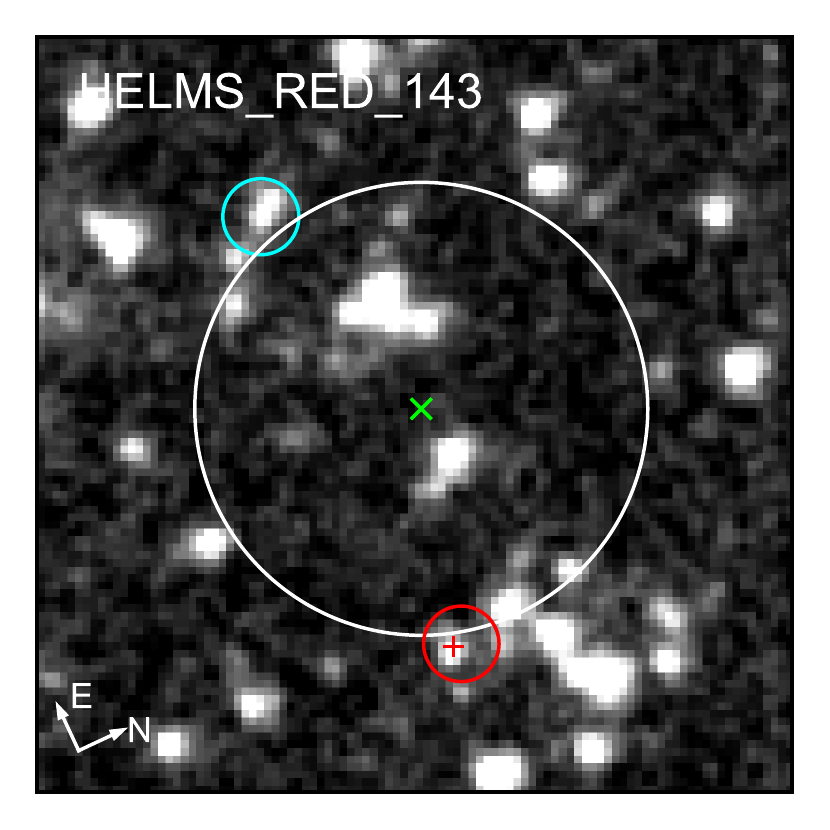}}
{\includegraphics[width=4.4cm, height=4.4cm]{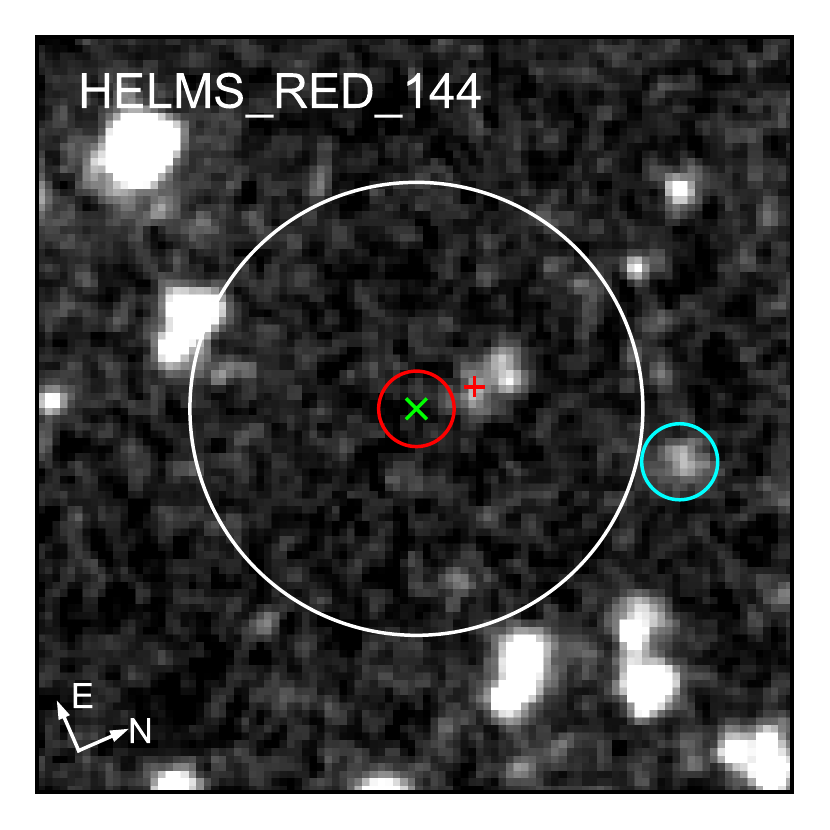}}
{\includegraphics[width=4.4cm, height=4.4cm]{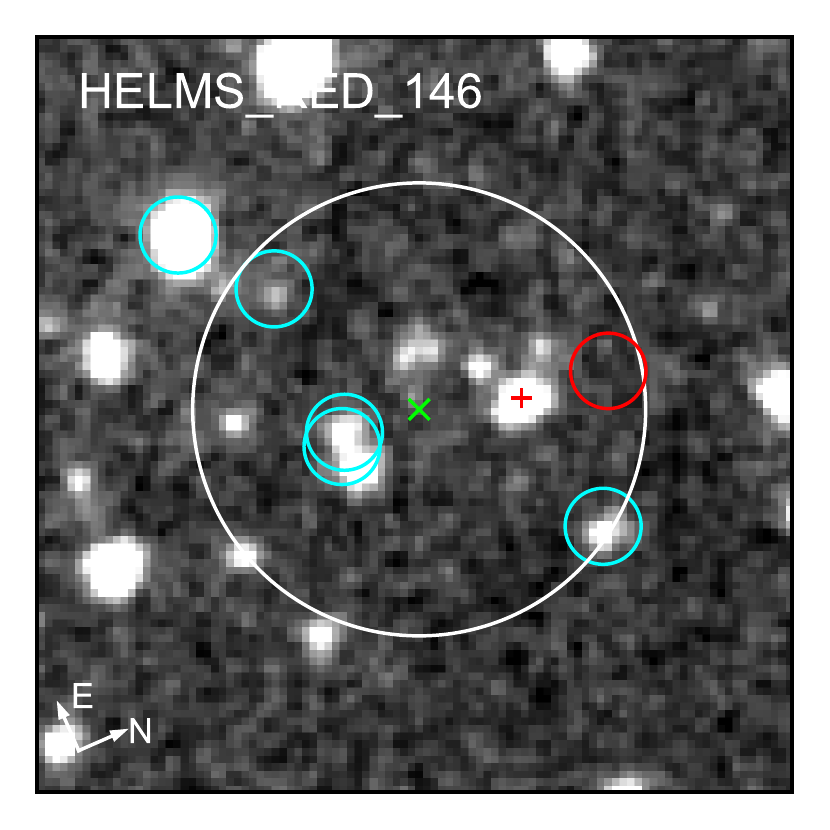}}
{\includegraphics[width=4.4cm, height=4.4cm]{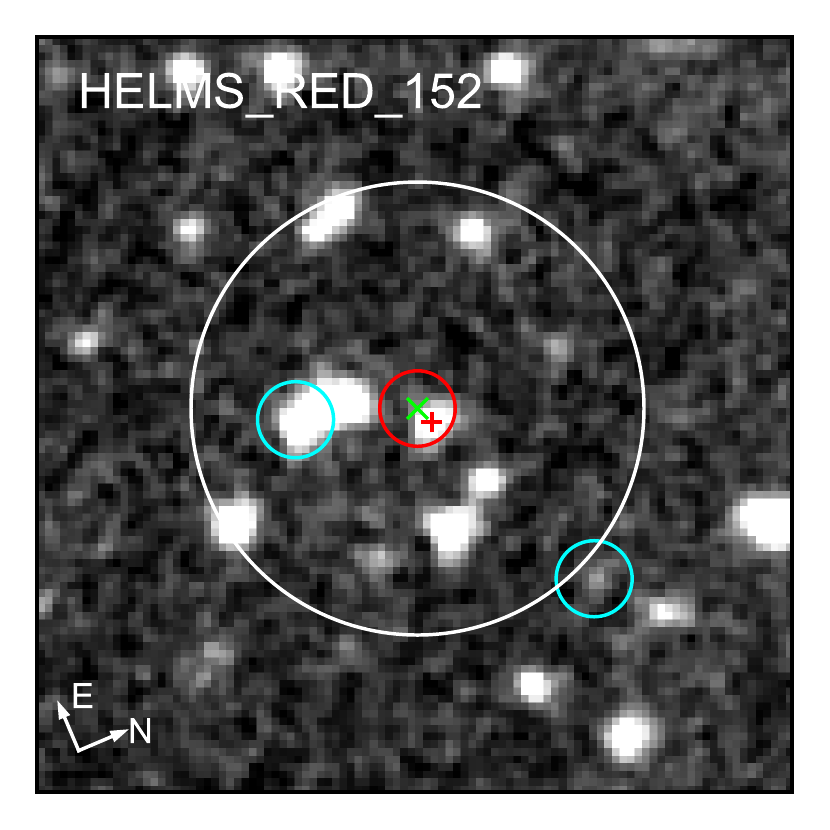}}
{\includegraphics[width=4.4cm, height=4.4cm]{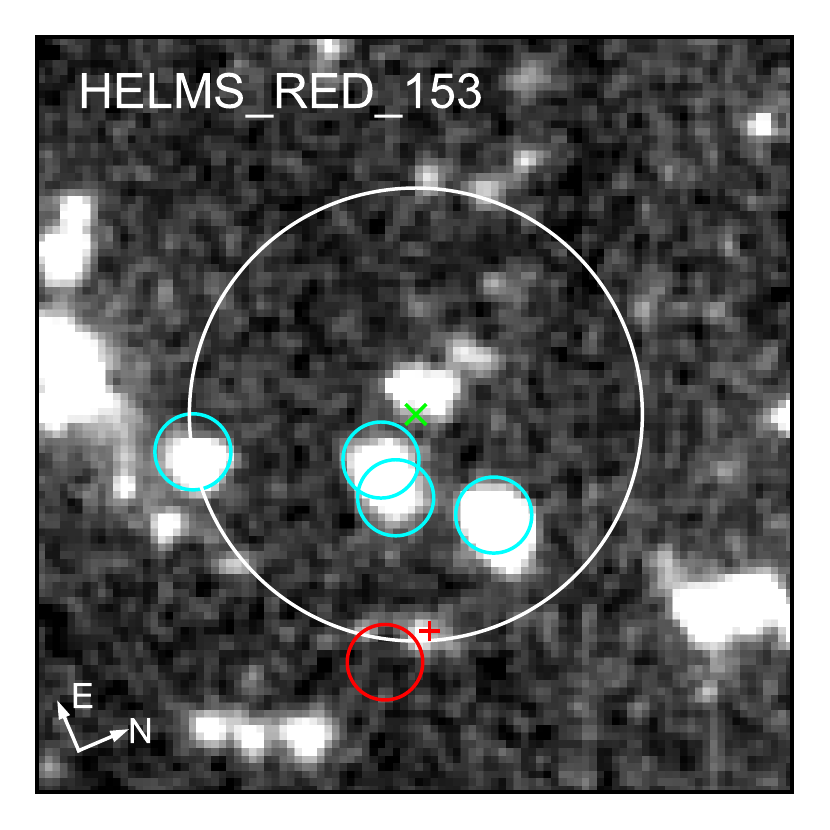}}
{\includegraphics[width=4.4cm, height=4.4cm]{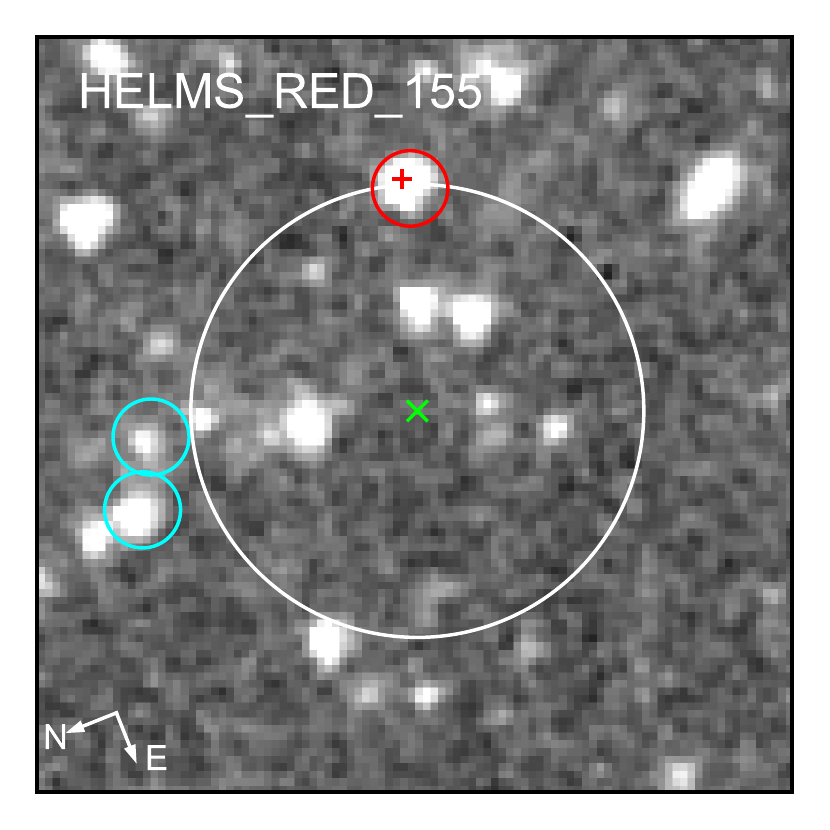}}
{\includegraphics[width=4.4cm, height=4.4cm]{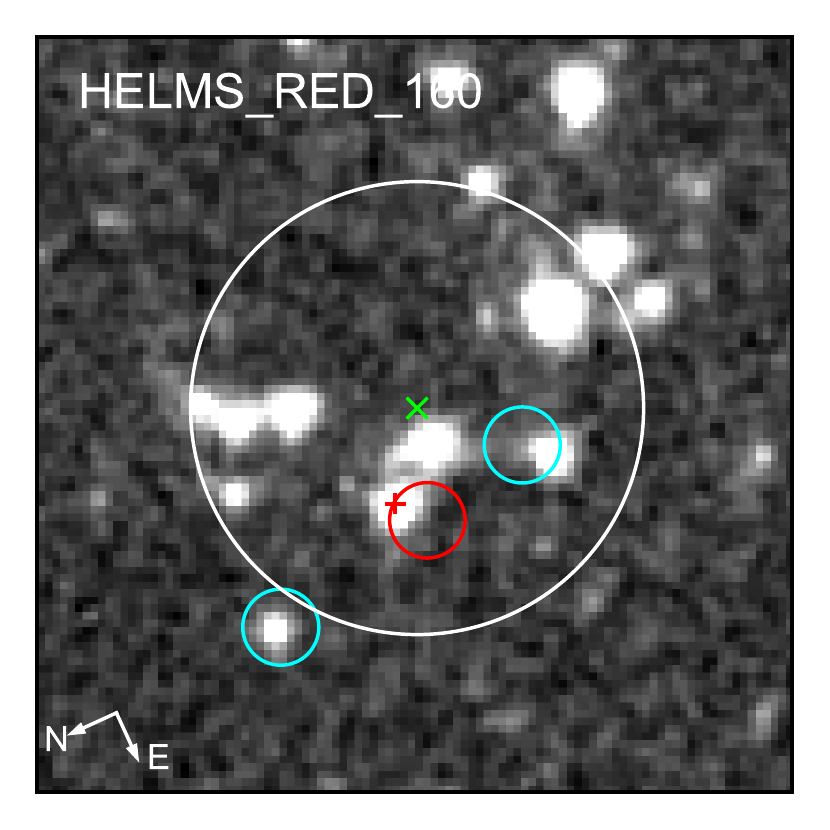}}
{\includegraphics[width=4.4cm, height=4.4cm]{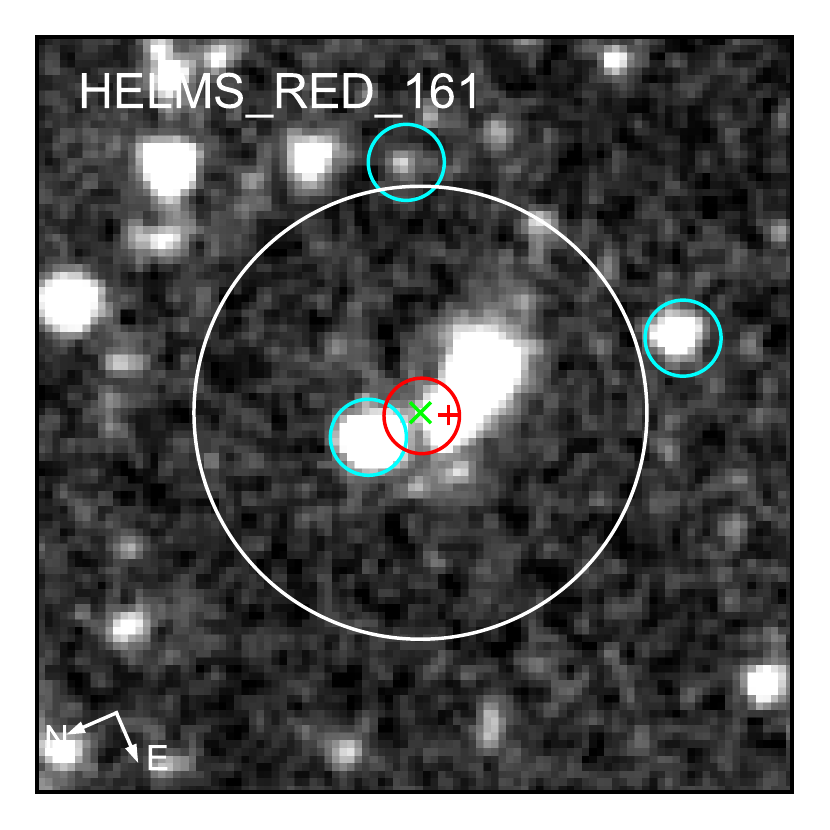}}
{\includegraphics[width=4.4cm, height=4.4cm]{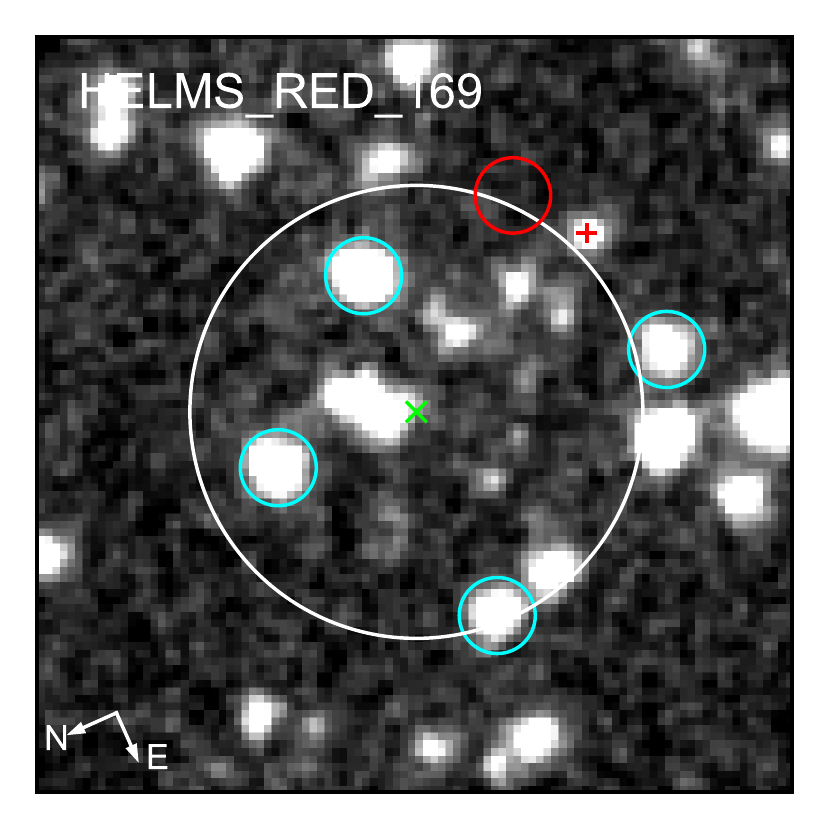}}
{\includegraphics[width=4.4cm, height=4.4cm]{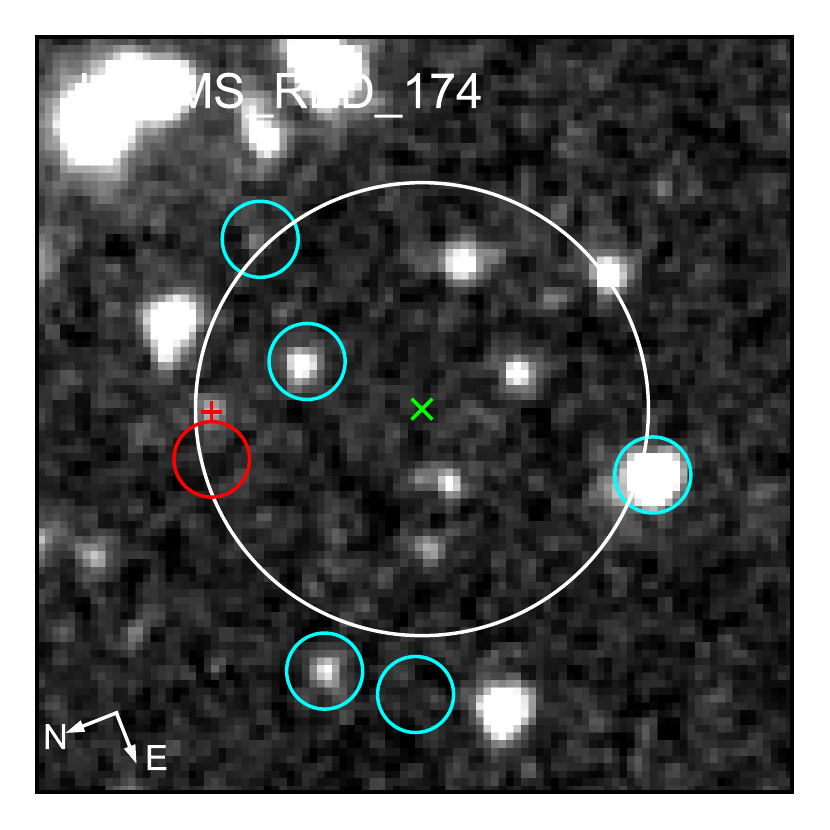}}
{\includegraphics[width=4.4cm, height=4.4cm]{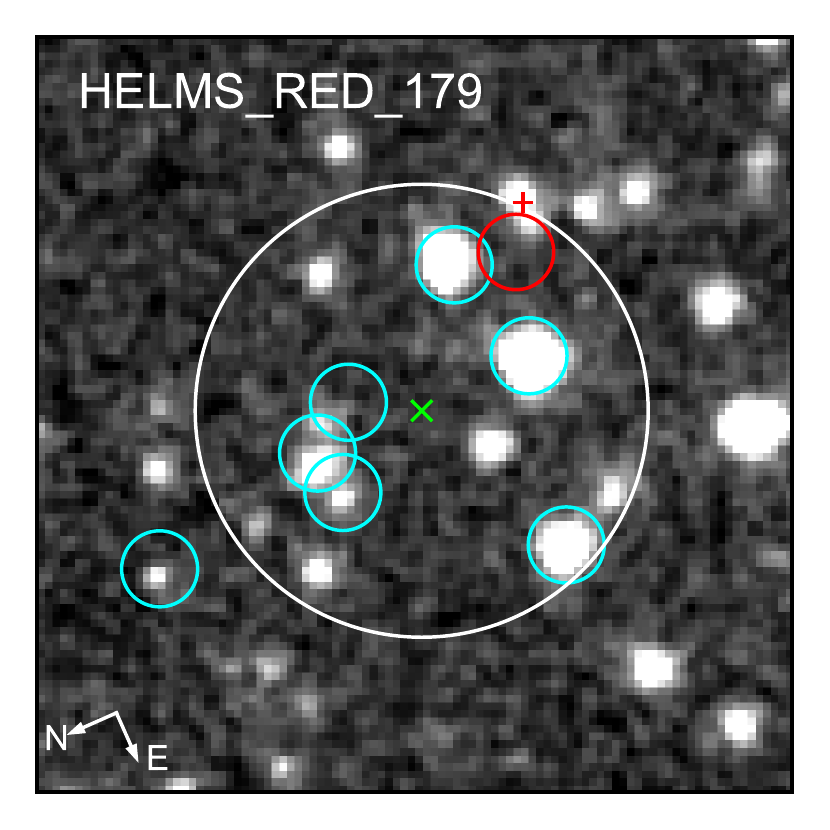}}
{\includegraphics[width=4.4cm, height=4.4cm]{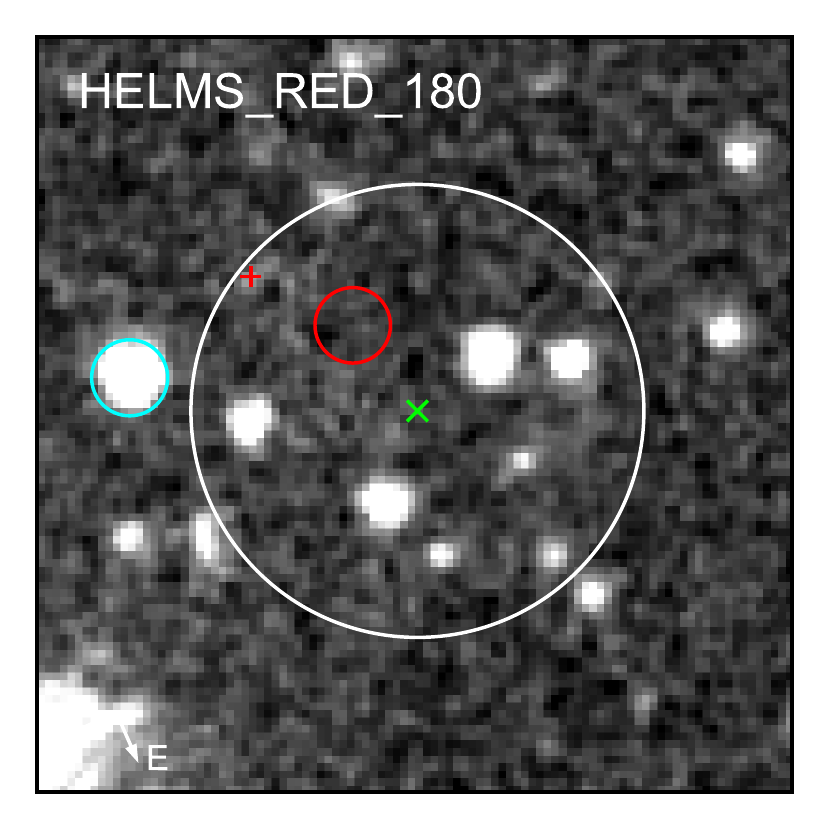}}
{\includegraphics[width=4.4cm, height=4.4cm]{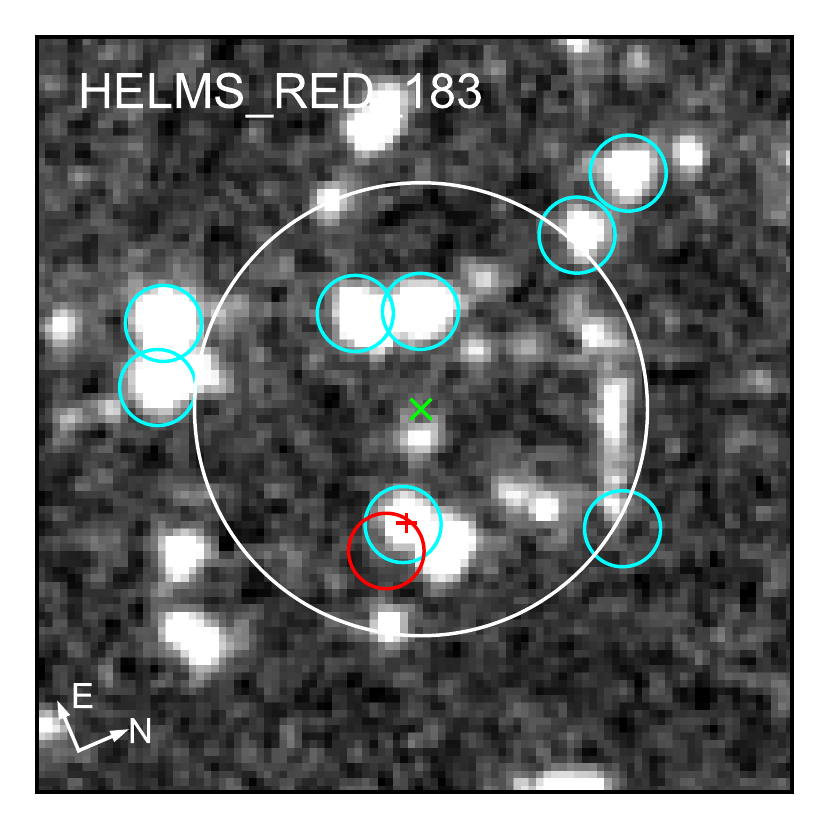}}
{\includegraphics[width=4.4cm, height=4.4cm]{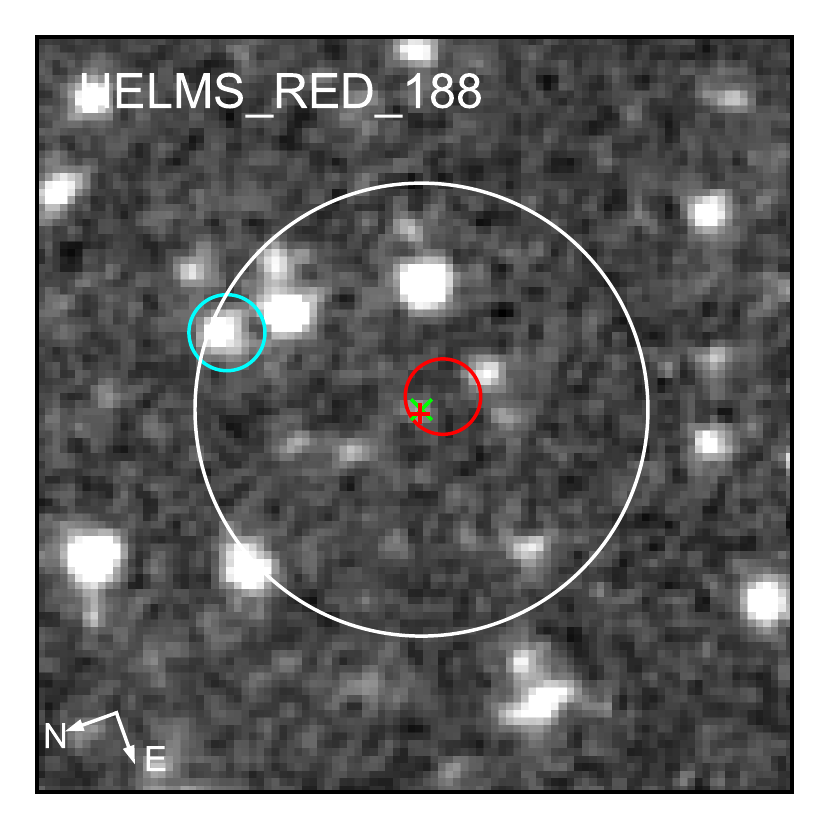}}
{\includegraphics[width=4.4cm, height=4.4cm]{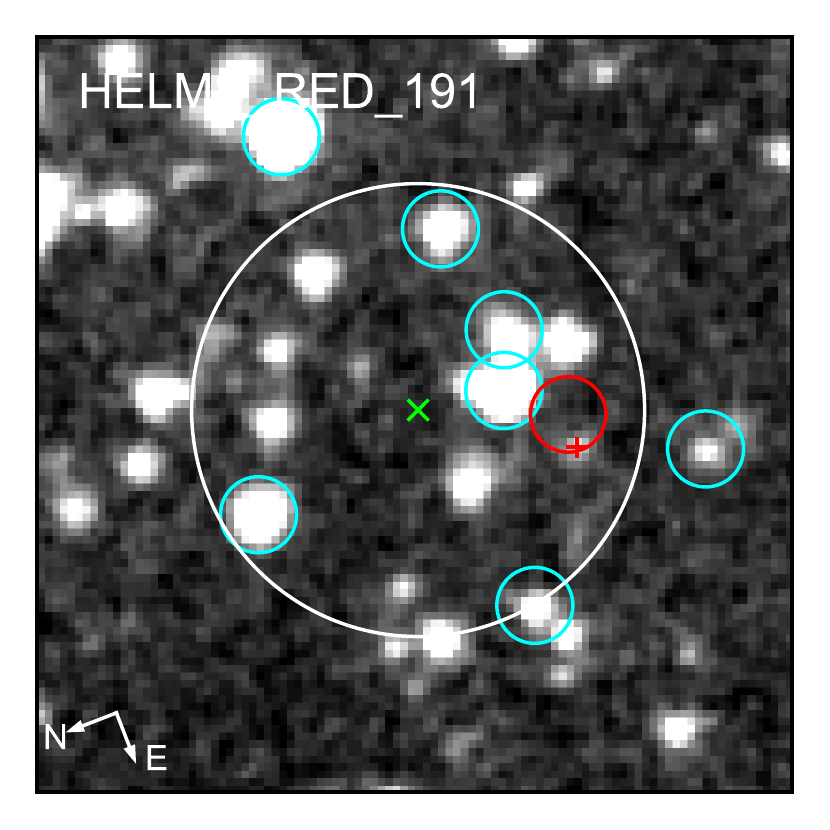}}
{\includegraphics[width=4.4cm, height=4.4cm]{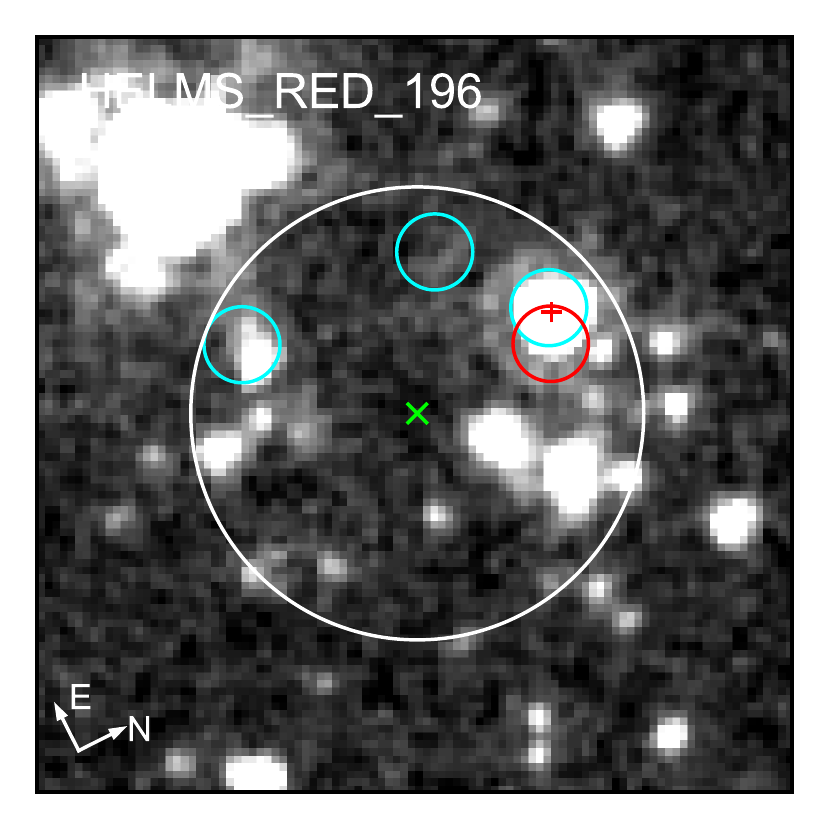}}
{\includegraphics[width=4.4cm, height=4.4cm]{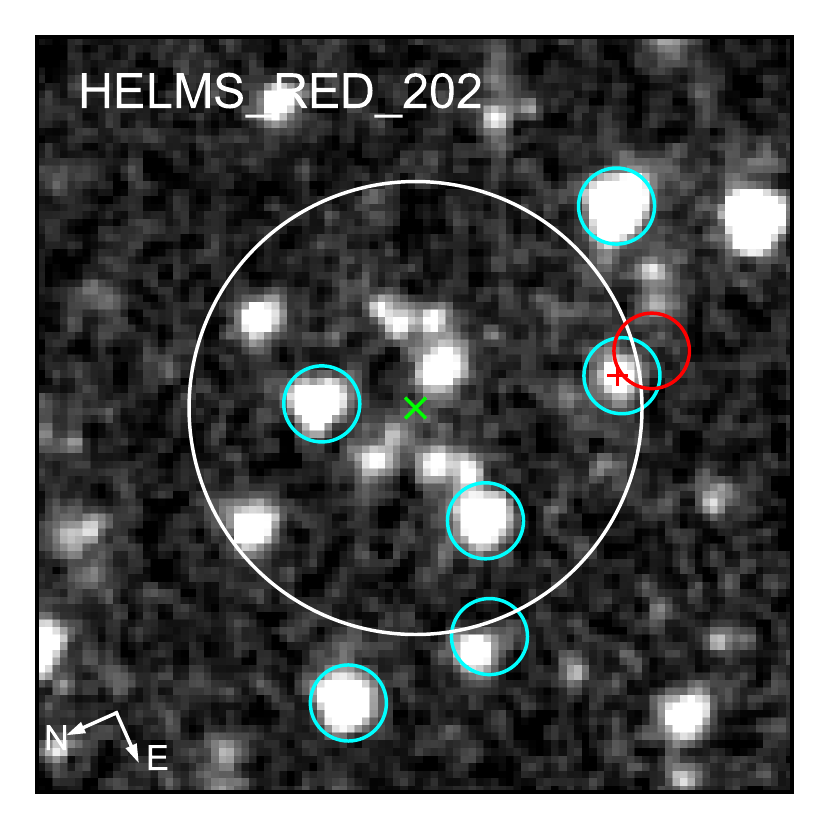}}
{\includegraphics[width=4.4cm, height=4.4cm]{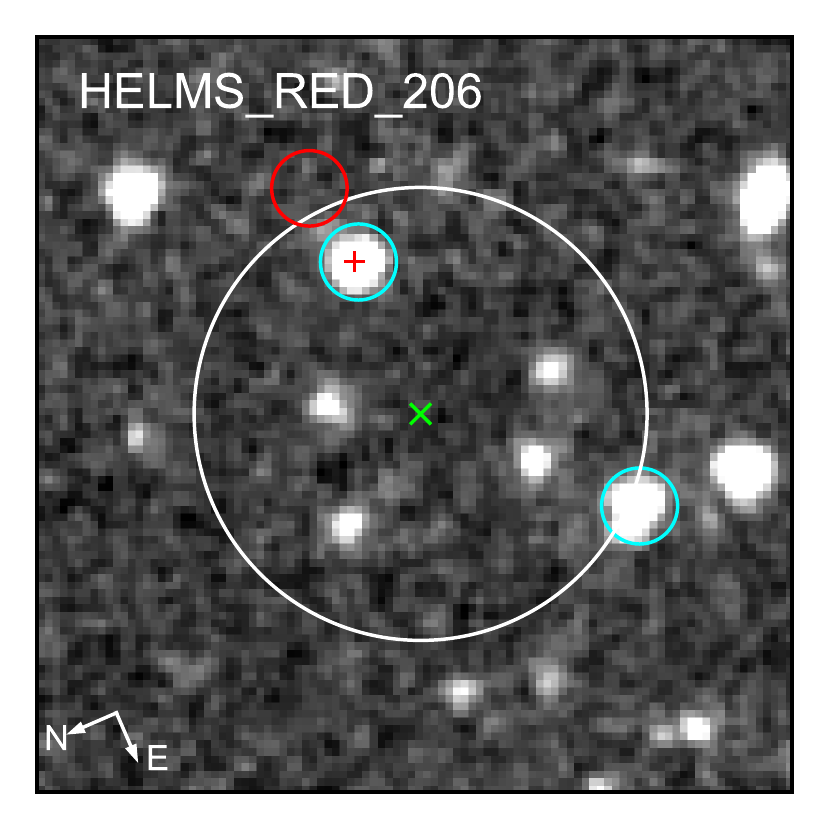}}
{\includegraphics[width=4.4cm, height=4.4cm]{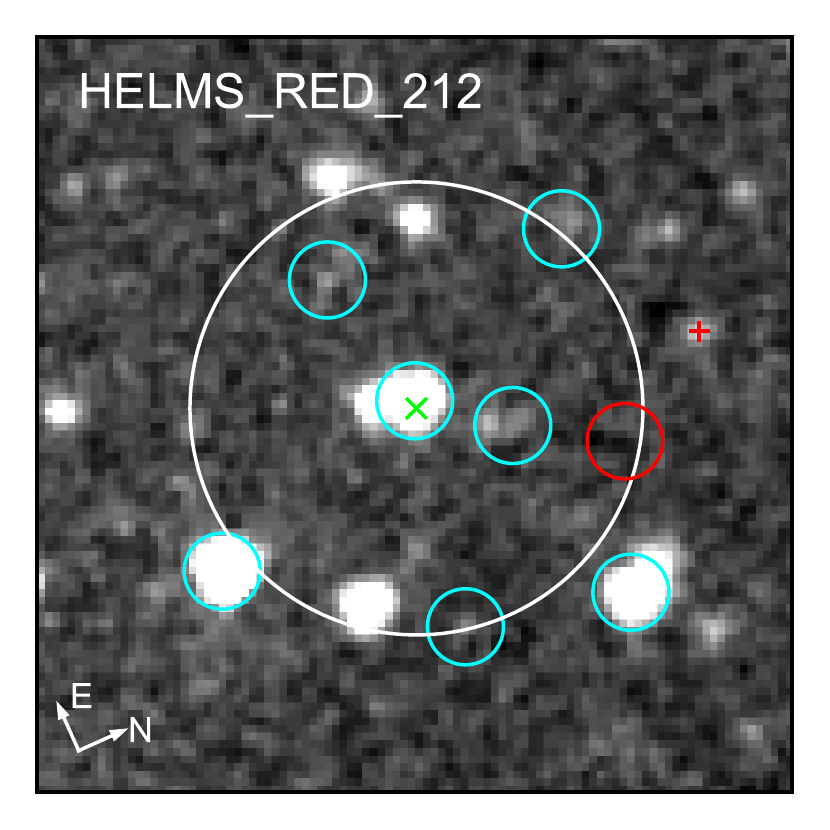}}
\caption{Continued 60$\arcsec$ $\times$ 60$\arcsec$ cutouts }
\label{fig:data}
\end{figure*}

\addtocounter{figure}{-1}
\begin{figure*}
\centering
{\includegraphics[width=4.4cm, height=4.4cm]{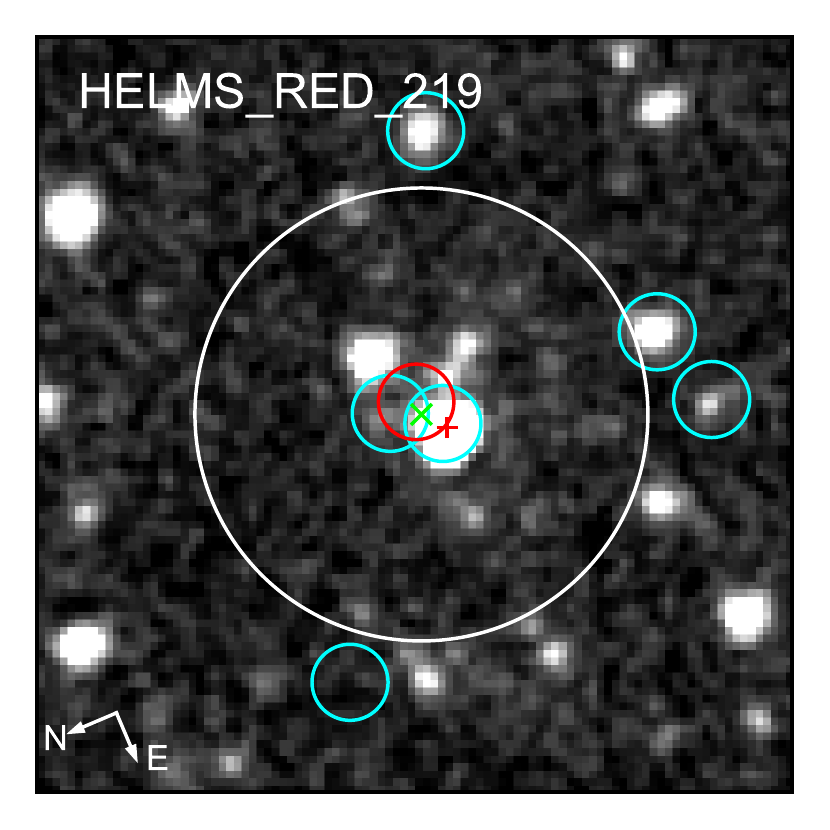}}
{\includegraphics[width=4.4cm, height=4.4cm]{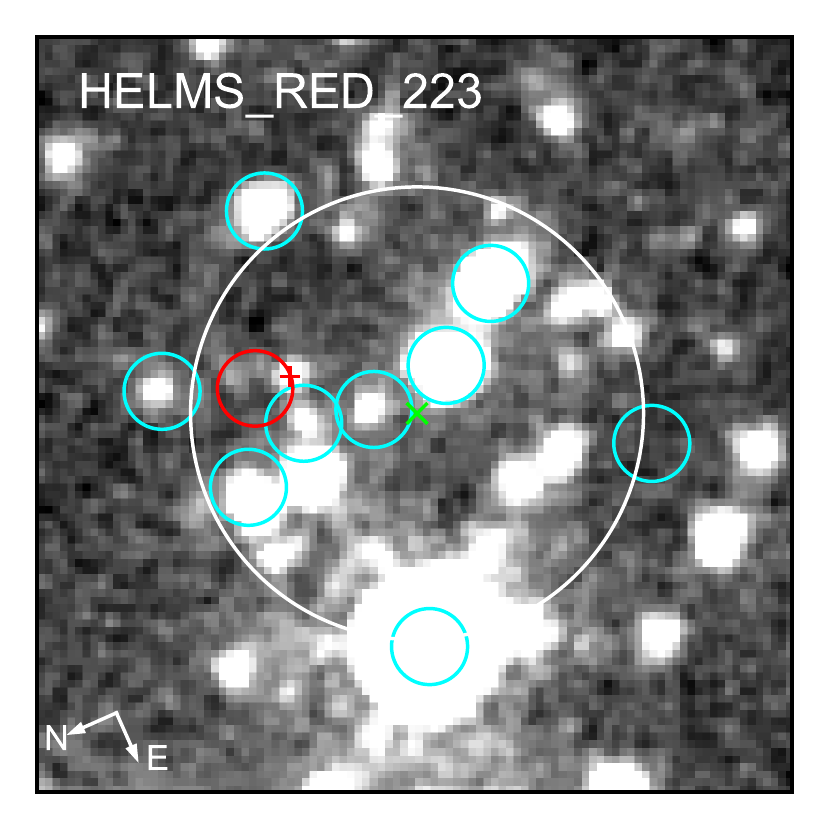}}
{\includegraphics[width=4.4cm, height=4.4cm]{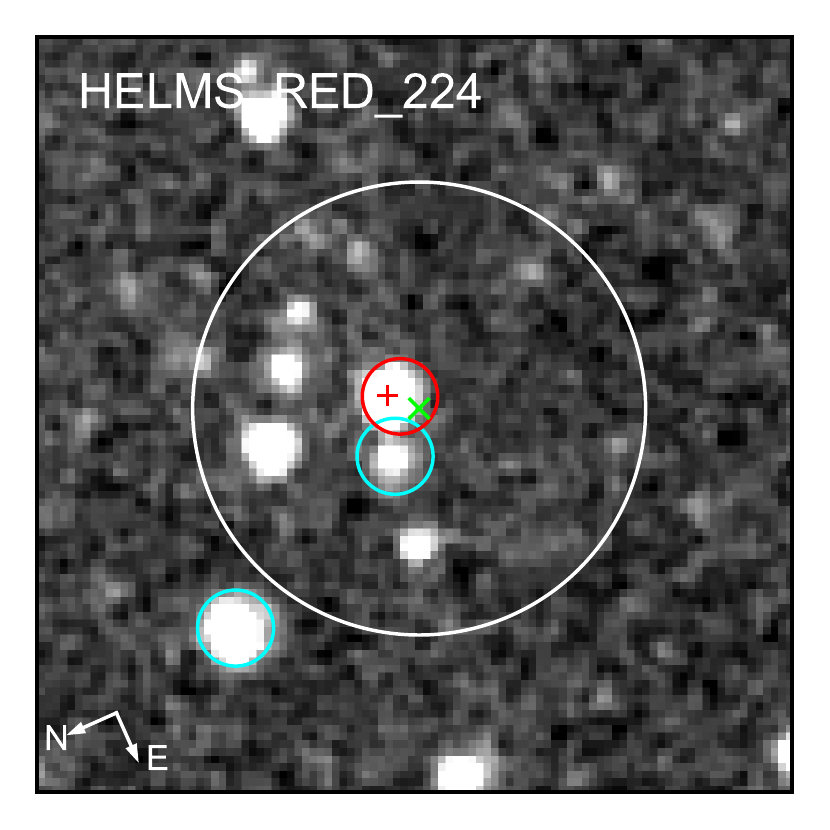}}
{\includegraphics[width=4.4cm, height=4.4cm]{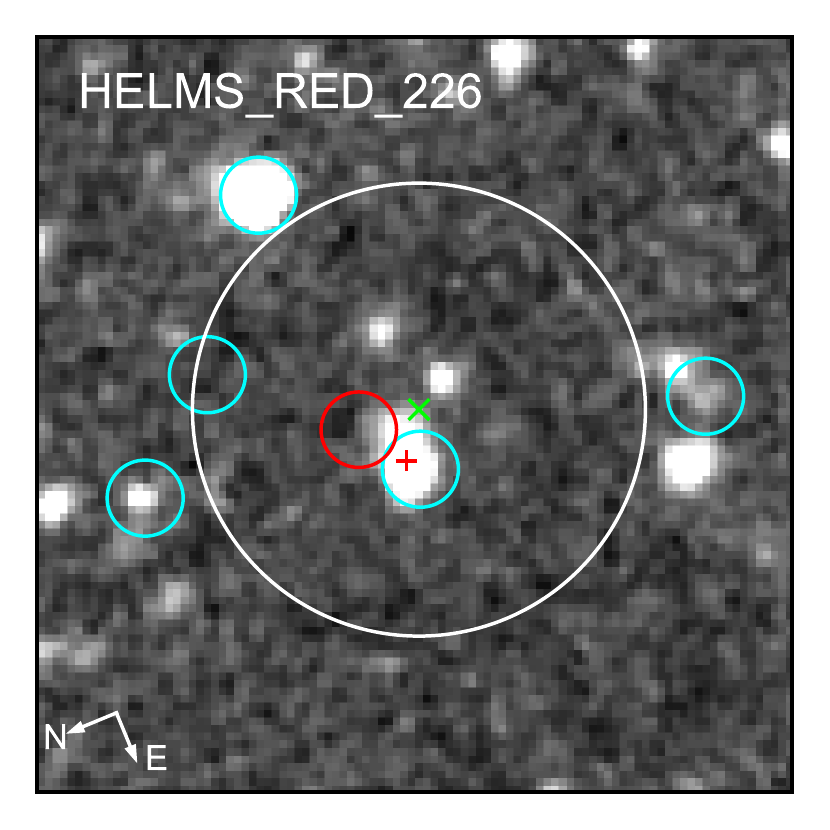}}
{\includegraphics[width=4.4cm, height=4.4cm]{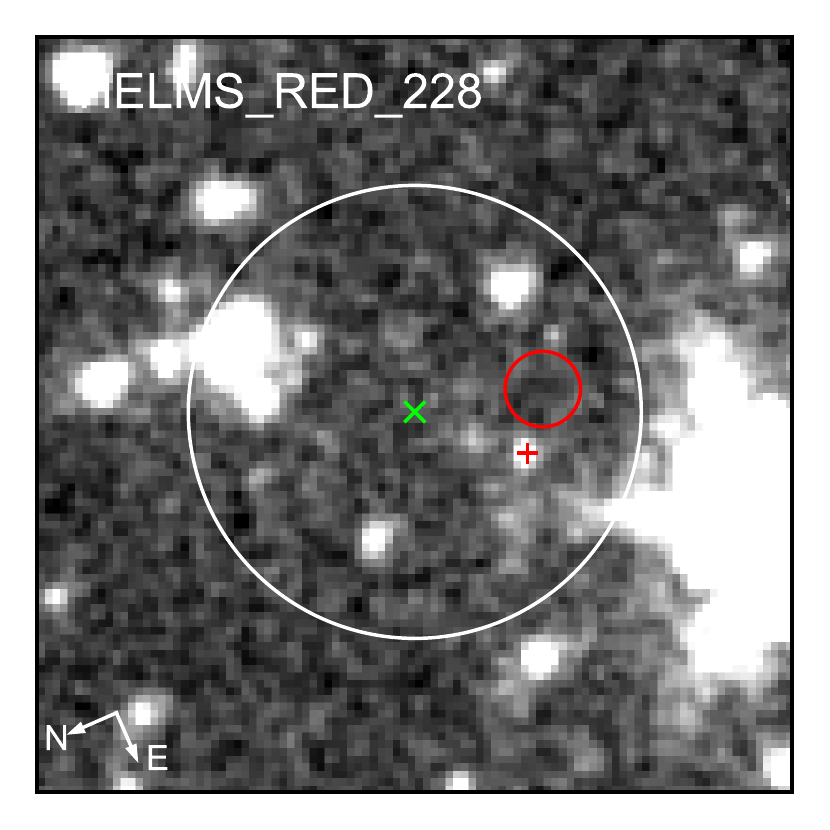}}
{\includegraphics[width=4.4cm, height=4.4cm]{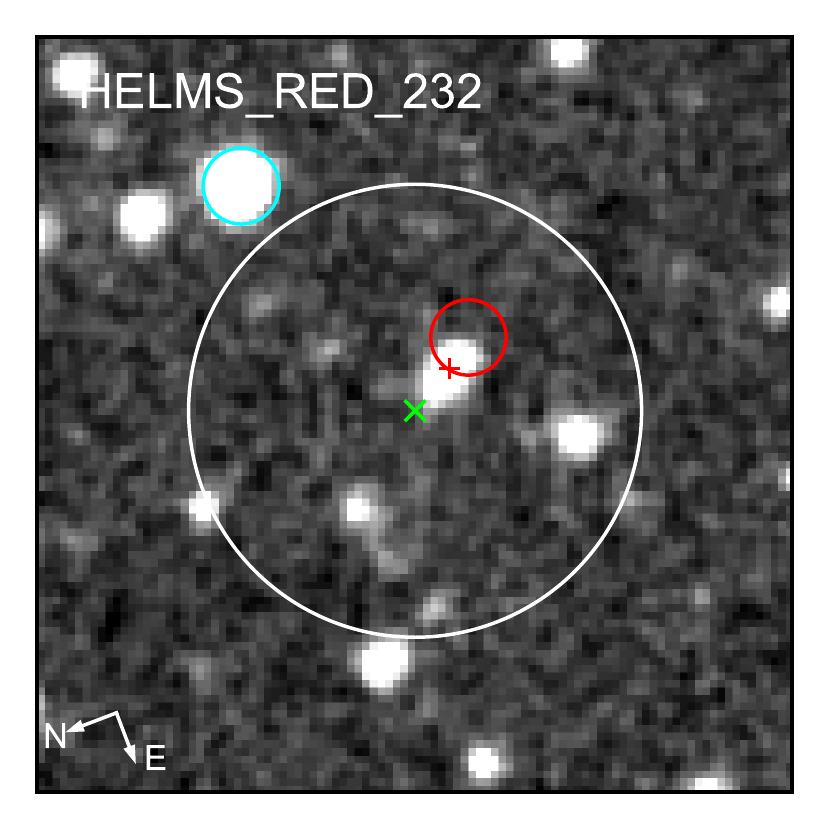}}
{\includegraphics[width=4.4cm, height=4.4cm]{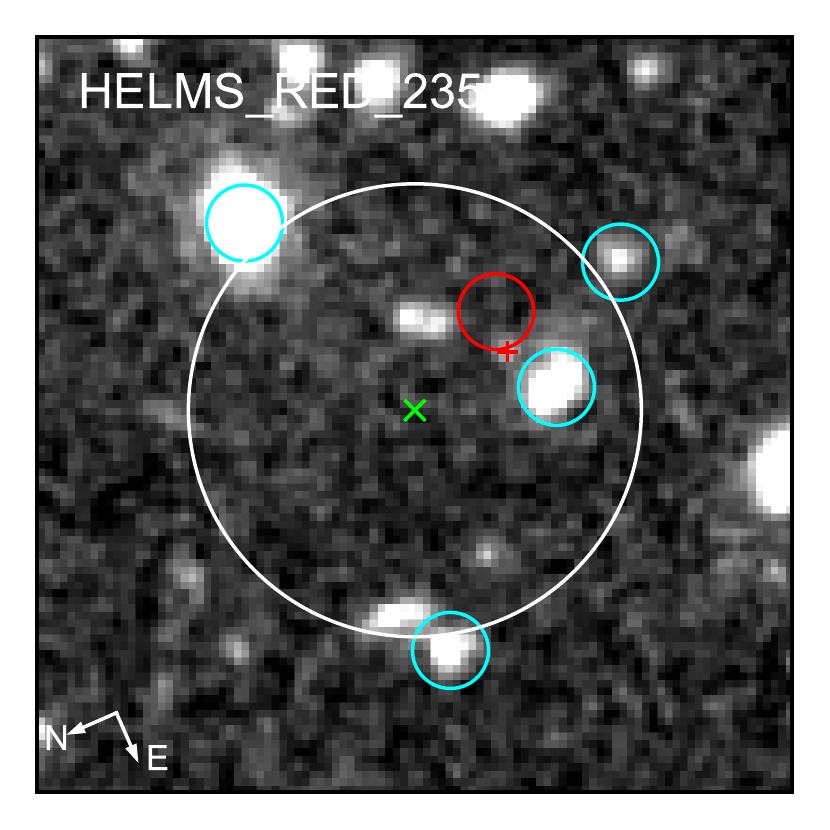}}
{\includegraphics[width=4.4cm, height=4.4cm]{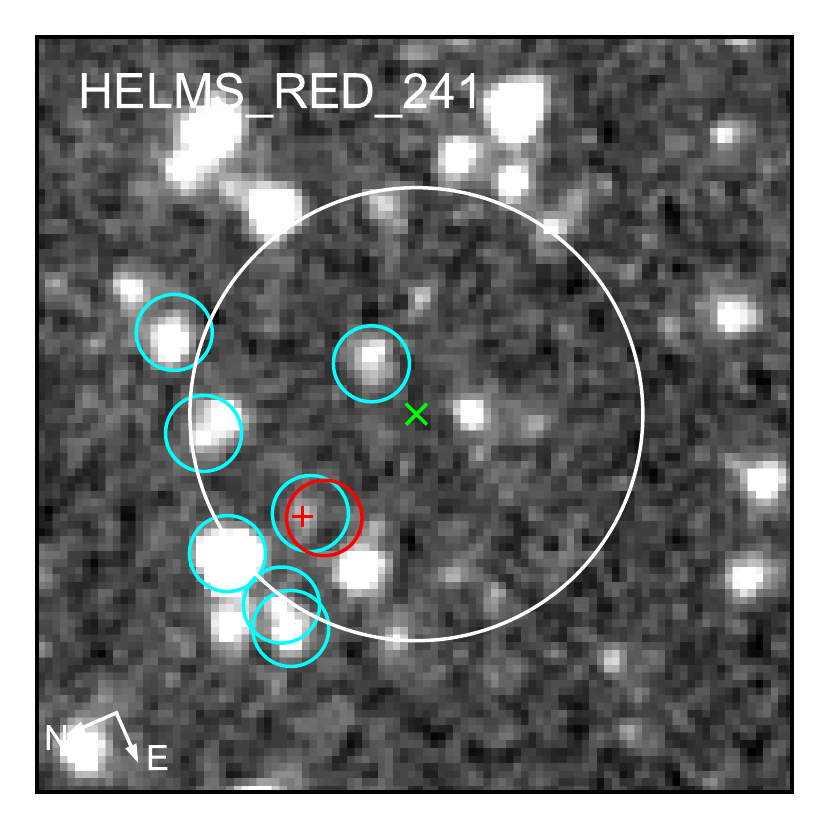}}
{\includegraphics[width=4.4cm, height=4.4cm]{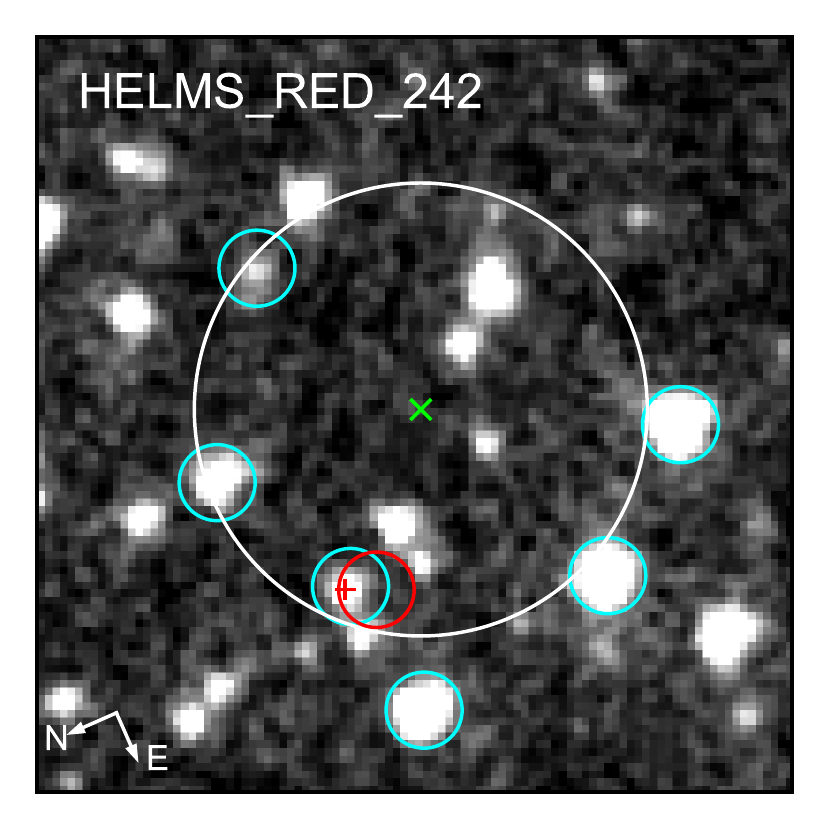}}
{\includegraphics[width=4.4cm, height=4.4cm]{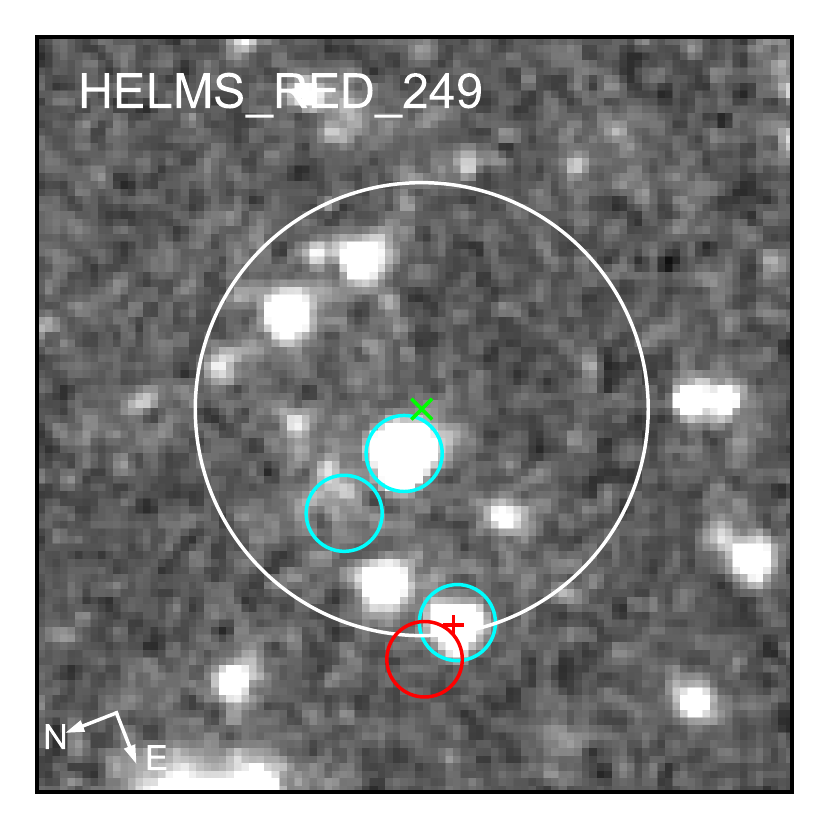}}
{\includegraphics[width=4.4cm, height=4.4cm]{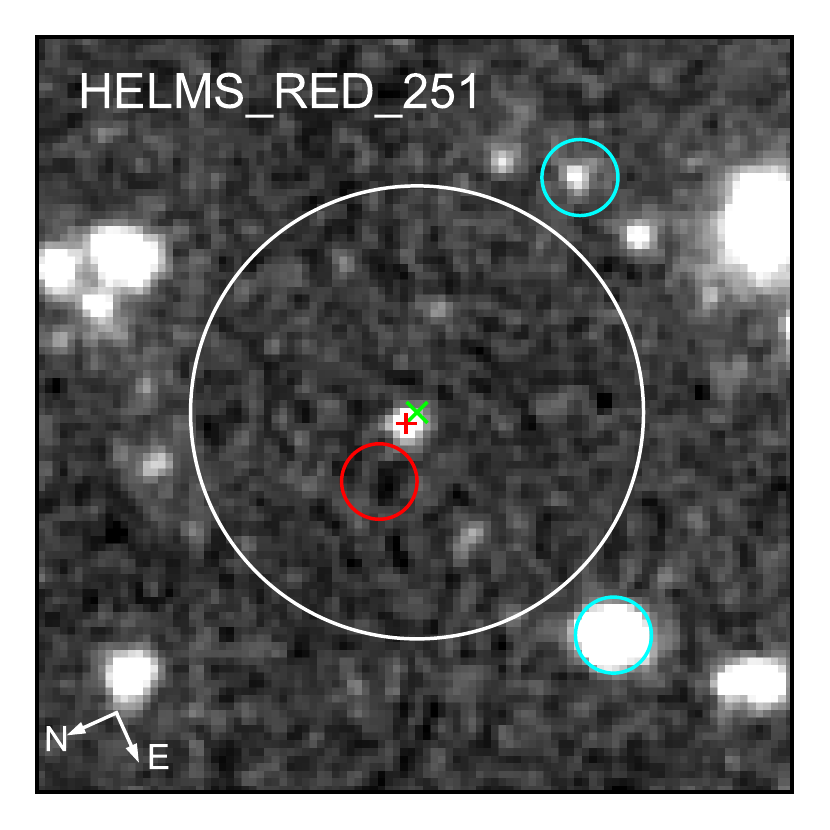}}
{\includegraphics[width=4.4cm, height=4.4cm]{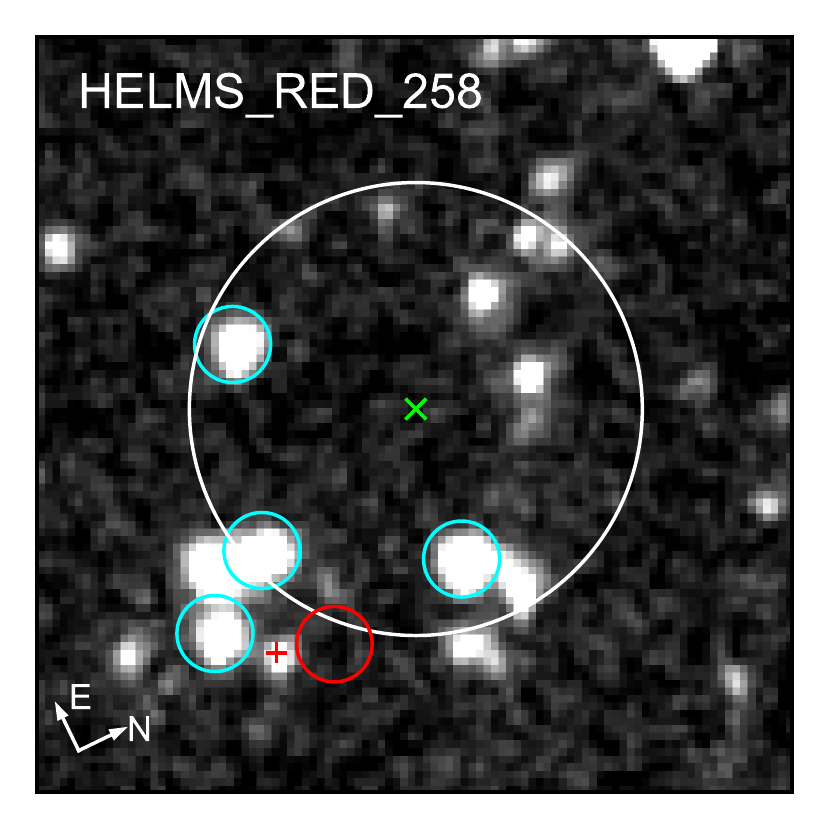}}
{\includegraphics[width=4.4cm, height=4.4cm]{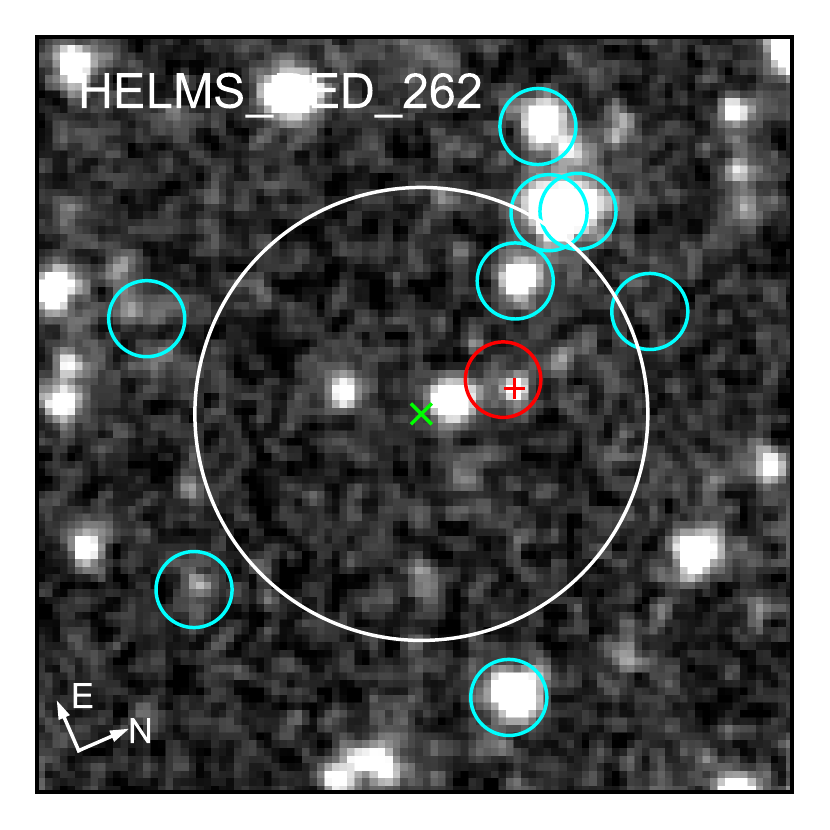}}
{\includegraphics[width=4.4cm, height=4.4cm]{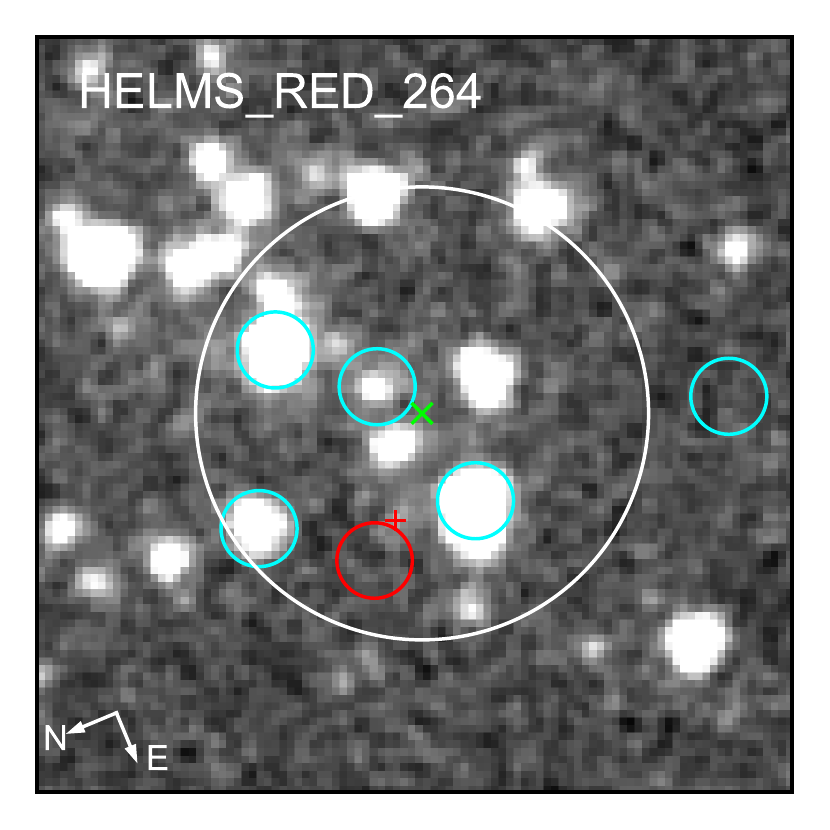}}
{\includegraphics[width=4.4cm, height=4.4cm]{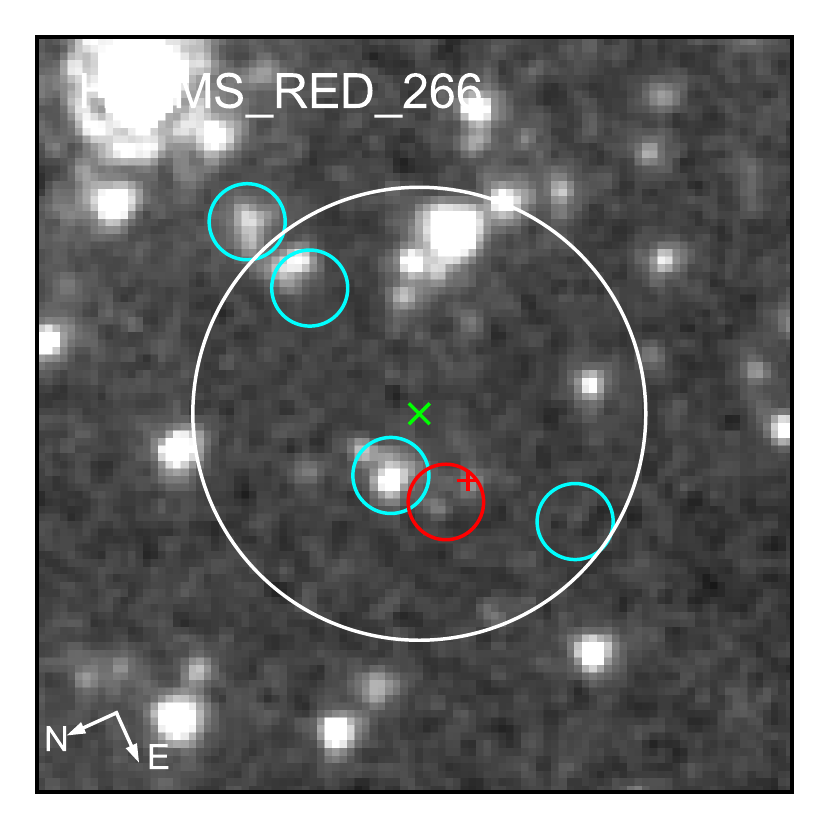}}
{\includegraphics[width=4.4cm, height=4.4cm]{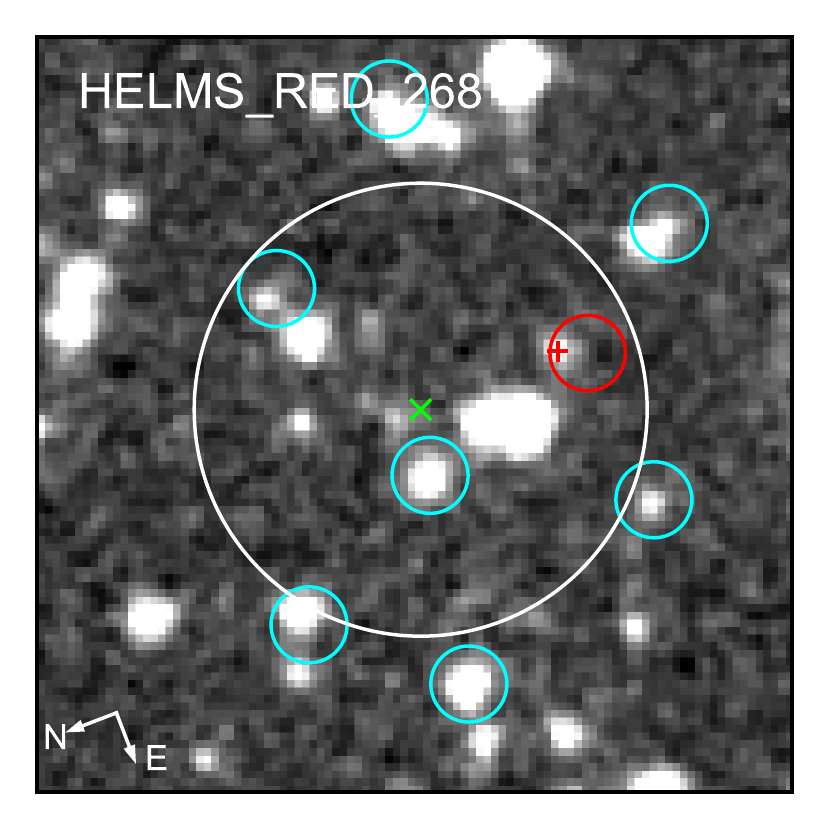}}
{\includegraphics[width=4.4cm, height=4.4cm]{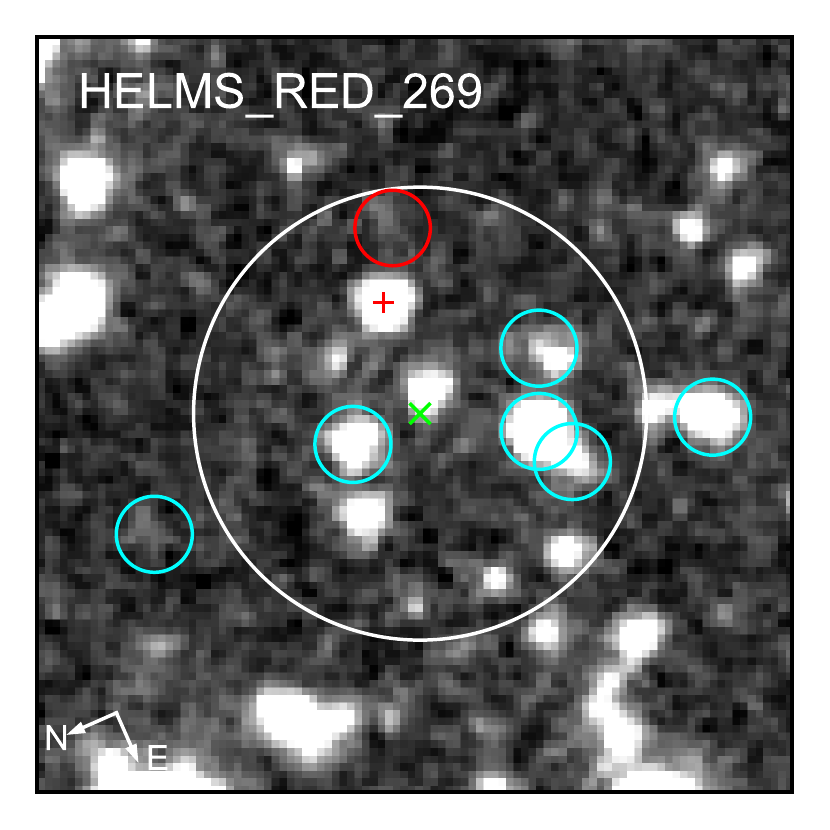}}
{\includegraphics[width=4.4cm, height=4.4cm]{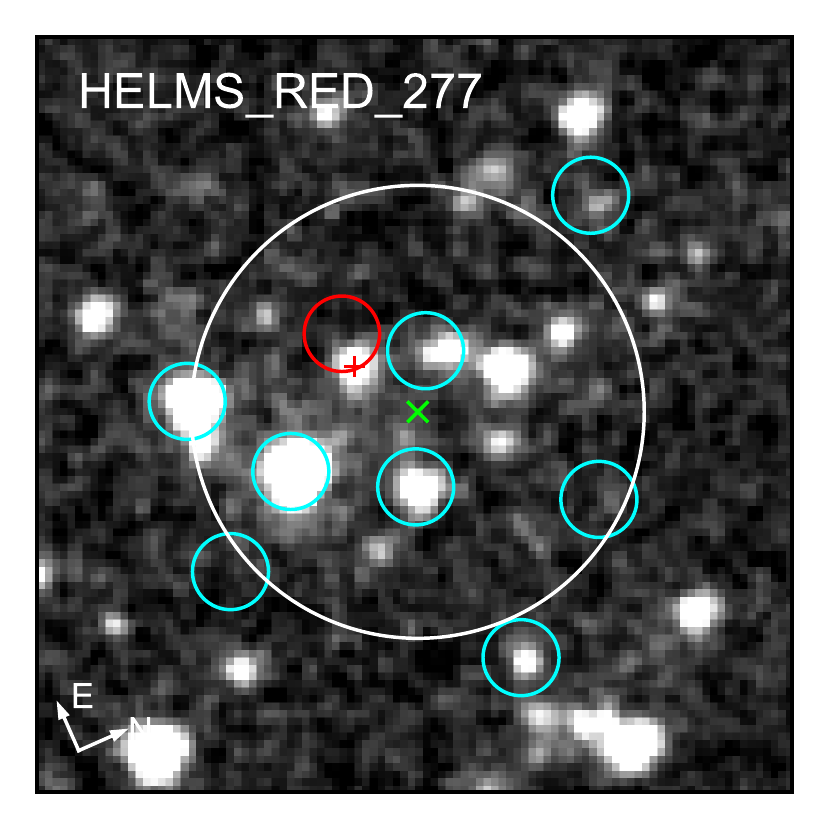}}
{\includegraphics[width=4.4cm, height=4.4cm]{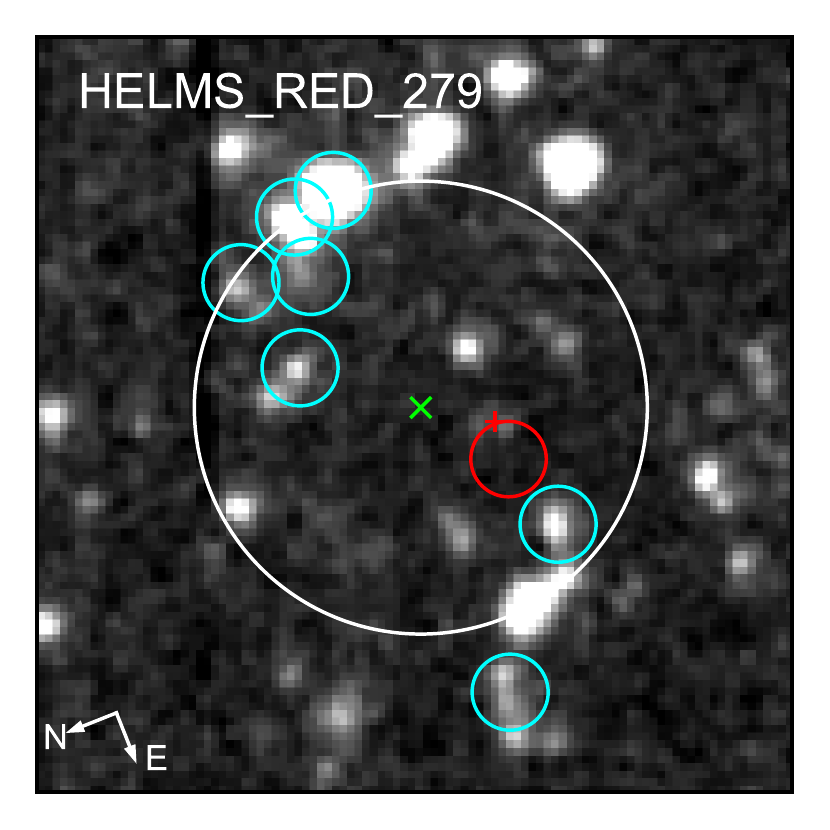}}
{\includegraphics[width=4.4cm, height=4.4cm]{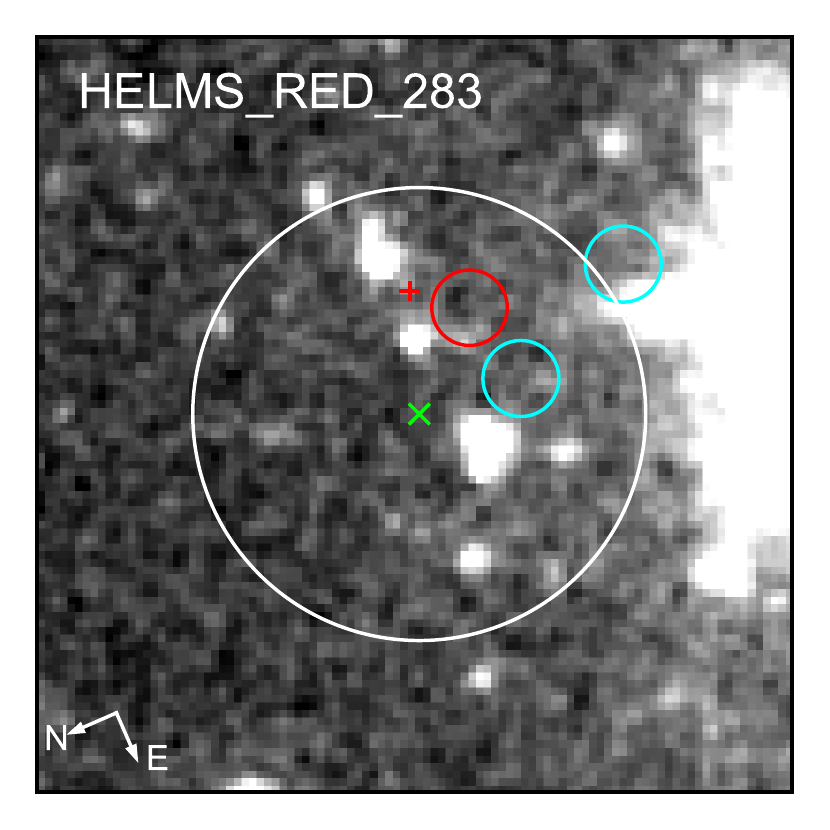}}
\caption{Continued 60$\arcsec$ $\times$ 60$\arcsec$ cutouts }
\label{fig:data}
\end{figure*}

\addtocounter{figure}{-1}
\begin{figure*}
\centering
{\includegraphics[width=4.4cm, height=4.4cm]{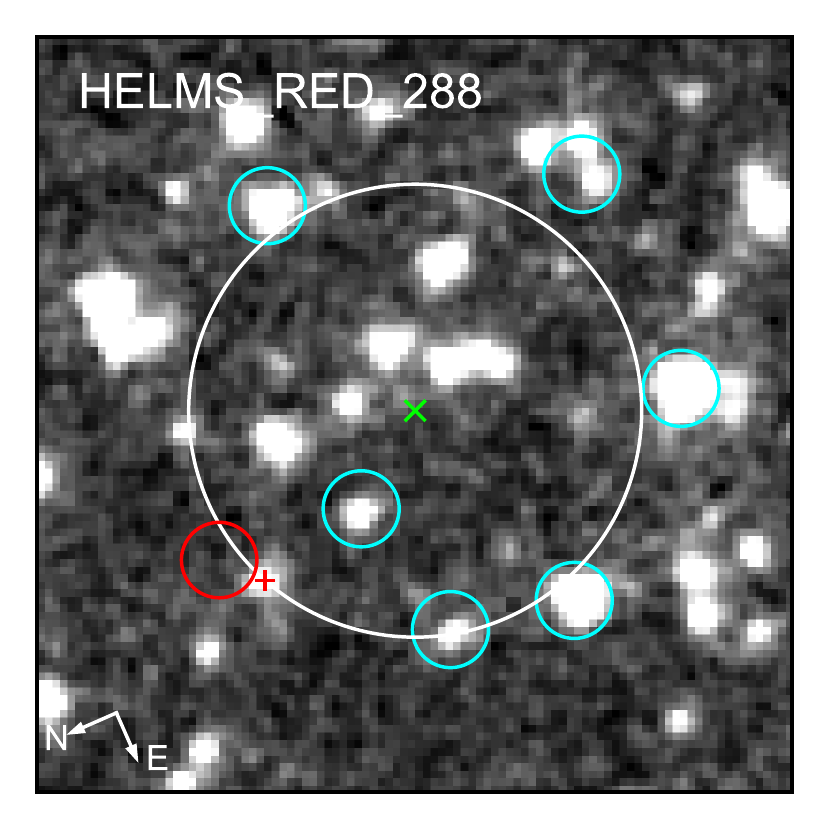}}
{\includegraphics[width=4.4cm, height=4.4cm]{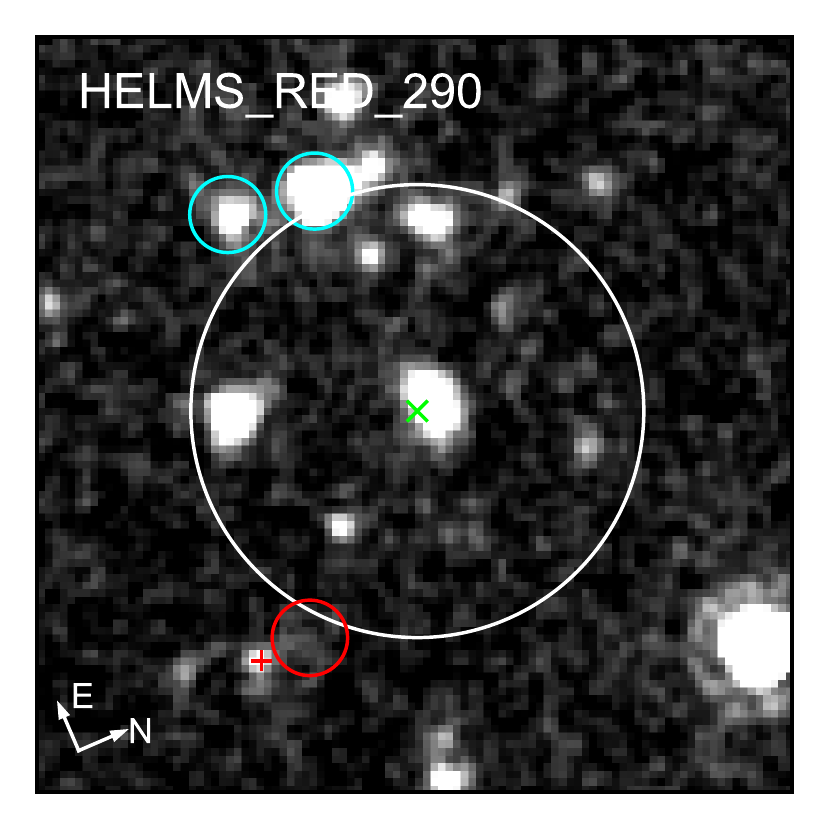}}
{\includegraphics[width=4.4cm, height=4.4cm]{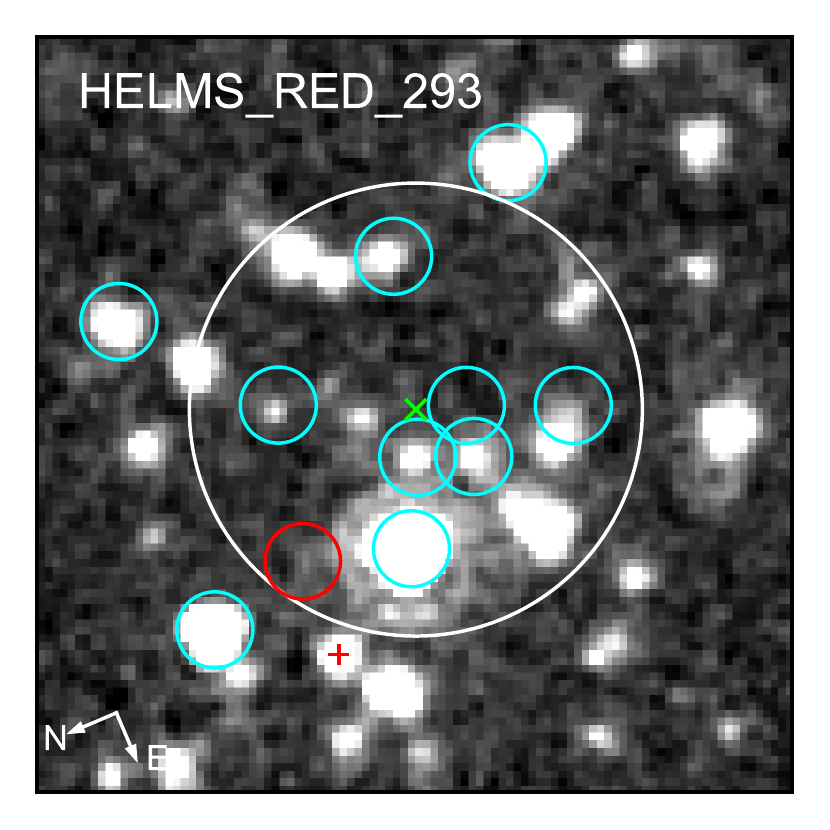}}
{\includegraphics[width=4.4cm, height=4.4cm]{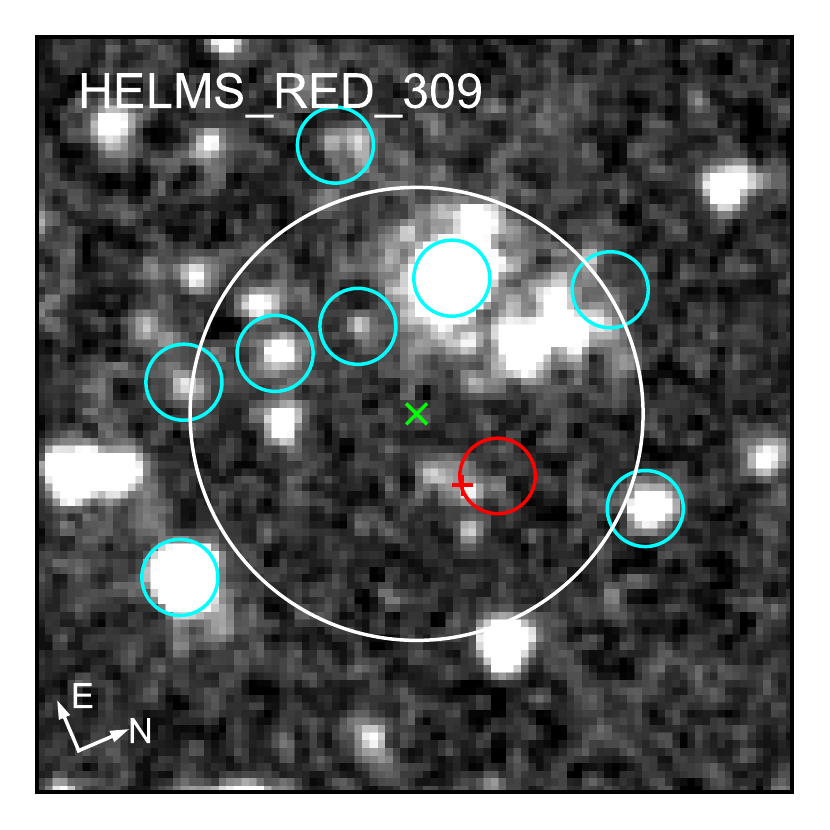}}
{\includegraphics[width=4.4cm, height=4.4cm]{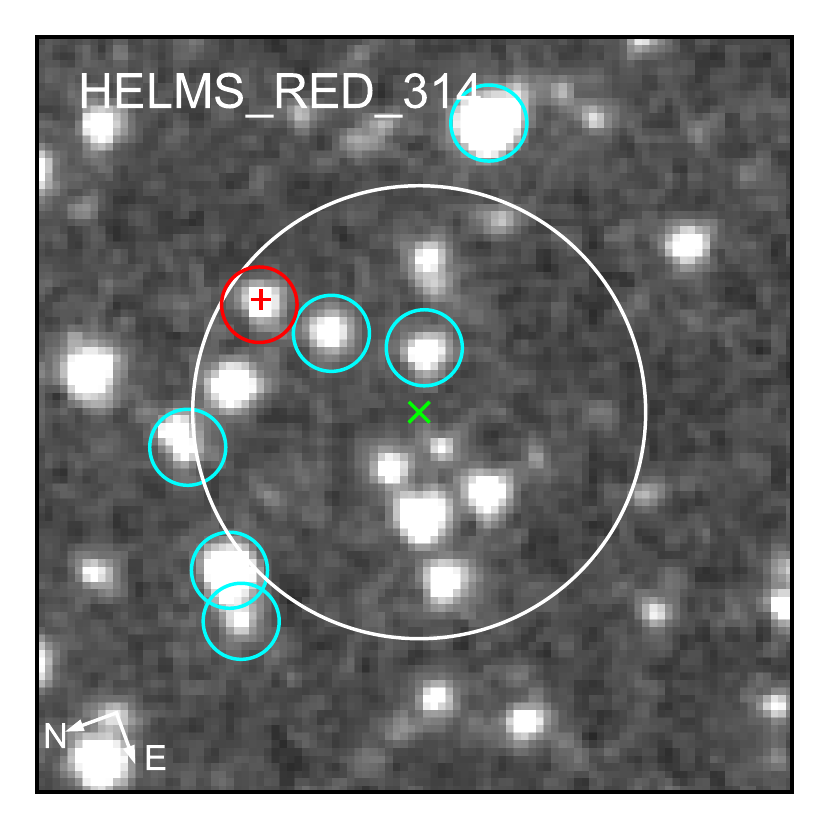}}
{\includegraphics[width=4.4cm, height=4.4cm]{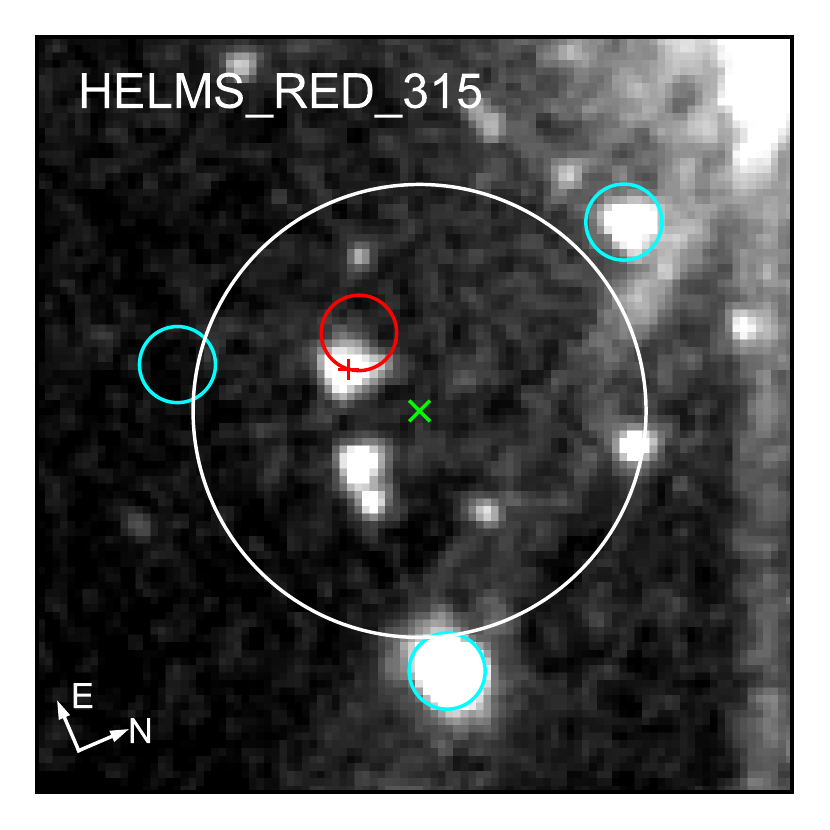}}
{\includegraphics[width=4.4cm, height=4.4cm]{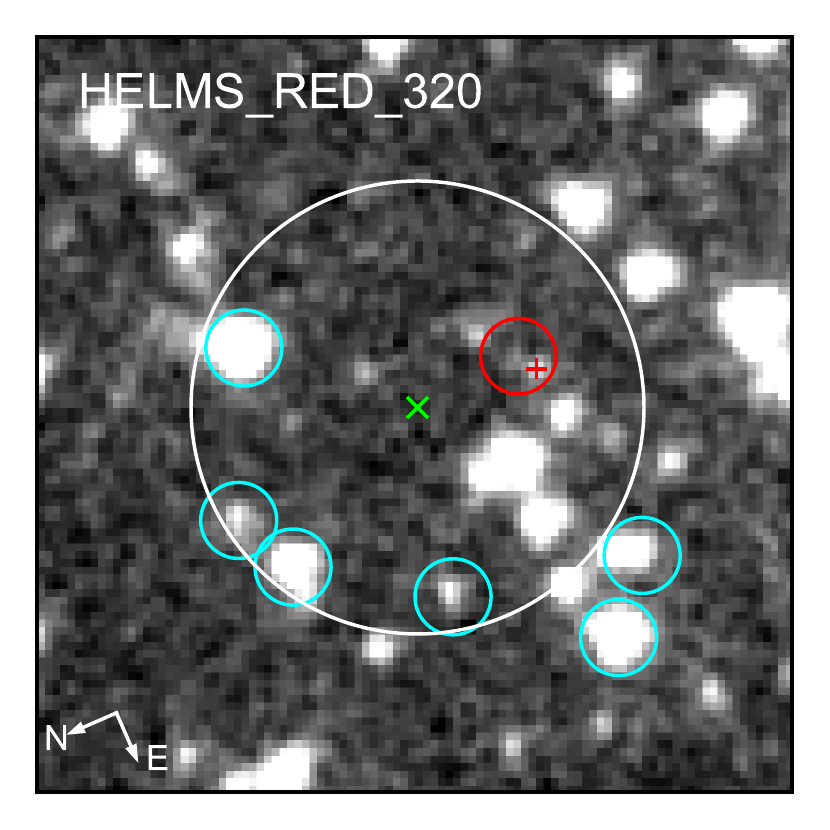}}
{\includegraphics[width=4.4cm, height=4.4cm]{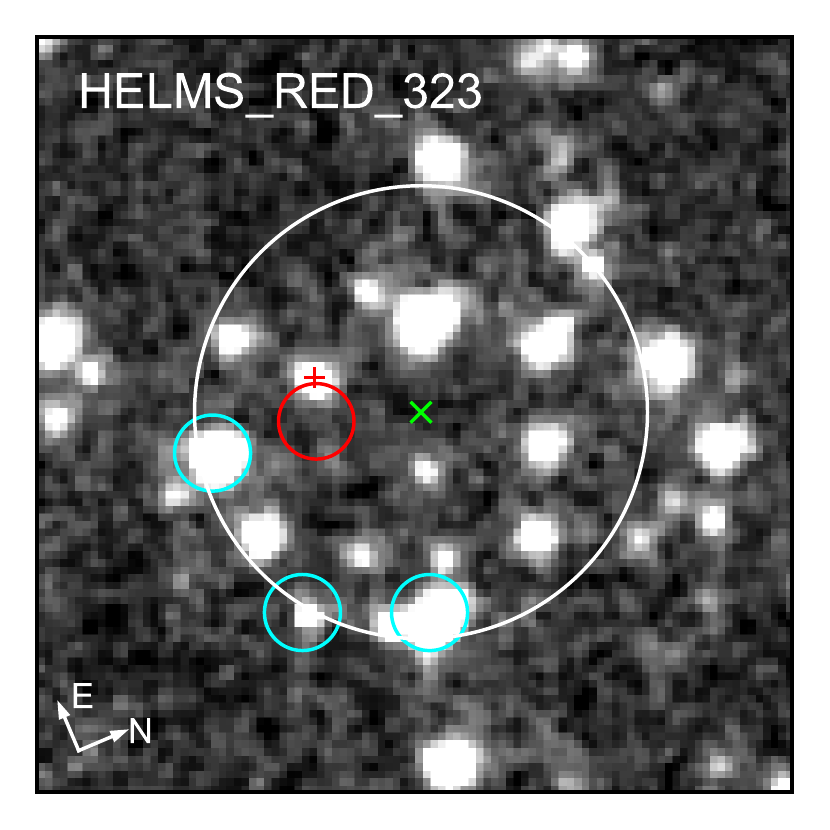}}
{\includegraphics[width=4.4cm, height=4.4cm]{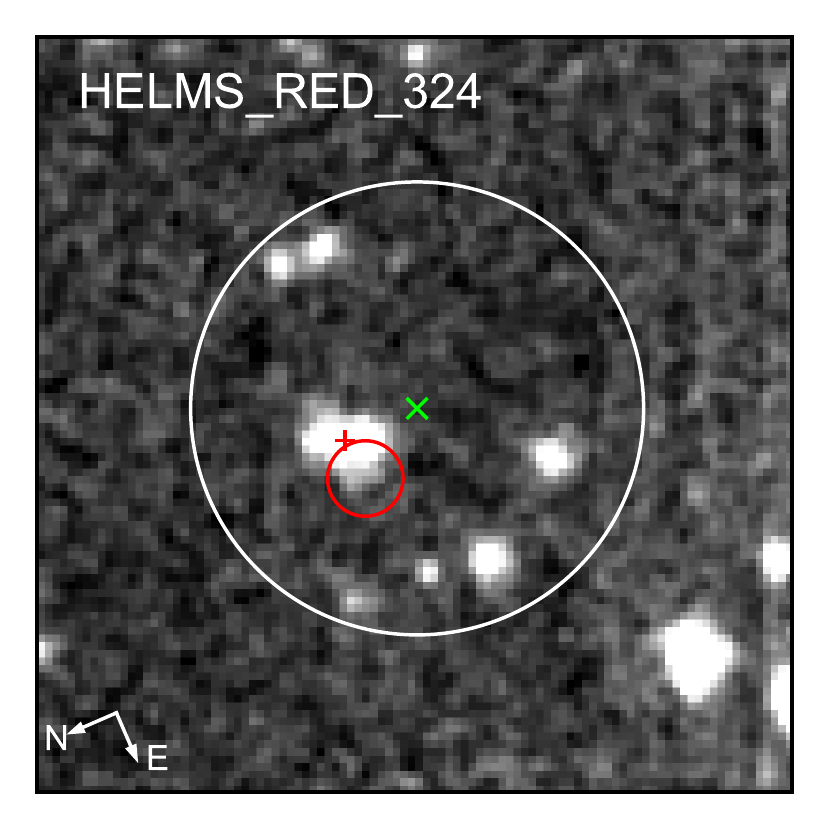}}
{\includegraphics[width=4.4cm, height=4.4cm]{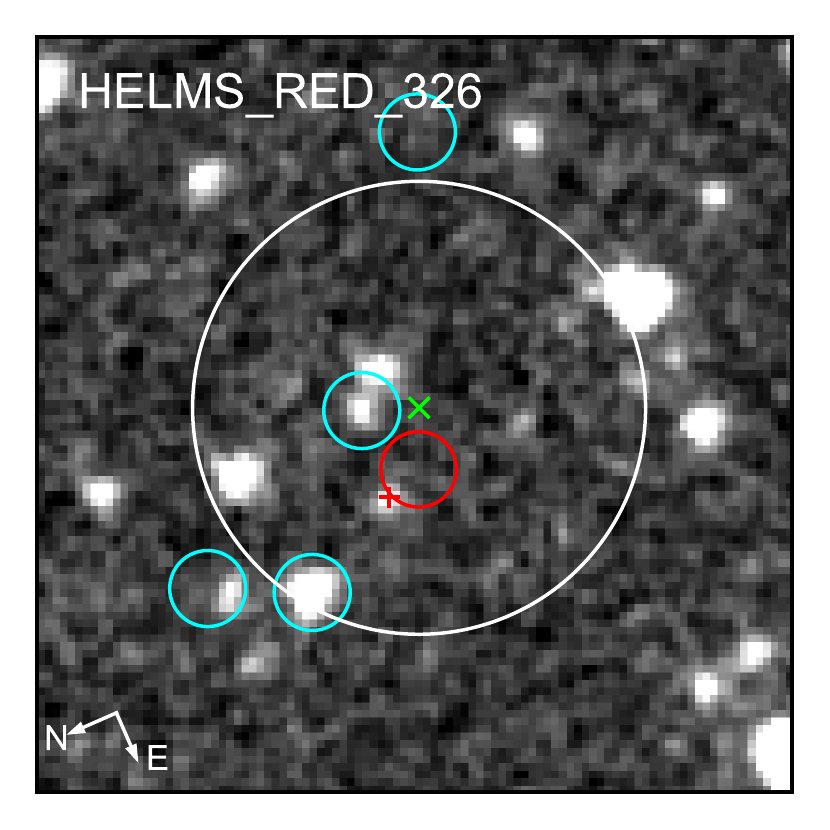}}
{\includegraphics[width=4.4cm, height=4.4cm]{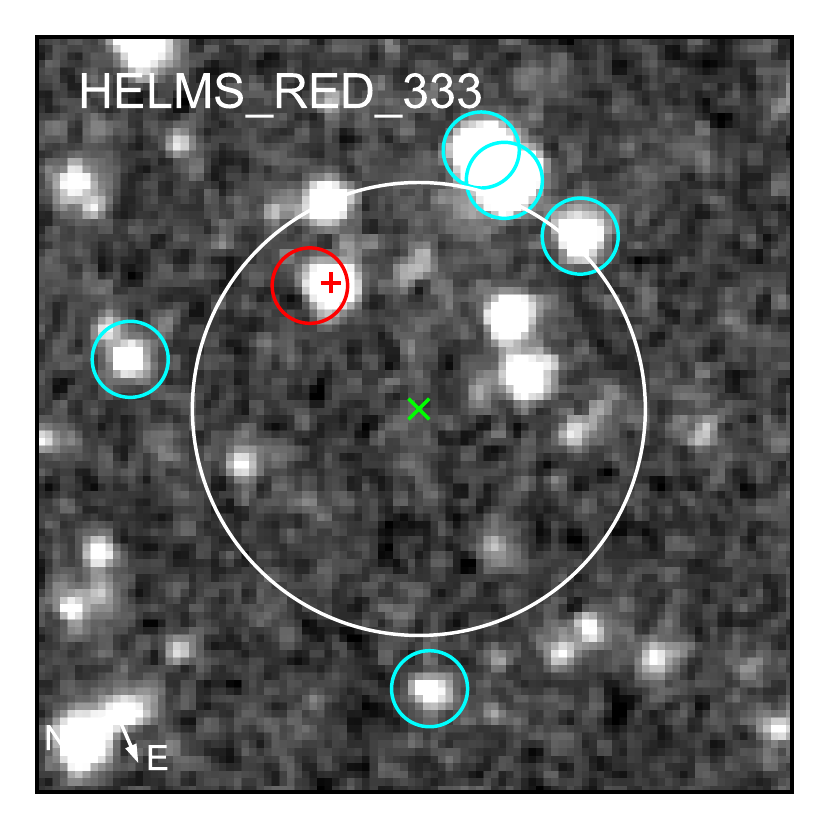}}
{\includegraphics[width=4.4cm, height=4.4cm]{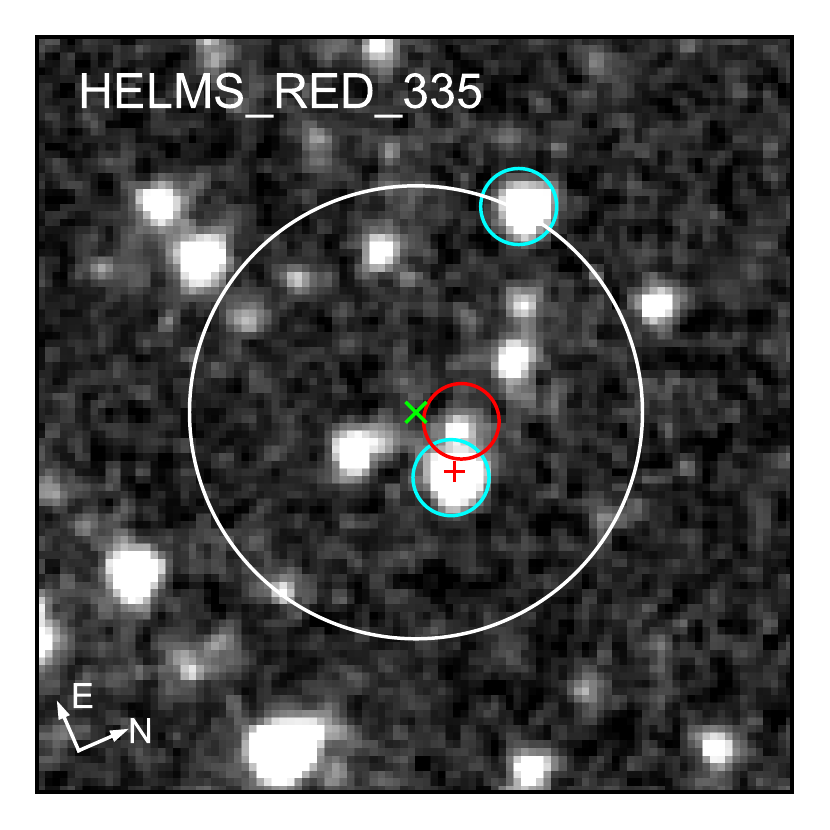}}
{\includegraphics[width=4.4cm, height=4.4cm]{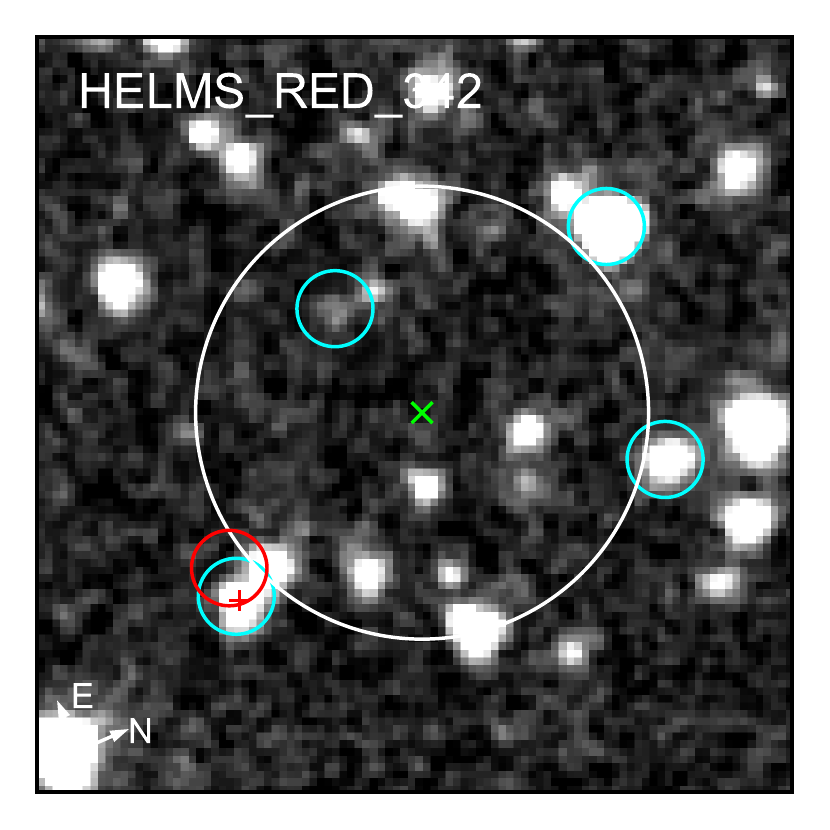}}
{\includegraphics[width=4.4cm, height=4.4cm]{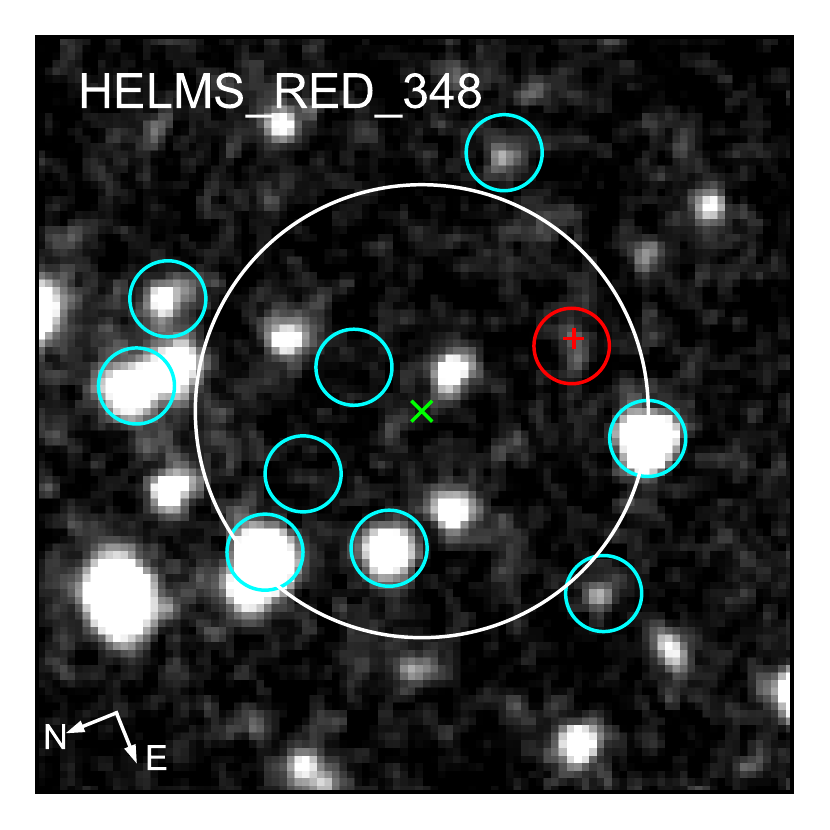}}
{\includegraphics[width=4.4cm, height=4.4cm]{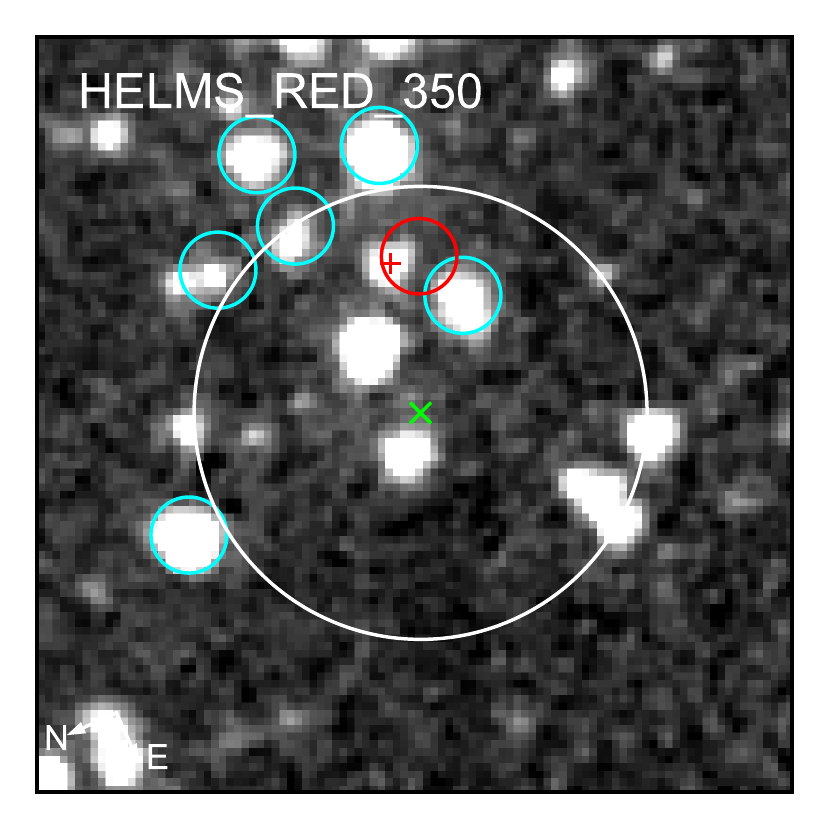}}
{\includegraphics[width=4.4cm, height=4.4cm]{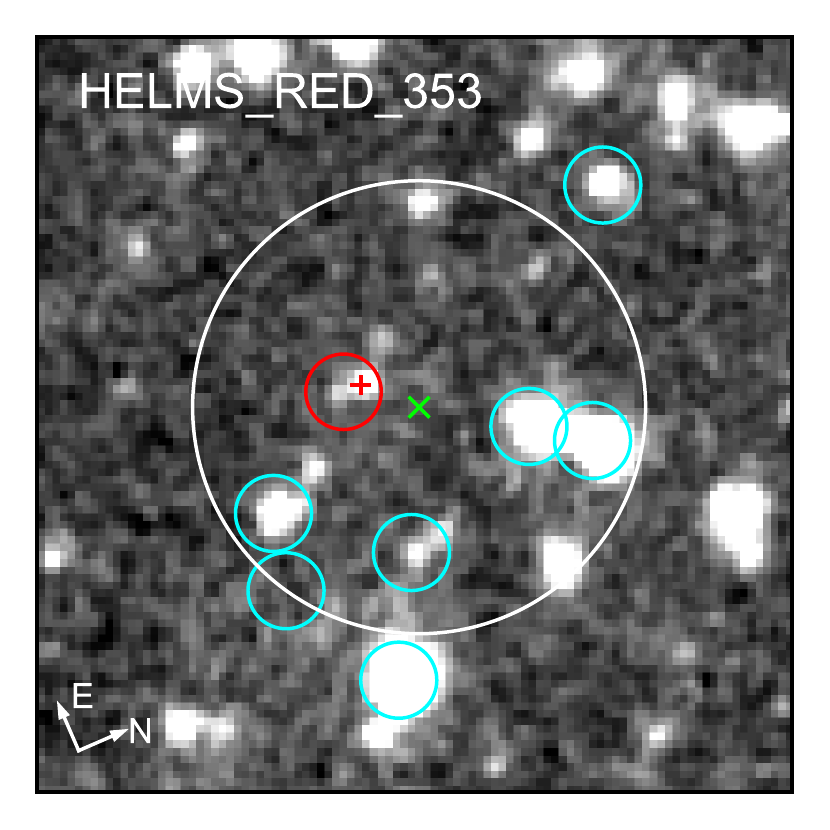}}
{\includegraphics[width=4.4cm, height=4.4cm]{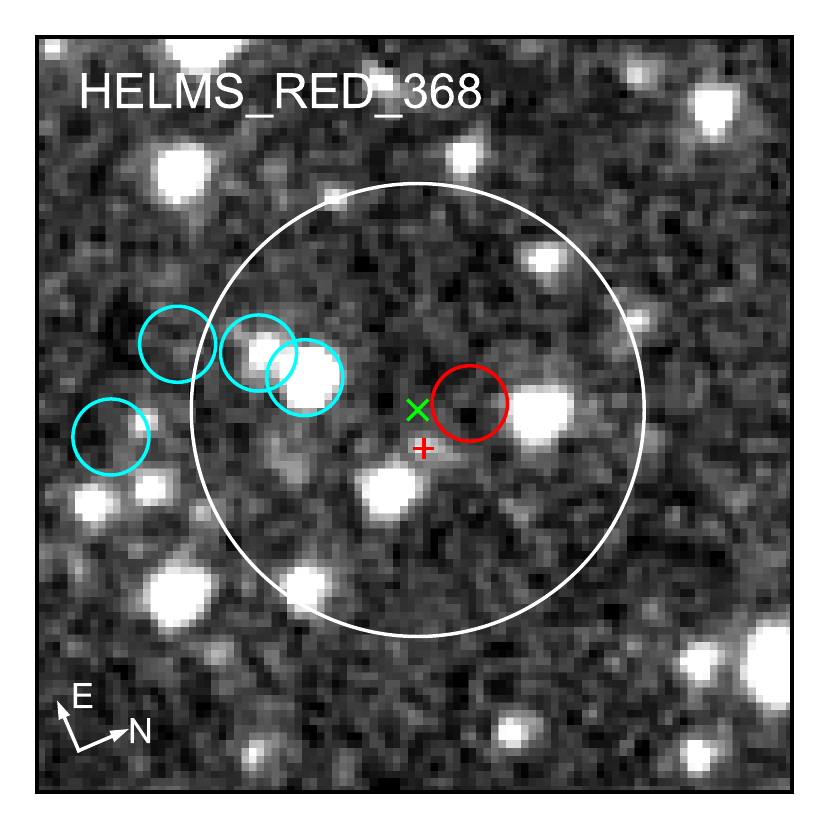}}
{\includegraphics[width=4.4cm, height=4.4cm]{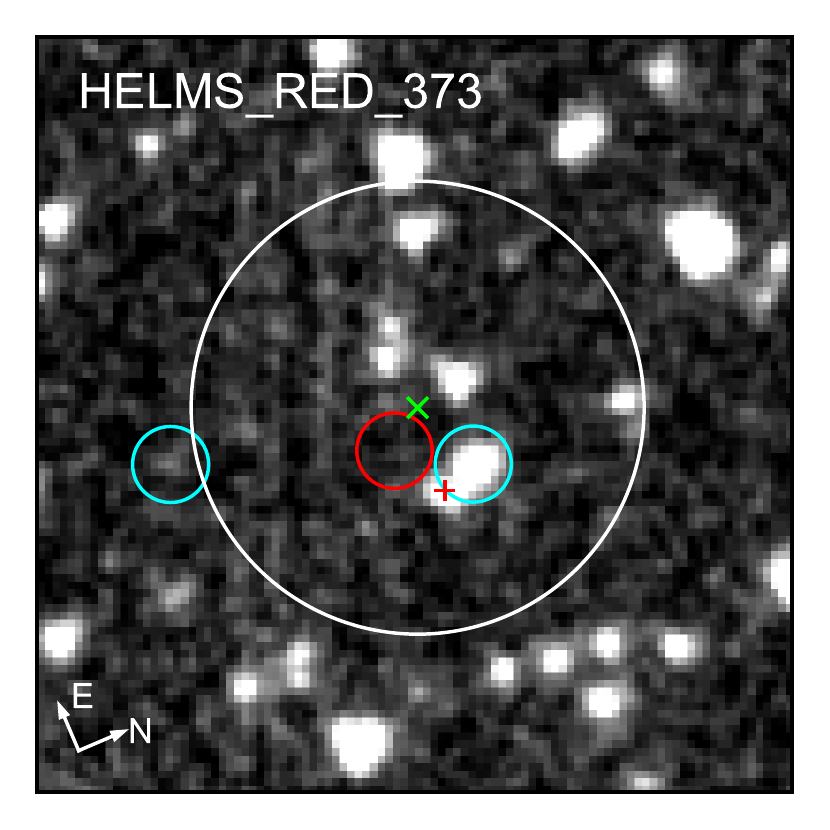}}
{\includegraphics[width=4.4cm, height=4.4cm]{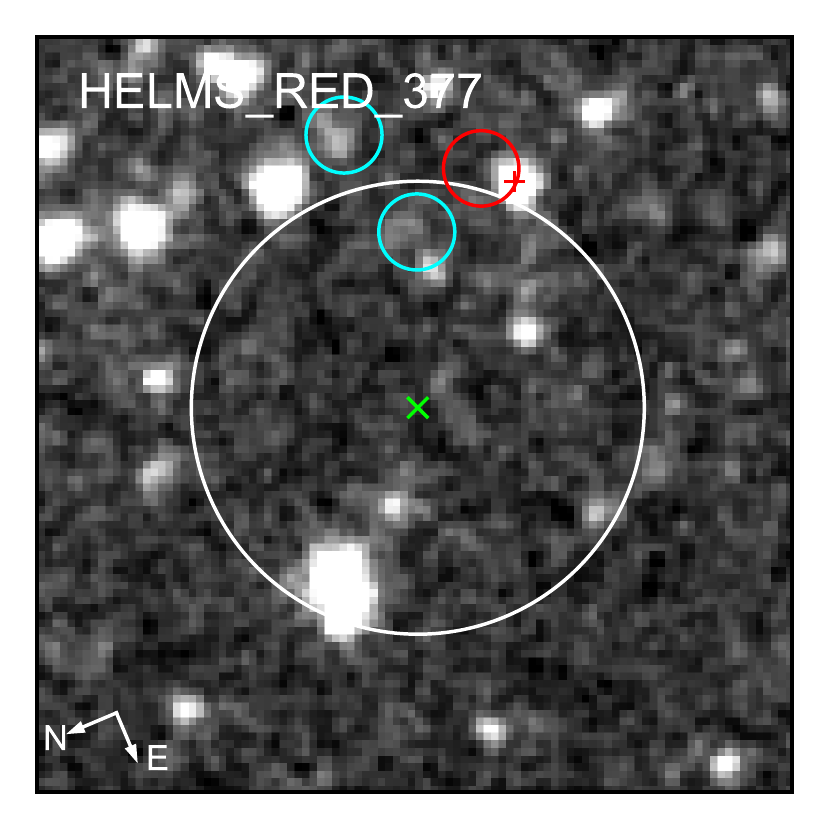}}
{\includegraphics[width=4.4cm, height=4.4cm]{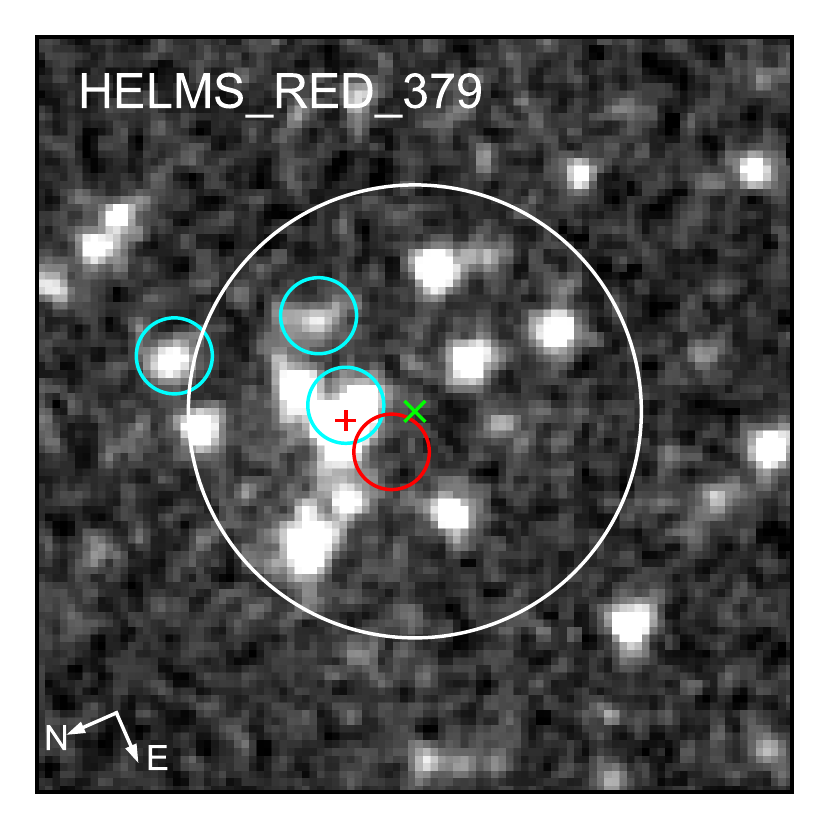}}
\caption{Continued 60$\arcsec$ $\times$ 60$\arcsec$ cutouts }
\label{fig:data}
\end{figure*}

\addtocounter{figure}{-1}
\begin{figure*}
\centering
{\includegraphics[width=4.4cm, height=4.4cm]{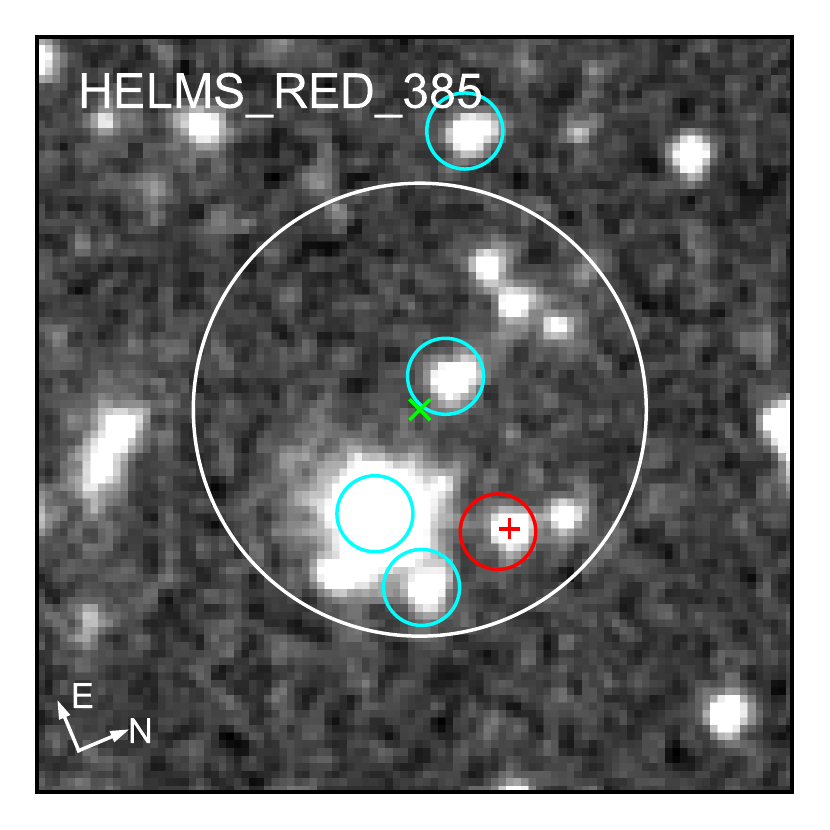}}
{\includegraphics[width=4.4cm, height=4.4cm]{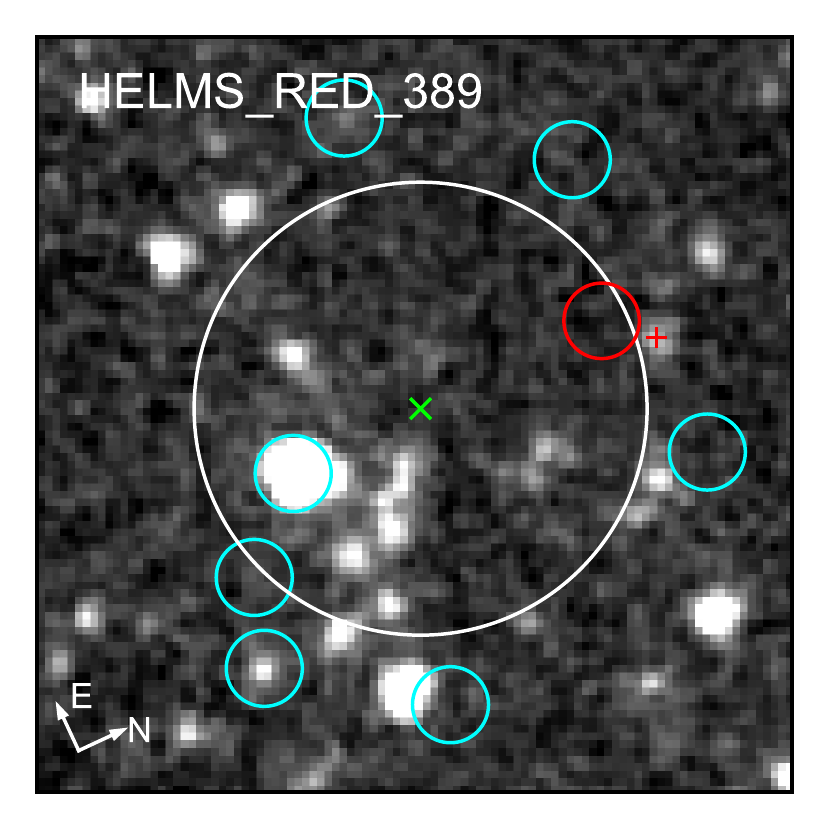}}
{\includegraphics[width=4.4cm, height=4.4cm]{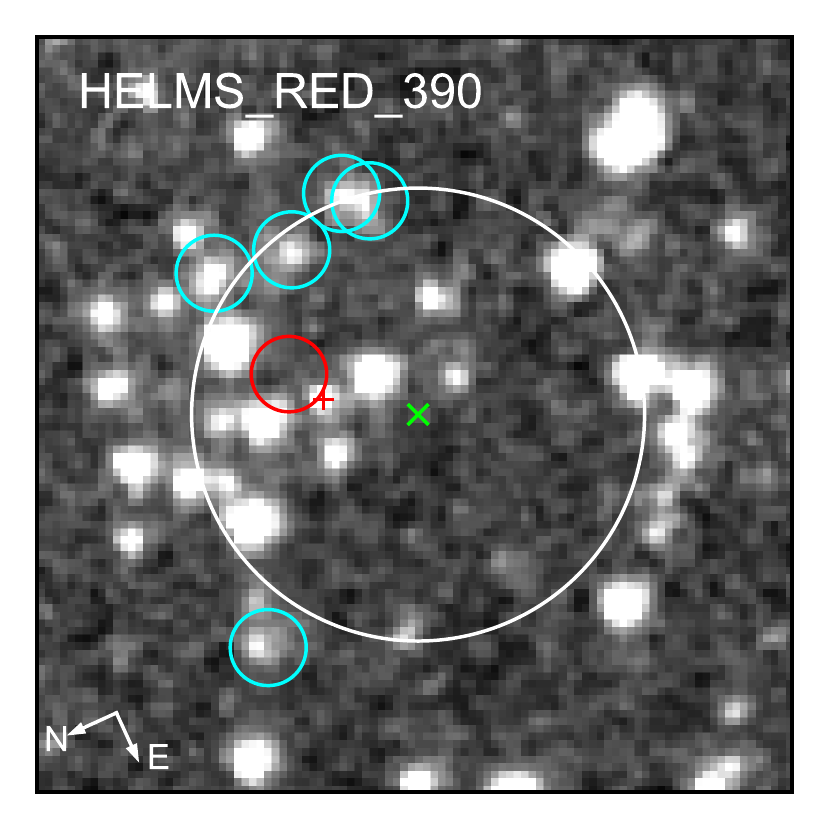}}
{\includegraphics[width=4.4cm, height=4.4cm]{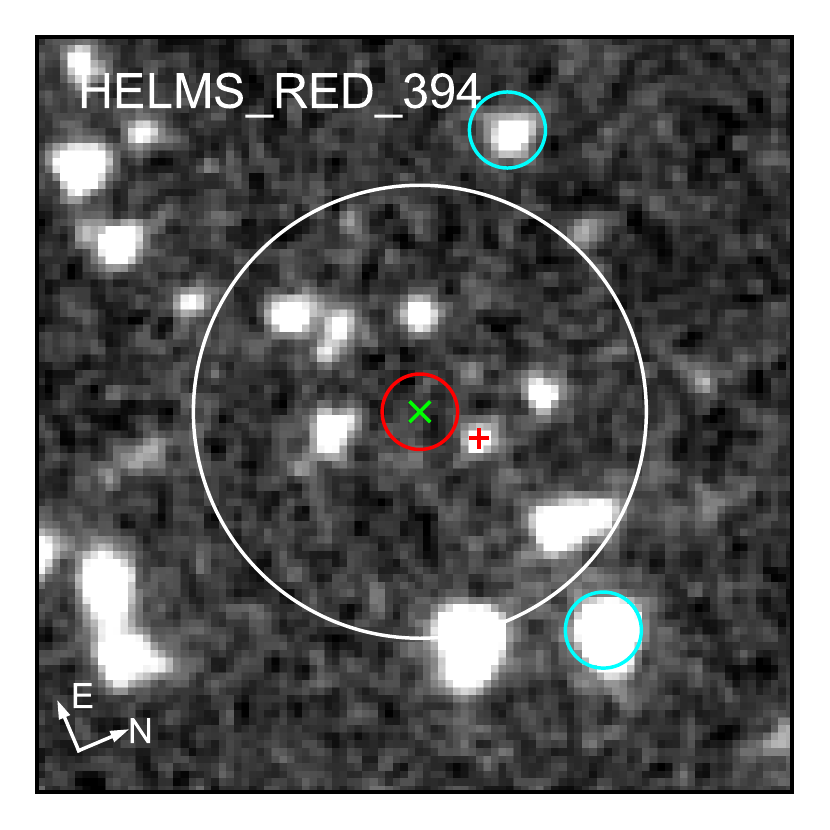}}
{\includegraphics[width=4.4cm, height=4.4cm]{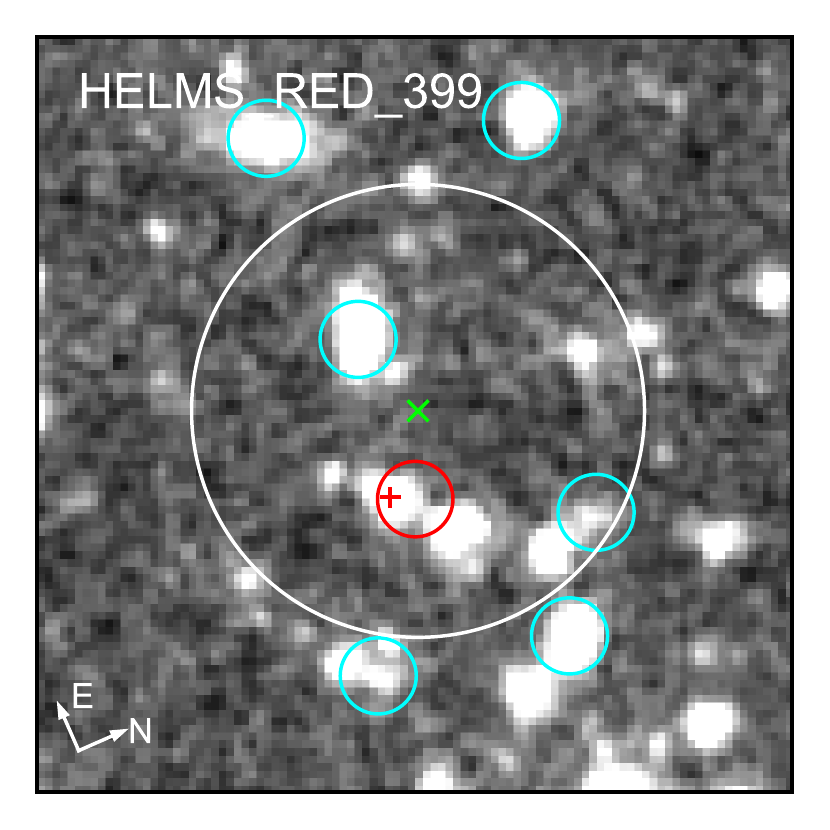}}
{\includegraphics[width=4.4cm, height=4.4cm]{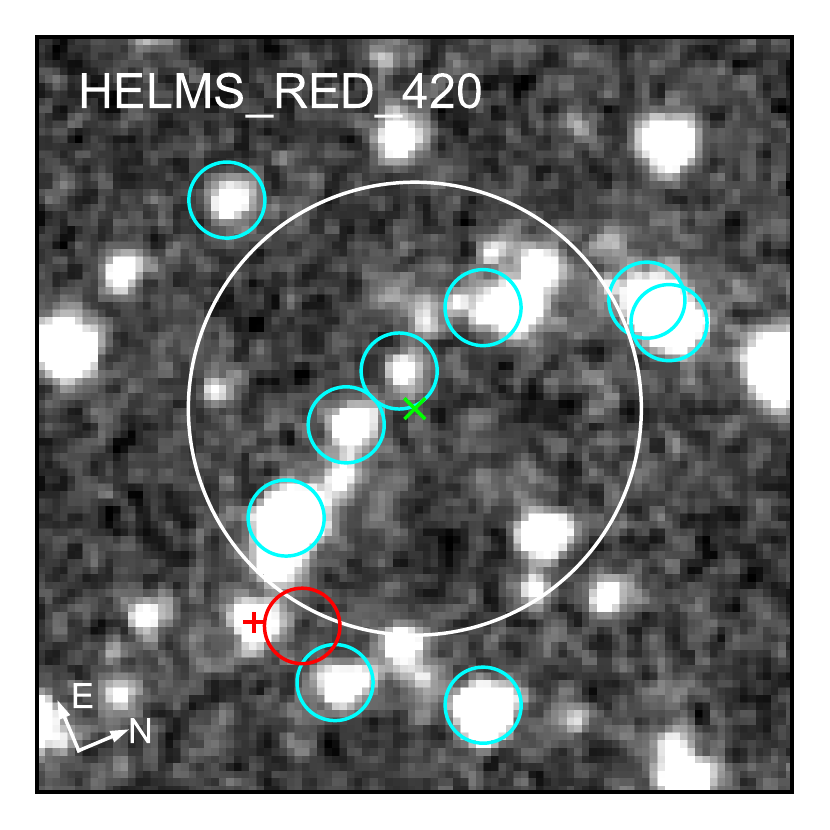}}
{\includegraphics[width=4.4cm, height=4.4cm]{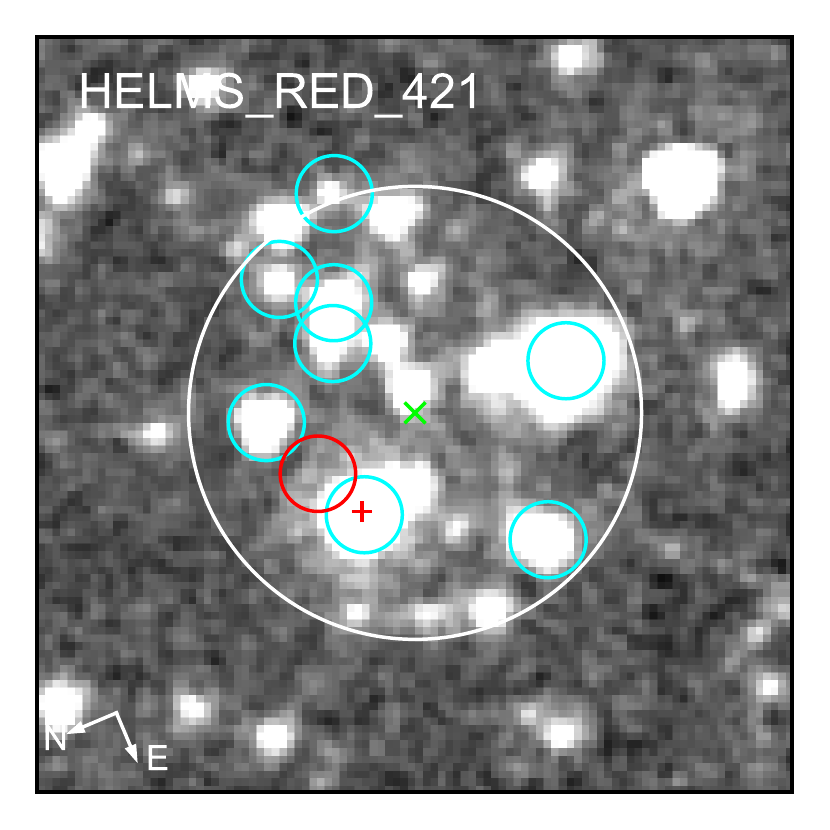}}
{\includegraphics[width=4.4cm, height=4.4cm]{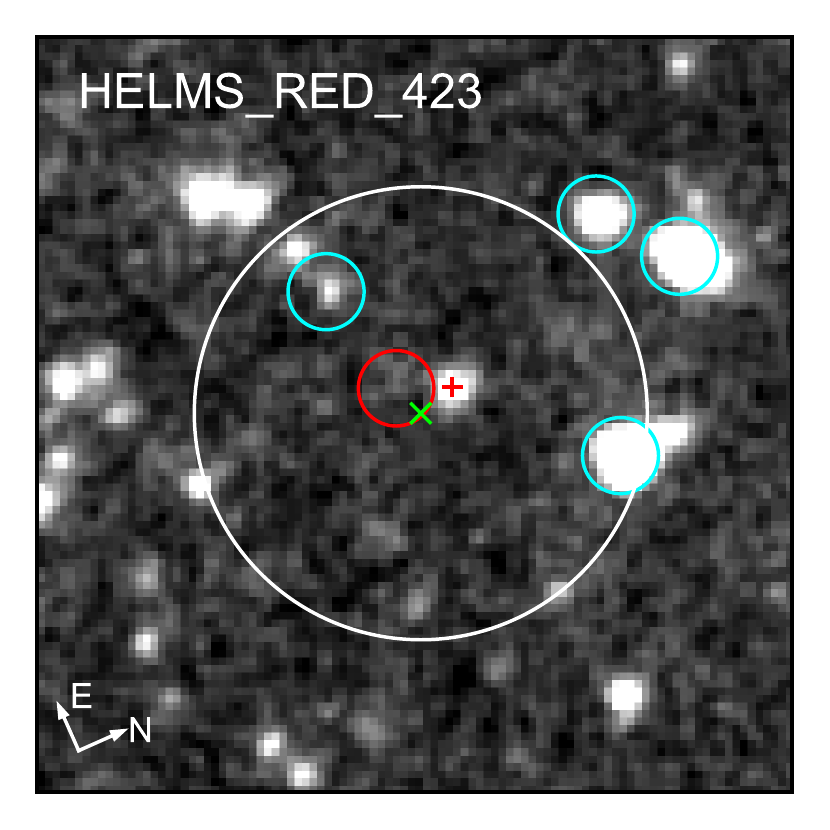}}
{\includegraphics[width=4.4cm, height=4.4cm]{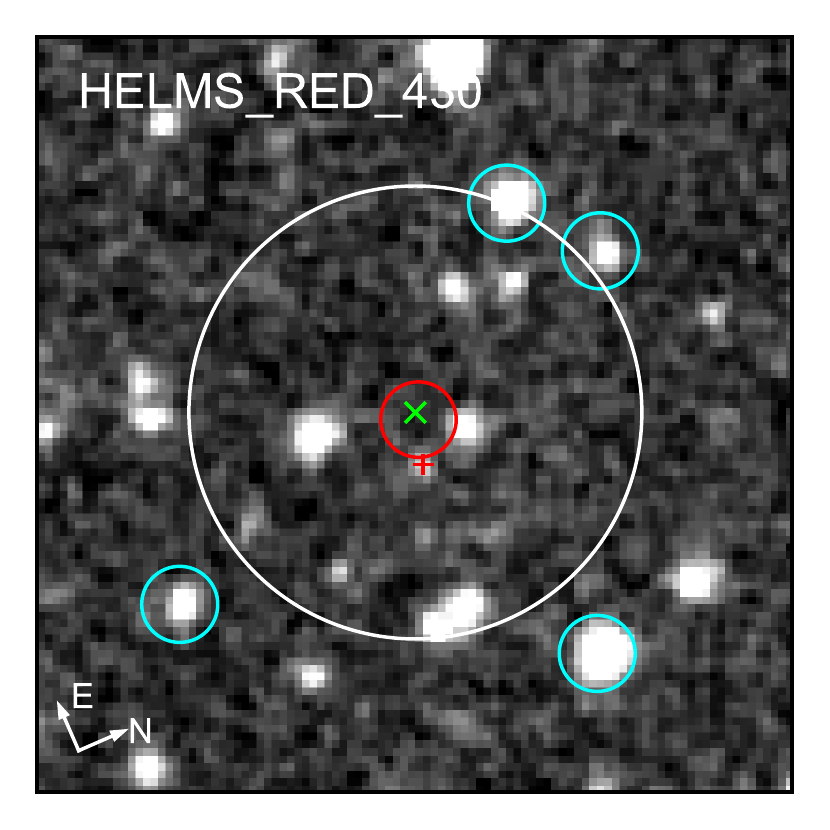}}
{\includegraphics[width=4.4cm, height=4.4cm]{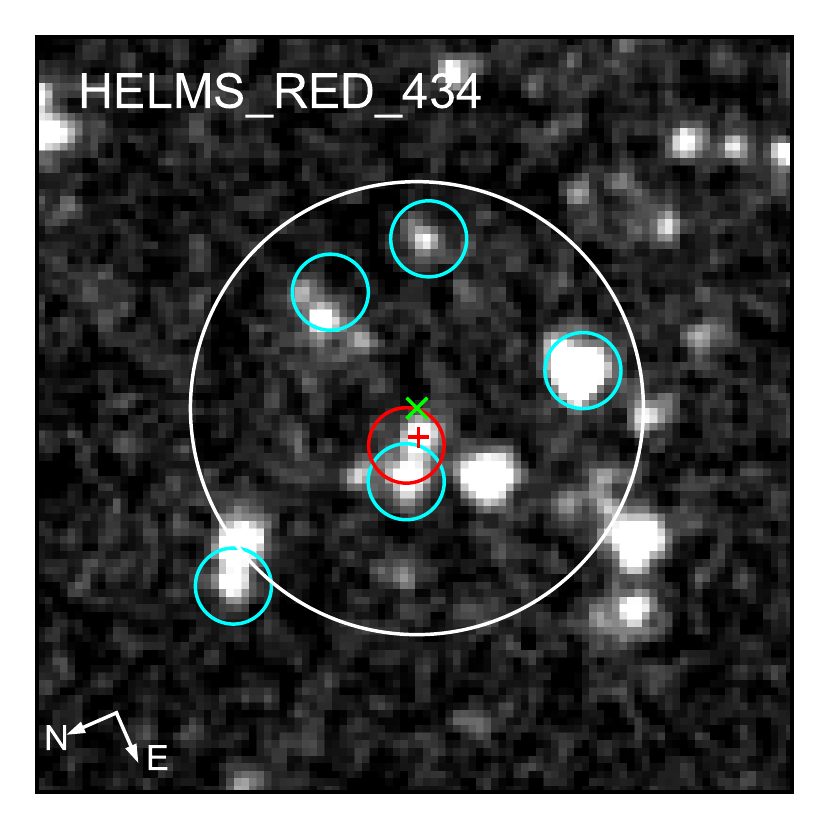}}
{\includegraphics[width=4.4cm, height=4.4cm]{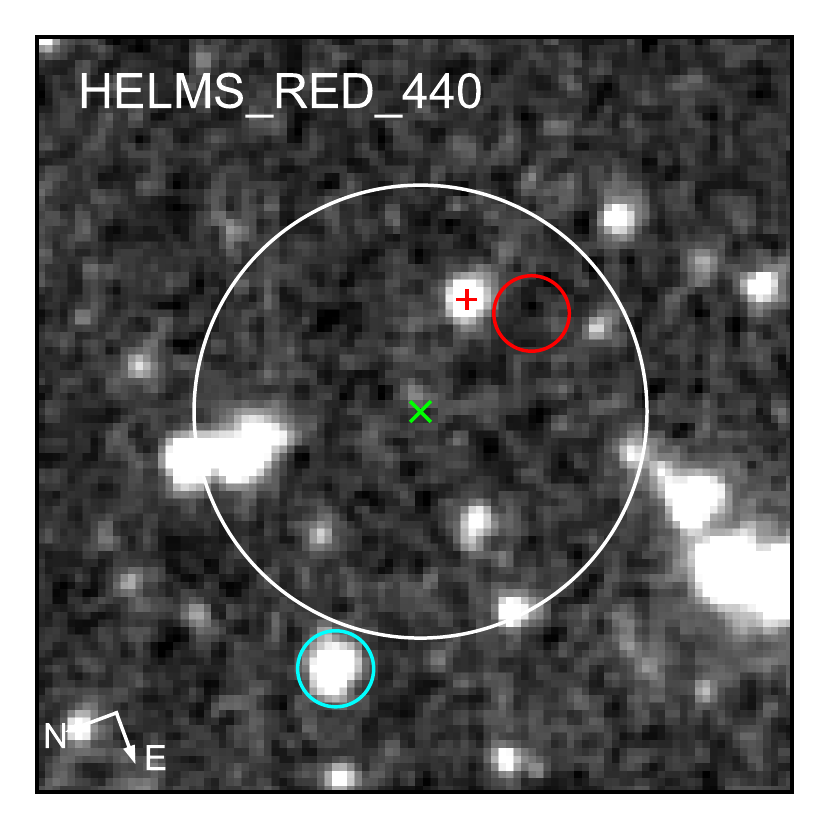}}
{\includegraphics[width=4.4cm, height=4.4cm]{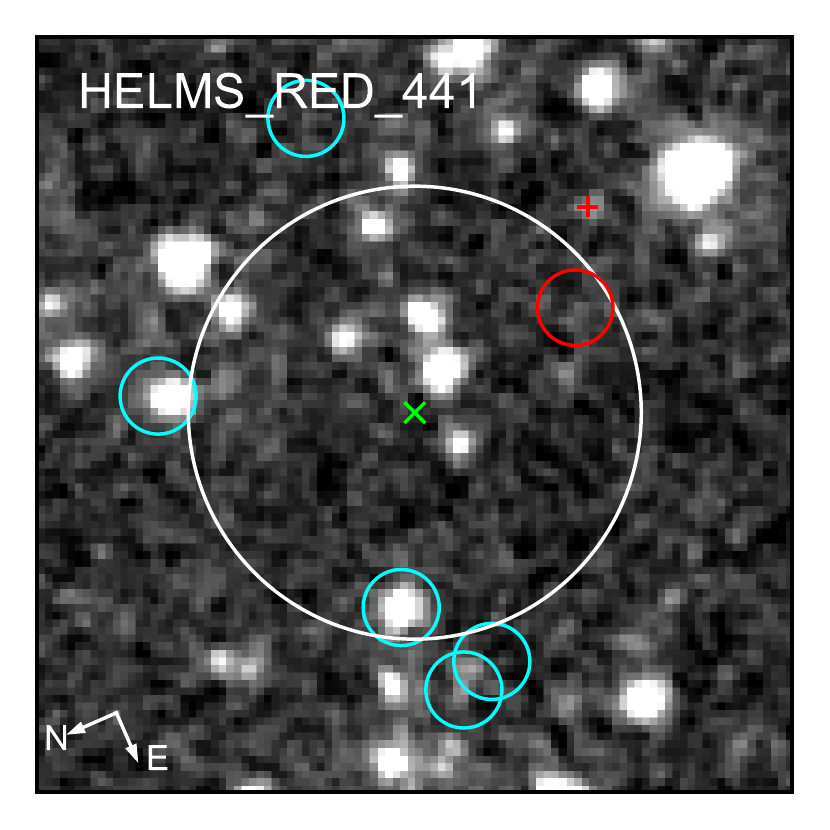}}
{\includegraphics[width=4.4cm, height=4.4cm]{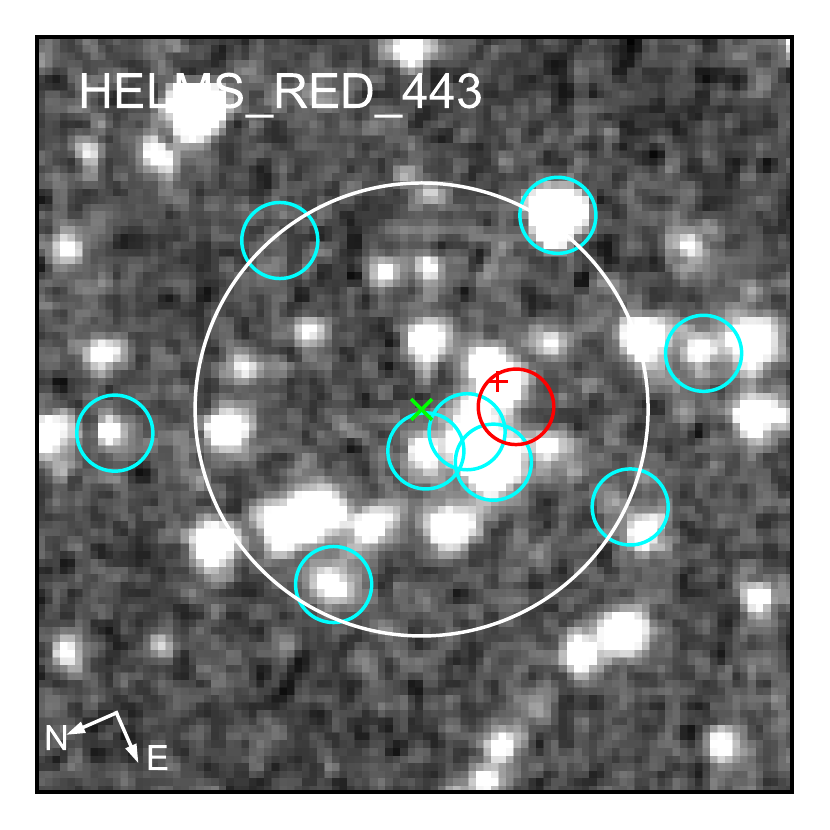}}
{\includegraphics[width=4.4cm, height=4.4cm]{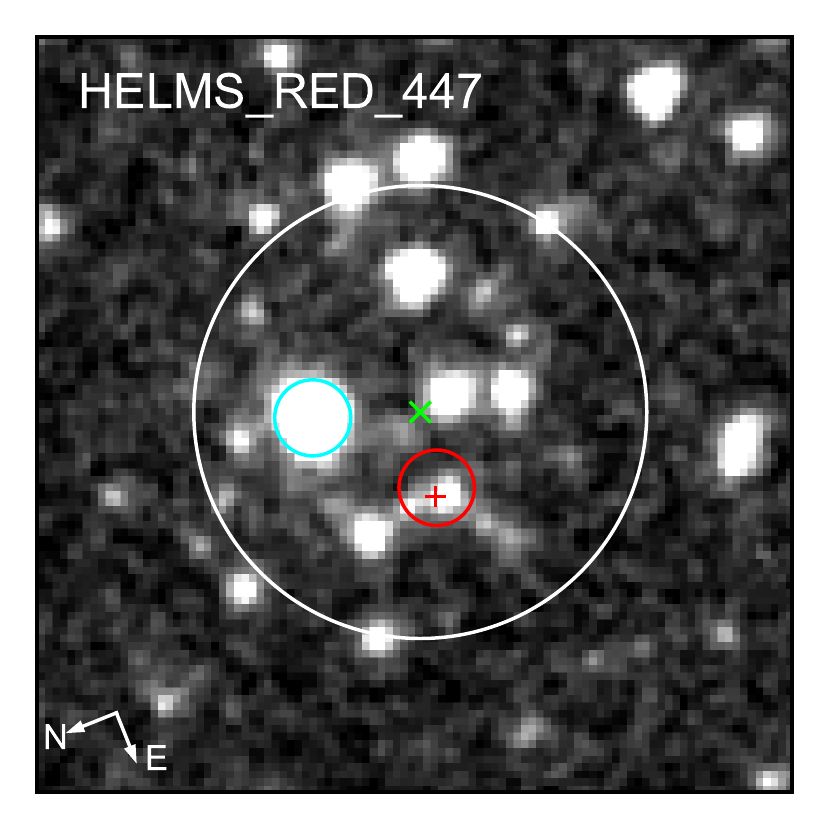}}
{\includegraphics[width=4.4cm, height=4.4cm]{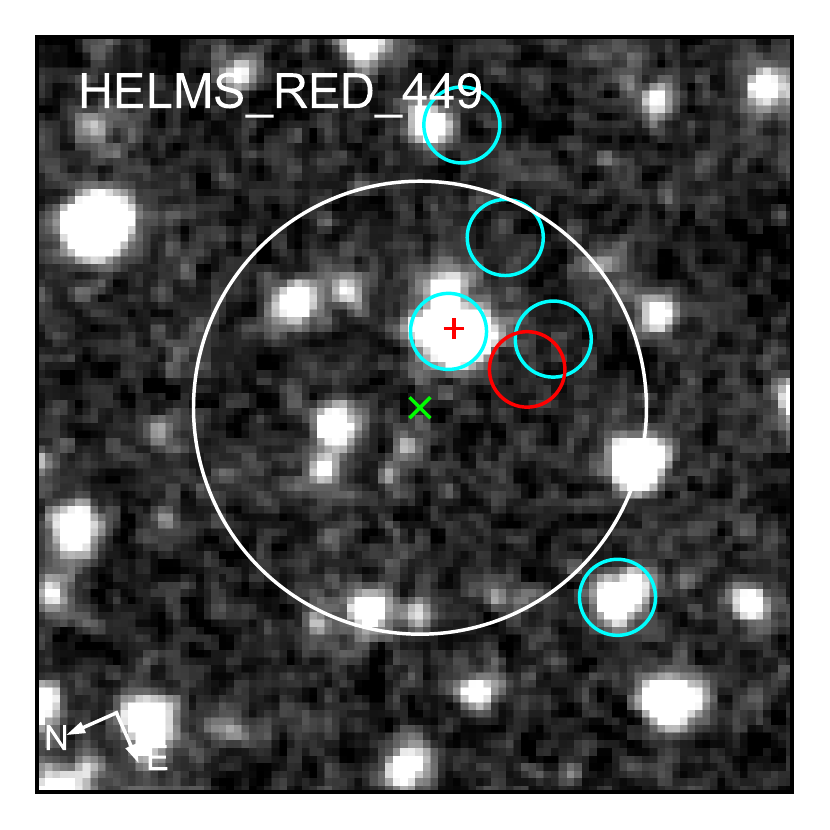}}
{\includegraphics[width=4.4cm, height=4.4cm]{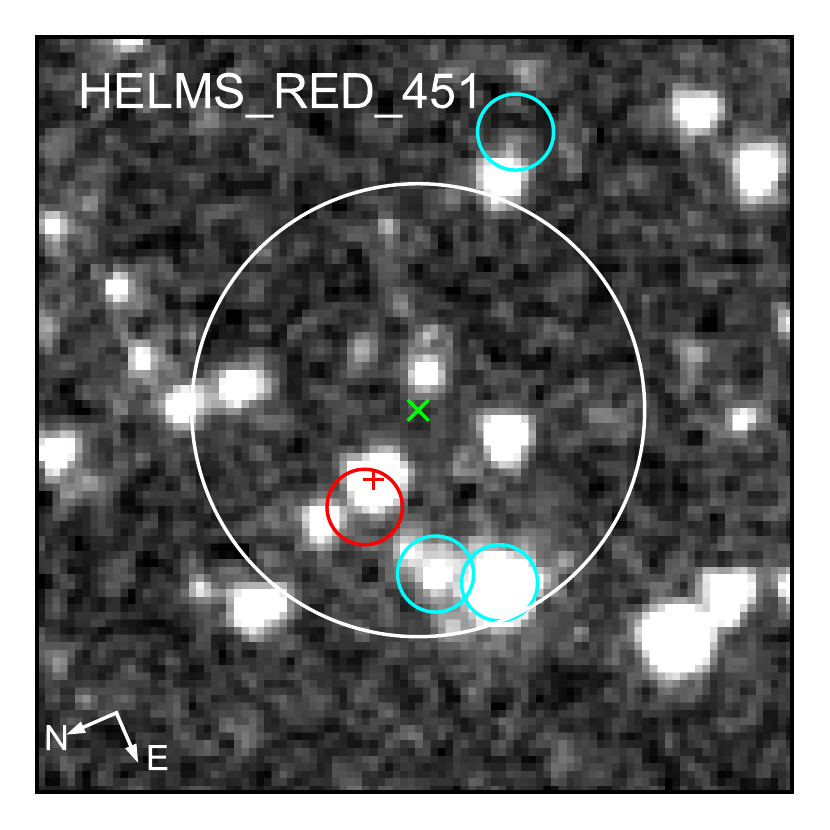}}
{\includegraphics[width=4.4cm, height=4.4cm]{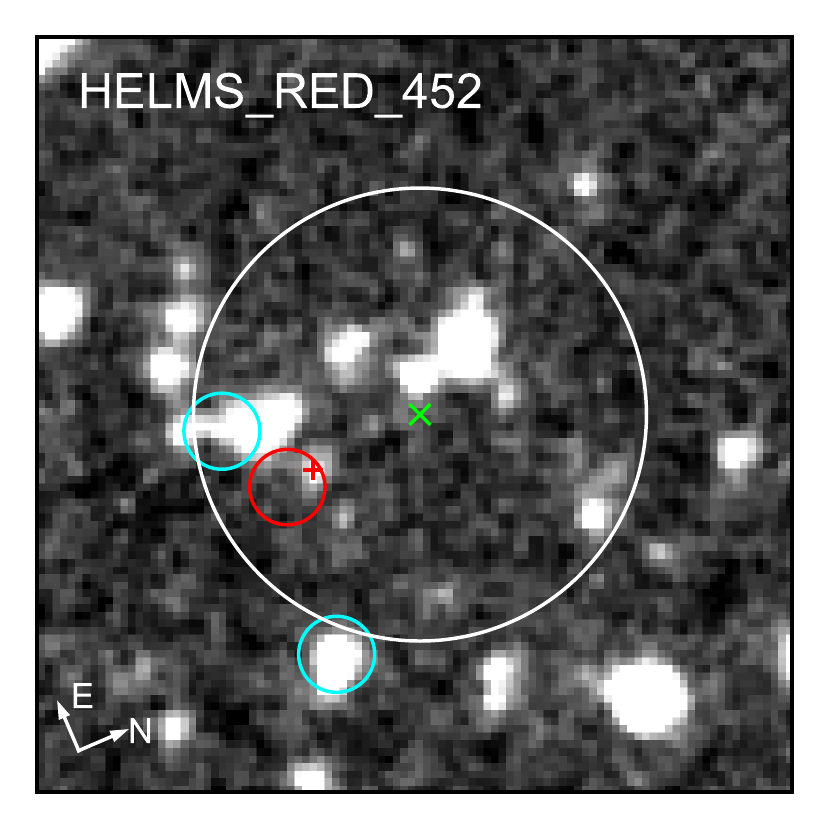}}
{\includegraphics[width=4.4cm, height=4.4cm]{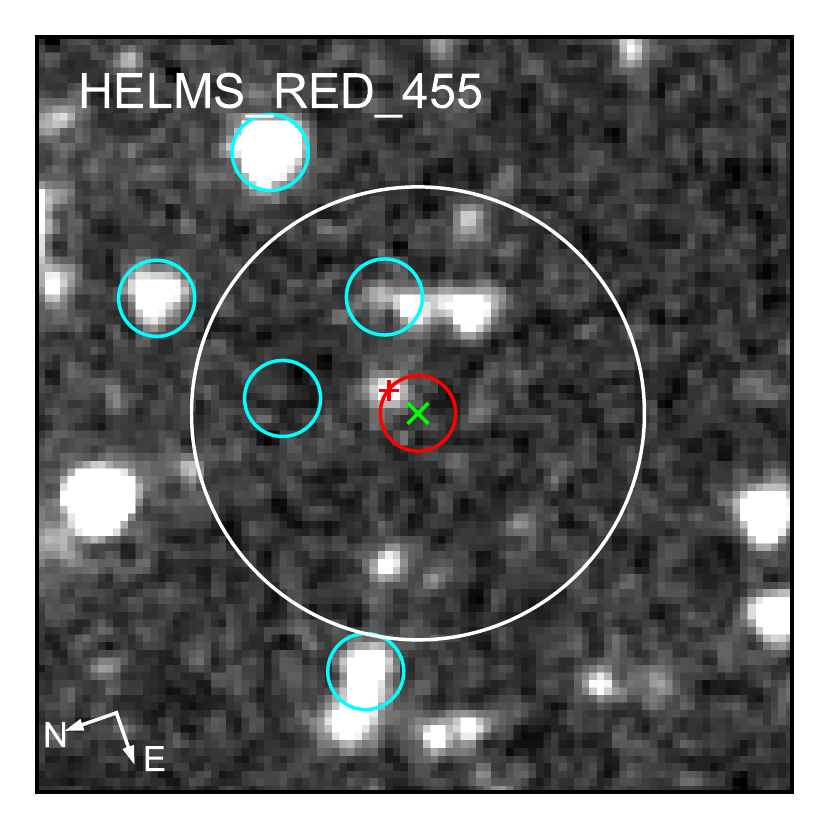}}
{\includegraphics[width=4.4cm, height=4.4cm]{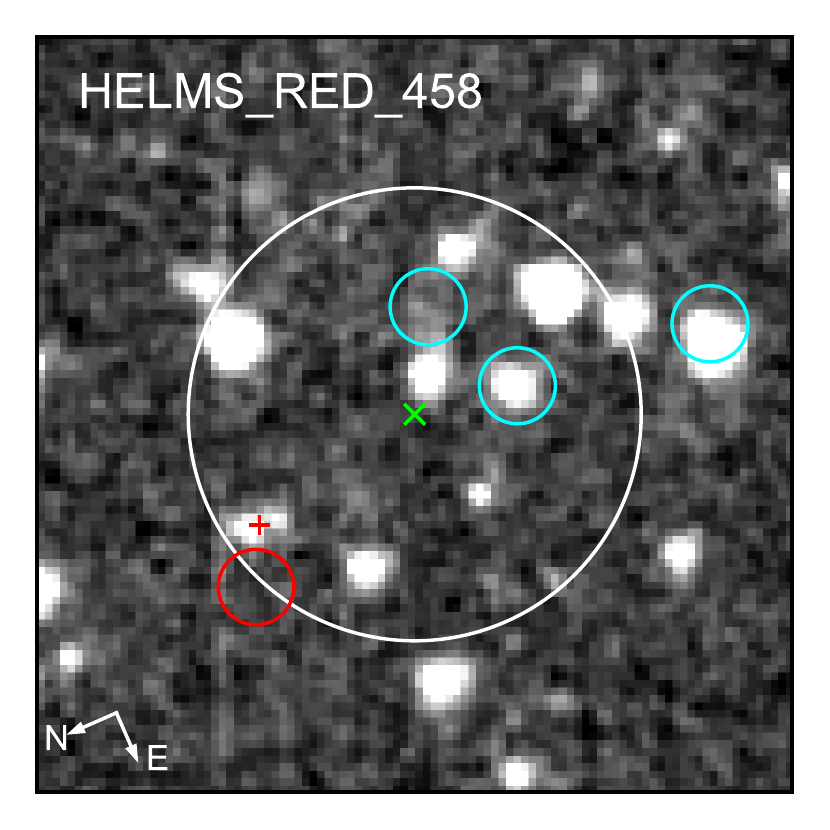}}
{\includegraphics[width=4.4cm, height=4.4cm]{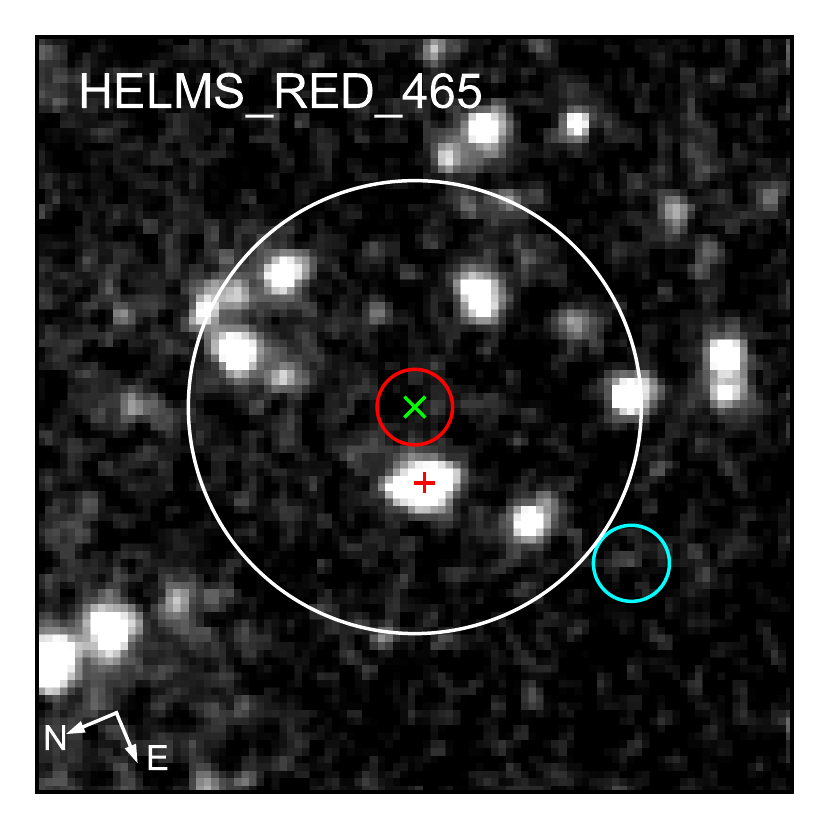}}
\caption{Continued 60$\arcsec$ $\times$ 60$\arcsec$ cutouts }
\label{fig:data}
\end{figure*}

\end{appendix}


\begin{thebibliography}{}

\bibitem[Alam et al.(2015)]{Alam2015} Alam, S., Albareti, F.~D., Allende Prieto, C., et al.\ 2015, \apjs, 219, 12.

\bibitem[Asboth et al.(2016)]{Asboth2016} Asboth, V., Conley, A., Sayers, J., et al.\ 2016, \mnras, 462, 1989 

\bibitem[Ashby et al.(2013)]{Ashby2013} Ashby, M.~L.~N., Stanford, S.~A., Brodwin, M., et al.\ 2013, \apjs, 209, 22 

\bibitem[Battisti et al.(2019)]{Battisti2019} Battisti, A., da Cunha, E., Grasha, K., et al. \ 2019, \apj, submitted

\bibitem[B{\'e}thermin et al.(2015)]{Bethermin2015} B{\'e}thermin, M., De Breuck, C., Sargent, M., et al.\ 2015, \aap, 576, L9

\bibitem[B{\'e}thermin et al.(2017)]{Bethermin2017} B{\'e}thermin, M., Wu, H.-Y., Lagache, G., et al.\ 2017, \aap, 607, A89 

\bibitem[Baugh et al.(2005)]{Baugh2005} Baugh, C.~M., Lacey, C.~G., Frenk, C.~S., et al.\ 2005, \mnras, 356, 1191 

\bibitem[Bertin \& Arnouts(1996)]{Bertin1996} Bertin, E., \& Arnouts, S.\ 1996, \aaps, 117, 393 

\bibitem[Bourne et al.(2017)]{Bourne2017} Bourne, N., Dunlop, J.~S., Merlin, E., et al.\ 2017, \mnras, 467, 1360.

\bibitem[Brammer et al.(2008)]{Brammer2008} Brammer, G.~B., van Dokkum, P.~G., \& Coppi, P.\ 2008, \apj, 686, 1503.

\bibitem[Bruzual \& Charlot(2003)]{Bruzual2003} Bruzual, G., \& Charlot, S.\ 2003, \mnras, 344, 1000 

\bibitem[Bussmann et al.(2013)]{Bussmann2013} Bussmann, R.~S., P{\'e}rez-Fournon, I., Amber, S., et al.\ 2013, \apj, 779, 25 

\bibitem[Ca{\~n}ameras et al.(2015)]{Canameras2015} Ca{\~n}ameras, R., Nesvadba, N.~P.~H., Guery, D., et al.\ 2015, \aap, 581, A105.

\bibitem[Carlstrom et al.(2011)]{Carlstrom2011} Carlstrom, J.~E., Ade, P.~A.~R., Aird, K.~A., et al.\ 2011, \pasp, 123, 568 

\bibitem[Carniani et al.(2013)]{Carniani2013} Carniani, S., Marconi, A., Biggs, A., et al.\ 2013, \aap, 559, A29 

\bibitem[Casey et al.(2012)]{Casey2012b} Casey, C.~M., Berta, S., B{\'e}thermin, M., et al.\ 2012b, \apj, 761, 140 

\bibitem[Casey et al.(2012)]{Casey2012a} Casey, C.~M., Berta, S., B{\'e}thermin, M., et al.\ 2012a, \apj, 761, 139 

\bibitem[Casey et al.(2014)]{Casey2014} Casey, C.~M., Narayanan, D., \& Cooray, A.\ 2014, \physrep, 541, 45 

\bibitem[Chabrier(2003)]{Chabrier2003} Chabrier, G.\ 2003, \apjl, 586, L133 

\bibitem[Chapin et al.(2011)]{Chapin2011} Chapin, E.~L., Chapman, S.~C., Coppin, K.~E., et al.\ 2011, \mnras, 411, 505 

\bibitem[Chapman et al.(2005)]{Chapman2005} Chapman, S.~C., Blain, A.~W., Smail, I., \& Ivison, R.~J.\ 2005, \apj, 622, 772 

\bibitem[Charlot \& Fall(2000)]{Charlot2000} Charlot, S., \& Fall, S.~M.\ 2000, \apj, 539, 718 

\bibitem[Combes et al.(2012)]{Combes2012} Combes, F., Rex, M., Rawle, T.~D., et al.\ 2012, \aap, 538, L4 

\bibitem[Cooray et al.(2014)]{Cooray2014} Cooray, A., Calanog, J., Wardlow, J.~L., et al.\ 2014, \apj, 790, 40 

\bibitem[Cowley et al.(2015)]{Cowley2015} Cowley, W.~I., Lacey, C.~G., Baugh, C.~M., \& Cole, S.\ 2015, \mnras, 446, 1784 

\bibitem[Cox et al.(2011)]{Cox2011} Cox, P., Krips, M., Neri, R., et al.\ 2011, \apj, 740, 63 

\bibitem[da Cunha et al.(2015)]{daCunha2015} da Cunha, E., Walter, F., Smail, I.~R., et al.\ 2015, \apj, 806, 110 

\bibitem[da Cunha et al.(2008)]{daCunha2008} da Cunha, E., Charlot, S., \& Elbaz, D.\ 2008, \mnras, 388, 1595 

\bibitem[Daddi et al.(2007)]{Daddi2007} Daddi, E., Dickinson, M., Morrison, G., et al.\ 2007, \apj, 670, 156.

\bibitem[Danielson et al.(2017)]{Danielson2017} Danielson, A.~L.~R., Swinbank, A.~M., Smail, I., et al.\ 2017, \apj, 840, 78 

\bibitem[Davidzon et al.(2017)]{Davidzon2017} Davidzon, I., Ilbert, O., Laigle, C., et al.\ 2017, \aap, 605, A70.

\bibitem[De Breuck et al.(2014)]{DeBreuck2014} De Breuck, C., Williams, R.~J., Swinbank, M., et al.\ 2014, \aap, 565, A59 

\bibitem[Diamond-Stanic et al.(2012)]{Diamond-Stanic2012} Diamond-Stanic, A.~M., Moustakas, J., Tremonti, C.~A., et al.\ 2012, \apj, 755, L26

\bibitem[Dole et al.(2006)]{Dole2006} Dole, H., Lagache, G., Puget, J.-L., et al.\ 2006, \aap, 451, 417 

\bibitem[Donevski et al.(2018)]{Donevski2018} Donevski, D., Buat, V., Boone, F., et al.\ 2018, \aap, 614, A33 

\bibitem[Dowell et al.(2014)]{Dowell2014} Dowell, C.~D., Conley, A., Glenn, J., et al.\ 2014, \apj, 780, 75 

\bibitem[Duivenvoorden et al.(2018)]{Duivenvoorden2018} Duivenvoorden, S., Oliver, S., Scudder, J.~M., et al.\ 2018, \mnras, 477, 1099 

\bibitem[Eales et al.(2010)]{Eales2010} Eales, S., Dunne, L., Clements, D., et al.\ 2010, \pasp, 122, 499 

\bibitem[Edge et al.(2013)]{Edge2013} Edge, A., Sutherland, W., Kuijken, K., et al.\ 2013, The Messenger, 154, 32 

\bibitem[Fazio et al.(2004)]{Fazio2004} Fazio, G.~G., Hora, J.~L., Allen, L.~E., et al.\ 2004, \apjs, 154, 10 

\bibitem[Fu et al.(2012)]{Fu2012} Fu, H., Jullo, E., Cooray, A., et al.\ 2012, \apj, 753, 134 

\bibitem[Fudamoto et al.(2017)]{Fudamoto2017} Fudamoto, Y., Ivison, R.~J., Oteo, I., et al.\ 2017, \mnras, 472, 2028 

\bibitem[Gonz{\'a}lez et al.(2011)]{Gonzalez2011} Gonz{\'a}lez, V., Labb{\'e}, I., Bouwens, R.~J., et al.\ 2011, \apj, 735, L34.

\bibitem[Gonz{\'a}lez et al.(2014)]{Gonzalez2014} Gonz{\'a}lez, V., Bouwens, R., Illingworth, G., et al.\ 2014, \apj, 781, 34.

\bibitem[Gralla et al.(2019)]{Gralla2019} Gralla, M.~B., Marriage, T.~A., Addison, G., et al.\ 2019, arXiv e-prints , arXiv:1905.04592.

\bibitem[Griffin et al.(2010)]{Griffin2010} Griffin, M.~J., Abergel, A., Abreu, A., et al.\ 2010, \aap, 518, L3 

\bibitem[Gruppioni et al.(2013)]{Gruppioni2013} Gruppioni, C., Pozzi, F., Rodighiero, G., et al.\ 2013, \mnras, 432, 23 

\bibitem[Gruppioni et al.(2017)]{Gruppioni2017} Gruppioni, C., Ciesla, L., Hatziminaoglou, E., et al.\ 2017, PASA, 34, e055.

\bibitem[Hayward et al.(2012)]{Hayward2012} Hayward, C.~C., Jonsson, P., Kere{\v{s}}, D., et al.\ 2012, \mnras, 424, 951.

\bibitem[Hayward et al.(2013)]{Hayward2013} Hayward, C.~C., Narayanan, D., Kere{\v s}, D., et al.\ 2013, \mnras, 428, 2529 

\bibitem[Hodge et al.(2013)]{Hodge2013} Hodge, J.~A., Karim, A., Smail, I., et al.\ 2013, \apj, 768, 91 

\bibitem[Holland et al.(2013)]{Holland2013} Holland, W.~S., Bintley, D., Chapin, E.~L., et al.\ 2013, \mnras, 430, 2513 

\bibitem[Hopwood et al.(2011)]{Hopwood2011} Hopwood, R., Wardlow, J., Cooray, A., et al.\ 2011, \apj, 728, L4.

\bibitem[Hurley et al.(2017)]{Hurley2017} Hurley, P.~D., Oliver, S., Betancourt, M., et al.\ 2017, \mnras, 464, 885 

\bibitem[Ikarashi et al.(2015)]{Ikarashi2015} Ikarashi, S., Ivison, R.~J., Caputi, K.~I., et al.\ 2015, \apj, 810, 133 

\bibitem[Ivison et al.(2007)]{Ivison2007} Ivison, R.~J., Greve, T.~R., Dunlop, J.~S., et al.\ 2007, \mnras, 380, 199 

\bibitem[Ivison et al.(2016)]{Ivison2016} Ivison, R.~J., Lewis, A.~J.~R., Weiss, A., et al.\ 2016, \apj, 832, 78 

\bibitem[Labb{\'e} et al.(2010)]{Labbe2010} Labb{\'e}, I., Gonz{\'a}lez, V., Bouwens, R.~J., et al.\ 2010, \apj, 708, L26.

\bibitem[Lee et al.(2012)]{Lee2012} Lee, K.-S., Ferguson, H.~C., Wiklind, T., et al.\ 2012, \apj, 752, 66.

\bibitem[Leja et al.(2015)]{Leja2015} Leja, J., van Dokkum, P.~G., Franx, M., et al.\ 2015, \apj, 798, 115

\bibitem[Ma et al.(2015)]{Ma2015} Ma, J., Gonzalez, A.~H., Spilker, J.~S., et al.\ 2015, \apj, 812, 88.

\bibitem[Ma et al.(2016)]{Ma2016} Ma, J., Gonzalez, A.~H., Vieira, J.~D., et al.\ 2016, \apj, 832, 114 

\bibitem[Madau, \& Dickinson(2014)]{Madau2014} Madau, P., \& Dickinson, M.\ 2014, \araa, 52, 415.

\bibitem[Maddox et al.(2018)]{Maddox2018} Maddox, S.~J., Valiante, E., Cigan, P., et al.\ 2018, \apjs, 236, 30.

\bibitem[Magdis et al.(2010)]{Magdis2010} Magdis, G.~E., Rigopoulou, D., Huang, J.-S., et al.\ 2010, \mnras, 401, 1521.

\bibitem[Magdis et al.(2011)]{Magdis2011} Magdis, G.~E., Daddi, E., Elbaz, D., et al.\ 2011, \apjl, 740, L15 

\bibitem[Magnelli et al.(2012)]{Magnelli2012} Magnelli, B., Lutz, D., Santini, P., et al.\ 2012, \aap, 539, A155 

\bibitem[Makovoz et al.(2006)]{Makovoz2006} Makovoz, D., Roby, T., Khan, I., \& Booth, H.\ 2006, \procspie, 6274, 62740C 

\bibitem[Marriage et al.(2011)]{Marriage2011} Marriage, T.~A., Baptiste Juin, J., Lin, Y.-T., et al.\ 2011, \apj, 731, 100.

\bibitem[Marrone et al.(2018)]{Marrone2018} Marrone, D.~P., Spilker, J.~S., Hayward, C.~C., et al.\ 2018, \nat, 553, 51 

\bibitem[Marsden et al.(2014)]{Marsden2014} Marsden, D., Gralla, M., Marriage, T.~A., et al.\ 2014, \mnras, 439, 1556.

\bibitem[Martin(2005)]{Martin2005} Martin, C.~L.\ 2005, \apj, 621, 227.

\bibitem[Mihos, \& Hernquist(1996)]{Mihos1996} Mihos, J.~C., \& Hernquist, L.\ 1996, \apj, 464, 641.

\bibitem[Micha{\l}owski et al.(2017)]{Michalowski2017} Micha{\l}owski, M.~J., Dunlop, J.~S., Koprowski, M.~P., et al.\ 2017, \mnras, 469, 492.

\bibitem[Mocanu et al.(2013)]{Mocanu2013} Mocanu, L.~M., Crawford, T.~M., Vieira, J.~D., et al.\ 2013, \apj, 779, 61.

\bibitem[Muzzin et al.(2013)]{Muzzin2013} Muzzin, A., Marchesini, D., Stefanon, M., et al.\ 2013, \apj, 777, 18.

\bibitem[Narayanan et al.(2010)]{Narayanan2010} Narayanan, D., Hayward, C.~C., Cox, T.~J., et al.\ 2010, \mnras, 401, 1613 

\bibitem[Nayyeri et al.(2014)]{Nayyeri2014} Nayyeri, H., Mobasher, B., Hemmati, S., et al.\ 2014, \apj, 794, 68.

\bibitem[Oliver et al.(2012)]{Oliver2012} Oliver, S.~J., Bock, J., Altieri, B., et al.\ 2012, \mnras, 424, 1614 

\bibitem[Oteo et al.(2018)]{Oteo2018} Oteo, I., Ivison, R.~J., Dunne, L., et al.\ 2018, \apj, 856, 72 

\bibitem[Oteo et al.(2016a)]{Oteo2016} Oteo, I., Ivison, R.~J., Dunne, L., et al.\ 2016a, \apj, 827, 34 

\bibitem[Oteo et al.(2016b)]{Oteo2016b} Oteo, I., Zwaan, M.~A., Ivison, R.~J., et al.\ 2016b, \apj, 822, 36.

\bibitem[Oteo et al.(2017)]{Oteo2017} Oteo, I., Ivison, R.~J., Negrello, M., et al.\ 2017, arXiv:1709.04191 

\bibitem[Pavesi et al.(2018)]{Pavesi2018} Pavesi, R., Riechers, D.~A., Sharon, C.~E., et al.\ 2018, \apj, 861, 43.

\bibitem[Pilbratt et al.(2010)]{Pilbratt2010} Pilbratt, G.~L., Riedinger, J.~R., Passvogel, T., et al.\ 2010, \aap, 518, L1 

\bibitem[Planck Collaboration et al.(2011)]{Planck2011} Planck Collaboration, Ade, P.~A.~R., Aghanim, N., et al.\ 2011, \aap, 536, A7.

\bibitem[Planck Collaboration et al.(2014)]{Planck2014} Planck Collaboration, Ade, P.~A.~R., Aghanim, N., et al.\ 2014, \aap, 571, A28.

\bibitem[Planck Collaboration et al.(2016)]{Planck2016} Planck Collaboration, Ade, P.~A.~R., Aghanim, N., et al.\ 2016, \aap, 594, A26.

\bibitem[Poglitsch et al.(2010)]{Poglitsch2010} Poglitsch, A., Waelkens, C., Geis, N., et al.\ 2010, \aap, 518, L2 

\bibitem[Riechers et al.(2013)]{Riechers2013} Riechers, D.~A., Bradford, C.~M., Clements, D.~L., et al.\ 2013, \nat, 496, 329 

\bibitem[Riechers et al.(2014)]{Riechers2014} Riechers, D.~A., Carilli, C.~L., Capak, P.~L., et al.\ 2014, \apj, 796, 84 

\bibitem[Riechers et al.(2017)]{Riechers2017} Riechers, D.~A., Leung, T.~K.~D., Ivison, R.~J., et al.\ 2017, \apj, 850, 1 

\bibitem[Romano et al.(2017)]{Romano2017} Romano, D., Matteucci, F., Zhang, Z.-Y., et al.\ 2017, \mnras, 470, 401

\bibitem[Rowan-Robinson et al.(2016)]{Rowan-Robinson2016} Rowan-Robinson, M., Oliver, S., Wang, L., et al.\ 2016, \mnras, 461, 1100.

\bibitem[Sayers et al.(2014)]{Sayers2014} Sayers, J., Bockstiegel, C., Brugger, S., et al.\ 2014, \procspie, 9153, 915304 

\bibitem[Scudder et al.(2016)]{Scudder2016} Scudder, J.~M., Oliver, S., Hurley, P.~D., et al.\ 2016, \mnras, 460, 1119 

\bibitem[Simpson et al.(2015)]{Simpson2015} Simpson, J.~M., Smail, I., Swinbank, A.~M., et al.\ 2015, \apj, 807, 128 

\bibitem[Simpson et al.(2014)]{Simpson2014} Simpson, J.~M., Swinbank, A.~M., Smail, I., et al.\ 2014, \apj, 788, 125 

\bibitem[Siringo et al.(2009)]{Siringo2009} Siringo, G., Kreysa, E., Kov{\'a}cs, A., et al.\ 2009, \aap, 497, 945.

\bibitem[Skrutskie et al.(2006)]{Skrutskie2006} Skrutskie, M.~F., Cutri, R.~M., Stiening, R., et al.\ 2006, \aj, 131, 1163 

\bibitem[Speagle et al.(2014)]{Speagle2014} Speagle, J.~S., Steinhardt, C.~L., Capak, P.~L., \& Silverman, J.~D.\ 2014, \apjs, 214, 15 

\bibitem[Spilker et al.(2016)]{Spilker2016} Spilker, J.~S., Marrone, D.~P., Aravena, M., et al.\ 2016, \apj, 826, 112 

\bibitem[Spilker et al.(2018)]{Spilker2018} Spilker, J.~S., Aravena, M., B{\'e}thermin, M., et al.\ 2018, Science, 361, 1016.

\bibitem[Spitler et al.(2014)]{Spitler2014} Spitler, L.~R., Straatman, C.~M.~S., Labb{\'e}, I., et al.\ 2014, \apj, 787, L36.

\bibitem[Stark et al.(2009)]{Stark2009} Stark, D.~P., Ellis, R.~S., Bunker, A., et al.\ 2009, \apj, 697, 1493.

\bibitem[Straatman et al.(2014)]{Straatman2014} Straatman, C.~M.~S., Labb{\'e}, I., Spitler, L.~R., et al.\ 2014, \apj, 783, L14.

\bibitem[Strandet et al.(2017)]{Strandet2017} Strandet, M.~L., Weiss, A., De Breuck, C., et al.\ 2017, \apjl, 842, L15 

\bibitem[Strandet et al.(2016)]{Strandet2016} Strandet, M.~L., Weiss, A., Vieira, J.~D., et al.\ 2016, \apj, 822, 80 

\bibitem[Su et al.(2017)]{Su2017} Su, T., Marriage, T.~A., Asboth, V., et al.\ 2017, \mnras, 464, 968 

\bibitem[Sutherland et al.(2015)]{Sutherland2015} Sutherland, W., Emerson, J., Dalton, G., et al.\ 2015, \aap, 575, A25 

\bibitem[Thompson et al.(2005)]{Thompson2005} Thompson, T.~A., Quataert, E., \& Murray, N.\ 2005, \apj, 630, 167 

\bibitem[Toft et al.(2014)]{Toft2014} Toft, S., Smol{\v{c}}i{\'c}, V., Magnelli, B., et al.\ 2014, \apj, 782, 68.

\bibitem[Swinbank et al.(2014)]{Swinbank2014} Swinbank, A.~M., Simpson, J.~M., Smail, I., et al.\ 2014, \mnras, 438, 1267 

\bibitem[Valiante et al.(2016)]{Valiante2016} Valiante, E., Smith, M.~W.~L., Eales, S., et al.\ 2016, \mnras, 462, 3146 

\bibitem[Vieira et al.(2010)]{Vieira2010} Vieira, J.~D., Crawford, T.~M., Switzer, E.~R., et al.\ 2010, \apj, 719, 763.

\bibitem[Vieira et al.(2013)]{Vieira2013} Vieira, J.~D., Marrone, D.~P., Chapman, S.~C., et al.\ 2013, \nat, 495, 344 

\bibitem[Walter et al.(2012)]{Walter2012} Walter, F., Decarli, R., Carilli, C., et al.\ 2012, \nat, 486, 233 

\bibitem[Walter et al.(2009)]{Walter2009} Walter, F., Riechers, D., Cox, P., et al.\ 2009, \nat, 457, 699 

\bibitem[Wang et al.(2013)]{Wang2013} Wang, R., Wagg, J., Carilli, C.~L., et al.\ 2013, \apj, 773, 44 

\bibitem[Wang et al.(2016)]{Wang2016} Wang, T., Elbaz, D., Schreiber, C., et al.\ 2016, \apj, 816, 84.

\bibitem[Wardlow et al.(2010)]{Wardlow2010} Wardlow, J.~L., Smail, I., Wilson, G.~W., et al.\ 2010, \mnras, 401, 2299 

\bibitem[Weinmann et al.(2011)]{Weinmann2011} Weinmann, S.~M., Neistein, E., \& Dekel, A.\ 2011, \mnras, 417, 2737.

\bibitem[Wei{\ss} et al.(2013)]{Weiss2013} Wei{\ss}, A., De Breuck, C., Marrone, D.~P., et al.\ 2013, \apj, 767, 88 

\bibitem[Whitaker et al.(2014)]{Whitaker2014} Whitaker, K.~E., Franx, M., Leja, J., et al.\ 2014, \apj, 795, 104

\bibitem[York et al.(2000)]{York2000} York, D.~G., Adelman, J., Anderson, J.~E., Jr., et al.\ 2000, \aj, 120, 1579 

\bibitem[Younger et al.(2008)]{Younger2008} Younger, J.~D., Fazio, G.~G., Wilner, D.~J., et al.\ 2008, \apj, 688, 59 

\bibitem[Yun et al.(2015)]{Yun2015} Yun, M.~S., Aretxaga, I., Gurwell, M.~A., et al.\ 2015, \mnras, 454, 3485 

\bibitem[Zavala et al.(2018)]{Zavala2018} Zavala, J.~A., Monta{\~n}a, A., Hughes, D.~H., et al.\ 2018, Nature Astronomy, 2, 56 

\bibitem[Zhang et al.(2018)]{Zhang2018} Zhang, Z.-Y., Romano, D., Ivison, R.~J., et al.\ 2018, \nat, 558, 260


\end{thebibliography}
\end{document}